\documentclass[12pt]{article}
\usepackage{graphicx}
\usepackage{hyperref}
\usepackage{amsmath}
\usepackage{amsfonts}
\usepackage{amssymb}

\begin{document}

\title{Gauge Invariant Lagrangian Formulations for Mixed
Symmetry Higher Spin Bosonic Fields in AdS Spaces}
\author{\textsc{A.A. Reshetnyak,}$^{a,b,c}$\thanks{%
e-mail: reshet@tspu.edu.ru}\,\,\, \textsc{P.Yu. Moshin,}$^{a}$\thanks{%
e-mail: moshin@phys.tsu.ru}\\
\textit{${}$}$^{a}${\normalsize National Research Tomsk State University,
634050, Tomsk, Russia}\\
$^{b}${\normalsize Tomsk State Pedagogical University, 634041, Tomsk, Russia}\\
$^{c}${\normalsize National Research Tomsk Polytechnic University, 634050,
Tomsk, Russia}}
\date{}
\maketitle

\begin{abstract}

We deduce a non-linear commutator higher-spin (HS) symmetry
algebra which encodes unitary irreducible representations of the
AdS group -- subject to a Young tableaux $Y(s_1,\ldots ,s_k)$ with
$k\geq 2$ rows -- in a $d$-dimensional anti-de-Sitter space.
Auxiliary representations for a  deformed non-linear HS
symmetry algebra in terms of a generalized Verma module, as
applied to additively convert a subsystem of second-class
constraints in the HS symmetry algebra into one with first-class
constraints, are found explicitly in the case of a $k=2$ Young
tableaux. An oscillator realization over the Heisenberg algebra
for the Verma module is constructed. The results generalize the
method of constructing auxiliary representations for the
symplectic $sp(2k)$ algebra used for mixed-symmetry HS fields in
flat spaces \cite{BRbos}. Polynomial deformations
of the $su(1,1)$ algebra related to the Bethe ansatz are studied as
a by-product. A nilpotent BRST operator for a non-linear HS
symmetry algebra of the converted constraints for $Y(s_1, s_2)$ is found, with
non-vanishing terms (resolving the Jacobi identities) of third
order in powers of ghost coordinates. A gauge-invariant
unconstrained reducible Lagrangian formulation for a free bosonic
HS field of generalized spin $(s_1,s_2)$ is deduced. Following the
results of \cite{BuchbinderReshetnyak, BRmasscub}, we develop a
BRST approach to constructing general off-shell local cubic
interaction vertices for irreducible massive higher-spin fields
(being candidates for massive particles in the Dark Matter
problem). A new  reducible gauge-invariant Lagrangian formulation
for an antisymmetric massive tensor field of spin $(1,1)$ is
obtained.

\end{abstract}

\section{Introduction}

The great interest in higher-spin field theory is mainly explained
by a hope for re-examining the problems of a unified description
for the variety of elementary particles and all of the known
interactions beyond the Standard Model, as well as by a possible
insight into the origin of Dark Matter \cite{LHC, LHC2} beyond the
scope of models with sterile neutrinos \cite{sterilen} or vector
massive fields \cite{vectbosDM}. Massive higher-spin fields in
constant-curvature spaces are legitimate candidates for Dark
Matter, providing a contribution to the total energy density at
the level of about $27 \%$ in a Universe having the geometry of a
4d de-Sitter space; see \cite{DM1, DM2} for a review. Such a model
calls for a study involving compactification of extra dimensions
after the Big Bang. In the case of a Big Bang singularity, this
indicates that the physics is no longer described by classical
gravity, but rather by a quantum version of gravity involving
higher-spin fields \cite{HSGRA}, or, perhaps, a superstring field
theory model is required. The problems of Dark Matter and Dark
Energy provide insight into the inflation theory, which
consistently describes the post-Planck classical early Universe
and plays a crucial role in explaining the late-time cosmic
acceleration; for a review, see, e.g., \cite{Odintsov1} and
references therein. Note that the simplest candidate for Dark
Energy is given by a positive-valued Einstein cosmological
constant $\Lambda$, constructed along with Cold Dark Matter, which
has proved to be quite successful in describing a large span of
observational data related to polarization anisotropies in Cosmic
Microwave Background Radiation; see, e.g., \cite{Odintsov2,
Odintsov3, Starobinsky}.

Should these expectations prove to be viable, the development of
higher-spin field theory appears to be of high relevance in view
of its close relation to superstring theory in constant-curvature
spaces, which operates an infinite set of massive and massless
bosonic and fermionic HS fields subject to a multi-row Young
tableaux (YT) $Y(s_1,...,s_k)$, $k \geq 1$; for a review, see,
e.g., \cite{reviews, reviews1, reviews2, reviewsV, Snowmass, Ponomarev, reviews3}
and references therein.

A description of such theories, aimed at the final purpose of
Lagrangian formulation, demands some advanced and sophisticated
group-theoretical techniques, related to constructing different
representations of the (super)algebras underlying the theories in
question. Whereas for a Lie (super)algebra -- being the case
relevant to HS fields in flat spaces -- the deduction and the
structure of such ingredients as Verma modules and generalized
Verma modules \cite{Vermamodule} are rather well established, the
similar case of non-linear algebraic and superalgebraic structures
-- which corresponds to HS fields in AdS spaces -- has not been
classified until now, with the exception of totally symmetric
bosonic \cite{BurdikNavratilPasnev, BKL} and fermionic
\cite{adsfermBKR, 0905.2705} HS fields.

There are two well-established approaches to constructing
gauge-invariant Lagrangians for higher-spin fields by using the
respective metric and frame-like formulations. In the first
approach, the fields remain unconstrained, i.e., all the
conditions (d'Alembert equation, Young symmetry, absence of trace
and divergence) that select an irreducible Poincar\'{e}
\cite{Wigner,Labastida, Labastida1} or (A)dS
\cite{metsaevmixirrep, Metsaev, metsaevmixirrep1} group
representation of a given mass and spin are implemented on equal
footing to obtain a Lagrangian. In the second approach, some of
the conditions (normally, those related to the trace) for the
fields and gauge parameters are implemented in a consistent manner
which is imposed by hand, outside the construction of a
Lagrangian. At the level of free fields, there exist two efficient
approaches to the above objectives, known as BRST methods
(initially developed to quantize gauge field theories \cite{BRST0, BRST1, BRST2},   constrained dynamical systems
using the BFV--BRST procedure \cite{BFV, BFV1, Henneaux}), with
respective complete (e.g., \cite{BuchbinderReshetnyak, PT,
symint-adsmassless}) and incomplete \cite{Alkalaev} BRST operators
(implied by String Field Theory \cite{SFT, SFT1}), whose
Lagrangian descriptions for one and the same higher-spin field in
a $d$-dimensional Minkowski space-time are shown to be equivalent
\cite{Reshetnyak_con}. For HS fields in AdS spaces, descriptions
of irreducible representations (in the metric-like formalism) for
the (A)dS group with both integer and half-integer spins are
rather different from those for the Poincar\'{e} group in a flat
space-time, even for free theories. In all of the known cases,
Lagrangian descriptions for one and the same higher-spin field
deduced using the constrained (incomplete) BRST approach with
additional holonomic constraints and the approach with a complete
BRST operator do not coincide. Successful Lagrangian descriptions
for integer and half-integer spins in AdS spaces using the
approach of a complete BRST operator have been suggested for
massless and massive particles of integer spins in \cite{BKL,
symint-adsmassless}, and for massive particles of half-integer
spins, in \cite{adsfermBKR}. Some problems of this approach have
not been solved (in view of the yet unknown form of holonomic
constraints consistent with an incomplete BRST operator) even for
free fields of a given higher spin.\footnote{For recent
developments in the ambient flat space-time techniques
of the incomplete BRST approach to totally-symmetric
tensor fields in AdS spaces, see \cite{BBGG}.}

The present article develops the research line initiated by
\cite{BRIOP2011}, and its objectives are achieved using a regular
method of constructing a Verma module and a Fock space for
quadratic algebras, whose negative- (or positive-) root vectors in
a Cartan-like triangular decomposition are entangled due to the
presence of a special parameter $r$, being the inverse square of
the AdS radius. The resulting ingredients allow one to obtain
Lagrangian formulations (LFs) for free mixed-symmetric integer HS
fields in a $d$-dimensional AdS space with $Y(s_1,...,s_k)$ by
using the Fronsdal metric-like formalism \cite{Fronsdal0, FangFronsdal, Fronsdal} in the
BRST approach. Such an LF is the starting point for an interacting
HS field theory in the scope of conventional Quantum Field Theory.
An application of the BRST construction with a complete BRST
operator to a free HS field theory in AdS spaces consists of
several stages and solves a \textit{problem being inverse} to that
of the method \cite{BFV, BFV1, Henneaux}, thus reflecting the
concept of BV--BFV duality \cite{GrigorievDamgaard, gmr, gmr1} as
a particular case of the AKSZ model\footnote{For the AKSZ model
used in higher-spin gravity, see, e.g., \cite{sharapovskvort}.}
\cite{AKSZ}. First of all, the conditions that determine
representations of a given mass and spin are regarded as a
topological (having no Hamiltonian) gauge system of mixed-class
operator constraints, $o_I$, $I=1,2,...$, in an auxiliary Fock
space $\mathcal{H}$. Second, the entire system of $o_I$, which
forms a quadratic commutator algebra, is additively converted (see
\cite{conversion, conversion1} for the conversion methods) within
a deformed (in powers of the parameter $r$) algebra of $O_I$,
$O_I=o_I+o'_I$, defined in a larger Fock space,
$\mathcal{H}\bigotimes \mathcal{H}'$, with first-class constraints
$O_\alpha$, $O_\alpha \subset O_I$. Third, one needs to find a
hermitian and nilpotent BRST operator $Q'$ for a non-linear
algebra of converted operators $O_I$, which contains a BRST
operator $Q$ for the subsystem $O_\alpha$. Fourth, a Lagrangian
action $\mathcal{S}$ is constructed for a given HS field, through
an inner product $\langle \ |K| \ \rangle$ in a total Hilbert
space $\mathcal{H}_{tot}$, $\mathcal{S} \sim \langle \chi |KQ
|\chi \rangle$, so that $\mathcal{S}$ obeys reducible gauge
transformations, $\delta|\chi \rangle = Q |\Lambda \rangle$, with
$|\chi \rangle$ containing the initial and auxiliary fields. As a
result, the corresponding equations of motion are to reproduce the
initial AdS group conditions.

The above algorithm, as applied to bosonic \cite{flatbos,
flatbosm} and fermionic \cite{flatferm, flatfermmix} HS fields in
flat spaces, does not encounter any problems in view of a linear
Lie structure of the initial constraint algebra $[o_I, o_J\} =
f_{IJ}^K o_K$ with structure constants $f_{IJ}^K$. Indeed, for the
same algebra of additional parts $o'_I$ it is sufficient to use
the construction of a Verma module (VM) for the integer HS
symmetry algebra $sp(2k)$ \cite{BRbos}, which is in
one-to-one correspondence with the Lorentz $so(1,d-1)$ algebra of
irreducible unitary representations subject to $Y(s_1,...,s_k)$,
$k\leq \left[\frac{d-2}{2}\right]$, for massless fields, due to
the Howe duality \cite{Howe1}. Then, an oscillator realization of
the symplectic algebra $sp(2k)$ in a Fock space $\mathcal{H}'$
\cite{BRbos} (orthosymplectic algebra $osp(1|2k)$
for half-integer massless field representations \cite{BRf}) has a
polynomial form, as compared to totally symmetric HS fields in AdS
spaces \cite{adsfermBKR, brst1}, where the (super)algebras of
$o_I$ are non-linear and different from those of $o'_I$
\cite{BurdikNavratilPasnev, BKL, 0905.2705}. A problem of equal
complexity arises in constructing a BRST operator $Q'$ for the
algebra of converted constraints $O_I$, which, in the transition
to AdS spaces, does not have a form being quadratic in ghost
coordinates $\mathcal{C}$ and requires a complete analysis of the
relations, starting from a resolution of the Jacobi identities;
see \cite{0812.2329} for the details of finding a relevant
operator $Q'$ and \cite{bl} for classical quadratic algebras.

The details of a Lagrangian description for mixed-symmetry
massless HS tensors on (A)dS backgrounds have been studied in a
``frame-like'' formulation \cite{AlkalaevVasiliev,
AlkalaevVasiliev1, AlkalaevVasiliev2, AlkalaevVasiliev3,
AlkalaevVasiliev4}, whereas such descriptions for mixed-symmetry
massive  bosonic fields with off-shell traceless constraints in
the (A)dS case are known for a Young tableaux with two rows
\cite{Zinoviev, Zinoviev1,  Zinoviev2}, the main result for
Lagrangian formulation being \cite{Zinoviev2}. In a metric-like
formalism, (non-Lagrangian) equations of motion from the viewpoint
of a BRST description with an incomplete BRST operator for
arbitrary massless HS fields has been discussed in \cite{UNIbrst1,
UNIbrst2}, whereas in the BRST--BV approach with an incomplete
BRST operator and also in the light-cone approach mixed-symmetric
HS fields have been examined in \cite{metsaads5, metsaadsBRSTBV,
metscontads5, metsLCads4}. Diverse aspects of mixed-symmetry HS
field Lagrangian dynamics in Minkowski spaces have been discussed
in \cite{Franciamix} and the references therein, and the case of
interacting mixed-symmetry HS fields in AdS spaces has been
considered in \cite{interlow, interlow2}. In the case of cubic
interaction (see \cite{Manvelyan, Manvelyan1, Joung, frame-like1,
Metsaev0712, BuchbinderTsulaia, Tsulaiai2009, BRST-BV3,
frame-like2, BKTW, Metsaev-mass} for the study of cubic vertices
in different approaches), irreducible higher-spin fields in flat
spaces have been classified by Metsaev \cite{Metsaev0512} using
the light-cone formalism. We intend to follow the application of
the BRST approach to Lagrangian cubic interaction vertices for
irreducible totally symmetric bosonic fields in flat spaces
\cite{BuchbinderReshetnyak, BRmasscub, BRcub1, BKStwis}, however,
as applied to HS fields (including mixed-symmetric ones) in AdS
spaces.

The article is devoted to solving the following problems. First,
we deduce an HS symmetry algebra for bosonic HS fields in a
$d$-dimensional AdS space subject to an arbitrary
$Y(s_1,...,s_k)$; second, we develop the Verma module construction
for an HS symmetry algebra, with a two-row YT $Y(s_1,s_2)$ and an
oscillator realization for a given non-linear algebra, as a formal
power series in the oscillators of the related Heisenberg algebra;
third, we construct a BRST operator for a converted HS symmetry
algebra, and an unconstrained LF for free bosonic HS fields in AdS
spaces subject to $Y(s_1,s_2)$; forth, we develop a procedure for
constructing cubic interaction vertices by using three copies of
massive HS fields in AdS spaces.

The article is organized as follows. In Section~\ref{HSfieldsAdS},
we examine bosonic HS fields, which includes, in
Subsection~\ref{HSalg}, the construction of an HS symmetry algebra
$\mathcal{A}(Y(k), AdS_d)$ for HS fields subject to a Young
tableaux with $k$ rows; then an auxiliary theorem on a deformation
of polynomial general commutator algebras under an additive
conversion procedure is presented in Subsection ~\ref{auxtheorem}.
In Section~\ref{HSauxY2b}, we deduce a manifest form of an HS
symmetry algebra of the additional parts for arbitrary HS fields,
formulate the problem of Verma module construction for the algebra
$\mathcal{A}'(Y(k), AdS_d)$, in Subsection~\ref{VMboconic}, and
solve it explicitly for a quadratic algebra $\mathcal{A}'(Y(2),
AdS_d)$ of the additional parts $o'_I$, in Subsection~\ref{VMY2b}.
We find a Fock space realization for $\mathcal{A}'(Y(2), AdS_d)$
in Subsection~\ref{oscVMY2b}. In Section~\ref{Lagrform}, we
manifestly deduce a non-linear algebra $\mathcal{A}_c(Y(k),
AdS_d)$ of converted operators $O_I$, in
Subsection~\ref{convalgg}; we also present the form of a BRST
operator for this algebra with $k=2$, in Subsection\ref{BRSTk2},
and develop an unconstrained Lagrangian formulation for massive
bosonic HS fields with a two-row Young tableaux $Y(s_1,s_2)$, in
Subsection~\ref{prLagrform}. The general concept of finding cubic
interaction vertices for massive HS fields in AdS spaces, using
the approach with a complete BRST operator, is examined in
Section~\ref{cubinter}. Some examples of fields with lower spins
are presented in Section~\ref{exg}. In Conclusion, we summarize
our results and discuss some open problems. In Appendix
\ref{proof}, we prove a certain proposition, and then, in Appendix
\ref{poltoy}, we consider its application to a polynomial algebra
related to the Bethe ansatz. In respective Appendices
\ref{Ap-AddPartsb} and \ref{oscVM2}, we present calculations for
the Verma module construction with the algebra $\mathcal{A}'(Y(2),
AdS_d)$, in terms of a series of lemmas, and also give its
oscillator representation. Finally, in Appendix~\ref{reductionC},
we prove that the resulting Lagrangian does indeed reproduce
appropriate conditions for a field to determine an irreducible
representation of the AdS group.

As a rule, we apply the notation of conventions of
\cite{BuchbinderReshetnyak}, and also the respective notation
$\epsilon(F)$, $gh(F)$, $[H,\,G\}$, $[x]$, $\vec{s}_k$,
$(\vec{s})^{3}$ for the Grassmann parity and ghost number of a
homogeneous quantity $F$, as well as for a supercommutator, the
integer part of a real-valued $x$, an integer-valued vector
$(s_1,s_2,...,s_k)$, and a triple
$(\vec{s}{}^1_{k_1},\vec{s}{}^2_{k_2},\vec{s}{}^3_{k_3})$.

\section{HS fields of integer spin in AdS spaces}\label{HSfieldsAdS}

In this section, we obtain a number of special HS symmetry non-linear
algebras which encode mixed-symmetry tensor fields as elements
of irreducible AdS group representations with a generalized spin
$\mathbf{s}=(s_1,\ldots , s_k)$ and mass $m$ in AdS${}_d$
space-times. We examine the problem of a Verma module
construction for one of the algebras and solve it explicitly
for a non-linear algebra with a two-row Young tableaux, as well as
for a polynomial algebra of order $(n-1)$. The construction
of a Fock space representation for a non-linear algebra,
with a resulting Verma module, finalizes the solution of the problem
under consideration.

\subsection{HS symmetry algebra $\mathcal{A}(Y(k),AdS_d)$
for mixed-symmetry tensor fields with $Y(s_1,...,s_k)$}\label{HSalg}

A massive generalized integer spin $\mathbf{s}=(s_1,...,s_k)$
($s_1 \geq s_2\geq ... \geq s_k>0$, $k \leq [d/2]$) AdS group
irreducible representation in an AdS${}_d$ space is realized
on mixed-symmetry tensors
$\Phi_{\mu^1(s_1),\mu^2(s_2),...,\mu^k(s_k)} \hspace{-0.2em}\equiv
\hspace{-0.2em}\Phi_{\mu^1_1\ldots\mu^1_{s_1},\mu^2_1\ldots\mu^2_{s_2},...,
\mu^k_1\ldots \mu^k_{s_k}}(x)$, which correspond to the Young tableaux
\begin{equation}\label{Young k2}
\Phi_{\mu^1(s_1),\mu^2(s_2),...,\mu^k(s_k)}
\hspace{-0.3em}\longleftrightarrow \hspace{-0.3em}
\begin{array}{|c|c|c c c|c|c|c|c|c| c|}\hline
  \!\mu^1_1 \!&\! \mu^1_2\! & \cdot \ & \cdot \ & \cdot \ & \cdot\  & \cdot\  & \cdot\ &
  \cdot\    &\!\! \mu^1_{s_1}\!\! \\
   \hline
    \! \mu^2_1\! &\! \mu^2_2\! & \cdot\
   & \cdot\ & \cdot  & \cdot &  \cdot & \!\!\mu^2_{s_2}\!\!   \\
  \cline{1-8} \cdot\ & \cdot\ & \cdot\
   & \cdot\ & \cdot  & \cdot &  \cdot & \cdot\!\!   \\
   \cline{1-8}
    \! \mu^k_1\! &\! \mu^k_2\! & \cdot\
   & \cdot\ & \cdot  & \cdot &   \!\!\mu^k_{s_k}\!\!   \\
   \cline{1-7}
   \end{array}
\end{equation}
subject to the Klein--Gordon (\ref{Eq-0b}), divergence-free
(\ref{Eq-1b}), tracelessness (\ref{Eq-2b}), and mixed-symmetry
(\ref{Eq-3b}) equations for $\beta = (2;3;...;k+1)
\Longleftrightarrow (s_1>s_2; s_1 = s_2>s_3;...;s_1 =
s_2=...=s_k)$ \cite{Metsaev}, namely,
\begin{eqnarray}
\label{Eq-0b} &&\{\nabla^2 +r[(s_1-\beta-1+ d)(s_1-\beta) -
\sum_{i=1}^ks_i]+m^2
\}\Phi_{\mu^1(s_1),\mu^2(s_2),...,\mu^k(s_k)}
 =0,
\\
&&\nabla^{\mu^i_{l_i}}\Phi_{\mu^1(s_1),\mu^2(s_2),...,\mu^k(s_k)} =0, \quad
i,j=1,...,k;\,\,\,\,\,\, l_i,m_i=1,...,s_i\,, \label{Eq-1b}
\\
&& g^{\mu^i_{l_i}\mu^i_{m_i}}\Phi_{\mu^1(s_1),\mu^2(s_2),...,\mu^k(s_k)}=
g^{\mu^i_{l_i}\mu^j_{m_j}}\Phi_{
\mu^1(s_1),\mu^2(s_2),...,\mu^k(s_k)} =0, \quad
 l_i<m_i,  \,,\label{Eq-2b}\\
&& \Phi_{
\mu^1(s_1),...,\{\mu^i(s_i)\underbrace{,...,\mu^j_{1}...}\mu^j_{l_j}\}...\mu^j_{s_j},...
\mu^k(s_k)}=0,\quad
i<j,\ 1\leq l_j\leq s_j, \label{Eq-3b}
\end{eqnarray}
where the underline brackets denote the fact that the indices inside do not take
part in symmetrization, i.e., the present symmetrization involves only
the indices $\mu^i(s_i), \mu^j_{l_j} $
in $\{\mu^i(s_i)\underbrace{,...,\mu^j_{1}...}\mu^j_{l_j}\}$.

To obtain the HS symmetry algebra of $o_I$ for a description of all integer
HS fields, we introduce, in a standard manner, a Fock space $\mathcal{H}$
generated by $k$ pairs of bosonic creation $a^i_{\mu^i}(x)$ and annihilation
$a^{j+}_{\nu^j}(x)$
operators,\footnote{This choice of oscillators corresponds to the
case of a symmetric basis, whereas there also exists another
realization of an auxiliary Fock space, generated by fermionic
oscillators (antisymmetric basis)
$\hat{a}^m_{\mu^m}(x)$, $\hat{a}^{\hat{n}+}_{\nu^n}(x)$, with the
anticommutation relations
$\{\hat{a}^m_{\mu^m},\hat{a}_{\nu^n}^{n+}\}
=-g_{\mu^m\nu^m}\delta^{mn}$, for $m, n = 1,..., s_1$. The
treatment below proceeds along the lines of \cite{brst1} for
totally antisymmetric tensors with $s_1=s_2=...=s_k=1$.}
$i,j =1,...,k, \mu^i,\nu^j  =0,1...,d-1$,
\begin{eqnarray}
 \label{comrels}
[a^i_{\mu^i}, a_{\nu^j}^{j+}]=-g_{\mu^i\nu^j}\delta^{ij}\,, \qquad
\delta^{ij} = diag(1,1,\ldots 1)\,,
\end{eqnarray}
and a set of constraints for an arbitrary string-like vector
$|\Phi\rangle \in \mathcal{H}$, which we call a basic vector,
\begin{eqnarray}
 \label{PhysState}
\hspace{-2ex}&& \hspace{-2ex} |\Phi\rangle  =
\sum_{s_1=0}^{\infty}\sum_{s_2=0}^{s_1}\cdots
\sum_{s_k=0}^{s_{k-1}}\frac{\imath^{s_1+...+s_k}}{s_1!...s_k!}
\Phi_{\mu^1(s_1),\mu^2(s_2),...,\mu^k(s_k)}(x)\,
\prod_{i=1}^k\prod_{l_i=1}^{s_i} a^{+\mu^i_{l_i}}_i|0\rangle,\\
 \label{l0}
\hspace{-2ex}&& \hspace{-3ex} {\tilde{l}}_0|\Phi\rangle
= \bigl(l_0+ \tilde{m}^2_b + r \bigl((g_0^1-2\beta-2)g_0^1 -
 \sum^k_{i=2}g_0^i \bigr)\bigr)|\Phi\rangle=0 , \quad l_0 = [D^2 -
r\textstyle\frac{d(d-2(k+1))}{4}],\\
 \label{lilijt}
\hspace{-2ex} && \hspace{-2ex} \bigl({l}^i, l^{ij},
t^{i_1j_1} \bigr)|\Phi\rangle  = \bigl(-i a^i_\mu D^\mu,
\textstyle\frac{1}{2}a^{i}_\mu a^{j\mu}, a^{i_1+}_\mu
a^{j_1\mu}\bigr) |\Phi\rangle=0,\quad i\leq j;\, i_1 < j_1,
\end{eqnarray}
with particle number operators, a central charge, and a covariant
derivative in $\mathcal{H}$,
respectively,\footnote{The operators $a^{a_i}_i$, $a^{b_j+}_j$
satisfy the usual (in the space $R^{1,d-1}$) commutation relations
$[a^{a_i}_i,a^{b_j+}_j]=-\eta^{a_ib_i}\delta_{ij}$
for $\eta^{ab}=\mathrm{diag}(+,-,\ldots,-)$.}
\begin{eqnarray}
 \label{numpart}
 && g_0^i = -\frac{1}{2}\{a^{i+}_\mu,
 a^{\mu{}i}\},\qquad \qquad \qquad\   \tilde{m}^2_b = {m}^2 + r\beta(\beta+1),\\
 && D_\mu = \partial_\mu - \omega_\mu^{ab}(x)\Bigl(\sum_{i}a_{i{}a}^+a_{i{}b}
 \Bigr), \qquad a^{\mu(+)}_i(x) = e^\mu_a(x)a^{a(+)}_i,
\end{eqnarray}
with a vielbein $e^\mu_a$, a spin connection $\omega_\mu^{ab}$, and
tangent indices $a,b$, $ a=0,1...,d-1$. The operator $D_\mu$ is
equivalent (as applied in $\mathcal{H}$) to the covariant
derivative $\nabla_{\mu}$ with the d'Alembertian $D^2 =
(D_a+\omega^{b}{}_{ba} )D^a$. The set of $k(k+1)$ primary
constraints (\ref{l0}), (\ref{lilijt}) with $\{o_\alpha\}$ =
$\bigl\{{\tilde{l}}_0,{l}^i, l^{ij}, t^{i_1j_1}\bigr\}$, is
equivalent to the equations (\ref{Eq-0b})--(\ref{Eq-3b}) for all
admissible values of spin and for the field
$\Phi_{\mu^1(s_1),\mu^2(s_2),...,\mu^k(s_k)}$ with a fixed spin
$\mathbf{s}=(s_1,s_2,\ldots , s_k)$, once, in addition to
(\ref{l0}), (\ref{lilijt}), we have to add $k$ more constraints
with $g_0^i$,
\begin{eqnarray}\label{g0iphys}
g_0^i|\Phi\rangle =(s_i+\frac{d}{2}) |\Phi\rangle.
\end{eqnarray}

The condition for the algebra of $o_\alpha$ to be closed under
the $[\ ,\ ]$-multiplication leads to an enlargement of $o_\alpha$
by adding the operators $g_0^i$ and the hermitian conjugates $o_\alpha^+$,
\begin{equation}\label{lilijt+}
 \bigl({l}^{i+},\
l^{ij+},\ t^{i_1j_1+} \bigr)  = \bigl(-i a^{i+}_\mu D^\mu,\
\textstyle\frac{1}{2}a^{i+}_\mu a^{j\mu+},\ a^{i_1}_\mu
a^{j_1\mu+}\bigr) ,\ i\leq j;\ i_1 < j_1,
\end{equation}
with respect to the inner product in $\mathcal{H}$,
\begin{eqnarray}
\label{sproduct} \langle{\Psi}|\Phi\rangle & =  & \int d^dx
\sqrt{|g|}\sum_{i=1}^k\sum_{s_i=0}^{s_{i-1}}\frac{(-1)^{s_1+...+s_k}}{s_1!...s_k!}
         \Psi^*_{\mu^1(s_1),\mu^2(s_2),...,\mu^k(s_k)}(x)
\Phi^{\mu^1(s_1),\mu^2(s_2),...,\mu^k(s_k)}(x).
\end{eqnarray}
This fact guarantees the hermiticity of a corresponding BFV--BRST
operator, with account taken of the self-conjugate operators
$(l_0^+,\ {g_0^i}^+) = (l_0,\ {g_0^i})$ [hence a real value
of the Lagrangian $\mathcal{L}$] for the system of all the operators
$\{o_\alpha, o_\alpha^+, g_0^i\}$. We call the algebra of these operators
an \emph{integer-higher-spin (integer HS) symmetry algebra in an AdS space,
with a Young tableaux having $k$
rows}\footnote{The term \emph{higher-spin symmetry algebra},
applied here to a free HS formulation, is not to be confused with
the algebraic structure known as a \emph{higher-spin algebra}
which arises in describing HS interactions; see, e.g.,
\cite{reviewsV}.}
and denote it as $\mathcal{A}(Y(k),AdS_d))$.

The maximal Lie subalgebra of the operators $l^{ij}, t^{i_1j_1},
g_0^i, l^{ij+},\ t^{i_1j_1+}$ is isomorphic to the symplectic
algebra $sp(2k)$ [see \cite{BRbos} for details;
later on, we refer to it as $sp(2k)$], whereas the only nontrivial
quadratic commutators in $\mathcal{A}(Y(k), AdS_d))$ are due to
the operators containing $D_{\mu}$: $l^i, \tilde{l}_0, l^{i+}$.
For the purpose of an LF construction, it is sufficient to have
a simplified (having no central charge $\tilde{m}^2_b$), so-called
\textit{modified}, HS symmetry algebra $\mathcal{A}_{m}(Y(k),AdS_d)$,
with the operator $l_0$ (\ref{l0}), instead of ${\tilde{l}}_0$,
so that the AdS mass term $\tilde{m}^2_b + r\bigl((g_0^1-2\beta-2)g_0^1
- \sum_{i=2}^kg_0^i \bigr)$ is restored later on, as usual, within
a conversion procedure and an appropriate LF construction.

The algebra $\mathcal{A}_{m}(Y(k), AdS_d))$ of the operators
$o_I$, from the viewpoint of a Hamiltonian analysis for dynamical
systems, contains the first-class constraint $l_0$, as well as the $2k$
differential $l_i, l_i^+ $ and $2k^2$ algebraic $t^{i_1j_1},
t^+_{i_1j_1}, l^{ij}, l_{ij}^+$ second-class constraints $o_{{a}}$,
and also the operators $g_0^i$, composing an invertible matrix
$\Delta_{ab}(g_0^i)$ for a topological (having a zero Hamiltonian)
gauge system, due to
\begin{eqnarray}
[o_a,\; o_b] = f^c_{ab} o_c + f^{cd}_{ab} o_co_d +
\Delta_{ab}(g_0^i),\ [o_a,\;l_o] = f^c_{a{}[l_0]}o_c +
f^{cd}_{a[l_0]} o_co_d . \label{inconstraintsd}
\end{eqnarray}
Here, $f^c_{ab},  f^{cd}_{ab}, f^c_{a{}[l_0]}, f^{cd}_{a[l_0]},
\Delta_{ab}$ are antisymmetric with respect to the permutations
of the lower indices; the constant quantities and the quantities
$\Delta_{ab}(g_0^i)$ form a non-vanishing $2k(k+1)\times
2k(k+1)$ matrix $\|\Delta_{ab}\|$ in the Fock space $\mathcal{H}$
on the surface $\Sigma \subset \mathcal{H}$,
$\|\Delta_{ab}\|_{|\Sigma} \ne 0 $, which is determined
by the equations $o_\alpha|\Phi\rangle = 0$. The set of $o_I$
satisfies non-linear relations, being additional to those
for $sp(2k)$ and given by the multiplication Table~\ref{table}.
\hspace{-1ex}{\begin{table}[t] {{
\begin{center}
\begin{tabular}{||c||c|c|c|c|c|c|c||}\hline\hline
$\hspace{-0.2em}[\; \downarrow, \rightarrow
\}\hspace{-0.5em}$\hspace{-0.7em}&
 $t^{i_1j_1}$ & $t^+_{i_1j_1}$ &
$l_0$ & $l^i$ &$l^{i{}+}$ & $l^{i_1j_1}$
&$l^{i_1j_1{}+}$  \\
\hline\hline $t^{i_2j_2}$
    & $A^{i_2j_2, i_1j_1}$ & $B^{i_2j_2}{}_{i_1j_1}$&$0$
   & \hspace{-0.3em}
    $\hspace{-0.2em}l^{j_2}\delta^{i_2i}$\hspace{-0.5em} &
    \hspace{-0.3em}
    $-l^{i_2+}\delta^{j_2 i}$\hspace{-0.3em}
    &\hspace{-0.7em} $\hspace{-0.7em}l^{\{j_1j_2}\delta^{i_1\}i_2}
    \hspace{-0.9em}$ \hspace{-1.2em}& \hspace{-1.2em}$\hspace{-0.9em}
    -l^{i_2\{i_1+}\delta^{j_1\}j_2}\hspace{-0.9em}$\hspace{-1.2em} \\
\hline $t^+_{i_2j_2}$
    & $-B^{i_1j_1}{}_{i_2j_2}$ & $A^+_{i_1j_1, i_2j_2}$&$0$
   & \hspace{-0.3em}
    $\hspace{-0.2em} l_{i_2}\delta^{i}_{j_2}$\hspace{-0.5em} &
    \hspace{-0.3em}
    $-l^+_{j_2}\delta^{i}_{i_2}$\hspace{-0.3em}
    & $l_{i_2}{}^{\{j_1}\delta^{i_1\}}_{j_2}$ & $-l_{j_2}{}^{\{j_1+}
    \delta^{i_1\}}_{i_2}$ \\
\hline $l_0$
    & $0$ & $0$&$0$
   & \hspace{-0.3em}
    $\hspace{-0.2em}-r{\mathcal{K}}^{i+}_1$\hspace{-0.5em} & \hspace{-0.3em}
    $r{\mathcal{K}}^{i}_1$\hspace{-0.3em}
    & $0$ & $0$  \\
\hline $l^j$
   & \hspace{-0.5em}$- l^{j_1}\delta^{i_1j}$ \hspace{-0.5em} &
   \hspace{-0.5em}$
   -l^{i_1}\delta^{j_1j}$ \hspace{-0.9em}  &\hspace{-0.3em}
    $\hspace{-0.2em}r{\mathcal{K}}^{j+}_1$\hspace{-0.5em} \hspace{-0.3em}&${W}^{ji}_b$ \hspace{-0.3em} & \hspace{-0.3em}
   ${X}^{ji}_b$\hspace{-0.3em}
    & $0$ & \hspace{-0.5em}$- \textstyle\frac{1}{2}l^{\{i_1+}\delta^{j_1\}j}$
    \hspace{-0.9em}  \\
\hline $l^{j+}$ & \hspace{-0.5em}$l^{i_1+}
   \delta^{j_1j}$\hspace{-0.7em} & \hspace{-0.7em}$l^{j_1+}\delta^{i_1j}$ \hspace{-1.0em} &\hspace{-0.3em}
    $\hspace{-0.2em}-r{\mathcal{K}}^{j}_1$\hspace{-0.5em}
   &\hspace{-0.3em}
   $-{X}^{ij}_b$\hspace{-0.3em}
   &\hspace{-0.5em} $- {W}^{ji+}_b$\hspace{-0.5em}
    &\hspace{-0.7em} $ \textstyle\frac{1}{2}l^{\{i_1}\delta^{j_1\}j}
    $\hspace{-0.7em} & $0$  \\
\hline $l^{i_2j_2}$
    & \hspace{-0.2em}$\hspace{-0.2em}-l^{j_1\{j_2}\delta^{i_2\}i_1}
    \hspace{-0.5em}$
    \hspace{-0.5em} &\hspace{-0.2em} $\hspace{-0.4em}
    -l^{i_1\{i_2}\delta^{j_2\}j_1}\hspace{-0.3em}$\hspace{-0.3em}&$0$
   & \hspace{-0.3em}
    $0$\hspace{-0.5em} & \hspace{-0.3em}
    $ \hspace{-0.7em}-\textstyle\frac{1}{2}l^{\{i_2}\delta^{j_2\}i}
    \hspace{-0.5em}$\hspace{-0.3em}
    & $0$ & \hspace{-0.7em}$\hspace{-0.3em}
L^{i_2j_2,i_1j_1}\hspace{-0.3em}$\hspace{-0.7em}  \\
\hline $l_{i_2j_2}^+$
    & $ l^{i_1+}_{\, \{i_2}\delta^{j_1}_{j_2\}}$  & $ l^{+}_{j_1\{j_2}
    \delta_{i_2\}i_1}$ & $0$
   & \hspace{-0.3em}
    $\hspace{-0.2em} \textstyle\frac{1}{2}l^{+}_{\{i_2}\delta^{i}_{j_2\}}$\hspace{-0.5em} & \hspace{-0.3em}
    $0$\hspace{-0.3em}
    & $-L^{i_1j_1,i_2j_2}$ & $0$  \\
\hline\hline $g^j_0$
    & $-F^{i_1j_1,j}$ & $F^{i_1j_1,j+}$&$0$
   & \hspace{-0.3em}
    $\hspace{-0.2em}-l^i\delta^{ij}$\hspace{-0.5em} & \hspace{-0.3em}
    $l^{i+}\delta^{ij}$\hspace{-0.3em}
    & \hspace{-0.7em}$\hspace{-0.7em}  -l^{j\{i_1}\delta^{j_1\}j}\hspace{-0.7em}$\hspace{-0.7em} & $ l^{j\{i_1+}\delta^{j_1\}j}$ \\
   \hline\hline
\end{tabular}
\end{center}}} \vspace{-2ex}\caption{The HS symmetry non-linear algebra
$\mathcal{A}_{m}(Y(k),AdS_d)$.\label{table}}
\end{table}

First of all, note that Table~\ref{table} does not include any columns
with $[\ ,\ \}$-products of all $o_I$ with $g_0^j$, which can be observed
from the rows containing $g_0^j$, as follows, $[b,o_I\}^+$ = $-[o^+_I ,b \}$,
with allowance for the closure of the HS algebra with respect to hermitian
conjugation. Second, the operators $t^{i_2j_2}, t^+_{i_2j_2}$ are defined
so as to satisfy the following properties:
\begin{equation}\label{thetasymb}
(t^{i_2j_2}, t^+_{i_2j_2}) \equiv
(t^{i_2j_2},t^+_{i_2j_2})\theta^{j_2i_2}, \ \theta^{j_2i_2} =
(1,0)\verb" for " (j_2> i_2, j_2\leq i_2),
\end{equation}
with the Heaviside
$\theta$-symbol\footnote{There is no summation with respect to
the indices $i_2, j_2$ in (\ref{thetasymb}), and the figure
brackets for the indices $i_1$, $i_2$ in the quantity $A^{\{ i_1
}B^{i_2\}i_3}\theta^{i_3i_2\}}$ imply symmetrization: $A^{\{i_1}
B^{i_2\}i_3}\theta^{i_3i_2\}}$
=$A^{i_1}B^{i_2i_3}\theta^{i_3i_2}+A^{i_2}B^{i_1i_3}\theta^{i_3i_1}$.
These indices are raised and lowered using the Euclidian metric
tensors $\delta^{ij}$, $\delta_{ij}$, $\delta^{i}_{j}$.}
$\theta^{ji}$. Third, the products $B^{i_2j_2}_{i_1j_1},A^{i_2j_2, i_1j_1},
F^{i_1j_1,i}, L^{i_2j_2,i_1j_1}$
are determined by the explicit relations
\begin{eqnarray}
  {}B^{i_2j_2}{}_{i_1j_1} &=&
  (g_0^{i_2}-g_0^{j_2})\delta^{i_2}_{i_1}\delta^{j_2}_{j_1} +
  (t_{j_1}{}^{j_2}\theta^{j_2}{}_{j_1} + t^{j_2}{}^+_{j_1}\theta_{
  j_1}{}^{j_2})\delta^{i_2}_{i_1}
  -(t^+_{i_1}{}^{i_2}\theta^{i_2}{}_{i_1} + t^{i_2}{}_{i_1}\theta_{i_1}{}^{
  i_2})
  \delta^{j_2}_{j_1}\,, \label{Bijkl}
  \\
  A^{i_2j_2, i_1j_1} &=&  t^{i_1j_2}\delta^{i_2j_1}-
  t^{i_2j_1}\delta^{i_1j_2}  , \qquad    F^{i_2j_2,i} \ = \
   t^{i_2j_2}(\delta^{j_2i}-\delta^{i_2i}),\label{Fijk} \\
  L^{i_2j_2,i_1j_1} &=&   \textstyle\frac{1}{4}\Bigl\{\delta^{i_2i_1}
 \delta^{j_2j_1}\Bigl[2g_0^{i_2}\delta^{i_2j_2} + g_0^{i_2} +
 g_0^{j_2}\Bigr]  - \delta^{j_2\{i_1}\Bigl[t^{j_1\}i_2}\theta^{i_2j_1\}} +t^{i_2j_1\}+}\theta^{j_1\}i_2}\Bigr] \nonumber \\
 && - \delta^{i_2\{i_1}\Bigl[t^{j_1\}j_2}\theta^{j_2j_1\}}
+t^{j_2j_1\}+}\theta^{j_1\}j_2}\Bigr] \Bigr\}\,. \label{Lklij}
\end{eqnarray}
They satisfy the obvious additional properties of antisymmetry and hermitian
conjugation
\begin{eqnarray}
  &&\hspace{-1em} A^{i_2j_2,  i_1j_1} = -A^{ i_1j_1, i_2j_2}
  ,\qquad  A^+_{i_1j_1,  i_2j_2}=(A_{i_1j_1,  i_2j_2})^+ = t^+_{i_2j_1}\delta_{j_2i_1}
   -   t^+_{i_1j_2}\delta_{i_2j_1},\\
  &&\hspace{-1em} ({L^{i_2j_2,i_1j_1}})^+ =  L^{i_1j_1, i_2j_2} ,\qquad  {F^{i_2j_2,i+}}
  =(F^{i_2j_2,i})^+= t^{i_2j_2+}(\delta^{j_2i}-\delta^{i_2i}),\\
\label{Bijk+}
  && \hspace{-1em} {B^{i_2j_2}{}_{i_1j_1}}^+ = (g_0^{i_2}-g_0^{j_2})\delta^{i_2}_{i_1}
   \delta^{j_2}_{j_1} +
  (t^+_{j_1}{}^{j_2}\theta^{j_2}{}_{j_1} + t^{j_2}{}_{j_1}
  \theta_{j_1}{}^{j_2})\delta^{i_2}_{i_1}
  -(t_{i_1}{}^{i_2}\theta^{i_2}{}_{i_1} +
  t^{i_2+}{}_{i_1}\theta_{i_1}{}^{i_2})\delta^{j_2}_{j_1}.
\end{eqnarray}
Fourth, the independent quantities ${\mathcal{K}}^{k}_1,{W}^{ki}_b, X^{ki}_b$
in Table~\ref{table} are quadratic in $o_I$:
\begin{eqnarray}
{} {W}^{ij}_b & = & [l^i,l^j]= 2r\left[(g_0^j-g_0^i)l^{ij} -
(\sum_{m} t^{m[j}\theta^{[jm}+
t^{[jm+}\theta^{m[j})l^{i]m}\right],
 \label{lilj}
\end{eqnarray}
\vspace{-1.5ex}
 \begin{eqnarray}
{\mathcal{K}}^{k}_1& =
&r^{-1}[l_0,l^{k+}]=\Bigl(4\sum_{i}l^{ki+}l^i +l^{k+}(2g_0^k-1)
-2(\sum_{i}l^{i+}t^{+ik}\theta^{ki}
+l^{i+}t^{ki}\theta^{ik})\Bigr)  \label{l'0li+} ,
\\
{} {X}^{ij}_b
  & =
 &\bigl\{{{l}}_{0}+ r\bigl( K^{0i}_0  +
\sum_{l=i+1}^k\mathcal{K}^{il}_0 +
\sum_{l=1}^{i-1}\mathcal{K}^{li}_0 \bigr)\bigr\}\delta^{ij}
      \label{lilj+b}\\
      && -
r\bigl[4\sum\nolimits_{l} l^{ jl+}l^{ li} -\sum_{l=1}^{j-1} t^{
lj+}t^{ li} -\sum_{l=i+1}^{k} t^{ il+}t^{jl}-\sum_{l=j+1}^{i-1}
t^{ li}t^{ jl} + \textstyle(g_0^{ j}+g_0^i
  -j-{1})t_{ji}\bigr]\theta^{ij}\nonumber\\
&& - r\bigl[4\sum\nolimits_{l} l^{ jl+}l^{ li} -\sum_{l=1}^{i-1}
t^{ lj+}t^{ li} -\sum_{l=j+1}^{k} t^{
il+}t^{jl}-\sum_{l=i+1}^{j-1} t^{ il+}t^{ lj+}+ t^+_{ij}(g_0^{
j}+g_0^{ i}-i
  -{1}
)\bigr]\theta^{ji}. \nonumber
\end{eqnarray}
In (\ref{lilj+b}), we have used the quantities
$K^{0i}_0,i,j=1,...,k$, $\mathcal{K}^{ij}_0$, composing a Casimir
operator $\mathcal{K}_0(k)$ for the $sp(2k)$ algebra:
\begin{equation}\label{Casimirsb}
\mathcal{K}_0(k) =
   \sum_{i}K_0^{0i} +
    2\sum_{i,j}\mathcal{K}_0^{ij}\theta^{ji} =
    \sum_{i}\bigl((g_0^i)^2-2g_0^i -4l^{ii+}l^{ii}\bigr)+2
    \bigl(\sum_{i=1}^k\sum_{j=i+1}^k(t^{ +}_{ij}t^{ij}-
    4l^{ +}_{ij}
      l^{ {ij}} -
    g_0^{ j}) \bigr).
\end{equation}
The algebra $\mathcal{A}_m(Y(k), AdS_d)$ may be regarded as a non-linear
deformation (in powers of $r$) of the integer HS symmetry algebra $\mathcal{A}(Y(k), {R} ^{1,d-1})$  \cite{BRbos}
in Minkowski space, namely,
\begin{equation}\label{identalg}
  \mathcal{A}_m(Y(k), AdS_d) = \mathcal{A}(Y(k), {R}^{1,d-1})(r)
  = \left(T^k \oplus T^{k*}\oplus l_0\right)(r) + \hspace{-1em} \supset sp(2k),
\end{equation}
for a $k$-dimensional commutative (in ${R}^{1,d-1}$) algebra $T^k
=\{l_i\}$ and its dual $T^{k*}=\{l^{i+}\}$, which represents a
semidirect sum of the symplectic algebra $sp(2k)$, being an
algebra of internal derivations of $(T^k \oplus T^{k*})$. For HS
fields with a single spin $s_1$, $k=1$, and a two-component spin
$(s_1,s_2)$, $k=2$, the algebra $\mathcal{A}_m(Y(k), AdS_d)$
coincides with the respective familiar HS symmetry algebras in AdS
spaces given by \cite{0812.2329} and \cite{BurdikNavratilPasnev}.

Now, we utilize the results of a special theorem, and then proceed to a conversion
of the algebra of $o_I$, so as to obtain the algebra of $O_I$ with first-class
constraints alone.

\subsection{On additive conversion for polynomial algebras}\label{auxtheorem}

In this subsection, having in mind the problem of an additive conversion
for non-linear algebras (superalgebras) with a subset of second-class constraints,
we need to use an important statement, based on the following (see
\cite{0905.2705} for a detailed description)

\noindent \textbf{\emph{Definition:}}
\emph{A non-linear commutator superalgebra $\mathcal{A}$ of basis elements
$o_I$, $I \in \Delta$ (with $\Delta$ being a finite or infinite set of indices)
is called a \textbf{polynomial superalgebra of order $n$}, $n \in \mathbf{N}$
if the set $\{{o}_I\}$ is subject to the $n$-th order polynomial supercommutator
relations}
\begin{eqnarray}\label{polynom}
&&
 [o_I, o_J\}  =
 F^{K}_{IJ}(o)o_{K},\
 F^{K}_{IJ} = f^{(1)K}_{IJ} + \sum_{n=2}^\infty
 f^{(n)K_1...K_{n-1}K}_{IJ}\prod_{i=1}^{n-1} o_{K_i}\,,
 \nonumber\\
 {}&&\qquad
 f^{(n)K_1...K_n}_{IJ} \neq 0, \, \mathrm{and} \,
 f^{(k)K_1...K_nK_{n+1}...K_k}_{IJ}=0,\, k>n,
\end{eqnarray}
\emph{with structure coefficients} $f^{(n)K_1...K_{n-1}K}_{IJ}$ of generalized
antisymmetry with respect to the permutations of lower indices,
$f_{IJ}^{(n)K_1\cdots K_n} = -(-1)^{\varepsilon(o_I)\varepsilon(o_J)}
f_{JI}^{(n)K_1\cdots K_n}$.

Now, we are in a position to formulate our basic statement
of Subsection~\ref{auxtheorem}, presented as the following

\noindent \textbf{\emph{Proposition:}} Let $\mathcal{A}$ be a polynomial
superalgebra of order $n$ with basis elements $o_I$ defined in a Hilbert space
$\mathcal{H}$. Then, for a set $\mathcal{A}'$ of elements $o'_I$ defined
in a new Hilbert space $\mathcal{H}'$ ($\mathcal{H} \bigcap \mathcal{H}'
= \emptyset$) and supercommuting with $o_I$, as well as for a direct sum
of sets $\mathcal{A}_{c}= \mathcal{A} + \mathcal{A}'$ of the operators
${O}_I$, ${O}_I = {o}_I + {o}'_I$, defined in the tensor product
$\mathcal{H}\otimes \mathcal{H}'$, the requirement to be in involution,
\begin{equation}\label{involrel}
 [{O}_I,{O}_J\} =
{F}_{IJ}^K({o}',{O}) {O}_K\,,
\end{equation}
implies that the sets $\{{o}'_I\}$ and $\{{O}_I\}$ form the respective
polynomial commutator superalgebra $\mathcal{A}'$ of order $n$
and the non-linear commutator superalgebra $\mathcal{A}_{c}$
with the composition laws
\begin{eqnarray}\label{addalg}
 && [\,o_I',o_J'\} = f_{IJ}^{(1)K_1}{o}'_{K_1}+ \sum_{m=2}^n(-1)^{
  m-1+\varepsilon_{K_{(m)}}}f_{IJ}^{(m)K_{m}\cdots
K_1}\prod_{s=1}^{m}{o}'_{K_s},\\ \label{conv-alg}
  && [\,{O}_I,{O}_J\} \hspace{-0,3ex}=\hspace{-0,3ex}
  \Bigl(f_{IJ}^{(1)K} + \sum_{m=2}^{n}
  F^{(m){}K}_{IJ}({o}',{O})\Bigr){O}_{K},
\end{eqnarray}
for the values $\varepsilon_{K_{(n)}}$,
$\varepsilon_{K_{(n)}}=\sum_{s=1}^{n-1}
\varepsilon_{K_s}\Bigl(\sum_{l=s+1}^n\varepsilon_{K_l}\Bigr)$. The
structure functions $F^{(m){}k}_{IJ}(o',O)$ in (\ref{addalg}) are
constructed with respect to the coefficients
$f^{(m)K_1...K_m}_{IJ} \equiv f^{K_1...K_m}_{IJ}$ used in
(\ref{polynom}), as follows:
\begin{eqnarray}
 \hspace{-1,5ex} && F^{(m){}K}_{IJ}
  \hspace{-0,3ex}=\hspace{-0,3ex}
f_{IJ}^{K_1\cdots K_{m}}\prod_{p=1}^{m-1}{O}_{K_p}
+\sum_{s=1}^{m-1}(-1)^{s+ \varepsilon_{K_{(s)}}}
f_{ij}^{\widehat{K_s\cdots K_1}\widehat{K_{s+1}\cdots K_{m}}}
\prod_{p=1}^{s}{o}'_{K_{p}}
\prod_{l=s+1}^{m-1}{O}_{K_{l}},\verb" where "\label{expfunc}
\\
 \hspace{-1,5ex}&& f_{ij}^{\widehat{K_s\cdots K_1}\widehat{K_{s+1}\cdots K_{m}}}
 = f_{ij}^{{K_s\cdots K_1}{K_{s+1}\cdots K_{m}}} +
f_{ij}^{K_s\cdots {K_{s+1}K_1}{K_{s+2}\cdots
K_{m}}}(-1)^{\varepsilon_{K_{s+1}}\varepsilon_{K_1}} +\cdots +
\nonumber \\
 \hspace{-1,5ex}{}&{}&  f_{ij}^{{K_{s+1}K_s\cdots K_1}{K_{s+2}\cdots
K_{m}}}(-1)^{\varepsilon_{K_{s+1}}\sum_{l=1}^{s}\varepsilon_{K_l}}
+\Bigl(f_{ij}^{K_{s+1}K_s\cdots {K_{s+2}K_1}{K_{s+3}\cdots
K_{m}}}(-1)^{\varepsilon_{K_{s+2}}\varepsilon_{K_1}} +\nonumber \\
 \hspace{-1,5ex}{}&{}&\cdots +
f_{ij}^{{K_{s+1}K_{s+2}K_s\cdots K_1}{K_{s+3}\cdots
K_m}}(-1)^{\varepsilon_{K_{s+2}}\sum_{l=1}^{s}\varepsilon_{K_l}}\Bigr)(-1)^{\varepsilon_{K_{s+1}}\sum_{l=1}^{s}\varepsilon_{K_l}
}+ \cdots
+\nonumber\\
 \hspace{-1,5ex}{}&{}&
(-1)^{\sum_{l=s+1}^{m}\varepsilon_{K_{l}}\sum_{p=1}^{s}\varepsilon_{K_p}
}f_{ij}^{{K_{s+1}\cdots K_m}K_s\cdots K_1},\label{sumcoeff}
\end{eqnarray}
where the sum in (\ref{sumcoeff}) contains $\frac{m!}{s!(m-s)!}$
terms, with all the possible ways of arrangement for
$(K_{s+1},..., K_{m})$ among the indices $(K_{s},..., K_{1})$ in
$f_{ij}^{{K_s\cdots K_1}{K_{s+1}\cdots K_{m}}}$, without changing
the separate ordering of the
indices\footnote{This superalgebra can be equally used for both bosonic HS fields
in an antisymmetric basis and fermionic HS fields in an AdS space.}
$K_{s+1},..., K_{m}$ and $K_{s},..., K_{1}.$

The correctness of the Proposition is examined in Appendix \ref{proof}.
Turning to the structure of the superalgebra $\mathcal{A}_c$,
we emphasize that, in contrast to $\mathcal{A}$ and $\mathcal{A}'$,
we call it a \emph{non-homogeneous polynomial superalgebra of order} $n$,
due to the form of relations (\ref{expfunc}); see Footnote~\ref{prop} in
Appendix \ref{proof} for remarks. In the case of a Lie superalgebra ($n=1$),
the structures of superalgebras $\mathcal{A}$, $\mathcal{A}'$
and $\mathcal{A}_c$ are identical, and used as the integer HS
symmetry algebra $\mathcal{A}(Y(k), {R}^{1,d-1})$
\cite{BRbos}. For quadratic algebras (n=2),
the algebraic relations for $\mathcal{A}^{n-1}$, $\mathcal{A}^{\prime n-1}$
and $\mathcal{A}_c^{n-1}$ do not coincide with each other, due to the presence
of structure functions $f^{(2)K_1K_{2}}_{IJ}$, which has been originally shown
for the algebra $\mathcal{A}(Y(1), AdS_d)$ with totally symmetric HS tensors
in an AdS space \cite{BKL}, having the form
\begin{eqnarray}
\hspace{-1em}&\hspace{-1em}&\hspace{-1em} [\,o_I,o_J]  =
f_{IJ}^{(1)K_1}{o}_{K_1}+
 f_{IJ}^{(2)K_{1}K_2}{o}_{K_1}{o}_{K_2},\quad [\,o'_I,o'_J] = f_{IJ}^{(1)K_1}{o}'_{K_1}-
f_{IJ}^{(2)K_{2}K_1}{o}'_{K_1}{o}'_{K_2}, \label{auxalg2}
 \\
 {}  \hspace{-1,7em}&\hspace{-1em}&\hspace{-1,7em} [\,{O}_I,{O}_J]  =
  \Bigl(f_{IJ}^{(1)K} +
  F^{(2){}K}_{IJ}({o}',{O})\Bigr){O}_{K},\quad
  F^{(2){}K}_{IJ}  = f_{IJ}^{(2)K_1K}{O}_{K_1} -
(f_{IJ}^{(2)K_1K} + f_{IJ}^{(2)KK_{1}}){o'}_{K_1}
\label{conv-alg2} .
\end{eqnarray}
The relations (\ref{auxalg2}) and (\ref{conv-alg2}) are sufficient
to determine the form of multiplication laws for the algebra of
the additional parts $o'_I$, $\mathcal{A}'(Y(k), AdS_d)$, and for
the converted, $O_I$, operator algebra $\mathcal{A}_c(Y(k),
AdS_d)$.

As a new result, we present an explicit form for the cubic commutator
algebras $\mathcal{A}'$, $\mathcal{A}_c$,
\begin{eqnarray}
 [\,o_I',o_J'\} & =& f_{IJ}^{(1)K_1}{o}'_{K_1}-f_{IJ}^{(2)K_{2}K_1}{o}'_{K_1}{o}'_{K_2}
+ f_{IJ}^{(3)K_{3}K_2 K_1}{o}'_{K_1}{o}'_{K_2}{o}'_{K_3},\nonumber \\
\label{conv-alg3}
   [\,{O}_I,{O}_J\} & = &
  \Bigl(f_{IJ}^{(1)K} + \sum_{l=2}^{n}
  F^{(l){}K}_{IJ}({o}',{O})\Bigr){O}_{K},\\
 F^{(3){}K}_{IJ}  &=&f_{IJ}^{(3)K_1K_{2}K}{O}_{K_1}{O}_{K_2}-
\bigl(f_{IJ}^{(3)K_1K_2K} +
f_{IJ}^{(3)K_2K_{1}K}\bigr){o'}_{K_1}{O}_{K_2}\nonumber\\
{}&& + \bigl(f_{IJ}^{(3)K_2K_1K} +
 f_{IJ}^{(3)K_2KK_{1}} +
f_{IJ}^{(3)KK_2K_{1}}\bigr){o'}_{K_1}{o'}_{K_2}\, ,
\end{eqnarray}
if the commutator relations for the initial algebra $\mathcal{A}$
are given by (\ref{polynom}) for $n=3$.

In Appendix \ref{poltoy}, we consider the example of a polynomial
$3$-parametric algebra $\mathcal{A}$ of order $n-1$, $n \in \mathbf{N}$,
being a polynomial deformation of the $su(1,1)$ algebra for $n>2$.


\section{Auxiliary HS symmetry algebra $\mathcal{A}'(Y(k),AdS_d)$}
\label{HSauxY2b}

The procedure of additive conversion for the non-linear HS symmetry
algebra $\mathcal{A}(Y(k), AdS_d)$ of the operators $o_I$ implies
establishing, first of all, an explicit form of the algebra
$\mathcal{A}'(Y(k), AdS_d)$ for the additional parts $o'_I$;
second, a representation of $\mathcal{A}'(Y(k), AdS_d)$ in terms
of appropriate Heisenberg algebra elements acting in a new
Fock space $\mathcal{H}'$. The structure of non-linear commutators
of the initial algebra implies a need for converting all the operators
$o_I$ in order to construct an unconstrained LF for a given HS field
$\Phi_{(\mu^1)_{s_1},(\mu^2)_{s_2},...,(\mu^k)_{s_k}}$.

The former step is based on determining a multiplication table
$\mathcal{A}'(Y(k), AdS_d)$ for the operators $o'_I$, following
the structure of the algebra $\mathcal{A}(Y(k), AdS_d)$, given by
(\ref{auxalg2}) and Table~\ref{table}.

As a result, the required composition law for $\mathcal{A}'(Y(k), AdS_d)$
is the same as the one for the algebra $\mathcal{A}(Y(k), AdS_d)$ in its linear
Lie part, i.e., for the $sp(2k)$ subalgebra of elements $(l^{\prime ij},
l^{\prime ij+},t^{\prime i_1j_1}, t^{\prime+}_{i_1j_1},g_0^{\prime i})$,
and is different in the non-linear part of Table~\ref{table}, determined by
the isometry group elements $l'_i, l^{\prime +}_j, l'_0$. The corresponding
non-linear submatrix of the multiplication matrix for $\mathcal{A}'(Y(k), AdS_d)$
has the form given by Table~\ref{table'}.
\begin{table}[t]{
\begin{center}
\begin{tabular}{||c||c|c|c||}\hline\hline
$[\,\downarrow\,,\to\}$&
 $l'_0$ &
$l^{\prime i}$ & $l^{\prime i{}+}$  \\
\hline\hline $l'_0$
    & $0$
   &
    $ r{\mathcal{K}}^{\prime bi+}_1$ & $-r{\mathcal{K}}^{\prime i}_1$\\
\hline $l^{\prime j}$
   &   $ -r{\mathcal{K}}^{\prime j+}_1$
   & $-{W}^{\prime ji}_b$  & ${X}^{\prime ji}_b$ \\
\hline $l^{\prime j+}$ &
   $r{\mathcal{K}}^{\prime j}_1$  &
   $-{X}^{\prime ij}_b$
   & $ {W}^{\prime ji+}_b$  \\
   \hline\hline
\end{tabular}\end{center}
} \caption{The non-linear part of the algebra $\mathcal{A}'(Y(k),
AdS_d)$.} \label{table'}
\end{table}
Here, the functions ${\mathcal{K}}^{\prime i+}_1$, ${\mathcal{K}}^{\prime i}_1$,
${W}^{\prime ji}_b$, ${W}^{\prime ji+}_b$ $(X^{\prime ij}_b-l'_0)$ have the same
definition as those in (\ref{lilj})--(\ref{lilj+b}) for the initial operators $o_I$, albeit
with an opposite sign for $(X^{\prime ij}_b-l'_0)$,
\begin{eqnarray}
{} && {W}^{\prime ij}_b  =  2r\left[(g_0^{\prime j}-g_0^{\prime
i})l^{\prime ij} - (\sum_{m} t^{\prime m[j}\theta^{[jm}+ t^{\prime
[jm+}\theta^{m[j})l^{\prime i]m} \right],
 \label{lilj'}\\
&& {\mathcal{K}}^{\prime j}_1 = \Bigl(4\sum_{i}l^{\prime
ji+}l^{\prime i} +l^{\prime j+}(2g_0^{\prime j}-1)
-2(\sum_{i}l^{\prime i+}t^{\prime +ij}\theta^{ji} +l^{\prime
i+}t^{\prime ji}\theta^{ij})\Bigr)  \label{l'0li+'} ,
\end{eqnarray}
\vspace{-1.5ex}
 \begin{eqnarray}
 {}&& {X}^{\prime ij}_b
   =
 \bigl\{l'_{0}- r\bigl( K^{\prime 0i}_0  +
\sum_{l=i+1}^k\mathcal{K}^{ \prime il}_0 +
\sum_{l=1}^{i-1}\mathcal{K}^{\prime li}_0 \bigr)\bigr\}\delta^{ij}
      \label{lilj+b'}\\
      &&\hspace{1em} +
r\bigl\{ 4\sum\nolimits_{l} l^{\prime jl+}l^{\prime li}
-\hspace{-0.2em}\sum_{l=1}^{j-1}\hspace{-0.1em} t^{\prime
lj+}t^{\prime li} -\hspace{-0.2em}\sum_{l=i+1}^{k}\hspace{-0.2em}
t^{\prime il+}t^{\prime jl}-\hspace{-0.2em}\sum_{l=j+1}^{i-1}
\hspace{-0.2em}t^{\prime li}t^{\prime jl} + \textstyle(g_0^{\prime
 j}+g_0^{\prime i}
  -j-{1})t'_{ ji}\bigr\}\theta^{ij}\nonumber\\
&& \hspace{1em} + r\bigl\{ 4\sum\nolimits_{l} l^{\prime
jl+}l^{\prime li} -\hspace{-0.2em}\sum_{l=1}^{i-1}\hspace{-0.1em}
t^{\prime lj+}t^{\prime li}
-\hspace{-0.2em}\sum_{l=j+1}^{k}\hspace{-0.2em} t^{\prime
il+}t^{\prime jl}-\hspace{-0.2em}\sum_{l=i+1}^{j-1}\hspace{-0.2em}
t^{\prime il+}t^{\prime lj+}+ t^{\prime +}_{ij}(g_0^{\prime
j}+g_0^{\prime i}-i
  -{1}
)\bigr\}\theta^{ji}. \nonumber
\end{eqnarray}
Now, we can outline the points leading to an oscillator representation
for the elements of an auxiliary HS symmetry algebra.

\subsection{On representations of $\mathcal{A}'(Y(k),AdS_d)$}\label{VMboconic}

Here, we assume that a generalization of the Poincar\'{e}--Birkhoff--Witt
theorem for the second-order algebra $\mathcal{A}'(Y(k),AdS_d)$
does indeed hold true (for a generalization of the PBW theorem involving
a quadratic algebra, see \cite{Vermamodule}), and so we begin
to construct a Verma module based on a Cartan-like
decomposition\footnote{One can choose $sp(2k)$ in a Cartan--Weyl basis
for a unified description; however, without loss of generality, the basis
elements and structure constants of the algebra under consideration
are chosen as in Table~\ref{table}.}
extended from that of  $sp(2k)$ ($i\leq j$, $l<m$),
\begin{equation}\label{Cartandecomp}
    \mathcal{A}'(Y(k),AdS_d) =  \{l^{\prime +}_{ij},
t^{\prime+}_{lm}, l^{\prime +}_i\} \oplus \{g_0^{\prime i}, l_0'\}
\oplus \{l^{\prime }_{ij}, t'_{lm}, l^{\prime }_i\} \equiv
\mathcal{E}^-_k\oplus H_k \oplus\mathcal{E}^+_k.
\end{equation}
As distinct from the case of a Lie algebra, the element $l'_0$
does not diagonalize the elements of the upper $\mathcal{E}^-_k$
(lower, $\mathcal{E}^+_k$), triangular subalgebra, due to the
quadratic relations (\ref{l'0li+'}), as in the case of totally
symmetric HS fields in AdS spaces \cite{BKL, BurdikNavratilPasnev}.
Besides, the negative-root vectors $l^{\prime +}_i, l^{\prime
+}_j$ do not commute.

Since the Verma module over a semi-simple finite-dimensional Lie algebra
$g$  (an induced module $\mathcal{U}(g)\bigotimes_{\mathcal{U}(b)} |0\rangle_V $
with a vacuum
vector\footnote{Here, the symbols $\mathcal{U}(g)$, $\mathcal{U}(b)$,
$\mathcal{U}(g^{-})$ denote the universal enveloping algebras, respectively,
for $g$, for the Borel subalgebra, and for the lower-triangular subalgebra $g^-$,
such as $\mathcal{E}^-_k$ in (\ref{Cartandecomp}).}
$|0\rangle_V$) is isomorphic, due to the PBW theorem, as a vector space
to a polynomial algebra $\mathcal{U}(g^{-})\bigotimes_{C} |0\rangle_V $,
it is clear that $g$ can be realized by first-order inhomogeneous differential
operators acting on these polynomials.

We examine a generalization of a \emph{Verma module for quadratic algebras}:
$g(r)$ presents a $g(r)$-deformation of a Lie algebra, $g$, in such a way that
$g(r=0) =g$. Thus, we consider a Verma module for such non-linear algebras,
assuming that the PBW theorem is valid for $g(r)$. The latter will be proved
later on, by an explicit construction of the Verma module.

Let us consider a quadratic algebra $\mathcal{A}'(Y(k),AdS_d)$ as
an $r$-deformation of the Lie algebra $\mathcal{A}'(Y(k),$
$R^{1,d-1})$, with account taken of (\ref{identalg}); see
\cite{BRbos} for details,
\begin{equation}\label{identalgaux}
\mathcal{A}'(Y(k),AdS_d) = \left(T^{\prime k} \oplus
T^{\prime k*}\oplus l'_0\right)(r) + \hspace{-1em} \supset
  sp(2k),\quad T^{\prime k}
= \bigl\{ l'_i\bigr\},\  T^{\prime k*} = \bigl\{ l^{\prime
+}_i\bigr\}.
 \end{equation}
Therefore, according to (\ref{Cartandecomp}) the basis vector of
the Verma module $|\vec{N}\rangle_V$ has the form
$|\vec{N}\rangle_V = \left|{\vec{n}}_{ij},({\vec{n}},
{\vec{p}}_{m})_l \rangle_V \right.$ $=$ $\left|
{\vec{n}}_{ij},{n}_{1}, {p}_{12},\ldots, {p}_{1k},n_2,
{p}_{23},\ldots, {p}_{2k},\ldots,p_{k-1k}, n_k \rangle_V \right.$,
\begin{equation}\label{VM}
   |\vec{N}\rangle_V \equiv  \prod_{i\leq j}^k\left(l^{\prime
+}_{ij}\right){}^{n_{ij}}\prod_{l}^k\Biggl[\prod_{m, m>l}
\left(\frac{l^{\prime +}_l}{m_l}\right){}^{ n_l}\left(t^{\prime
+}_{lm}\right){}^{p_{lm}}\Biggr] |0\rangle_V, \
\mathcal{E}^+_k|0\rangle_V=0,
\end{equation}
with the (vacuum) highest-weight vector $|0\rangle_V$, non-negative
integers $n_{ij}, n_l, p_{lm}$, and arbitrary constants $m_l$
with the dimensionality of mass.

Now, we are in a position to construct a Verma module for the algebra
$\mathcal{A}'(Y(2), AdS_d)$, thus leaving the general case of an arbitrary
Young tableaux with $k>2$ rows out of the scope of the paper.
The solution of a similar problem for the auxiliary polynomial algebra
$\mathcal{A}'$ is given by Appendix~\ref{poltoy}.

\subsection{Verma module for quadratic algebra $\mathcal{A}'(Y(2),AdS_d)$
with $k=2$ rows}\label{VMY2b}

Let us first specify a commutator multiplication law for the algebra
$\mathcal{A}'(Y(2), AdS_d)$, belonging to $\mathcal{A}'(Y(k), AdS_d)$.
To do so, we can choose $\theta^{ij}= \delta^{i2}\delta^{j1}$, and therefore
the only surviving operators are $t^{ij} = t^{12}$, $t^+_{ij} =  t^{\prime +}_{12}$.

As a result, the Lie part of Tables~\ref{table},~\ref{table'} for $k=2$ is the same as
in the case of a bosonic Lie subalgebra in \cite{flatfermmix}, once the following
expressions for the non-vanishing operators  $B^{\prime 12}{}_{12}$,
$A^{\prime 12{}12}$, $F^{\prime 12,j}$, $F^{\prime 12,j+}$,  $L^{\prime i_1j_1,i_2j_2}$,
\begin{eqnarray}
&& B^{\prime 12}{}_{12} = (g_0^{1} - g_0^{2}),\quad A^{\prime
 12,12} =0, \quad  (F^{\prime 12,j},\equiv  F^{\prime j}= t^{\prime
 12}(
 \delta^{j2}-\delta^{j1}),  \label{B1212} \\
&& L^{\prime i_1j_1,i_2j_2} =
 \textstyle\frac{1}{4}\Bigl\{\delta^{i_2i_1}
 \delta^{j_2j_1}\Bigl[2g_0^{i_2}\delta^{i_2j_2} + g_0^{i_2} +
g_0^{j_2}\Bigr]  - \delta^{j_2\{i_1}\Bigl[t^{12}
 \delta^{j_1\}1}\delta^{i_22}
 +t^{12+}\delta^{j_1\}2}\delta^{i_21}\Bigr] \label{Lklij2} \\
&& \hspace{5em} -  \delta^{i_2\{i_1}\Bigl[t^{12}\delta^{j_22}\delta^{j_1\}1}
 +t^{12+}\delta^{j_1\}2}\delta^{j_21}\Bigr] \Bigr\}
 \,,\nonumber
\end{eqnarray}
are identical to the same operators $o_I$ from the initial algebra
$\mathcal{A}(Y(2), AdS_d)$. Since the non-linear part of
$\mathcal{A}'(Y(2), AdS_d)$ is determined by the same functions
${W}^{\prime ji}_b, {\mathcal{K}}^{\prime i}_1,{W}^{\prime ji+}_b,
{\mathcal{K}}^{\prime i+}_1 {X}^{\prime ij}_b$ as those for
arbitrary $k$, its manifest expression is implied directly by
(\ref{lilj'})--(\ref{lilj+b'}), as follows:\footnote{For the first
time \cite{0812.2329}, the treatment of quadratic ``primed'' quantities has
been presented in the case of $Y(s_1.s_2)$,
so that the following properties hold true: $\left({X}^{\prime
ii}_b\right)^+ = {X}^{\prime ii}_b$, $\left({X}^{\prime
12}_b\right)^+ = {X}^{\prime 21}_b$.}
\begin{eqnarray}
 {} {W}^{\prime ij}_b & = & 2r\epsilon^{ij} \left[(g_0^{\prime
 2}-g_0^{\prime 1})l^{\prime 12} - t^{\prime 12}l^{\prime 11}+
 t^{\prime 12+}l^{\prime 22} \right],\
  \label{lilj'2}
 \\
 {\mathcal{K}}^{\prime j}_1& = &\Bigl(4\sum_{i}l^{\prime ji+}
 l^{\prime i} +l^{\prime j+}(2g_0^{\prime j}-1) -2l^{\prime
 2+}t^{\prime +12}\delta^{j1} -2l^{\prime 1+}t^{\prime
 12}\delta^{j2}\Bigr)  \label{l'0li+'2} ,
\\
 {} {X}^{\prime ij}_b
  & = &
  \bigl\{l'_{0}- r\bigl( K^{\prime 0i}_0  +\mathcal{K}^{ \prime 12}_0\bigr)\bigr\}\delta^{ij}
  +r\bigl\{ \bigl[4\sum\nolimits_{l} l^{\prime 1l+}l^{\prime l2}
  +\textstyle(g_0^{\prime
 2}+g_0^{\prime 1}
  -2)t'_{ 12}\bigr]\delta^{i2}\delta^{j1}\label{lilj+b'2}\\
 && + r\bigl\{ \bigl[4\sum\nolimits_{l} l^{\prime l2+}l^{\prime 1l}
  +t^{\prime +}_{12}(g_0^{\prime 2}+g_0^{\prime 1}-2  )\bigr]\delta^{i1}\delta^{j2}.
  \nonumber
\end{eqnarray}
with the totally antisymmetric $sp(2)$-invariant tensor
$\epsilon^{ij}$, $\epsilon^{12}=1$, and the operator
$\mathcal{K}^{12}_0 = \bigl(t^{ +}_{12}t^{12}- 4l^{ +}_{12} l^{
{12}} - g_0^{ 2} \bigr)$, deduced from a Casimir operator for the
$sp(4)$ algebra in (\ref{Casimirsb}) for $k=2$.

All the Lie part quantities of Table~\ref{table}, albeit for
$o'_I$, and those of Table~\ref{table'} are specified by
(\ref{B1212})--(\ref{lilj+b'2}) used in the HS symmetry algebra
$\mathcal{A}'(Y(2), AdS_d)$ of the additional parts $o'_I$.

The Cartan-like decomposition (\ref{Cartandecomp}) and the Verma module
basis vector (\ref{VM}) of the algebra $\mathcal{A}'(Y(k),AdS_d)$ are reduced
to those for $\mathcal{A}'(Y(2),AdS_d)$ in the form
\begin{eqnarray}\label{Cartandecomp2}
 \hspace{-1em}&&  \hspace{-1em} \mathcal{A}'(Y(2),AdS_d) =  \{l^{\prime +}_{ij},
t^{\prime+}_{12}, l^{\prime +}_i\} \oplus \{g_0^{\prime i}, l_0'\}
\oplus \{l^{\prime }_{ij}, t'_{12}, l^{\prime }_i\} \equiv
\mathcal{E}^-_2\oplus H_2 \oplus\mathcal{E}^+_2,\\
\hspace{-1em}&&\hspace{-1em} \label{VM2}
   |\vec{N}(2)\rangle_V = \left| {\vec{n}}_{ij},{n}_{1}, {p}_{12},n_2\rangle_V
\right. \equiv \prod_{i\leq j}^2\textstyle\bigl(l^{\prime
+}_{ij}\bigr){}^{n_{ij}}\left(\frac{l^{\prime
+}_1}{m_1}\right){}^{ n_1}\bigl(t^{\prime
+}_{12}\bigr){}^{p_{12}}\left(\frac{l^{\prime
+}_2}{m_2}\right){}^{ n_2} |0\rangle_V, \
\mathcal{E}^+_2|0\rangle_V=0.
\end{eqnarray}
Here, in contrast to the case of a Lie algebra with totally
symmetric HS fields in AdS spaces \cite{adsfermBKR, flatfermmix,
BurdikNavratilPasnev}, the negative-root vectors $l^{\prime +}_1,
t^{\prime+}, l^{\prime +}_2$ do not commute, which means that
$|\vec{N}(2)\rangle_V$ is not an eigenvector for the operators
$t^{\prime+}_{12}, l^{\prime  +}_2$, thereby featuring substantial
peculiarities in constructing a Verma module for
$\mathcal{A}'(Y(2),AdS_d)$.

By definition, the highest-weight vector $|0\rangle_V$ is an eigenvector
of the Cartan-like subalgebra $H_2$,
\begin{equation}\label{nullV}
   (g_0^{\prime i}, l_0')|0\rangle_{V} =
 (h^i, m_0^2) |0\rangle_{V} ,
\end{equation}
with certain real numbers $h^1, h^2, m_0$, whose values are to be
determined later on, at the end of LF construction, in order for
the Lagrangian equations of motion to reproduce the initial
AdS group irrep conditions (\ref{Eq-0b})--(\ref{Eq-3b}).

Then, we determine the action of negative-root vectors from the subspace
$\mathcal{E}^-_2$ on the basis vector $|\vec{N}(2)\rangle_V$, which
does not present the action of the raising operators in a manifest form,
and reads as
\begin{eqnarray}
 l^{\prime+}_{lm}\left|\vec{N}(2)\rangle_V \right. & = &
 \left|\vec{n}_{ij}+ \delta_{ij,lm}, \vec{n}_s
 \rangle_V \right. \,,
 \label{l'+ijb} \\
 \label{l'+lb}
 l^{\prime  +}_l\left|\vec{N}(2)\rangle_V
 \right. & = & \delta^{l2}
 \prod_{i\leq j}^2\textstyle\bigl(l^{\prime
 +}_{ij}\bigr){}^{n_{ij}} l^{\prime +}_2\left|
 \vec{0}_{ij},\vec{n}_s\rangle_V\right.  +\delta^{l1}
 m_1\left|\vec{n}_{ij}, n_1 +1,
 p_{12}, n_2\rangle_V\right. \,,\ l=1,2,
 \\
 t^{\prime+}_{12}  \left|\vec{N}(2)\rangle_V
 \right. & = & \prod_{i\leq j}^2\textstyle\bigl(l^{\prime
 +}_{ij}\bigr){}^{n_{ij}}t^{\prime
 +}_{12}|\vec{{0}}_{ij},\vec{n}_s\rangle_V -
 2n_{11}\left| {n}_{11}-1, n_{12}+1, n_{22}, \vec{n}_s\rangle_V
 \right. \label{t'+b}\\
 &&  - n_{12}\left| {n}_{11}, {n}_{12}-1, {n}_{22}+1,
 \vec{n}_s\rangle_V
 \right. .
 \nonumber
\end{eqnarray}
Here, we have used the following notation: first, $\vec{n}_s
\equiv (n_1, p_{12}, n_2)$, $\vec{0}_{ij} \equiv (0,0,0)$, in
accordance with the definition of $|\vec{N}(2)\rangle_V$; second,
$\delta_{ij,lm}=\delta_{il}\delta_{jm}$, for $i\leq j, l\leq m$,
so that the vector $\left|\vec{N} +
\delta_{ij,lm}\rangle_V\right.$ in (\ref{l'+ijb}) is subject to
the definition (\ref{VM2}), increasing the coordinate $n_{ij}$ in
the vector $|\vec{N}\rangle_V$, for $i=l, j=m$, by a unit, while
leaving intact the values of the remaining ones.

In turn, the action of Cartan-like generators on the vector $|\vec{N}(2)\rangle_V $
is given by the relations
\begin{eqnarray}
 g_0^{\prime l} \left|\vec{N}(2)\rangle_V
 \right. & = & \Bigl( 2n_{ll}+n_{12} + n_l + (-1)^l p_{12} + h^l\Bigr)
 \left|\vec{N}(2)\rangle_V
 \right.,\ l=1,2, \label{g'0ib}\\
 \label{l'0auxb}
 l'_0\left|\vec{N}(2)\rangle_V
 \right. & =& \prod\nolimits_{i\leq j}^2\textstyle\bigl(l^{\prime
+}_{ij}\bigr){}^{n_{ij}}l'_0\left| \vec{0}_{ij},
\vec{n}_s\rangle_V\right..
\end{eqnarray}
To obtain the representations (\ref{l'+ijb})--(\ref{l'0auxb}),
we have used a formula for the product of operators $A$, $B$,
\begin{eqnarray}
\label{product} &&    AB^n = \sum^{n}_{k=0} \frac{n!}{k!(n-k)!}
B^{n-k}\mathrm{ad}^k_B{}A\,,   \   \mathrm{ad}^k_B{}A=
[[...[A,\stackrel{ k{\,} {\rm times}}{
\overbrace{B\},...\},B}\}},\verb" for " n\geq 0.
\end{eqnarray}
Finally, the action of positive-root vectors from the subspace
$\mathcal{E}^+_2$ on the vector $|\vec{N}(2)\rangle_V$,
according to the rule (\ref{product}), reads as follows:
\begin{eqnarray}
\label{t'auxb}
  t^{\prime}_{12}
\left|\vec{N}(2)\rangle_V
 \right. &=&  \hspace{-0.3em}  \prod_{i\leq j}^2\textstyle\bigl(l^{\prime
+}_{ij}\bigr){}^{n_{ij}}\hspace{-0.3em}\left(\frac{l^{\prime
+}_1}{m_1}\right)\hspace{-0.2em}{}^{ n_1}\bigl(t^{\prime
+}_{12}\bigr){}^{p_{12}} t'_{12}\left| \vec{0}_{ij},
0,0,n_2\rangle_V \right. -\sum\limits_l
ln_{l2}\hspace{-0.2em}\left|{\vec{n}}_{ij}-\delta_{ij,l2}+
\delta_{ij,1l}, \vec{n}_s\rangle_V
 \right. \nonumber \\
  &&+
p_{12}(h^1-h^2-n_2-p_{12}+1)\left|\vec{n}_{ij}, n_1, p_{12}-1,
{n}_2\rangle_V
 \right.
\,,\end{eqnarray}
\vspace{-1.5ex}
 \begin{eqnarray}
\nonumber
 l^{\prime 1}\left|\vec{N}(2)\rangle_V
 \right. & =& -m_1 n_{11}\bigl|{\vec{n}}_{ij}-\delta_{ij,11},
  n_1+1, p_{12}, {n}_2\rangle_V \\
{}&{}&+\Bigl\{ \prod_{i\leq j}^2\bigl(l^{\prime
+}_{ij}\bigr){}^{n_{ij}}l^{\prime 1}- \frac{n_{12}}{2}\prod_{i\leq
j}^2\bigl(l^{\prime +}_{ij}\bigr){}^{n_{ij}-\delta_{ij,12}}
l^{\prime  +}_2 \Bigr\}\left|\vec{{0}}_{ij},
\vec{n}_s\rangle_V\right.,
\label{l'1auxb}
 \\
 l^{\prime 2}\left|\vec{N}(2)\rangle_V
 \right. & =& -m_1 \frac{n_{12}}{2}\bigl| {\vec{n}}_{ij}-\delta_{ij,12},
   n_1+1, p_{12}, {n}_2\rangle_V \nonumber \\
{}&{}& + \Bigl\{ \prod_{i\leq j}^2\bigl(l^{\prime
+}_{ij}\bigr){}^{n_{ij}}l^{\prime 2} -{n_{22}} \prod_{i\leq
j}^2\bigl(l^{\prime +}_{ij}\bigr){}^{n_{ij}-\delta_{ij,22}}l^{\prime +}_2
\Bigr\}\left|\vec{{0}}_{ij}, \vec{n}_s\rangle_V\right.,
\label{l'2auxb} \\
{}
 l^{\prime 11}
\hspace{-0.2em}\left|\vec{N}(2)\rangle_V
 \right. \hspace{-0.2em}&\hspace{-0.2em}=\hspace{-0.2em}&
 \hspace{-0.2em}  {n}_{11}({n}_{11}+{n}_{12}+n_1- p_{12} -1 + h^1)
 \left|{\vec{n}}_{ij} -\delta_{ij,11}, \vec{n}_s\rangle_V
 \right.
\nonumber \\
&&+ \frac{{n}_{12}({n}_{12}-1)}{4}
 \left| {\vec{n}}_{ij} -2\delta_{ij,12}+\delta_{ij,22},\vec{n}_s\rangle_V
 \right.\label{l'11auxb}\\
 && \hspace{-0.2em} +\Bigl\{\prod_{i\leq j}^2\bigl(l^{\prime  +}_{ij}\bigr){}^{n_{ij}}l^{\prime 11} -
 \frac{{n}_{12}}{2}\prod_{i\leq j}^2\bigl(l^{\prime  +}_{ij}\bigr){}^{n_{ij}-\delta_{ij,12}}t^{\prime +}_{12}  \Bigr\}\left|\vec{{0}}_{ij},  \vec{n}_s\rangle_V\right.,
\nonumber\\
  l^{\prime 12}  \left|\vec{N}(2)\rangle_V  \right. &=&  \frac{{n}_{12}}{4}
\Bigl({n}_{12}+ \displaystyle\sum_{l}\hspace{-0.1em}(2{n}_{ll}+n_l+h^l)-1\Bigr)\left|
 {\vec{n}}_{ij} - \delta_{ij,12}, \vec{n}_s\rangle_V \right.
 \nonumber \\
&& + \frac{1}{2} p_{12} n_{11}(h^2-h^1 + n_2+ p_{12} - 1)\left|
 {\vec{n}}_{ij}-\delta_{ij,11}, n_1, p_{12}-1, {n}_2\rangle_V \right.
 \nonumber \\
 &&  + n_{11}n_{22}\left|
 {\vec{n}}_{ij}-\delta_{ij,11} -\delta_{ij,22}+ \delta_{ij,12}, \vec{n}_s\rangle_V
 \right.
\nonumber \\
 && +\Bigl\{\prod_{i\leq j}^2\bigl(l^{\prime
 +}_{ij}\bigr){}^{n_{ij}}l^{\prime 12}-
 \frac{n_{22}}{2}\displaystyle\prod_{i\leq j}^2\bigl(l^{\prime
 +}_{ij}\bigr){}^{n_{ij}-\delta_{ij,22}}t^{\prime
 +}_{12}\Bigr\}\left|\vec{{0}}_{ij}, \vec{n}_s\rangle_V\right. \nonumber
 \\
 && -\frac{n_{11}}{2}\displaystyle\prod_{i\leq j}^2\bigl(l^{\prime
 +}_{ij}\bigr){}^{n_{ij}-\delta_{ij,11}}\textstyle\left(\frac{l^{\prime
 +}_1}{m_1}\right){}^{ n_1}\bigl(t^{\prime
 +}_{12}\bigr){}^{p_{12}}t^{\prime
 }_{12}\left|\vec{{0}}_{ij},0,0, {n}_2\rangle_V\right. ,
\label{l'12auxb} \\
 l^{\prime 22}
 \left|\vec{N}(2)\rangle_V
 \right. &=&  n_{22}(n_{12}+ n_{22} + p_{12}  + n_2  -1 + h^2)
 \left|{\vec{n}}_{ij}-\delta_{ij,22}, \vec{n}_s\rangle_V
 \right. \nonumber \\
 && + \frac{n_{12}p_{12}}{2}(p_{12}-1 + h^2 - h^1+n_2)\left|
 {\vec{n}}_{ij}-\delta_{ij,12},
 n_1, p_{12}-1, {n}_2\rangle_V
 \right.
 \nonumber \\
 && + \frac{n_{12}(n_{12}-1)}{4}\left|{\vec{n}}_{ij}+\delta_{ij,11}
 -2\delta_{ij,12},
 \vec{n}_s\rangle_V
 \right.
\nonumber \\
 && +\prod_{i\leq j}^2\bigl(l^{\prime
 +}_{ij}\bigr){}^{n_{ij}}\hspace{-0.3em}
 \textstyle\left(\frac{l^{\prime
 +}_1}{m_1}\hspace{-0.3em}\right){}^{ n_1}l^{\prime
 22}\left|\vec{0}_{ij}, 0,p_{12}, {n}_2\rangle_V\right.
 \nonumber\\
 &&- \displaystyle\frac{n_{12}}{2}\prod_{i\leq j}^2\bigl(l^{\prime
 +}_{ij}\bigr){}^{n_{ij}-\delta_{ij,12}}
 \textstyle\left(\frac{l^{\prime +}_1}{m_1}\right){}^{
 n_1}\bigl(t^{\prime +}_{12}\bigr){}^{p_{12}}t^{\prime}_{12}
 \left|\vec{0}_{ij}, 0,0, {n}_2\rangle_V\right.\,.
\label{l'22auxb}
\end{eqnarray}
It is easy to see that, in order to complete the calculation in
(\ref{l'+lb}), (\ref{t'+b}), (\ref{l'1auxb})--(\ref{l'22auxb}), we
need to find the action of the positive-root vectors $ l^{\prime
m}, l^{\prime 1m}$, $m=1,2$, the Cartan-like vector $l^{\prime
}_0$, and the negative-root vectors $l^{\prime +}_2, t^{\prime
+}_{12}$ on $|\vec{{0}}_{ij},\vec{n}_s\rangle_V$, and also
the action of the remaining operators $t'_{12},  l^{\prime 22}$
on an arbitrary $|\vec{0}_{lm},0,0,{n}_2\rangle_V$
in terms of linear combinations of certain vectors. We solve
this rather non-trivial technical problem explicitly in Appendix
\ref{addalgebra}, with an introduction of some auxiliary quantities,
which we call a \emph{primary block operator}, $\widehat{t}'_{12}$,
(\ref{t'ntot}), (\ref{Liepart}) and \emph{derived block
operators}, $\widehat{t}^{\prime +}_{12}, \widehat{l}^{\prime
+}_{2},\widehat{l}^{\prime }_{0}, \widehat{l}^{\prime }_{m},
\widehat{l}^{\prime }_{m2}, m=1,2$, (\ref{t+fin}), (\ref{ll+fin}),
(\ref{l0nn2})--(\ref{l22nn2}), (\ref{l11ns}), presenting the
result in the form of

\vspace{1ex} \noindent \textbf{\emph{Theorem 1.}} A Verma module
for the non-linear second-order algebra $\mathcal{A}'(Y(2),AdS_d)$ does
exist, is determined by the relations (\ref{l'+ijb}), (\ref{g'0ib}),
(\ref{t12fin})--(\ref{l22totnV}), and is expressed using the primary
$\widehat{t}'_{12}$ and derived block operators
$\widehat{t}^{\prime +}_{12}, \widehat{l}^{\prime +}_{2},
\widehat{l}^{\prime }_{0} , \widehat{l}^{\prime }_{m},
\widehat{l}^{\prime }_{m2}, m=1,2$, (\ref{t+fin}), (\ref{ll+fin}),
(\ref{l0nn2})--(\ref{l22nn2}), (\ref{l11ns}) in the final form
\begin{eqnarray}
\label{t12fin} t'_{12}\left|\vec{N}(2)\rangle_V\right. & =&
 p_{12}(h^1-h^2-n_2-p_{12}+1)\left|\vec{n}_{ij}, n_1, p_{12}-1, {n}_2\rangle_V
 \right. \nonumber \\
  && -\sum_l ln_{l2}\hspace{-0.2em}\left|{\vec{n}}_{ij} -
\delta_{ij,l2}+ \delta_{ij,1l}, \vec{n}_s\rangle_V
 \right.
+ \hat{t}'_{12} \left|\vec{N}(2)\rangle_V\right.,\\
{}
 {{t}^{\prime +}_{12}\left|\vec{N}(2)\rangle_V\right.} & =& -
 \sum_l (3-l) n_{1l}\left| {\vec{n}}_{ij}-\delta_{ij,1l} +\delta_{ij,l2},
 \vec{n}_s\rangle_V
 \right. +\hat{t}^{\prime +}_{12}\left|\vec{N}(2)\rangle_V\right.,
 \label{t+fin2}\\
{{l}^{\prime
 +}_l\left|\vec{N}(2)\rangle_V\right.} & = &\delta^{l1}
 m_1\left|\vec{N}(2) + \delta_{s,1}\rangle_V\right. +
 \delta^{l2}\hat{l}^{\prime +}_{2}\left|\vec{N}(2)\rangle_V\right.,
 \label{ll+fin2}\\
  {l^{\prime  }_0\left|\vec{N}(2)\rangle_V\right.}
 &=& {{l}^{\prime  }_0\left|\vec{0}_{ij},\vec{n}_s\rangle_V\right.}\vert_{[\vec{0}_{ij}\to
 \vec{n}_{ij}]},   \label{l0finnV}\\
 {l^{\prime  }_1\left|\vec{N}(2)\rangle_V\right.}
 & =& {l^{\prime
 }_1\left|\vec{0}_{ij},\vec{n}_s\rangle_V\right.}\vert_{[
 \vec{0}_{ij}\to
 \vec{n}_{ij}]}  -m_1 n_{11}\left|\vec{n}_{ij}-\delta_{ij,11},
 \vec{n}_s+\delta_{s,1}\rangle_V\right. \nonumber \\
 {}&{}&- \frac{n_{12}}{2}\widehat{l}^{\prime
 +}_2\left|\vec{n}_{ij}-\delta_{ij,12},\vec{n}_s\rangle_V\right. ,
 \label{l1finnV}
 \\
 {l^{\prime }_2\left|\vec{N}(2)\rangle_V\right.} & = & {l^{\prime  }_2\left|\vec{0}_{ij},\vec{n}  _s\rangle_V\right.}\vert_{ [\vec{0}_{ij}\to  \vec{n}_{ij}]} -m_1 \frac{n_{12}}{2}
 \left| {\vec{n}}_{ij}-\delta_{ij,12},    \vec{n}_s +\delta_{s,1}\rangle_V\right.
 \nonumber \\
 {}&{}& - {n_{22}} {\widehat{l}^{\prime +  }_2\left|\vec{n}_{ij}-\delta_{ij,22},\vec{n}_s\rangle_V\right.},  \label{l2finnV}
 \\
 {l^{\prime
 }_{11}\left|\vec{N}(2)\rangle_V\right.} &=&
 {n}_{11}({n}_{11}+{n}_{12}+n_1- p_{12} -1 + h^1)
 \left|\vec{n}_{ij} -\delta_{ij,11}, \vec{n}_s\rangle_V
 \right. \nonumber\\
 && + \frac{{n}_{12}({n}_{12}-1)}{4}
 \left| \vec{n}_{ij} -2\delta_{ij,12}+\delta_{ij,22},
\vec{n}_s\rangle_V \right.+l^{\prime}_{11}\left|\vec{0}_{ij},\vec{n}_s\rangle_V\right.\vert_{[\vec{0}_{ij}
\to \vec{n}_{ij}]}
 \nonumber \\
 &&
 - \frac{{n}_{12}}{2}\widehat{t}^{\prime
 +}_{12}\left|\vec{n}_{ij}-\delta_{ij,12},\vec{n}_s\rangle_V
 \right.,
 \label{l11nVfin}
 \\
 {l^{\prime }_{12}\left|\vec{N}(2)\rangle_V\right.}& =
 &\frac{{n}_{12}}{4} \Bigl({n}_{12}+
 \displaystyle\sum_{l}\hspace{-0.1em}(2{n}_{ll}+n_l+h^l)-1
\Bigr)\left|
 {\vec{n}}_{ij} - \delta_{ij,12},  \vec{n}_s\rangle_V
 \right.\nonumber
 \\
 && + \frac{1}{2} p_{12} n_{11}(h^2-h^1 + n_2+ p_{12} - 1)\left|
 {\vec{n}}_{ij}-\delta_{ij,11}, n_1, p_{12}-1, {n}_2\rangle_V \right.
 \nonumber \\
 &&  + n_{11}n_{22}\left|
 {\vec{n}}_{ij}-\delta_{ij,11} -\delta_{ij,22}+ \delta_{ij,12}, \vec{n}_s\rangle_V
 \right. + l^{\prime
 }_{12}\left|\vec{0}_{ij},\vec{n}_s\rangle_V\right.\vert_{[\vec{0}_{ij}
 \to
 \vec{n}_{ij}]} \nonumber\\
 && -\frac{n_{22}}{2}\widehat{t}^{\prime
 +}_{12}\left|\vec{n}_{ij}-\delta_{ij,22},\vec{n}_s\rangle_V\right.-\frac{n_{11}}{2}\widehat{t}^{\prime
 }_{12}\left|\vec{n}_{ij}-\delta_{ij,11},\vec{n}_s\rangle_V\right.,
 \label{l12nVfin}
 \\
 {l^{\prime }_{22}\left|\vec{N}(2)\rangle_V\right.} &= &
 n_{22}(n_{12} + p_{12} + n_2 + n_{22} -1 + h^2)
 \left|\vec{n}_{ij} -\delta_{ij,22}, \vec{n}_s\rangle_V
 \right. \nonumber\\
  && +  \frac{n_{12}p_{12}}{2}(p_{12}-1 + h^2 - h^1+n_2)
 \left| \vec{n}_{ij}-\delta_{ij,12},
 \vec{n}_s -\delta_{s,12}\rangle_V
 \right.
 \nonumber \\
 && +  \frac{n_{12}(n_{12}-1)}{4}\left|\vec{n}_{ij}+\delta_{ij,11}
 -2\delta_{ij,12}, \vec{n}_s\rangle_V\right.+\widehat{l}^{\prime
 }_{22}\left|\vec{N}(2)\rangle_V\right. \nonumber\\
 &&
 -\frac{n_{12}}{2}\widehat{t}'_{12}\left|{\vec{n}}_{ij}-\delta_{ij,12},
 \vec{n}_s\rangle_V\right. \label{l22totnV}.
\end{eqnarray}
To obtain these relations, we have applied the obvious rule
\begin{equation}\label{convnsN2}
\prod_{i\leq j}^2\bigl(l^{\prime +}_{ij}\bigr){}^{n_{ij}} (l'_0,
l'_n, l'_{lm})\left|\vec{0}_{ij},\vec{n}_s\rangle_V\right. \equiv
(l'_0, l'_n, l'_{lm})\left|\vec{0}_{ij},\vec{ n}_s
\rangle_V\right.\vert_{[\vec{0}_{ij}\to \vec{n}_{ij}]},\quad
l,m,n=1,2, \ l\leq m,
\end{equation}
when the multipliers $\bigl(l^{\prime +}_{ij}\bigr){}^{n_{ij}}$
act as the only raising operators for the vectors
$(l'_0, l'_k,l'_{km})\left|\vec{0}_{ij},\vec{n}_s\rangle_V\right.$.

The above result has obvious corollaries: first of all, in the case of
$\mathcal{A}'(Y(2), AdS_d)$ reduced to the quadratic
algebra $\mathcal{A}'(Y(1), AdS_d)$, given in \cite{BurdikNavratilPasnev} 
for the vanishing components $n_{l2}=n_2=p_{12}=0$,
$l=1,2$, of an arbitrary VM vector
$\left|\vec{N}(1)\rangle_V\right. \equiv \left|n_{11},
0,0,n_1,0,0\rangle_V\right.$, and, secondly, in the case of the AdS space reduced
to the Minkowski space $R^{1,d-1}$, when the non-linear
algebra $\mathcal{A}'(Y(2), AdS_d)$ at $r=0$ turns to the Lie
algebra $\left(T^{\prime 2} \oplus T^{\prime 2*}\oplus
l'_0\right)(0) + \hspace{-1em} \supset sp(4)$ for $k=2$ in
(\ref{identalgaux}). In the former case, we obtain the
Verma module \cite{BurdikNavratilPasnev}, familiar from the results
of Theorem~1, whereas in the latter case we obtain a new Verma
module (see Appendix \ref{addalgebra}), with the above Lie algebra
being different from that for $sp(4)$ in \cite{flatbos}, and also in
\cite{BRbos} for $k=2$.

\subsection{Fock space realization of $\mathcal{A}'(Y(2),AdS_d)$}\label{oscVMY2b}

In this section, we obtain, on a basis of the constructed Verma module,
a realization of $\mathcal{A}'(Y(2), AdS_d)$ as a formal power series
in the creation and annihilation operators $(B_a,B^+_a) = \bigl((b_i,
b_{ij}, d_{12}), (b_i^+, b_{ij}^+, d_{12}^+)\bigr)$ in $\mathcal{H}'$,
whose number coincides with that of second-class constraints among
$o'_I$, i.e., with $\dim(\mathcal{E}^-_2\oplus\mathcal{E}^+_2)$.
This task is solved by following the results of \cite{Liealgebra} and
the algorithms suggested in \cite{0001195}, \cite{BurLeites}, initially elaborated
for a simple Lie algebra, and then enlarged to a non-linear quadratic
algebra\footnote{In the case of a quadratic superalgebra
$\mathcal{A}^{\prime f}(Y(1), AdS_d)$, elaborated for a totally
symmetric fermionic HS field oscillator realization, this has been
established in \cite{adsfermBKR, 0905.2705}.}
$\mathcal{A}(Y(1), AdS_d)$ \cite{BurdikNavratilPasnev}. To this end, we
make use of the following mapping for an arbitrary basis vector in
the Verma module and the vector $\left|\vec{n}_{ij},
\vec{n}_s\rangle\right.$ in a new Fock space $\mathcal{H}'$:
\begin{eqnarray}\label{map}
   & \left|\vec{n}_{ij}, \vec{n}_s\rangle_V \right.
    \longleftrightarrow \left|\vec{n}_{ij}, \vec{n}_s\rangle \right.
 = \prod_{i\leq j}^2\bigl(b^+_{ij}\bigr){}^{n_{ij}}\prod_{l=1}^2\bigl(b^{+}_l\bigr){}^{
 n_{l}}\bigl(d^{+}_{12}\bigr){}^{p_{12}}|0\rangle\,, &\\
&  \mathrm{for}\ b_{ij}|0\rangle = b_l|0\rangle = d_{12} |0\rangle
= 0,
 \quad |\vec{N}(2)\rangle {}_{[\vec{N}=\vec{0}]} \equiv |0\rangle\,.
&\nonumber
\end{eqnarray}
Here, the vectors $\left|\vec{n}_{ij}, \vec{n}_s\rangle \right.$
for non-negative integers $n_{ij}, n_l, p_{12}$ are the basis vectors
of the Fock space $\mathcal{H}'$, generated by six pairs of bosonic
operators $(B,B^+)$, being the basis elements of the Heisenberg algebra
$A_{6}$ with the standard (non-vanishing) commutation relations
\begin{equation}\label{commrelations}
 [B_a,B^+_b] =
 \delta_{ab} \Longleftrightarrow [b_{ij},b^+_{lm}]
=\delta_{il,jm},\ [b_{i},b^+_{m}]=\delta_{im},\ [d_{12},d^+_{12}]=1 .
\end{equation}
To present the elements of the Verma module as a formal power
series in the generators of the algebra $A_{6}$, we  use an
additive correspondence between the special Verma module vector
$\left|A_{\vec{0}_{ij},0,0,n_2}\rangle_V\right.$ (\ref{An2}) and
the Fock space $\mathcal{H}'$ vector
$\left|\vec{0}_{ij},0,0,n_2\rangle\right.$; see (\ref{exmap}) in
Appendix~\ref{oscVM2}. Summarizing, we present our basic result
in the form of Theorem 2, given in the same Appendix~\ref{oscVM2}
as polynomials for the  trivial negative-root vectors $l^{\prime  +}_1 $,
as well as for the particle number operators $g_0^{\prime i}$ (\ref{l'+ijF}),
the remaining root vectors, and the Cartan-like vector ${l^{\prime }_0}$
from $\mathcal{A}'(Y(2), AdS_d)$, as power series in oscillators
$B_a, B^+_b$. The latter ones are given by (\ref{t'Lf})--(\ref{l22fin}).
To obtain the above  representations, we use
the \emph{primary block operator} $\widehat{t}'_{12}$
(\ref{hattosc}) and the \emph{derived block operators}
$\widehat{t}^{\prime +}_{12}, \widehat{l}^{\prime }_{0},
\widehat{l}^{\prime }_{m}, \widehat{l}^{\prime }_{m2}, m=1,2$.

The additional parts $o^{\prime}_I(B,B^+)$, as formal power
series in the oscillators $(B,B^+)$, do not obey the usual properties,
\begin{equation}
 \left(l^{\prime }_{lm}\right)^+\neq l^{\prime +}_{lm}\,,
 \qquad
\left(t^{\prime }_{12}\right)^+\neq t^{\prime +}_{12},\qquad
\left(l^{\prime }_{0}\right)^+\neq l^{\prime }_{0}\,, \qquad
\left(l^{\prime }_{m}\right)^+\neq l^{\prime +}_{m}\,,\ l\leq m,
\label{hermcong}
\end{equation}
if one uses the standard rules of hermitian conjugation for the new creation
and annihilation operators, $(B_a)^+ = B_a$. The standard hermitian conjugation
properties for $o'_I$ are restored by changing the inner product in $\mathcal{H}'$
as follows:
\begin{eqnarray}
\langle{\Phi}_1|\Phi_2\rangle_{\mathrm{new}} =
\langle{\Phi}_1|K'|\Phi_2\rangle\,, \label{newsprod}
\end{eqnarray}
for any vectors $|\Phi_1\rangle, |\Phi_2\rangle$ with a certain non-degenerate operator
$K'$. This operator is determined by the condition that all the operators of the algebra
should have the standard hermitian properties with respect to the new inner product:
\begin{equation}\label{HCproperty}
 \langle{\Phi}_1|K'E^{- \prime\alpha}|\Phi_2\rangle  =
 \langle{\Phi}_2|K'E^{\prime\alpha}|\Phi_1\rangle^* ,  \
 \langle{\Phi}_1|K'G^{\prime }|\Phi_2\rangle =
 \langle{\Phi}_2|K'G^{\prime }|\Phi_1\rangle^*,
\end{equation}
for  $(E^{\prime\alpha};E^{-\prime\alpha}) = (l^{\prime }_{lm},t^{\prime }_{12}, l^{\prime }_{m};
l^{\prime +}_{lm}, t^{\prime +}_{12}, l^{\prime +}_{m})$, $G^{\prime } = (g^{\prime i}_0,l^{\prime }_0)$.
The relations (\ref{HCproperty}) determine an operator $K'$ which is hermitian with respect
to the standard inner product $\langle\, |\, \rangle$,
\begin{eqnarray}\label{explicit K}
K'=Z^+Z, \qquad
Z=\sum_{(\vec{n}_{lm},\vec{n}_s)=(\vec{0},\vec{0})}^{\infty}
\left|\vec{N}(2)\rangle_V\right.\frac{1}{(\vec{n}_{lm})!(\vec{n}_{s})!}\langle
0|\prod_{r=1}^2b_r^{n_r}{d}_{12}^{p_{12}}\prod_{l,m \geq
l}{b}_{lm}^{n_{lm}},
\end{eqnarray}
where $(\vec{n}_{lm})! ={n}_{11}!{n}_{12}!{n}_{22}!$,
$(\vec{n}_{s})! = {n}_{1}!{n}_{2}!{p}_{12}!$,
and the normalization ${}_V\langle0|0\rangle_V = 1$ is assumed.

Theorem 2 has the same consequences as those of Theorem 1, which
concerns, first of all, a flat-space limit of the algebra
$\mathcal{A}'(Y(2), AdS_d)$, and therefore a new representation
for the Lie algebra $\left(T^{\prime 2} \oplus T^{\prime 2*}\oplus
l'_0\right)(0) + \hspace{-1em} \supset sp(4)$. Second, modulo the
oscillator pairs $b_{m2}, b^+_{m2}, b_2, b^+_{2}, d_{12},
d^+_{12}, m=1,2$, the deduced representation coincides with the
one for a totally symmetric HS field quadratic algebra
$\mathcal{A}'(Y(1), AdS_d)$ in \cite{BurdikNavratilPasnev} and in
\cite{BKL} for the Weyl ordering of quadratic combinations  of $O_I$.

The set of equations that constitute the results of Theorems 1, 2 presents a general
solution of the second problem (mentioned in Introduction) of an LF construction
for mixed-symmetry HS tensors in AdS spaces with a given mass and spin
$\mathbf{s}=(s_1,s_2)$.

\section{Construction of Lagrangian actions}\label{Lagrform}

In order to construct a general Lagrangian formulation for an HS tensor field
of fixed generalized spin $\mathbf{s}=(s_1,\ldots,s_k)$, we should explicitly
determine a composition law for the deformed algebra $\mathcal{A}_{c}(Y(k), AdS_d)$,
then we have to specify this to the case of a YT with $k=2$ rows, find a BRST
operator for the non-linear algebra $\mathcal{A}_{c}(Y(2), AdS_d)$,
and, finally, reproduce a correct gauge-invariant Lagrangian formulation
for the basic bosonic field $\Phi_{(\mu)_{s_1},(\nu)_{s_2}}$.

\subsection{Explicit form of $\mathcal{A}_c(Y(k),AdS_d)$}\label{convalgg}

As in the case of the algebra $\mathcal{A}'(Y(k),AdS_d)$
of $o'_I$, it is only the multiplication law for the quadratic part
of the initial algebra $\mathcal{A}(Y(k), AdS_d)$ that is modified,
whereas the linear part is given by the same $sp(2k)$ algebra as
for the maximal Lie subalgebra, $\mathcal{A}(Y(k), AdS_d)$ and
$\mathcal{A}'(Y(k), AdS_d)$, with the same form of commutators
$[O_a, O_I]$, $O_a \in sp(2k)$. From (\ref{auxalg2}),
(\ref{conv-alg2}) and Table~\ref{table}, the non-linear part of
the algebra  $\mathcal{A}_c(Y(k), AdS_d)$ can be recovered using
Table~\ref{tablec},
\begin{table}[t]{
\begin{center}
\begin{tabular}{||c||c|c|c||}\hline\hline
$[\,\downarrow\,,\to]$&
 $L_0$ &
$L^{ i}$ & $L^{i{}+}$  \\
\hline\hline $L_0$
    & $0$
   &
    $- r\hat{\mathcal{K}}^{ i+}_1$ & $r{\hat{\mathcal{K}}}^{ i}_1$\\
\hline $L^{j}$
   &   $ r\hat{\mathcal{K}}^{ j+}_1$
   & ${\hat{W}}^{ji}_b$  & ${\hat{X}}^{ji}_b$ \\
\hline $L^{ j+}$ &
   $-r\hat{\mathcal{K}}^{ j}_1$  &
   $-{\hat{X}}^{ij}_b$
   & $- {\hat{W}}^{ ji+}_b$  \\
   \hline\hline
\end{tabular}\end{center}
} \caption{The non-linear part of the converted algebra
$\mathcal{A}_c(Y(k), AdS_d)$.} \label{tablec}
\end{table}
Here, the functions $\hat{\mathcal{K}}^{ i}_1$, ${\hat{W}}^{ ji}_b$,
$\hat{X}^{ ij}_b$ (hence, the hermitian conjugate quantities
$\hat{\mathcal{K}}^{ i+}_1$, ${\hat{W}}^{ ji+}_b$) are given by
\begin{eqnarray}
{} {\hat{W}}^{ij}_b & = & 2r\left\{\bigl[G_0^{
j}-G_0^{i}-(g_0^{\prime j}-g_0^{\prime i})\bigr] L^{ij} -
l^{\prime ij}(G_0^{ j}-G_0^{i})- \sum\nolimits_{m}
\Bigl[(T^{ m[j}-t^{\prime m[j})\theta^{[jm}L^{ i]m}  \right.\nonumber\\
{}&&\left.   - l^{\prime [im}T^{ mj]}\theta^{j]m} +
(T^{[jm+}-t^{\prime[jm+})\theta^{m[j}L^{ i]m}- l^{\prime
[im}T^{j]m+} \theta^{mj]}\Bigr] \right\},
 \label{LiLj}\\
 \hat{\mathcal{K}}^{ j}_1& = &
4\sum\nolimits_{i=1}^k\Bigl\{\bigl(L^{ ji+} - l^{\prime
ji+}\bigl)L^{ i}-l^{\prime i}L^{ ji+}\Bigr\} +
2\bigl(L^{j+}-l^{\prime j+}\bigr)G_0^{j} -
2\textstyle\bigl(g_0^{\prime j}+\frac{1}{2}\bigr)L^{ j+}  \nonumber\\
{}&& -2\sum\nolimits_{i=1}^k\Bigl\{\bigl[(L^{ i+}- l^{\prime
i+})T^{ ij+}- t^{\prime ij+}L^{i+}\bigr]\theta^{ji} +\bigl[(L^{
i+}-l^{\prime i+})T^{ ji}- t^{\prime ji}L^{
i+}\bigr]\theta^{ij}\Bigr\} \label{L0Li+} ,
\\
 {} {\hat{X}}^{ij}_b
  & =
 &\Bigl\{L_{0}+ r\Bigl[ \hat{K}^{ 0i}_0 - 2g^{\prime i}_0 G_0^i +4 l^{\prime
ii+}L^{ii} +4 l^{\prime ii}L^{ii+}    + \sum_{l=i+1}^k
\Bigl(\hat{\mathcal{K}}^{il}_0 - t^{\prime il+}T^{ il}- t^{\prime
il}T^{ il+}
  \nonumber
   \\
{}&& +4 l^{\prime il+}L^{il} +4 l^{\prime il}L^{il+}\Bigr) +
\sum_{l=1}^{i-1}\Bigl(\hat{\mathcal{K}}^{ li}_0 -t^{\prime li+}T^{
li}-t^{\prime li}T^{ li+} +4 l^{\prime li+}L^{li}+4l^{\prime
li}L^{li+} \Bigr)\Bigr]\Bigr\}\delta^{ij}
  \nonumber
\\
  && -
r\Bigl\{ 4\sum\nolimits_{l} \Bigl[(L^{jl+}-l^{\prime jl+})L^{ li}
- l^{\prime li}L^{ jl+} \Bigr] -\sum_{l=1}^{j-1}
\Bigl[(T^{lj+}-t^{\prime lj+})T^{li}-
t^{\prime li}T^{lj+} \Bigr] \nonumber\\
{} && - \sum_{l=i+1}^{k} \Bigl[(T^{ il+}- t^{\prime il+})T^{ jl}
- t^{\prime jl}T^{ il+}\Bigr]
-\sum_{l=j+1}^{i-1} \Bigl[(T^{ li}-t^{\prime li})T^{jl} -t^{\prime jl}T^{li} \Bigr] \nonumber \\
{} &&+ \textstyle\Bigl(G_0^{
 j}+G_0^{i}-(g_0^{\prime
 j}+g_0^{\prime i})
  -j-{1}\Bigr)T^{ ji} - t^{\prime ji}\Bigl(G_0^{
 j}+G_0^{i}\Bigr) \Bigr\}\theta^{ij}\nonumber\\
&& - r\Bigl\{ 4\sum\nolimits_{l} \Bigl[(L^{jl+}-l^{\prime jl+})
L^{li} -l^{\prime li}L^{jl+}\Bigr] -\sum_{l=1}^{i-1}
\Bigl[(T^{lj+}- t^{\prime lj+})T^{li} -
t^{\prime li}T^{lj+}\Bigr] \nonumber \\
{}&&- \sum_{l=j+1}^{k} \Bigl[(T^{ il+}-t^{\prime
il+})T^{jl}-t^{\prime jl}T^{il+}\Bigr] -\sum_{l=i+1}^{j-1}
\Bigl[(T^{il+}-t^{\prime il+}) T^{lj+}- t^{\prime lj+}
T^{il+}\Bigr]
\nonumber \\
{} && + \Bigl(T^{ +}_{ij}-t^{\prime +}_{ij}\Bigr)\Bigl(G_0^{
j}+G_0^{ i} \Bigr)- \Bigl(g_0^{\prime i}+ g_0^{\prime
j}+i+1\Bigr)T^{ +}_{ij}\Bigl\}\theta^{ji}. \label{LiLj+b}
\end{eqnarray}
Here, the quantities $ \hat{K}^{ 0i}_0$, $\hat{\mathcal{K}}^{
ij}_0$, $\hat{K}^{ 1i}_0$ are the same as those of
(\ref{Casimirsb}), albeit expressed in terms of $O_I$.

Let us follow our experience in the study of the (super)algebra
$\mathcal{A}_c(Y(1), AdS_d)$ \cite{BurdikNavratilPasnev, BKL, adsfermBKR},
when, in order to find an exact BRST operator, we choose the Weyl
(symmetric) ordering for the quadratic combinations of $O_I$ in
the r.h.s. of (\ref{LiLj})--(\ref{LiLj+b}) as $O_IO_J =
\frac{1}{2}(O_IO_J+O_JO_I)+ \frac{1}{2}[O_I\,,O_J]$. As a result,
Table~\ref{tablec} for this kind of ordering must contain certain
quantities ${\hat{W}}^{ij}_{b{}W}$, $\hat{\mathcal{K}}^{
j}_{1{}W}$, ${\hat{X}}^{ij}_{bW}$ (and ${\hat{W}}^{ij+}_{b{}W}$,
$\hat{\mathcal{K}}^{ j+}_{1{}W}$), which have the form
\begin{eqnarray}
{} {\hat{W}}^{ij}_{b{}W} & = & r\left\{\bigl[\mathcal{G}_0^{
j}-\mathcal{G}_0^{i}\bigr] L^{ij} + \mathcal{L}^{ij}\bigl[G_0^{
j}-G_0^{i}\bigr]- \sum\nolimits_{m=1}^k\Bigl[
\mathcal{T}^{ m[j}\theta^{[jm}L^{ i]m} + \mathcal{L}^{ [im}
T^{ mj]}\theta^{j]m} \right.\nonumber\\
{}&&\left.  + \mathcal{T}^{[jm+}\theta^{m[j}L^{ i]m} +
\mathcal{L}^{ [im}T^{j]m+} \theta^{mj]}\Bigr] \right\},
 \label{LiLjW}\\
\hat{\mathcal{K}}^{ j}_{1{}W} & = &
2\sum\nolimits_{i=1}^k\Bigl\{\mathcal{L}^{ ji+} L^{ i}+
\mathcal{L}^{ i}L^{ ji+}\Bigr\} + \mathcal{L}^{j+}G_0^{j}+
\mathcal{G}_0^{j}L^{ j+}
-\sum\nolimits_{i=1}^k\Bigl\{\bigl[\mathcal{L}^{ i+}T^{ ij+}
 \nonumber\\
{}&& + \mathcal{T}^{ ij+}L^{i+}\bigr]\theta^{ji}
+\bigl[\mathcal{L}^{ i+}T^{ ji}+\mathcal{T}^{ ji}L^{
i+}\bigr]\theta^{ij}\Bigr\} \label{L0Li+W} ,
\\
 {} {\hat{X}}^{ij}_{bW}
  & =
 &\Bigl\{L_{0}+ r\Bigl[
 \mathcal{G}^{\prime i}_0 G_0^i -2\mathcal{L}^{ii+}L^{ii} -2\mathcal{L}^{ii}L^{ii+}
 + \frac{1}{2} \sum\nolimits_{l=i+1}^k
 \Bigl({\mathcal{T}}^{  il}T^{ il+}+{\mathcal{T}}^{  il+}T^{ il}
  \nonumber\\
 {}&& -4 \mathcal{L}^{il+}L^{il} -4 \mathcal{L}^{il}L^{il+}\Bigr) +
 \frac{1}{2}\sum\nolimits_{l=1}^{i-1}\Bigl( {\mathcal{T}}^{  li}T^{ li+}
 +{\mathcal{T}}^{  li+}T^{ li}-4 \mathcal{L}^{li+}L^{li} -4
\mathcal{L}^{li}L^{li+}\Bigr)\Bigr]\Bigr\}\delta^{ij}\nonumber
\end{eqnarray}
\begin{eqnarray}
      && -
r\Bigl\{ 2\sum\nolimits_{l} \bigl[\mathcal{L}^{jl+}L^{
li}+\mathcal{L}^{ li}L^{ jl+} \bigr] -\frac{1}{2}\sum_{l=1}^{j-1}
\bigl[\mathcal{T}^{lj+}T^{li}+\mathcal{T}^{li}T^{lj+} \bigr] -
\frac{1}{2}\sum_{l=i+1}^{k} \bigl[\mathcal{T}^{
il+}T^{ jl}\nonumber
\\
{} && \ + \mathcal{T}^{ jl}T^{ il+}\bigr]
-\frac{1}{2}\sum_{l=j+1}^{i-1} \bigl[\mathcal{T}^{ li} T^{jl} +
\mathcal{T}^{jl}T^{li} \bigr] + \frac{1}{2}\bigl(\mathcal{G}_0^{
 j}+\mathcal{G}_0^{i}\bigr)T^{ ji} + \frac{1}{2}\mathcal{T}^{ ji}\bigl(
 G_0^{
 j}+G_0^{i}\bigr) \Bigr\}\theta^{ij}\nonumber\\
&& - r\Bigl\{ 2\sum\nolimits_{l} \{\mathcal{L}^{jl+} L^{li}
+\mathcal{L}^{li}L^{jl+}\} -\frac{1}{2}\sum_{l=1}^{i-1}
\bigl[\mathcal{T}^{lj+}T^{li} +
\mathcal{T}^{ li}T^{lj+}\bigr] - \frac{1}{2}\sum_{l=j+1}^{k}
\bigl[\mathcal{T}^{ il+}T^{jl}\nonumber \\
{}&& \ + \mathcal{T}^{ jl}T^{il+}\bigr]
-\frac{1}{2}\sum_{l=i+1}^{j-1} \bigl[\mathcal{T}^{+}_{il}
T^{+}_{lj} + \mathcal{T}^{\prime +}_{lj} T^{+}_{il}\bigr]
 + \frac{1}{2}\Bigl[\mathcal{T}^{ +}_{ij}\bigl(G_0^{ j}+G_0^{ i} \bigr)+
\bigl(\mathcal{G}_0^{ i}+ \mathcal{G}_0^{ j}\bigr)T^{
+}_{ij}\Bigr]\Bigr\}\theta^{ji}, \label{LiLj+bW}
\end{eqnarray}
with the notation  $\mathcal{O}_I$ for the quantity
$\mathcal{O}_I= ({O}_I - 2o'_I)$. Note that the ordering in
(\ref{LiLjW})--(\ref{LiLj+bW}) does not contain any linear terms
(except for $L_0$ in the r.h.s. of the last relation), as compared
to (\ref{LiLj})--(\ref{LiLj+b}).

Thus, we deduce the algebra of converted operators $O_I$ underlying an HS field
subject to an arbitrary unitary irreducible AdS group representation with spin
$\mathbf{s} = (s_1,..,s_k)$ in an AdS space, so that the problem is to find a BRST
operator for $\mathcal{A}_c(Y(k), AdS_d)$.

\subsection{BRST operator for converted algebra $\mathcal{A}_c(Y(2), AdS_d)$}
\label{BRSTk2}

In this section, we turn to HS fields subject to a YT with two
rows. The non-linear algebra no longer has a closed form, due to
the operator functions $F^{(2){}K}_{IJ}({o}',{O})$ in
(\ref{conv-alg2}), and, as shown in \cite{0812.2329}, this
leads to an emergence of higher-order structure functions, due to
the quadratic algebraic relations (\ref{LiLjW})--(\ref{LiLj+bW})
and their hermitian conjugation, corresponding to the quantities
$F^{(2){}K}_{IJ}({o}',{O})$ for $i,j=1,2$. In \cite{0812.2329},
new structure functions have been found,
$F_{IJK}^{RS}(O)$, being of third order \cite{Henneaux} and
implied by a resolution of the Jacobi identities $[[{O}_I,
{O}_J],{O}_K] + cycl. perm. (I,J,K)=0$, as follows
(\ref{sumcoeff}):
\begin{eqnarray}\label{Jid}
     \Bigl\{ F^{M}_{IJ} F^{P}_{MK}  + [F^{(2){}P}_{IJ},
   {O}_K] + cycl.perm.(I,J,K)\Bigr\} =
   F_{IJK}^{RS}\Bigl({O}_R\delta^P_S       -
   \textstyle\frac{1}{2} F^{P}_{RS}\Bigr) ,
   \end{eqnarray}
   for  $F^{M}_{IJ}\equiv \bigl(f^{M}_{IJ}+
   F^{(2){}M}_{IJ}\bigr)$.
The structure functions $F_{IJK}^{RS}({o}',{O})$ are antisymmetric
with respect to a permutation of any two of the lower indices $(I,J,K)$
and the upper ones $R,S$, and do exist because of non-trivial Jacobi
identities for the $k(2k-1)$ triples $(L_i, L_j, L_0)$, $(L_i^+, L_j^+, L_0)$,
$(L_i, L_j^+, L_0)$.

The construction of a BFV--BRST operator ${Q}'$ for $\mathcal{A}_{c}(Y(2),
AdS_d)$ is studied in \cite{0812.2329} and has the general form
\begin{eqnarray}\label{genQ'}
    {Q}'  = \mathcal{C}^I\bigl[{{O}}_I  + \textstyle\frac{1}{2}
    \mathcal{C}^{J}(f^{P}_{JI}+
   F^{(2){}P}_{JI})\mathcal{P}_{P}
    +\frac{1}{12}
    \mathcal{C}^{J}\mathcal{C}^{K}
    F^{RP}_{KJI}\mathcal{P}_{R}\mathcal{P}_{P}\bigr],
\end{eqnarray}
with the $(\mathcal{C}\mathcal{P})$-ordering for the ghost coordinates $\mathcal{C}^I$ = $\{\eta_0$,
$\eta^i_G$ $\eta_i^+$, $\eta_i$, $\eta_{ij}^+$, $\eta_{ij}$, $\vartheta_{12}$, $\vartheta^+_{12}\}$,
and their conjugate
momenta\footnote{The momenta satisfy the independent non-vanishing anticommutation
relations $\{\vartheta_{12},\lambda^+_{12}\}= 1,   \{\eta_i,{\cal{}P}_j^+\}= \delta_{ij}\,, \
 \{\eta_{lm},{\cal{}P}_{ij}^+\}= \delta_{li}\delta_{jm}\,,
 \{\eta_0,{\cal{}P}_0\}= \imath,\{\eta^i_{\mathcal{G}}, {\cal{}P}^j_{\mathcal{G}}\}
 = \imath\delta^{ij}$,  as well as possess the standard ghost number distribution
$gh(\mathcal{C}^I)$ = $ - gh(\mathcal{P}_I)$ = $1$, providing
the property  $gh({Q}')$ = $1$, and have the hermitian conjugation
properties of zero-mode pairs, $\left( \eta_0, \eta^i_{{G}},  {\cal{}P}_0,
{\cal{}P}^i_{G} \right)^+  =  \left( \eta_0, \eta^i_{G},  -
{\cal{}P}_0, -{\cal{}P}^i_{G}\right)$.}
$\mathcal{P}_I$ = $\{{\cal{}P}_0$, ${\cal{}P}^i_G$, ${\cal{}P}_i$, ${\cal{}P}_i^+$,
${\cal{}P}_{ij}$, ${\cal{}P}_{ij}^+$, $\lambda_{12}^+$, $\lambda_{12}\}$.
Explicitly, ${Q'}$ is given by
\begin{eqnarray}\label{explQ'}
   {Q'}\hspace{-0.5em}
& = \hspace{-0.5em}& Q'_1 + Q'_2 +
r^2\left\{\eta_0\sum\nolimits_{i,j}\eta_i\eta_j\varepsilon^{ij}\Bigl[
\textstyle \frac{1}{2}\sum_m\Bigl(G^m_0\bigl[\lambda_{12}
\mathcal{P}^{+}_{22}
 - \lambda_{12}^+
\mathcal{P}^{+}_{11}+ {i}
\mathcal{P}^{+}_{12}\sum_l(-1)^l\mathcal{P}_G^l\bigr]
   \right. \nonumber\\
&& \textstyle -
  {i}({L}^{+}_{11}\lambda_{12}^+-{L}^{+}_{22}\lambda_{12})
  \mathcal{P}_G^m + 4L^{mm}\mathcal{P}^{+}_{m2}
\mathcal{P}^{+}_{1m}\Bigr)
 - {L}^{+}_{12}\mathcal{P}_G^1\mathcal{P}_G^2  +2
L^{12}\mathcal{P}^{+}_{22}\mathcal{P}^{+}_{11}\Bigr]
\nonumber\\
&& \textstyle + \eta_0\sum_{i,j}\eta^+_i\eta_j\Bigl[
\sum_{m}\Bigl((-1)^m\frac{i}{2}G^m_0 \sum_l\mathcal{P}_G^l +
2(L^{+}_{22}\mathcal{P}^{22}-L^{11}\mathcal{P}^{+}_{11})\Bigr)
\lambda_{12}\delta^{1j} \delta^{2i} \nonumber
\end{eqnarray}
\begin{eqnarray}
&& +
\varepsilon^{\{1j}\delta^{2\}i}\Bigl(\imath\textstyle\sum_m\Bigl[
\frac{1}{2}T^{12}\lambda_{12}^+ - 2 L^{12}\mathcal{P}^{+}_{12}
(-1)^m\Bigr]\mathcal{P}_G^m  +2\Bigl[
L^{12}\lambda_{12}-T^{12}\mathcal{P}^{12}\Bigr]\mathcal{P}^{+}_{22}
\nonumber\\
&&  +
2\Bigl[T^+_{12}\mathcal{P}^{12}-L^{12}\lambda_{12}^+\Bigr]\mathcal{P}^{+}_{11}
\Bigr) -T^{12}\Bigl[\mathcal{P}_G^1\mathcal{P}_G^2\delta^{2i}
\delta^{1j}+2\mathcal{P}^{11}\mathcal{P}^{+}_{22}\delta^{1i}\delta^{2j}\Bigr]
\nonumber\\
&& \left.\hspace{-1em} - \textstyle
2\sum_{m}(-1)^m\hspace{-0.2em}\Bigl[(G^m_0\mathcal{P}^{11}+\imath{L}^{11}\mathcal{P}_G^m)
\delta^{1i}\delta^{2j}- (G^m_0\mathcal{P}^{22} +
\imath{L}^{22}\mathcal{P}_G^m) \delta^{2i}\delta^{1j} \Bigr]
\mathcal{P}^{+}_{12}
  \Bigr] + h. c.\right\}\hspace{-0.2em},
\end{eqnarray}
for $\varepsilon^{ij}=-\varepsilon^{ji}, \varepsilon^{12}=1$, with the standard form
of the linear $Q_1'$ and quadratic $Q_2'$ terms in the ghosts $\mathcal{C}^I$; for details, see \cite{0812.2329}. The hermiticity of the  nilpotent operator
$Q'$ in the total Hilbert space $\mathcal{H}_{tot}$ = $\mathcal{H}\bigotimes
\mathcal{H}'\bigotimes \mathcal{H}_{gh}$ is defined by the rule
 \begin{equation}\label{tK}
  Q^{\prime +}K = K Q', \mathrm{for} K = \hat{1} \otimes K' \otimes \hat{1}_{gh},
\end{equation}
with the operator $K'$ given by (\ref{explicit K}), which provides
the hermiticity of $o'_I$ in $\mathcal{H}'$.

\subsection{Lagrangian formulation}\label{prLagrform}

The construction of Lagrangians for bosonic HS fields in AdS${}_d$ spaces
can be achieved by partially following the algorithm of \cite{BKL, symint-adsmassless}
(see also \cite{BRbos}), which is a particular case of our construction,
corresponding to $s_2 = 0$. As a first step, we extract the dependence of the BRST
operator $Q'$ (\ref{explQ'}) on the ghosts $\eta^i_{G}, {\cal{}P}^i_{G}$,
\begin{eqnarray}
\label{Q'} {Q}' \hspace{-0.4em} &=& Q + \eta^i_{G}(\sigma^i+h^i)+
\mathcal{B}^i \mathcal{P}^i_{G},
\end{eqnarray}
with certain (non-essential later on) operators $\mathcal{B}^i$
and a BRST
operator\footnote{Here, in the coefficients depending on $o'_I,
O_I$, we use the convention for $\mathcal{O}_I$ adopted in
(\ref{LiLjW})--(\ref{LiLj+bW}).}
$Q$ which corresponds only to the converted first-class constraints, $\{O_I\} \setminus \{G^i_0\}$,
\begin{eqnarray}
{Q}  &=& \textstyle
 \frac{1}{2}\eta_0L_0+\sum_{i}\eta_i^+L_i
+\sum_{l\leq m}\eta_{lm}^+L_{lm} + \vartheta^+_{12}T_{12}
 + \frac{\imath}{2}\sum_l\eta_l^+\eta_l{\cal{}P}_0
\nonumber \\
&& - \vartheta_{12}^+\sum\nolimits_{n}(1+\delta_{1n})\eta^+_{1n}
\mathcal{P}_{n2}+ \vartheta_{12}^+
\sum\nolimits_{n}(1+\delta_{n2})\eta_{n2} \mathcal{P}^+_{1n}+
\textstyle\frac{1}{2}\sum_{n }\eta^+_{n2}\eta_{1n}\lambda_{12}
\nonumber\\
&&
 - \textstyle\frac{1}{2}\sum\limits_{l\leq
m}(1+\delta_{lm})\eta_m\eta_{lm}^+\mathcal{P}_l  -
\bigl[\vartheta_{12} \eta^+_{2}\mathcal{P}_1 +\vartheta^+_{12} \eta^+_{1}\mathcal{P}_2 \bigl]
\nonumber\\
&& + r\left\{
 \eta_0 \sum\nolimits_{i}\eta^+_i\Bigl[
 2\mathcal{L}_{ ii}\mathcal{P}_i^+ + 2\mathcal{L}^+_{ i}\mathcal{P}_{ii}
  + \mathcal{G}^{i}_0  \mathcal{P}_{ i }+2
  \bigl(\mathcal{\mathcal{L}}_{12 }
  \mathcal{P}^+_{\{1} +
\mathcal{L}^+_{ \{1}\mathcal{P}_{12 }\bigr)\delta_{2\}i }
 \right.\nonumber \\
&&\left. \quad -\delta^{1i}\bigl(\mathcal{L}_{2}\lambda^+_{12}
 + \mathcal{T}^+_{12}\mathcal{P}_{2}\bigr) - \delta^{2i}
\bigl(\mathcal{L}_{ 1}\lambda_{12} +
\mathcal{T}_{12}\mathcal{P}_{1}\bigr)\Bigr]\right.\nonumber
\\
&& \left. - \textstyle\frac{1}{2}\sum_{i,j}\eta^+_i \eta^+_{j}
\varepsilon^{ij}\Bigl[\sum_m(-1)^m\mathcal{G}_0^m \mathcal{P}_{12}
- \bigl( \mathcal{T}_{12} \mathcal{P}_{11} +
\mathcal{L}_{11}\lambda_{12}\bigr)  + \mathcal{T}^+_{12}
\mathcal{P}_{22} + \mathcal{L}_{22}\lambda^+_{12} \Bigr]\right. \nonumber \\
&  & \left. +\textstyle 2\eta^+_i \eta_{j}\Bigl[\sum_{m}
\mathcal{L}^+_{jm}\mathcal{P}_{im} -\frac{1}{8}\bigl(
\mathcal{T}^+_{12} \lambda_{12} +\mathcal{T}_{12}\lambda^+_{12}
\bigr)\delta^{ij} + \frac{1}{4}\sum_{m}\mathcal{G}_0^m
\lambda_{12} \delta^{j1}\delta^{i2} \Bigr] \right\}\nonumber\\
&&
 +
r^2\hspace{-0.3em}\left\{\eta_0\sum\nolimits_{i,j}\hspace{-0.3em}\eta_i\eta_j\varepsilon^{ij}\Bigl[
\textstyle \frac{1}{2}\sum_m\Bigl(G^m_0\bigl[\lambda_{12}
\mathcal{P}^{+}_{22}
 - \lambda_{12}^+
\mathcal{P}^{+}_{11}\bigr]
    + 4L_{mm}\mathcal{P}^{+}_{m2}
\mathcal{P}^{+}_{1m}\Bigr)  +2
L_{12}\mathcal{P}^{+}_{22}\mathcal{P}^{+}_{11}\Bigr]
 \right. \nonumber\\
&& \textstyle \left. + \eta_0\sum_{i,j}\eta^+_i\eta_j\Bigl[
2\sum_{m}\bigl(
L^{+}_{22}\mathcal{P}_{22}-L_{11}\mathcal{P}^{+}_{11})\bigr)
\lambda_{12}\delta^{1j} \delta^{2i}
-2T_{12}\mathcal{P}_{11}\mathcal{P}^{+}_{22}\delta^{1i}\delta^{2j}\right.
 \nonumber\\
&& \left.\hspace{1em}+ 2\varepsilon^{\{1j}\delta^{2\}i}\Bigl(\bigl(
L_{12}\lambda_{12}-T_{12}\mathcal{P}_{12}\bigr)\mathcal{P}^{+}_{22}
  + \bigl(T^+_{12}\mathcal{P}_{12}-L_{12}\lambda_{12}^+\bigr)
\mathcal{P}^{+}_{11} \Bigr)\right.
\nonumber\\
&& \left.\hspace{1em} - \textstyle
2\sum_{m}(-1)^m\hspace{-0.2em}\Bigl(G^m_0\mathcal{P}_{11}\delta^{1i}\delta^{2j}-
G^m_0\mathcal{P}_{22} \delta^{2i}\delta^{1j} \Bigr)
\mathcal{P}^{+}_{12}
\Bigr] \right\}+ h. c.\,.
\label{Q}
\end{eqnarray}
The generalized spin operator $\vec{\sigma}_2 =(\sigma^1,\sigma^2)$,
extended by ghost Wick-pair variables,
\begin{eqnarray}
\label{sigmai}
  \sigma^i = G_0^i - h^i   - \eta_i \mathcal{P}^+_i +
   \eta_i^+ \mathcal{P}_i + \sum_{
m}(1+\delta_{im})(
\eta_{im}^+{\cal{}P}_{im}-\eta_{im}{\cal{}P}^+_{im}) +
(\vartheta^+_{12} \lambda_{12} -
\vartheta_{12}\lambda^+_{12})(-1)^i\,,
\end{eqnarray}
supercommutes with  $Q$. The nilpotency $Q'{}^2=0$ entails the system of equations
 \begin{align}\label{geneq}
   & Q^2 =  2 \imath \sum_i  \mathcal{B}^i \sigma^i ,\ &&  [Q,\, \sigma^i\} =0; \\
   & Q^+K =  KQ ,  \ && (\sigma^i)^+K =  K\sigma^i.  \label{geneq1}
\end{align}
We choose a representation for the Hilbert space $\mathcal{H}_{tot}$
to be coordinated with the decomposition (\ref{Q'}), so that the operators
$(\eta_i,\eta_{ij}, \vartheta_{12}, \mathcal{P}_0, \mathcal{P}_i,
\mathcal{P}_{ij}, \lambda_{12}, \mathcal{P}^{i}_G)$ annihilate
the vacuum vector $|0\rangle$, and suppose that the field vectors
$|\chi \rangle$, as well as the gauge parameters $|\Lambda \rangle$,
do not depend on the ghosts $\eta^{i}_G$,
\begin{eqnarray}
|\chi \rangle &=& \sum_n \prod\nolimits_{l}^2 ( b_l^+
)^{n_{l}}\prod\nolimits_{i\le j}^2( b_{ij}^+ )^{n_{ij}}( d_{12}^+
)^{p_{12}}( \eta_0^+ )^{n_{f 0}}  \prod^2_{i, j, l\le m, n\le o}(
\eta_i^+ )^{n_{f i}} ( \mathcal{P}_j^+ )^{n_{p j}} ( \eta_{lm}^+
)^{n_{f lm}} ( \mathcal{P}_{no}^+ )^{n_{pno}} \nonumber
\\
&&{}\times ( \vartheta_{12}^+)^{n_{f 12}} ( \lambda_{12}^+
)^{n_{\lambda 12}} |\Phi(a^+_i)^{n_{f 0} (n)_{f i}(n)_{p j}(n)_{f
lm} (n)_{pno}(n)_{f 12}(n)_{\lambda
12}}_{(n)_{l}(n)_{ij}p_{12}}\rangle \,. \label{chi}
\end{eqnarray}
The brackets $(n)_{f i},(n)_{p j}, (n)_{p no}$ in (\ref{chi})
imply, e.g., for $(n)_{p no}$, the set of indices $(n_{p 11},n_{p
12},n_{p 22})$. The above sum runs over $n_{l}$, $n_{ij}$,
$p_{12}$, in the range from $0$ to infinity, as well as over the
remaining $n$ from $0$ to $1$. The Hilbert space
$\mathcal{H}_{tot}$ decomposes into a direct sum of Hilbert
subspaces with definite ghost numbers: $\mathcal{H}_{tot} =
\bigoplus_{k=-6}^6 \mathcal{H}_k$. Denote by $|\chi^k\rangle \in
\mathcal{H}_{-k}$ the state (\ref{chi}) with the ghost number
$-k$, i.e., $gh(|\chi^k\rangle)=-k$. Thus, the physical state
having the ghost number zero is $|\chi^0\rangle$, and the gauge
parameters $|\Lambda \rangle$ having the ghost number $-1$ are
$|\chi^1\rangle$, and so on. For the vanishing of all the
auxiliary creation operators $B^+$ and ghost variables $\eta_0,
\eta^+_i, \mathcal{P}^+_i,...$, the vector $|\chi^0\rangle$ must
contain only the physical string-like vector $|\Phi\rangle =
|\Phi(a^+_i)^ {(0)_{f o}(0)_{f i}(0)_{p j}(0)_{f lm}
(0)_{p{}no}(0) _{f 12}(0)_{\lambda 12}}_{(0)_l
(0)_{ij}0_{12}}\rangle$,
\begin{eqnarray}\label{decomptot}
|\chi^0\rangle&=&|\Phi\rangle+  |\Phi_A\rangle ,\quad
|\Phi_A\rangle\Big|{}_{[B^+=\eta_0= \eta^+_i=
\mathcal{P}^+_i=\eta_{lm}^+ = \mathcal{P}_{no}^+ =
\vartheta_{12}^+ = \lambda_{12}^+=0]} = 0.
\end{eqnarray}
One can show, using some of the equations of motion and gauge transformations,
that the vector $|\Phi_A\rangle$ can be completely removed.

The equation for the physical state ${Q}'|\chi^0\rangle=0$ and the tower of reducible
gauge transformations $\delta|\chi\rangle$ = $Q'|\chi^1\rangle$, $\delta|\chi^1\rangle =
Q'|\chi^2\rangle$, $\ldots$, $\delta|\chi^{(s-1)}\rangle =Q'|\chi^{(s)}\rangle$ lead
to the following relations:
\begin{eqnarray}
\label{Qchi}  Q|\chi\rangle & = &0,\qquad
\qquad(\sigma^i+h^i)|\chi\rangle=0, \qquad\left(\epsilon,
{gh}\right)(|\chi\rangle)=(0,0),
\\
 \delta|\chi\rangle & = &Q|\chi^1\rangle, \qquad
(\sigma^i+h^i)|\chi^1\rangle=0, \qquad\left(\epsilon,
{gh}\right)(|\chi^1\rangle)=(1,-1), \label{QLambda}
\\
 \ldots && \qquad\ldots  \qquad \ldots \nonumber\\
\delta|\chi^{s-1}\rangle &=& Q|\chi^{s}\rangle, \qquad
(\sigma^i+h^i)|\chi^{s}\rangle=0, \qquad \left(\epsilon,
{gh}\right)(|\chi^{s}\rangle)= (s\, mod\, 2 ,-s). \label{QLambdai}
\end{eqnarray}
Here, $s=5$ is the maximal stage of reducibility for a massive bosonic HS field,
due to the subspaces $\mathcal{H}_{k} = \emptyset$, for all integer $k\leq -
7$. The middle set of equations (\ref{Qchi})--(\ref{QLambdai}) determines
the possible values of the parameters $h^i$ and the eigenvectors of the operators
$\sigma^i$. Solving the spectral problem, we obtain a set of eigenvectors,
$|\chi^0\rangle_{\vec{n}_2}$, $|\chi^1\rangle_{\vec{n}_2}$, $\ldots$,
$|\chi^{s}\rangle_{\vec{n}_2}$, $n_1 \geq n_2 \geq 0$,
and a set of eigenvalues,
\begin{eqnarray}
\label{hi} \sigma_i | \chi \rangle_{\vec{n}_2} = \Bigl(\textstyle
n^i+\frac{d-1-4i}{2} \Bigr) | \chi \rangle_{\vec{n}_2},\,\, -h^i =
n_i+\frac{d-1-4i}{2} \,\,\, \ i=1,2,\,\, n_1 \in \mathbf{Z},\,\, n_2 \in
\mathbf{N}_0\,.
\end{eqnarray}
It is easy to see that, in order to construct a Lagrangian for a
field corresponding to a certain Young tableaux (\ref{Young k2}),
the numbers $n_i$ must be equal to the numbers of boxes in the
$i$-th row of the corresponding Young tableaux, i.e., $n_i=s_i$.
Thus, the state $|\chi\rangle_{\vec{s}_2}$ contains the physical
field (\ref{PhysState}) and all of its auxiliary fields. Here, we
fix some values of $n_i=s_i$, and after the substitution $h^i \to
h^i(s_i)$ the operator $Q_{\vec{s}_2} \equiv Q_{\vert h^i \to
h^i(s_i)} $ becomes nilpotent in each subspace
$\mathcal{H}_{tot{}\vec{s}_2}$ whose vectors satisfy (\ref{Qchi})
for (\ref{hi}). Having in mind the Lagrangian equations of motion
(those corresponding to (\ref{Eq-0b})--(\ref{Eq-3b}) for $k=2$),
the sequence of reducible gauge transformations acquires the form
\begin{eqnarray}
\hspace{-1em}&& Q_{\vec{s}_2}|\chi^0\rangle_{\vec{s}_2}=0, \quad
\delta|\chi^{s} \rangle_{\vec{s}_2} = Q_{\vec{s}_2}|\chi^{s+1}
\rangle_{\vec{s}_2}, \ s = 0,...,5.
\label{LEoM}
\end{eqnarray}
By analogy with the totally symmetric bosonic HS fields \cite{BKL, symint-adsmassless},
one can show that the Lagrangian action for a fixed spin $\vec{n}_2=\vec{s}_2$ is defined
up to an overall factor, as follows:
\begin{eqnarray}
\hspace{-1em}&&  \mathcal{S}^{m}_{0|\vec{s}_2} = \int d \eta_0 \;
{}_{\vec{s}_2}\langle \chi^0 |K_{\vec{s}_2} Q_{\vec{s}_2}| \chi^0
\rangle_{\vec{s}_2},\,\, \rm{for}\,\, |\chi^0\rangle\equiv
|\chi\rangle, \label{S}
\end{eqnarray}
where the usual inner product for the creation and annihilation
operators is assumed, with the measure $d^dx\sqrt{|{g}|}$ over the
AdS space. The vector $| \chi^0 \rangle_{(s)_2}$ and the operator
$K_{\vec{s}_2}$ in (\ref{S}) are, respectively, the vector $|\chi
\rangle$ (\ref{chi}), subject to the spin distribution relations
(\ref{hi}) for the HS tensor field
$\Phi_{(\mu^1)_{s_1},(\mu^2)_{s_2}}(x)$, and the operator $K$
(\ref{tK}), where the substitution $h_i\to-(n_i+\frac{d-1-4i}{2})$
has been made. The corresponding LF for the bosonic field of spin
$\mathbf{s}$, subject to $Y(s_1,s_2)$, is a reducible gauge theory
of $L=6$ stage of reducibility at the most.

One can prove that the equations of motion (\ref{LEoM}) reproduce
only the basic conditions (\ref{Eq-0b})--(\ref{Eq-3b}) for HS fields
of a given spin $\vec{s}_2$ and mass $m$. Indeed, the corresponding
analysis  is presented in the Appendix~\ref{reductionC} and  repeats a similar proof of the paper \cite{BRbos}
for arbitrary bosonic HS fields in flat spaces (with the AdS space being
the only specific one), and allows one to gauge away all the auxiliary
fields, so that the remaining physical vector satisfies only the AdS group
irreducible conditions (\ref{Eq-0b})--(\ref{Eq-3b}). Therefore, the resulting
equations of motion, due to the representation (\ref{decomptot}), acquire
the form
\begin{equation}\label{fincompEoM}
 L_0|\Phi\rangle_{\vec{s}_2} = (l_0 + m_0^2)|\Phi\rangle_{\vec{s}_2}, \
 (l_i, l_{ij}, t_{i_1j_1})|\Phi\rangle_{\vec{s}_2} = (0,0,0), \
 i\leq j,\ i_1 < j_1.
\end{equation}
The above relations allow one, in a unique way, to determine the parameter
$m_0$ in terms of~$h^i(s_i)$,
\begin{equation}\label{m02}
 \textstyle  m_0^2  =
 m^2+r\Bigl\{\beta(\beta+1) + \frac{d(d-6)}{4} + \Bigl(h^1-\frac{1}{2}
 + 2\beta\Bigr)\Bigl(h^1-\frac{5}{2}\Bigr) + \Bigl(h^2-\frac{9}{2}\Bigr)\Bigr\},
\end{equation}
whereas the values of parameters $m_1, m_2$ remain arbitrary, and can be
used to provide the characteristic properties of a Lagrangian for a given
HS field.

The general action (\ref{S}) presents a direct recipe to obtain
a Lagrangian for any component field $\Phi_{(\mu^1)_{s_1},(\mu^2)_{s_2}}(x)$,
starting from a general vector $|\chi^0\rangle_{\vec{s}_2}$, since we only need
to calculate the vacuum expectation values of products for certain creation
and annihilation operators. Examples of applying the general action are given below in Section~\ref{exg}. We emphasize that Lagrangian formulations for massive bosonic HS
fields described by a Young tableaux with two rows subject to traceless and Young constraints in  (A)dS spaces have only been known
in a frame-like form: see \cite{Zinoviev2}.

\section{BRST approach to cubic interaction vertices}\label{cubinter}

Here, we follow our results on constructing general cubic vertices
for massless \cite{BuchbinderReshetnyak, BRcub1} and massive \cite{BRmasscub}
totally symmetric fields of integer higher spin in Minkowski spaces
by using two approaches. The first one is based on an ambient formalism
of embedding an AdS${}_d$ space in a $(d+1)$-dimensional Minkowski space
\cite{0607248_AdS_amb}; see also \cite{BBGG} and references therein.
Now, we use the second of the above approaches to a description
of AdS${}_d$ spaces: with no reference to flat space-times.

For completeness, we draw the reader's attention to the results
of Lagrangian cubic vertices construction in AdS${}_d$ spaces
\cite{JoungLopezTaronna, FranciaMonacoMkrtchyan},
using various methods, basically for reducible interacting
totally symmetric HS fields, Maxwell-like Lagrangians,
and the $d=3$ case \cite{ProkushkinVasiliev}, including
interaction with gravity for a partially massless spin-2
field \cite{JoungMkrtchyanPoghosyan}.
Also, cubic vertices deduced from
the Fradkin--Vasiliev recipe \cite{FradkinVasiliev,FradkinVasiliev1}
are shown to have a smooth flat limit for AdS vertices
\cite{BoulangerLeclercqSundell, JoungTaronna}.

To construct a cubic vertex, we consider three copies of vectors
$|\chi^{(i)}\rangle_{\vec{s}{}^i_2}$ and gauge parameters
$|\chi^{(i)1}\rangle_{\vec{s}{}^i_2}$,
$\ldots$, $|\chi^{(i)6}\rangle_{\vec{s}{}^i_2}$
from a corresponding Hilbert space $\mathcal{H}^{(i)}_{tot}$, with
the respective vacuum vectors $|0\rangle^i$ and oscillators for $i=1,2,3$.
This allows one to obtain a deformed action and a sequence of deformed
reducible gauge transformations for a massive triple
$(m)_3\equiv (m_1, m_2, m_3) $,  with accuracy up to the first order
in the parameter $g$:
\begin{eqnarray}\label{S[n]}
 && S^{(m)_3}_{[1]|(\vec{s})^3}[\chi^{(1)},\chi^{(2)}, \chi^{(3)}] \
 = \ \sum_{i=1}^{3} \mathcal{S}^{m_i}_{0|\vec{s}{}^i_2}
 + g \int \prod_{e=1}^{3} d\eta^{(e)}_0  \Big( {}_{\vec{s}{}^{e}_{2}}\langle
 \chi^{(e)} K^{(e)} \big| V^{(3)}\rangle^{(m)_3}_{(\vec{s})^3_{2}}+h.c.\Big),
\\
 && \delta_{[1]} \big| \chi^{(i)} \rangle_{\vec{s}{}^i_2}
 = Q^{(i)} \big| \chi^{(i)1} \rangle_{\vec{s}{}^i_2}
 - g \int \prod_{e=1}^{2} d\eta^{(i+e)}_0  \Big( {}_{\vec{s}{}^{i+1}_2}\langle
 \chi^{({i+1}1)}K^{(i+1)}\big|{}_{\vec{s}{}^{i+2}_2} \langle
 \chi^{({i+2})}K^{(i+1)}\big| \label{cubgtr}\\
 &&
 \ \ \phantom{\delta_{[1]} \big| \chi^{(i)} \rangle_{s_i}}
 + (i+1 \leftrightarrow i+2)\Big)
 \big|\widetilde{V}{}^{(3)}_0\rangle^{(m)_3}_{(s){}^{3}_2}, \nonumber\\
 && \delta_{[1]} \big| \chi^{(i)1} \rangle_{\vec{s}{}^i_2} = Q^{(i)}
 \big| \chi^{(i)2} \rangle_{\vec{s}{}_2^i} -g \int \prod_{e=1}^{2}
 d\eta^{(i+e)}_0  \Big( {}_{\vec{s}{}^{i+1}_2}\langle
 \chi^{(i+1)2}K^{(i+1)}\big|{}_{\vec{s}{}^{i+2}_2}
 \langle \chi^{({i+2})}K^{(i+1)}\big| \label{cubggtr0}\\
 &&
 \ \ \phantom{\delta_{[1]} \big| \chi^{(i)} \rangle_{s_i}}
 + (i+1\leftrightarrow i+2)\Big)
 \big|\widetilde{V}{}^{(3)}_1\rangle^{(m)_3}_{(\vec{s}){}^{3}_2}\,\, ,
 \nonumber\\
 && \ \ \ldots \ \ \ldots \ \ \ldots \ \ \ldots \ \ \ldots \ \
 \ldots \ \ \ldots \ \  \ldots \ \   \ldots \ \ \ldots \ \ \ldots \ \
 \ldots \ \ \ldots \ \ \ldots \ \ \ldots \ \ \ldots \ \ \nonumber\\
 && \delta_{[1]} \big| \chi^{(i)5} \rangle_{\vec{s}{}^i_2} =
 Q^{(i)} \big| \chi^{(i)6} \rangle_{\vec{s}{}_2^i} -g  \int
 \prod_{e=1}^{2} d\eta^{(i+e)}_0 \Big( {}_{\vec{s}{}^{i+1}_2}\langle
 \chi^{(i+1)6}K^{(i+1)}\big|{}_{\vec{s}{}^{i+2}_2}
 \langle \chi^{({i+2})}K^{(i+1)}\big|  \label{cubggtr}\\
 && \ \   \phantom{\delta_{[1]} \big| \chi^{(i)} \rangle_{s_i}} +
 (i+1 \leftrightarrow i+2)\Big)
 \big|\widetilde{V}{}^{(3)}_5\rangle^{(m)_3}_{(\vec{s}){}^{3}_2} \nonumber ,
\end{eqnarray}
with some unknown three-vectors
$\big| V^{(3)}\rangle^{(m)_3}_{(\vec{s}){}^{3}_2},
\,\big|\widetilde{V}{}^{(3)}_l\rangle^{(m)_3}_{(\vec{s}){}^{3}_2}$,
$l=0,1,...,5$, defined in a total Hilbert space
$\bigotimes_{i=1}^3\mathcal{H}^{(i)}_{tot}$. Here,
$\mathcal{S}^{m_i}_{0|\vec{s}{}^i_2}$ is the free action (\ref{S})
for the field $\big| \chi^{(i)} \rangle_{\vec{s}{}^i_2}$; $Q^{(i)}$
is the BRST charge (\ref{Q}) corresponding to the spin
$\vec{s}{}^{i}_2,\, i=1,2,3$; $K^{(i)}$ is the operator $K$ (\ref{tK})
related to the spin $\vec{s}{}^i_2,\, i=1,2,3$ for a massive field;
$g$ is a deformation parameter, usually called a coupling constant.
We also use the convention $[i+3\simeq i]$.

The construction of a cubic interaction involves finding some
3-vectors $\big| V^{(3)}\rangle^{(m)_3}_{(\vec{s}){}^{3}_2}$,
$\big|\widetilde{V}{}^{(3)}_l\rangle^{(m)_3}_{(\vec{s}){}^{3}_2}$.
For this purpose, we can apply the set of fields, constraints, and ghost
operators related to the spins $\vec{s}^1_2,\vec{s}^2_2, \vec{s}^3_2$,
as well as the respective conditions of gauge invariance for the deformed
action under the deformed gauge transformations, and also the preservation
of the form of gauge transformations for the fields
$\big|\chi^{(i)}\rangle_{\vec{s}^i_2}$ under the first-level gauge
transformations $\delta_{[1]}\big|\chi^{(i)1}\rangle_{\vec{s}^i_2}$,
having the form of $l$-th level gauge transformations for the gauge
parameters $\big|\chi^{(i)l}\rangle_{\vec{s}^i_2 }$ under the $(l+1)$-th
level gauge transformations $\delta_{[1]}\big|\chi^{(i)l+1}
\rangle_{\vec{s}^i_2}$ for $l=1,...,5$ at the first power in $g$:
\begin{eqnarray}
&&  g \int \prod_{e=1}^{3} d\eta^{(e)}_0   {}_{\vec{s}^j_2}\langle
\chi^{(j)1}K^{(j)} \big| {}_{\vec{s}^{j+1}_2}\langle
\chi^{(j+1)}K^{(j+1)}\big|{}_{\vec{s}^{j+2}_2}\langle
\chi^{(j+2)}K^{(j+2)}\big| \mathcal{Q}(V^3,\widetilde{V}_0^3) = 0,
\label{g1L}
\end{eqnarray}
\begin{eqnarray}
&& \hspace{-1em}   g \int \prod_{e=1}^{2} d\eta^{(e)}_0 {}_{\vec{s}^{j+1}_2}
\langle \chi^{(j+1)2}K^{(j+1)}
\big| {}_{\vec{s}^{j+2}_2}\langle \chi^{(j+2)}K^{(j+2)}\big|
\Big(\mathcal{Q}(\widetilde{V}_0^3,\widetilde{V}_1^3)
- Q^{(j+2)} |\widetilde{V}{}_1^{(3)}\rangle\Big) = 0, \label{g1L1}\\
&& \ \ \ldots \ \ \ldots \ \ \ldots \ \ \ldots \ \ \ldots \ \ \ldots \ \
\ldots \ \ \ldots \ \  \ldots \ \ \ldots \ \ \ldots \ \ \ldots \ \
\ldots \ \ \ldots \ \ \ldots \ \ \ldots \ \ \nonumber\\
&& \hspace{-1em} g \int\prod_{e=1}^{2} d\eta^{(e)}_0
{}_{\vec{s}^{j+1}_2} \langle \chi^{(j+1)6}K^{(j+1)} \big|{}_{\vec{s}^{j+2}_2}
\langle \chi^{(j+2)}K^{(j+2)}\big| \Big(\mathcal{Q}(\widetilde{V}_4^3,
\widetilde{V}_5^3) - Q^{(j+2)} |\widetilde{V}{}_5^{(3)}\rangle\Big) = 0,
\label{g1Ll1}
\end{eqnarray}
where
\begin{eqnarray}
&& \mathcal{Q}(\widetilde{V}_l^3,\widetilde{V}_{l+1}^3) = \sum_{k=1}^3
Q^{(k)}\big |\widetilde{V}{}_{l+1}^{(3)}\rangle^{(m)_3}_{(\vec{s}){}^{3}_2}
+Q^{(j)}\Big( \big|{\widetilde{V}}{}_l^{(3)}\rangle^{(m)_3}_{(\vec{s}){}^{3}_2}
-\big|\widetilde{V}{}_{l+1}^{(3)}\rangle^{(m)_3}_{(\vec{s}){}^{3}_2}\Big),
\ j=1,2,3,
\label{g1operV3}
\end{eqnarray}
for $l=-1,...,4$ and $\widetilde{V}_{-1}^3\equiv {V}^3$.

Following our results \cite{BuchbinderReshetnyak, BRcub1}, we assume coincidence
for the vertices $\big|{{V}}{}^{(3)}\rangle$
= $\big|{\widetilde{V}}{}_0^{(3)}\rangle$
= $\big|\widetilde{V}{}_e^{(3)}\rangle$, $e=1,...,5$, which provides
the validity of the operator generating equations at the first order in $g$
(the highest orders being necessary to find the quartic and higher vertices),
\begin{equation}\label{g1Lmod}
 Q^{tot}\big|{V}{}^{(3)}\rangle^{(m)_3}_{(\vec{s}){}^{3}_2} =0, \qquad
 \vec{\sigma}{}_2^{(i)}\big|{V}{}^{(3)}
 \rangle^{(m)_3}_{(\vec{s}){}^{3}_2}\ =\ 0,
\end{equation}
along with the spin conditions, as a consequence of the generalized spin equation (\ref{hi})
for each sample (with $|\chi^{(i)}\rangle_{\vec{s}^{i}_2},\, |\chi^{(i)1}
\rangle_{\vec{s}^{i}_2},...,|\chi^{(i)6}\rangle_{\vec{s}^{i}_2}$),
providing the nilpotency of a total BRST operator $Q^{tot}\equiv\sum_i Q^{(i)}$
when evaluated on the vertex, due to the equations (\ref{geneq}) and
$\{Q^{(i)},Q^{(j)}\}=0$ for $i\ne j$.

A local dependence on the space-time coordinates in the vertices
(given by Figure \ref{m0m0m0}) $\big |V^{(3)}\rangle$, $\big
|\widetilde{V}{}_e^{(3)}\rangle$ implies
\begin{equation}\label{xdep}
 \big |V^{(3)}\rangle^{(m)_3}_{(\vec{s}){}^{3}_2}
 = \int d^dx \sqrt{|g|} \prod_{i=1}^3 \delta^{(d)}\big(x -
 x_{i}\big) V^{(3)|}{}_{(\vec{s}){}^{3}_2}^{(m)_3}\prod_{j=1}^3
 \eta^{(j)}_0 |0\rangle , \ \ \ |0\rangle\equiv\otimes_{e=1}^3|0\rangle^{e},
\end{equation}
for $(\varepsilon,gh)V^{(3)|}{}_{(\vec{s}){}^{3}_2}^{(m)_3} = (0,0)$.

\vspace{15ex}
\begin{figure}[h]
{\footnotesize\begin{picture}(10,4)
\put(135.5, -30.5){\includegraphics[scale=0.35]{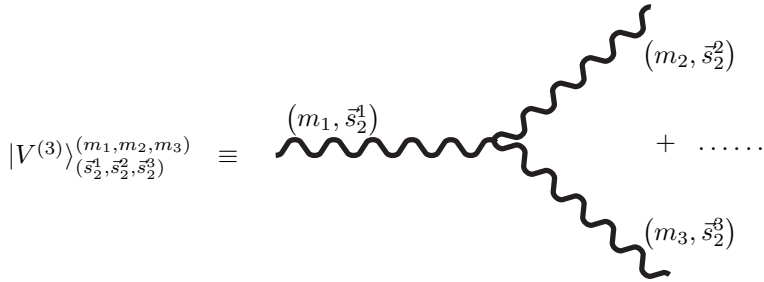}}
\put(45.5, 25.5){$ |{V}{}^{(3)}\rangle^{(m_1,m_2,m_3)}_{(\vec{s}{}^1_2, \vec{s}{}^2_2,  \vec{s}{}^3_2)}\ \ \equiv $ }
\put(150.5, 39.5){$ \big(m_1, \vec{s}{}^1_2\big)$ }
\put(285.5, 63.5){$\big(m_2, \vec{s}{}^2_2\big)$ }
\put(285.5, -3.5){$\big(m_3, \vec{s}{}^3_2\big)$}
\put(290.5, 30.5){$+ \ \ \ldots\ldots $ }
\end{picture}}
\vspace{4ex} \caption{The interaction vertex
$|{V}{}^{(3)}\rangle^{({m})_3}_{(\vec{s})^3_2}$  for  massive
fields $\Phi^{(i)}_{\mu^1(s^i_1)\mu^2(s^i_2)}$  of  masses $m_i$
and generalized spins $\vec{s}{}^i_2$  for  $i=1,2,3$. The terms
in ``$\ldots$'' correspond to the auxiliary fields of
$|\Phi^{(i)}\rangle_{\vec{s}{}^i_2}$.}\label{m0m0m0}
\end{figure}~Here, we have a conservation law $\sum_{i=1}^3p^{(i)}_\mu  = 0$
for the momenta associated with all the vertices. Once again, as in the case
of flat space-times \cite{BuchbinderReshetnyak, BRmasscub}, the deformed gauge
transformations form a closed algebra, which implies, after a simple calculation,
\begin{eqnarray}
  && \big[\delta^{\chi^{1}_1}_{[1]},\delta^{\chi^{1}_2}_{[1]}\big]  |\chi^{(i)} \rangle \  =
  \ - g \delta^{\chi^{1}_3}_{[1]}  |\chi^{(i)} \rangle
  \nonumber, \\
 && \big|\chi^{(i)1}_3 \rangle \sim  \int \prod_{e=1}^{2}
 d\eta^{(i+e)}_0  \Big(\langle\chi^{(i+1)1}_2K\big|\langle\chi^{(i+2)1}_1\big|K
 +\big(i+1\leftrightarrow i+2\big)\Big) - \big({\chi^1_1} \leftrightarrow
 {\chi^1_2}\big) \bigg\} \big|{V}{}^{(3)}\rangle ,
 \label{closuregtr}
\end{eqnarray}
with the Grassmann-odd gauge parameter $\chi^1_3$ being a function of the parameters
$\chi^1_1,\chi^1_2:$ $\chi^1_3 = \chi^1_3(\chi^1_1,\chi^1_2)$.

The equations (\ref{g1Lmod}) for identical vertices,
$V{}^{(3)}=\widetilde{V}{}_0^{(3)}=...=\widetilde{V}{}_5^{(3)}$,
along with the commutator of gauge transformations (\ref{closuregtr}),
determine the cubic interaction vertices for irreducible massive mixed-symmetric,
$Y^{(i)}(2)\equiv Y(s^{(i)}_1,s^{(i)}_2)$, higher-spin fields
in an AdS${}_d$ space. To  solve the  generating equations (\ref{xdep}),
we should find BRST-closed forms, e.g., having,
in the flat-space limit $r \to 0$, the following expressions, in a so-called
\emph{minimal derivative scheme} for different masses $m_i\ne m_j$, $i\ne j$,
\begin{eqnarray}\label{LrZ}
 &&\hspace{-1 cm} L^{(i)}_t \ = \  \widehat{p}{}^{(i)}_{\mu}a^{(i)\mu+}_t
 + \frac{m^2_{i+1}-m^2_{i+2}}{m_i}b_t^{(i)+}
 - \imath \widehat{\mathcal{P}}{}^{(i)}_{0|t}\eta_t^{(i)+}
 +\mathcal{O}(r) , \ \ \  i=1,2,3,\ t=1,2; \\
 && \hspace{- 1 cm} L^{(i{}i+1)+}_{st} \ = \ a^{(i)\mu+}_s a^{(i+1)+}_{\mu|t}
 - \frac{1}{2}\mathcal{P}^{(i)+}_s\eta_t^{(i+1)+}
 - \frac{1}{2}\mathcal{P}^{(i+1)+}_t\eta_s^{(i)+}+\frac{b_s^{(i)+}}{2m_i}L^{(i+1)}_t -\frac{b_t^{(i+1)+}}{2m_{i+1}}L^{(i)}_s
 +\mathcal{O}(r); \label{Lrr+1} \\
 && \hspace{- 1 cm} L^{(ii)+}_{st} \ = \ \frac{1}{2}a^{(i)\mu+}_s a^{(i)+}_{\mu|t}
 - (1/2)b^{(i)+}_s b^{(i)+}_t + b^{(i)+}_{st}
 + \mathcal{P}^{(i)+}_{s}\eta^{(i)+}_{t}+\mathcal{P}^{(i)+}_{t}\eta^{(i)+}_{s}
 +\mathcal{O}(r),\label{Lrr+2}
\end{eqnarray}
for $s,t=1,2$, $\widehat{p}{}^{(i)}_{\mu} = {p}{}^{(i+1)}_{\mu}
- {p}{}^{(i+2)}_{\mu}$, ${p}{}^{(i)}_{\mu} = - \imath D^{(i)}_{\mu}$
and $ \widehat{\mathcal{P}}{}^{(i)}_{0|t}= \mathcal{P}{}^{(i)}_{0|t}
- \mathcal{P}{}^{(i+2)}_{0|t}$. The above operators are such as those suggested
in the case of a Minkowski space-time for reducible mixed-symmetry integer spins
\cite{BRST-BV3}, with an incomplete BRST operator and without the vanishing
of non-dynamical algebraic (trace and mixed symmetry) constraints,
as the operator is evaluated on the vertices (see \cite{BRmasscub}
for details), being, at the same time, augmented by the trace and mixed-symmetry
operators, according to our presentation \cite{BRmasscub} of cubic vertices
for irreducible totally symmetric higher-spin fields in $\mathbb{R}^{1,d-1}$
with a complete BRST operator.

A solution for the cubic vertex (\ref{xdep}) satisfying the generating equations (\ref{g1Lmod})
requires extending the operators (\ref{LrZ})--(\ref{Lrr+2}) by means of mixed-symmetric
oscillators, and adding some new BRST-closed trace and mixed-symmetric operators, which poses
a separate problem outside the scope of the present article.

\section{Examples}\label{exg}

Here, we apply the general prescriptions of our Lagrangian formulation to the case
of free mixed-symmetry bosonic fields at the lowest values of rows and spins.

\subsection{Spin-$(s_1,0)$ totally symmetric tensor field}

Let us consider a totally symmetric field $\Phi_{\mu^1(s_1)}$
corresponding to spin $(s_1,0)$, i.e., to the  Young tableaux $Y(s_1)$. We suppose that our result should reduce to the one
examined in \cite{BKL} for totally symmetric massive bosonic fields
in an AdS space. According to our procedure, we have $n_1=s_1$, $n_2=0$.
It can be shown that in the case $n_2=0$ (\ref{chi}) all the components related
to the second row in the Young tableaux are equal to zero, namely,
\begin{eqnarray}\label{restr1}
n_{a2}=n_{b2}=n_{b22}=n_{b21}=n_b=n_{f2}=n_{p2} =n_{f22}=n_{p22}
=n_{f12}=n_{p12} =n_{fg12}=n_{pg12}=0 \,.
\end{eqnarray}
Therefore, the state vector reduces to
\begin{eqnarray}
|\chi \rangle &=& \sum_n ( a_1^{ \mu_1+} \cdots a_n^{ \mu_n+} ) ( b_1^+ )^{n_{b 1}} ( b_{11}^+ )^{n_{b 11}} ( \eta_0^+
)^{n_{f 0}} ( \eta_1^+ )^{n_{f 1}} ( \mathcal{P}_1^+ )^{n_{p 1}} (
\eta_{11}^+ )^{n_{f 11}} ( \mathcal{P}_{11}^+ )^{n_{p 11}} \nonumber
\\
&&{}\times \chi ( x )_{\mu_1 \cdots \mu_{n_{a 1}}0n_{p 1}0n_{p11}
000}^{n_{b_1}0n_{b11}000n_{f0}n_{f1}0n_{f11}000} |0\rangle\,,
\label{x1}
\end{eqnarray}
which corresponds to the one of \cite{BKL}, with the Weyl ordering in the right-hand sides
of non-linear supercommutators for the HS symmetry algebra $\mathcal{A}_c(Y(1),
AdS_d)$. Then the BRST and spin operators (supercommuting with each other) are reduced to
$\widetilde{Q}$ and $\widetilde{\sigma}$:
\begin{eqnarray}
{\widetilde{Q}}  &=& \textstyle
 \frac{1}{2}\eta_0L_0+\sum_{i}\eta_i^+L^i
+ \eta_{11}^+\big(L_{11}+ \eta_1 \mathcal{P}_1\big)+ \frac{\imath}{2}\eta_1^+\eta_1{\cal{}P}_0
\nonumber \\
&&
 + r\left\{
 \eta_0 \eta^+_1\Bigl[
 2\mathcal{L}_{ 11}\mathcal{P}_1^+ + 2\mathcal{L}_1^{ +}\mathcal{P}_{11}
  + \mathcal{G}^{1}_0  \mathcal{P}_{1}
\Bigr] +\textstyle 2\eta^+_1 \eta_{1}
\mathcal{L}^+_{11}\mathcal{P}_{11}  \right\}\nonumber+ h. c.;\\
\label{sigma1}
  \widetilde{\sigma} &= &  G_0^1 - h^1   - \eta_1 \mathcal{P}^+_1 +
 \eta_1^+ \mathcal{P}_1 + 2(
\eta_{11}^+{\cal{}P}_{11}-\eta_{11}{\cal{}P}^+_{11}) .
\label{Q1}
\end{eqnarray}

The corresponding Lagrangian formulation for a free massive field of higher integer spin
is given by the action
\begin{eqnarray}
\label{PhysStatetot} \mathcal{S}^m_{0|s_1}[\phi,\phi_1,...]=
\mathcal{S}^m_{0|s_1}[|\chi\rangle_{s_1}] = \int d\eta_0 {}_{s_1}\langle\chi|
K\widetilde{Q}|\chi\rangle_{s_1},
\end{eqnarray}
which is invariant under the reducible gauge transformations
\begin{eqnarray}
\label{gauge trasnform}
\delta|\chi\rangle_{s_1} =  \widetilde{Q}|\chi^{1}\rangle_{s_1} , \ \ \delta |\chi^1\rangle_{s_1} = \widetilde{Q}|\chi^2\rangle_{s_1}
,  \ \ \delta |\chi^2\rangle_{s_1} =0,
\end{eqnarray}
and thus determines, at most, a first-stage reducible abelian gauge theory.

A cubic vertex for three copies of massive interacting totally symmetric fields
$(m_i; s_i)$, for $i=1,2,3$, being determined as a solution of the generating equations
\begin{equation}
\label{g1Lmodts}
  \widetilde{Q}{}^{tot}
\big|{V}{}^{(3)}\rangle^{(m)_3}_{({s}){}^{3}} =0, \qquad \widetilde{\sigma}{}^{(i)}\big|{V}{}^{(3)}\rangle^{(m)_3}_{({s}){}^{3}}\ =\ 0,
\end{equation}
for $ \widetilde{Q}{}^{tot}= \sum_i\widetilde{Q}{}^{(i)}$, may have various
structures, due to various relations among the masses, and thus poses a separate
problem. For non-coincident masses, however, the vertex may have a structure
generated by the monomials (\ref{LrZ})--(\ref{Lrr+2}), for $t=s=1$.

We emphasize that for massless totally symmetric fields of integer helicities $\lambda^{(i)}$
the corresponding BRST and spin operators are simplified to the form presented
in \cite{symint-adsmassless}, so that, in addition to the above restrictions (\ref{restr1}),
there are no oscillators $b^{(i)}, b^{(i)+}$, and thus the additional parts $o'_I$ become finite polynomials.
The cubic vertices -- for three massless fields, for one massless and two massive totally symmetric HS fields, as well as for two massless and one massive totally symmetric HS fields in an AdS space-time
-- can be found according to our approach \cite{BuchbinderReshetnyak, BRmasscub, BRcub1}.

\subsection{Spin-$(1,1)$ antisymmetric tensor field}\label{antsymm}

Our next example is the simplest mixed-symmetry case, namely, a spin-$(1,1)$ rank-2 totally
antisymmetric tensor field. Below, we denote all the gauge parameters by primed characters,
and all of the gauge parameters of first level for the gauge transformations, by double-primed
characters. The values of the parameters $h_i$ are equal to $h_i=-(\frac{d+1-4i}{2})$.

\subsubsection{Action}

Let us decompose the state vector (\ref{chi}), having
a zero ghost number\footnote{The states (\ref{chi}) in the case of spin-$(1,1)$
have the lowest ghost number $-2$ due to the restriction (\ref{QLambdai}),
(\ref{hi}) at $n_1=n_2=1$.}
and obeying (\ref{hi}) at $n_1=n_2=1$, in the ghost oscillators,
as well as in the auxiliary creation operators $b_{11}^+$,
$b_{22}^+$, $b_{12}^+$, $d^+_{12}$,
\begin{eqnarray}
|\chi^0\rangle_{1,1} &=& |\widetilde{\Phi}\rangle_{1,1} +\mathcal{P}_1^+\eta_1^+|\widetilde{\phi}_3\rangle_{-1,1}
+\mathcal{P}_1^+\eta_2^+|\phi_4\rangle_{0,0}
+\mathcal{P}_2^+\eta_1^+|\phi_5\rangle_{0,0} \nonumber
\end{eqnarray}
\begin{eqnarray}
&&{} +\vartheta_{12}^{+}\mathcal{P}_1^+|A_1\rangle_{1,0}
+\vartheta_{12}^{+}\mathcal{P}_{11}^+|\phi_6\rangle_{0,0}
+\lambda_{12}^{+}\eta_1^+|A_2\rangle_{1,0}
+\lambda_{12}^{+}\eta_{11}^+|\phi_7\rangle_{0,0} \nonumber
\\
&&+\eta_0\Big( \mathcal{P}_1^+
|\widetilde{H}\rangle_{0,1}
+\mathcal{P}_2^+|A_4\rangle_{1,0} +\mathcal{P}_{11}^+|\widetilde{\phi}_8\rangle_{-1,1}
+\mathcal{P}_{12}^+|\phi_{10}\rangle_{0,0} \nonumber
\\
&&{} +\lambda_{12}^{+} |\widetilde{T}_2\rangle_{2,0}
+\lambda_{12}^{+}\mathcal{P}_1^+\eta_1^+|\phi_{11}\rangle_{0,0} \Big),
\label{x0}
\end{eqnarray}
where the states $|...\rangle$ in the right-hand sides depend on $a_{i,\mu}^+$,
 $b_i^+$, $d^+_{12}$, $b_{1i}^+$, and are expanded according to
\begin{eqnarray}
|\widetilde{\Phi}\rangle_{1,1}&=&
|\Phi\rangle_{1,1} + d^+_{12}|T_1\rangle_{2,0} +
b_{12}^+|\phi_1\rangle_{0,0} + b_{11}^+d^+_{12} |\phi_2\rangle_{0,0},\\
|\widetilde{H}\rangle_{0,1}&=&  |H\rangle_{0,1}+d^+_{12}|A_3\rangle_{1,0}, \qquad   |\widetilde{T}_2\rangle_{2,0}\ =\  |T_2\rangle_{2,0} +
b_{11}^+|\phi_9\rangle_{0,0},\\
|\Phi\rangle_{1,1} &=& \big(- a_1^{+\mu}a_2^{+\nu}  \Phi_{\mu\nu}(x)
-i  b_1^+  a_2^{+\mu}  H_{(2) \mu} (x) -i  b_2^+
a_1^{+\mu}  A_{(7) \mu} (x) +  b_1^+ b_2^+
\phi_{(19)}(x)\big)|0\rangle ,
\\
|T_{j}\rangle_{2,0} &=&  \big(- a_1^{+\mu}a_1^{+\nu}
T_{(j)\mu\nu}(x) -i  b_1^+  a_1^{+\mu}  A_{(4+j) \mu}
(x) +( b_1^+)^2 \phi_{(16+j)}(x) \big)|0\rangle ,
\\
|\widetilde{H}\rangle_{0,1} & = &  \big(-i a_2^{+\mu}  H_\mu (x) + b_2^+
 \phi_{(16)} (x)\big)|0\rangle ,
\\
|A_{j}\rangle_{1,0} & = & \big( -i a_1^{+\mu} |0\rangle A_{(j)\mu} (x) +
b_1^+  \phi_{(11+j)} (x)\big)|0\rangle,\ j=1,2,3,4,
\\
|\widetilde{\phi}_k\rangle_{-1,1}&=&d^+_{12} |{\phi}_k\rangle_{0,0}, \ k=3,8,\\
|\phi_{i}\rangle_{0,0} &=&  \phi_{(i)}(x)|0\rangle .
\end{eqnarray}

Then the relation (\ref{S}) gives the  Lagrangian action in a ghost-independent form,
\begin{eqnarray}
&&{\cal S}^m_{(1, 1)}\left[|\chi^0\rangle_{1,1}\right] \ = \ {}_{1,1}\langle\widetilde{\Phi}|  K_{1.1}\Big\{\frac{1}{2}L_0|\widetilde{\Phi}\rangle_{1,1} -L_1^+|\widetilde{H}\rangle_{0,1}-L_2^+|A_4\rangle_{1,0}-L_{11}^+|\widetilde{\phi}_8\rangle_{-1,1}-L_{12}^+|{\phi}_{10}\rangle_{0,0}
\nonumber
\\
&&\phantom{{\cal S}^m_{(1, 1)} }-T_{12}^+|\widetilde{T}_2\rangle_{2,0} +2r \Big[(l_{11}^+-l^{\prime +}_{11}) |\widetilde{\phi}_3\rangle_{-1,1}+ (l_{12}^+-l^{\prime +}_{12})( |{\phi}_4\rangle_{0,0}+|{\phi}_5\rangle_{0,0})
\nonumber
\\
&&\phantom{{\cal S}^m_{(1, 1)} }+\frac{1}{2}(l_{2}^+-l^{\prime +}_{2})|A_1\rangle_{1,0}\Big]\Big\}+{}_{-1,1}\langle\widetilde{\phi}_3|  K_{1.1}\Big\{\frac{1}{2}(L_0-2r[g_0^1-g_0^{\prime 1}] ) |\widetilde{\phi}_3\rangle_{-1,1}-L_1\widetilde{H}\rangle_{0,1}
\nonumber
\\
&&\phantom{{\cal S}^m_{(1, 1)} }+T_{12}^+|{\phi}_{11}\rangle_{0,0}- |\widetilde{\phi}_{8}\rangle_{-1,1}+r (t_{12}^+-t^{\prime +}_{12})( |{\phi}_4\rangle_{0,0}+|{\phi}_5\rangle_{0,0})\Big\} +{}_{0,0}\langle{\phi}_4|  K_{1.1}\Big\{(L_0-r\sum_i (g_0^i\nonumber
\\
&&\phantom{{\cal S}^m_{(1, 1)} }-g_0^{\prime i}) )|{\phi}_5\rangle_{0,0} - L_1 |A_4\rangle_{1,0}
- |{\phi}_{11}\rangle_{0,0} +\frac{1}{2} |{\phi}_{10}\rangle_{0,0}\Big\}   +{}_{0,0}\langle{\phi}_5|  K_{1.1}\Big\{ - L_2|\widetilde{H}\rangle_{0,1}\nonumber
 \\
&&\phantom{{\cal S}^m_{(1, 1)} }-r(l_1-l_1^{\prime})|A_1\rangle_{1,0} +\frac{1}{2} |{\phi}_{10}\rangle_{0,0} - |{\phi}_{11}\rangle_{0,0}  \Big\}+{}_{1,0}\langle{A}_1|  K_{1.1}\Big\{(L_0+r[g_0^1-g_0^{\prime 1}] ) |A_2\rangle_{1,0}+2r(l_1^+\nonumber
\end{eqnarray}
 \begin{eqnarray}
&&\phantom{{\cal S}^m_{(1, 1)} } -l_1^{\prime+})|{\phi}_{7}\rangle_{0,0}+L_1|\widetilde{T}\rangle_{2,0}+L_1^+ |{\phi}_{11}\rangle_{0,0}+\frac{r}{2}
 (t_{12}-t^{\prime}_{12})|\widetilde{H}\rangle_{0,1}-\frac{r}{2}\sum_m[g_0^m-g_0^{\prime m}] ) |A_4\rangle_{1,0} \Big\} \nonumber
\\
&&\phantom{{\cal S}^m_{(1, 1)} } +{}_{0,0}\langle{\phi}_6|  K_{1.1}\Big\{L_0 |{\phi}_{7}\rangle_{0,0}+L^+_{11}|\widetilde{T}\rangle_{2,0} - |{\phi}_{11}\rangle_{0,0}-\frac{1}{2}|{\phi}_{10}\rangle_{0,0}\Big\} \nonumber
\\
&&\phantom{{\cal S}^m_{(1, 1)} } -{}_{1,0}\langle{A}_2|  K_{1.1}T_{12}|\widetilde{H}\rangle_{0,1}  -{}_{0,0}\langle{\phi}_7|  K_{1.1}|{\phi}_{10}\rangle_{0,0}\Big\} \nonumber
\\
&&\phantom{{\cal S}^m_{(1, 1)} }-\frac{1}{2}\Big( {}_{0,1}\langle\widetilde{H}|  K_{1.1}|\widetilde{H}\rangle_{0,1}+ {}_{1,0}\langle A_4|  K_{1.1}|A_4\rangle_{1,0}\Big) +h.c.\,,
\label{S(1,1)g}
\end{eqnarray}
and a tensor form (for the leading field term),
\begin{eqnarray}
{\cal S}^m_{(1, 1)} &=&\int d^dx \sqrt{|g|}\Big\{ \Phi_{\mu \nu} (\nabla^2-\frac{d(d-6)}{4}+m_0^2) \Phi^{\mu \nu} - 4
T_{(2) \left(\mu \nu \right)} \Phi^{\mu \nu} \nonumber \\
&& + 2 A_{(4) \mu}\nabla_{\nu} \Phi^{\mu \nu} + 2 H_{\nu}
\nabla_{\mu} \Phi^{\mu \nu}
+\phi_{(10)} \Phi_{\mu}^{\mu} + \mathrm{more} \Big\}.
\label{S(1,1)}
\end{eqnarray}
Here, the mass parameter $m_0^2$ (\ref{m02}) equals to
\begin{equation}\label{mass11}
  m_0^2  =  m^2+r\Bigl\{4+\frac{d(d-10)}{2} \Bigr\}.
\end{equation}

Let us now consider the set of gauge transformations.

\subsubsection{Zero-level gauge transformations}

We decompose the vector (\ref{chi}), satisfying (\ref{hi}),
for $n_1=n_2=1$, and having the ghost number $-1$, as follows:
\begin{eqnarray}
|\chi^{1}\rangle_{1,1} &=& \mathcal{P}_1^+ \left( |H^\prime
\rangle_{0,1}+d^+_{12}|A_1^\prime \rangle_{1,0} \right)
+\mathcal{P}_2^+|A_2^\prime \rangle_{1,0} +\mathcal{P}_{11}^+d^+_{12}|\phi_1^\prime
\rangle_{0,0} +\mathcal{P}_{12}^+|\phi_{2}^\prime \rangle_{0,0} \nonumber
\\
&&{} +\lambda_{12}^{+} \left(|T_1^\prime \rangle_{2,0} +
b_{11}^+|\phi_3^\prime \rangle_{0,0} \right)
+\lambda_{12}^{+}\mathcal{P}_1^+\eta_1^+|\phi_{4}^\prime \rangle_{0,0} \nonumber
\\
&&{} +\eta_0 \Big( \mathcal{P}_1^+\mathcal{P}_2^+|\phi_5^\prime \rangle_{0,0}
+\mathcal{P}_1^+\lambda_{12}^{+}|A_3^\prime \rangle_{1,0}
+\mathcal{P}_{11}^+\lambda_{12}^{+}|\phi_6^\prime \rangle_{0,0} \Big),
\label{x-1}
\end{eqnarray}
where
\begin{eqnarray}
|T^\prime\rangle_{2,0} &=& \big(- a_1^{+\mu}a_1^{+\nu}
T^\prime_{\mu\nu}(x) -i  b_1^+  a_1^{+\mu}  A^\prime_{(4)
\mu} (x) +( b_1^+)^2 \phi^\prime_{(11)}(x)\big)|0\rangle ,
\\
|H^\prime\rangle_{0,1} & = & \big(-i a_2^{+\mu}  H^\prime_\mu
(x) + b_2^+ \phi^\prime_{(10)} (x)\big)|0\rangle ,
\\
|A^\prime_{j}\rangle_{1,0} & = &\big( -i a_1^{+\mu}
A^\prime_{(j)\mu} (x) + b_1^+ \phi^\prime_{(6+j)}(x)\big)|0\rangle
\ j=1,2,3,\\
|\phi^\prime_{i}\rangle_{0,0} &=& \phi^\prime_{(i)}(x)|0\rangle
\,.
\end{eqnarray}
Substituting (\ref{x0}), (\ref{x-1}) into the relations
(\ref{QLambda}) on the left, we obtain the following gauge
transformations for the fields:
\begin{eqnarray}
\delta \Phi_{\mu\nu} &=& - 2 {T^\prime_{\mu  \nu }}   + {\nabla_{\mu
}}  {H^\prime_{\nu }} + {\nabla_{\nu }}  {A^\prime_{{(2)} \mu }}
-  \frac{{{\eta }_{\mu  \nu }}}{2} {\phi^\prime }_{(2)}\, ,
\\
\delta T_{{(1)}\mu\nu} &=&{T^\prime_{\mu  \nu }} + {\nabla_{\{\mu
}}  {A^\prime_{(1)\nu\} }} - \frac{{{\eta }_{\mu  \nu
}}}{2}{{\phi^\prime }_{(1)}}\, ,
\\
\delta T_{{(2)}\mu\nu}&=&  \left(\nabla^2-r\frac{d(d-6)}{4}+m_0^2\right)  {T^\prime_{\mu  \nu
}}    -{\nabla_{\{\mu }}  {A^\prime_{{(3)} \nu\} }} \\
&&  + \frac{{{\eta
}_{\mu  \nu }}}{2} \left({\phi^\prime }_{(6)}+2r {\phi^\prime }_{(4)}+\frac{r}{2} {\phi^\prime }_{(5)}\right) -r{\nabla_{\{\mu }}  {A^\prime_{{(2)} \nu\} }}, \nonumber
\\
\delta A_{{(1)}\mu} &=& -2  {A^\prime_{(1)\mu }} -
{A^\prime_{{(2)} \mu }} - {H^\prime_{\mu }}\, ,
\\
\delta A_{{(2)}\mu} &=& -  2 \nabla^{\nu }  {T^\prime_{\mu  \nu
}} - \frac{1}{m_1} \big(m_0^2  - r\left[\frac{d^2}{4}-\frac{17}{4} \right]\big) {A^\prime_{(4)  \mu }} - {\nabla_{\mu }}  {{\phi^\prime
}_{(4)}} +  {A^\prime_{{(3)} \mu }}
\\
&&+\frac{r}{2}\left( {H^\prime_{\mu }}-2  {A^\prime_{(1)\mu }}+{A^\prime_{(2)\mu }}(1+2d)\right),
\nonumber \\
\delta A_{{(3)}\mu} &=&  \left(\nabla^2-r\frac{(d+4)(d-6)}{4}+r(d-\frac{3}{2})+m_0^2\right)  {A^\prime_{(1)\mu }}
+ {A^\prime_{{(3)} \mu }}+r {A^\prime_{{(2)} \mu }}\, ,
\\
\delta A_{{(4)}\mu} &=&  -{\nabla_{\mu }}  {{\phi^\prime
}_{(5)}}  + {\left(\nabla^2-r\frac{d(d-6)}{4}+m_0^2+r(d-\frac{7}{2})\right)}  {A^\prime_{{(2)} \mu }} -
{A^\prime_{{(3)} \mu }}+r({H^\prime_{\mu }} -2 {A^\prime_{(1)\mu }}),
\end{eqnarray}
\begin{eqnarray}
\delta A_{(5)\mu} &=&  m_1 {A^\prime_{(1)\mu }} + {\nabla_{\mu }}
{{\phi^\prime }_{(7)}} + {A^\prime_{(4)  \mu }}\, ,
\\
\delta A_{(6)\mu}   &=&    -m  {A^\prime_{{(3)} \mu }} -
{\nabla_{\mu }}  {{\phi^\prime }_{(9)}} + \left(\nabla^2-r\frac{d(d-6)}{4}+m_0^2\right)
{A^\prime_{(4)  \mu }}\\
&& -r ({\nabla_{\mu }}  {{\phi^\prime }_{(8)}} -  {A^\prime_{{(2)} \mu }}+(3-d){A^\prime_{{(4)} \mu }}), \nonumber
\\
\delta A_{(7)\mu}  &=&   m_2  {A^\prime_{{(2)} \mu }} +
{\nabla_{\mu }}  {{\phi^\prime }_{(10)}} - \frac{m_2}{m_1}  {A^\prime_{(4)  \mu
}}\, ,
\\
\delta H_{\mu}  &=&   {\nabla_{\mu }}  {{\phi^\prime }_{(5)}} +
{\left(\nabla^2-r\frac{d(d-6)}{4}+m_0^2+r(d-\frac{3}{2})\right)}  {H^\prime_{\mu }} - {A^\prime_{{(3)} \mu }}+r {A^\prime_{{(2)} \mu }}\, ,
\\
\delta H_{(2)\mu}  &=&  m_1  {H^\prime_{\mu }} + {\nabla_{\mu }}
{{\phi^\prime }_{(8)}} -    {A^\prime_{(4)  \mu }}\, ,
\\
\delta \phi_{{(1)}} &=&  {{\phi^\prime }_{(2)}} - 2 {{\phi^\prime
}_{(3)}}\, ,
\\
\delta \phi_{{(2)}}  &=& {{\phi^\prime }_{(1)}} + {{\phi^\prime
}_{(3)}}\, ,
\\
\delta \phi_{{(3)}}&=& -  \nabla^{\mu }  {A^\prime_{(1)\mu }} -
\frac{1}{m_1}  \big(m_0^2  - r\left[\frac{d(d+4)}{4}-\frac{29}{4} \right]\big)   {{\phi^\prime }_{(7)}}  + {{\phi^\prime }_{(1)}} +
{{\phi^\prime }_{(4)}}\, ,
\\
\delta \phi_{{(4)}}   &=& -  \nabla^{\mu }  {H^\prime_{\mu }}
+\frac{1}{m_2}  \big(m_0^2  - r\left[\frac{d(d-8}{4}+\frac{14}{4} \right]\big)  {{\phi^\prime }_{(10)}}+2r(3-d){{\phi^\prime }_{(7)}}\, ,\\
 && + \frac{{1}}{2} {\phi^\prime }_{(2)}-
{{\phi^\prime }_{(4)}} -  {{\phi^\prime }_{(5)}},  \nonumber
\\
\delta \phi_{{(5)}}  &=& -  \nabla^{\mu }  {A^\prime_{{(2)} \mu
}} -
\frac{1}{m_1}  \big(m_0^2  - r\left[\frac{d^2}{4}-5 \right]\big)   {{\phi^\prime }_{(8)}} + \frac{{1}}{2} {\phi^\prime
}_{(2)}- {{\phi^\prime }_{(4)}} + {{\phi^\prime }_{(5)}}\, ,
\\
\delta \phi_{{(6)}}  &=&   -2  {{\phi^\prime }_{(1)}} -
{{\phi^\prime }_{(2)}},\\
\delta \phi_{(7)} &=& \frac{d-3}{2} \phi^\prime_{(3)}
+T^\prime_\mu{}^\mu
+\phi^\prime_{(4)} -\frac{1}{m_1^2}\big(m_0^2  - r\left[\frac{d^2-17}{4}\right]\big)  \phi^\prime_{(11)}+\frac{1}{2}\phi^\prime_{(2)},
\\
\delta \phi_{(8)} &=&{\left(\nabla^2-r\frac{d(d-6)}{4}+m_0^2\right)}  {{\phi^\prime }_{(1)}} +
{{\phi^\prime }_{(6)}}+r {{\phi^\prime }_{(5)}} -2r \nabla^{\mu }  {A^\prime_{{(1)} \mu }}+\frac{2r^2}{m_2}(3-d){{\phi^\prime }_{(10)}} \\
\phantom{\delta \phi_{(8)}} &&-2r\left(m_0^2 - r\left[\frac{d(d+4)-29}{4}\right]\big)\right){{\phi^\prime }_{(7)}}\, , \nonumber
\\
\delta \phi_{(9)} &=&  \left(\nabla^2-r\frac{d(d-6)}{4}+m_0^2\right)  {{\phi^\prime }_{(3)}}
- {{\phi^\prime }_{(6)}},-2r  ({\phi^\prime }_{(4)}\\
&& +r  ({\phi^\prime }_{(8)}+\frac{r}{2} {\phi^\prime }_{(5)}-2(4-d){\phi^\prime }_{(11)-}),
 \nonumber\\
\delta \phi_{(10)} &=&  {\left(\nabla^2-r\frac{d(d-6)}{4}+m_0^2\right)} {{\phi^\prime }_{(2)}}
- 2 {{\phi^\prime }_{(6)}}+2r{{\phi^\prime }_{(5)}}+2r\Big\{  \nabla^{\mu }  {A^\prime_{{(2)} \mu }}-\frac{2r}{m_1}(3-d){{\phi^\prime }_{(7)}}
\\
\phantom{\delta \phi_{(8)}} &&
-\frac{1}{m_1}\left(m_0^2 - r\left[\frac{d^2-17}{4}\right]
\right){{\phi^\prime }_{(8)}}+ \nabla^{\mu }  {H^\prime_{{} \mu }}+ \left(m_0^2 - r\left[\frac{d(d+8)+7}{4}\right]\right){{\phi^\prime }_{(10)}} \Big\}\, ,  \nonumber
\\
\delta \phi_{(11)}  &=&   \nabla^{\mu }  {A^\prime_{{(3)} \mu }}
+ \frac{1}{m_1}\left(m_0^2 - r\left[\frac{d^2-17}{4}\right]
\right) {{\phi^\prime }_{(9)}} +  {\left(\nabla^2-r\frac{d(d-6)}{4}+m_0^2\right)}  {{\phi^\prime
}_{(4)}} - {{\phi^\prime }_{(6)}} \\
\phantom{\delta \phi_{(8)}} &&+ r(2d-3){{\phi^\prime }_{(4)}}+r\left(d-\frac{5}{2}\right){{\phi^\prime }_{(5)}}
+ 2r\left(T^{\prime \mu}{}_\mu+ \frac{1}{m_1^2}\left(m_0^2 - r\left[\frac{d^2-17}{4}\right]
\right) {{\phi^\prime }_{(11)}}\right.\nonumber \\
&& \left.+\frac{3-d}{2}{{\phi^\prime }_{(3)}}+ \frac{1}{2}\nabla^{\mu }  {H^\prime_{} \mu }+\frac{1}{2m_2}\left(m_0^2 - r\left[\frac{d(d-8)+7}{4}\right]\right)\phi^{\prime }_{(10)}  + \frac{r}{m_1}(3-d)\phi^{\prime }_{(7)}\right),\nonumber
\end{eqnarray}
\begin{eqnarray}
\delta \phi_{(12)}  &=&  -2 \phi^{\prime }_{(7)} -\phi^{\prime }_{(8)} -\frac{m_1}{m_2}\phi^{\prime }_{(10)}\, ,
\\
\delta \phi_{(13)} &=& - \nabla^\mu A^\prime_{(4)\mu} + m_1
(\phi^\prime_{(3)}-\phi^\prime_{(4)})  + \frac{2}{m_1} \big(m_0^2  - r\left[\frac{d(d-4)}{4}-\frac{13}{4} \right]\big) \phi^\prime_{(11)} + \phi^\prime_{(9)}
\\
&&+\frac{r}{2}\left(-\frac{m_2}{m_1} \phi^{\prime }_{(10)}+2 \phi^{\prime }_{(7)}+{\phi^\prime_{(8)}}(2d-1)\right), \nonumber
\\
\delta \phi_{(14)} &=&  {\left(\nabla^2-r\frac{d(d-10)-8}{4}+r(d-\frac{3}{2})+m_0^2\right)}  \phi^{\prime }_{(7)}
+ \phi^{\prime }_{(9)}+r \phi^{\prime }_{(7)}+2r\frac{m_1}{m_2}\phi^{\prime }_{(10)}\, ,
\\
\delta \phi_{(15)} &=&
{\left(\nabla^2-r\frac{d(d-6)}{4}+m_0^2+r(2d-\frac{9}{2})\right)}  \phi^{\prime }_{(8)} - \phi^{\prime }
_{(9)}-r\left(2\phi^{\prime }_{(7)}+\frac{m_1}{m_2}\phi^{\prime }_{(10)}\right)\, ,
\\
\delta \phi_{(16)}  &=&     m_2 {{\phi^\prime }_{(5)}} +
{\left(\nabla^2-r\frac{d(d-6)}{4}+r(d-\frac{3}{2})+m_0^2\right)}  {{\phi^\prime }_{(10)}} -\frac{m_2}{m_1}( {{\phi^\prime
}_{(9)}}\\
&& +r {{\phi^\prime }_{(8)}})+r\frac{m_2}{2m_1}{{\phi^\prime }_{(7)}}, \nonumber
\\
\delta \phi_{(17)} &=& m_1 \phi^\prime_{(7)}  +\phi^\prime_{(11)}\, ,
\\
\delta \phi_{(18)} &=& -m\phi^\prime_{(9)} +
\left(\nabla^2-r\frac{d(d-6)}{4}+m_0^2\right) \phi^\prime_{(11)} + \frac{1}{2}\phi^\prime_{(6)}\, ,
\\
\delta \phi_{(19)}&=&  m_2  {{\phi^\prime }_{(8)}} + m_1
{{\phi^\prime }_{(10)}}-2\frac{m_2}{m_1}{{\phi^\prime }_{(11)}}.
\end{eqnarray}

Let us now turn to the gauge transformation for the gauge parameters.

\subsubsection{First-level gauge transformations}

We decompose the vector (\ref{chi}), satisfying (\ref{hi}) at
$n_1=n_2=1$ and having the ghost number $-2$, according to
\begin{eqnarray}
|\chi^{2}\rangle_{1,1} &=& \mathcal{P}_1^+\mathcal{P}_2^+|\phi_1^{\prime \prime}
\rangle_{0,0} +\mathcal{P}_1^+\lambda_{12}^{+}|A^{\prime \prime} \rangle_{1,0}
+\mathcal{P}_{11}^+\lambda_{12}^{+}|\phi_2^{\prime \prime} \rangle_{0,0}\,,
\label{x-2}
\end{eqnarray}
 where
\begin{eqnarray}
|A^{\prime \prime}\rangle_{1,0}  = \big(-i a_1^{+\mu}
A^{\prime \prime}_\mu (x) + b_1^+ \phi^{\prime
\prime}_{(3)} (x)\big)|0\rangle \,, &\qquad& |\phi^{\prime
\prime}_{i}\rangle_{0,0} = \phi^{\prime \prime}_{(i)}(x)|0\rangle
\,.
\end{eqnarray}

Substituting (\ref{x-1}), (\ref{x-2}) into the relations
(\ref{QLambdai}) on the right, we find the gauge transformations
for the gauge parameters:
\begin{align}
& \delta T_{\mu\nu}^\prime =
\nabla_{\{\mu}  A^{\prime \prime}_{\nu\} }
-\frac{1}{2}\eta_{\mu\nu}(\phi^{\prime\prime}_{(2)}+r\phi^{\prime\prime}_{(1)}) , && \delta
A_{(1)\mu}^\prime  = - A^{\prime\prime}_\mu,
\\
& \delta A_{(2)\mu}^\prime =
\nabla_\mu\phi^{\prime\prime}_{(1)} - A^{\prime\prime}_\mu,
&& \delta A_{(3)\mu}^\prime = (\nabla^2 -r \frac{d(d-6)-34}{4} +m_0^2)
A^{\prime\prime}_\mu,\\
& \delta A_{(4)\mu}^\prime = m_1 A^{\prime\prime}_{\mu}
+\nabla_{\mu}\phi^{\prime\prime}_{(3)}, && \delta H_\mu^\prime =
-\nabla_\mu \phi^{\prime\prime}_{(1)} +A^{\prime\prime}_{\mu},
\\
& \delta\phi_{(1)}^\prime = -\phi^{\prime\prime}_{(2)}+r\phi^{\prime\prime}_{(1)}, &&
\delta\phi_{(2)}^\prime  = 2\phi^{\prime\prime}_{(2)}-2r\phi^{\prime\prime}_{(1)},
\\
& \delta\phi_{(3)}^\prime  =\phi^{\prime\prime}_{(2)}-r\phi^{\prime\prime}_{(3)}, &&\delta\phi_{(5)}^\prime =
(\nabla^2 -r \frac{d(d-6)-40}{4} +m_0^2)\phi^{\prime\prime}_{(1)}
,
\end{align}
\begin{align}
&\delta\phi_{(4)}^\prime = -\nabla^\mu A^{\prime\prime}_\mu -
\frac{1}{m_1}\big(m_0^2  - r\frac{d^2-14}{4} \big)\phi^{\prime\prime}_{(3)}  ,  &&
\delta\phi_{(6)}^\prime =
(\nabla^2 -r \frac{d(d-6)-20}{4} +m_0^2)\phi^{\prime\prime}_{(2)}+ 2r\nabla^\mu A^{\prime\prime}_\mu
\nonumber \\
& \phantom{\delta\phi_{(5)}^\prime}+\phi^{\prime\prime}_{(2)} +\frac{5}{2}r\phi^{\prime\prime}_{(1)},  && \phantom{\delta\phi_{(5)}^\prime =} - \frac{1}{m_1}\big(m_0^2  - r\frac{d(d+2)-23}{2} \big)\phi^{\prime\prime}_{(3)}-10r^2\phi^{\prime\prime}_{(1)} ,
\\
& \delta\phi_{(7)}^\prime = -\phi^{\prime\prime}_{(3)}, &&
\delta\phi_{(8)}^\prime = m_1 \phi^{\prime\prime}_{(1)} +
\phi^{\prime\prime}_{(3)},
\\
& \delta\phi_{(9)}^\prime =
(\nabla^2 -r \frac{d(d-6)-34}{4} +m_0^2)\phi^{\prime\prime}_{(3)}, && \delta
\phi_{(10)}^\prime = - m_2  \phi^{\prime \prime}_{(1)} -\frac{m_2}{m_1}  \phi
^{\prime \prime}_{(3)},
\\
& \delta \phi_{(11)}^\prime = m_1\phi^{\prime\prime}_{(3)} -
\frac{1}{2}\phi^{\prime\prime}_{(2)}.
\end{align}

\subsubsection{Gauge-fixing and partial use of equations of motion}

We now fix the gauge symmetry completely, using the gauge-fixing
conditions (\ref{G1}), obtained in our general consideration.
It is easy to see that the following fields are gauged away:
\begin{equation}
\phi_{(1)},\phi_{(2)},\phi_{(9)},\phi_{(13)},\phi_{(15)},\phi_{(17)},
\phi_{(18)},\phi_{(19)},A_{(5)},A_{(6)},A_{(7)},H_{(2)},T_{(1)}
\longrightarrow 0.
\end{equation}
Then, using the equations of motion for all the fields, except the antisymmetric
part of the basic field $\Phi_{[\mu\nu]}$, we find that we are left only with
$\Phi_{[\mu\nu]}$, so that the corresponding Lagrangian action (\ref{S(1,1)g}),
up to a total derivative of a certain quantity $f^\mu$, has the form
\begin{eqnarray}\label{S011}
{\cal S}^m[\Phi_{[\mu\nu]}] &=& \int d^dx \sqrt{|g|}\Big\{
 \Phi_{[ \mu \nu ]} (\nabla^2+m^2) \Phi^{[ \mu \nu ]} + 2 (\nabla_{\mu} \Phi^{[ \mu \lambda ]} ) ( \nabla^{\nu} \Phi_{[ \nu \lambda ]})
 \\
&=& \int d^dx \sqrt{|g|}\Big\{- \frac{1}{3}F_{\mu \nu \lambda} F^{\mu \nu \lambda} + m^2
\Phi_{[ \mu \nu ]}\Phi^{[ \mu \nu ]} +\nabla_\mu f^\mu \Big\},\nonumber
\end{eqnarray}
where $F_{\mu \nu \lambda}$ stands for the field strength of $\Phi_{[ \mu \nu ]}$, namely,
\begin{eqnarray}
F_{\mu \nu \lambda} =
\nabla_{\lambda} \Phi_{[ \mu \nu ]}
+ \nabla_{\mu} \Phi_{[ \nu\lambda ]} + \nabla_{\nu} \Phi_{[ \lambda
\mu ]} .
\end{eqnarray}
As a result, we have obtained the gauge-invariant action (\ref{S(1,1)g})
for a massive rank-2 antisymmetric higher integer spin field interacting
with an AdS background field $g_{\mu\nu}$, containing a complete set
of auxiliary fields and gauge parameters, as well as the action of a non-gauge
formulation (\ref{S011}).  Note that an equivalent form of the Lagrangian
formulation for the field $\Phi_{[ \mu \nu ]}$ has been constructed
in \cite{brst1}, using an antisymmetric basis of the initial oscillators.

\section{Conclusion}

In this paper, we have obtained quadratic non-linear HS symmetry
algebras for a description of ar\-bi\-trary integer HS fields
defined in $d$-dimensional AdS space and subject to a $k$-row
Young tableaux $Y(s_1, \ldots, s_k)$. We have shown that the
difference of resulting algebras $\mathcal{A}(Y(k),AdS_d)$,
$\mathcal{A}'(Y(k),AdS_d)$ and $\mathcal{A}_c(Y(k),AdS_d)$,
corresponding, respectively, to the initial set of operators,
their additional parts, and the converted set of operators within
an additive conversion procedure, is due to purely non-linear
parts, related to the AdS radius $(\sqrt{r})^{-1}$, in the set of
AdS space isometry operators.

To obtain the above algebras, we start by embedding the bosonic HS
fields into the vectors of an auxiliary Fock space, regarding the
fields as coordinates of Fock space vectors, and reformulate the
theory in these terms. We realize the conditions that determine an
irreducible AdS group representation of a given mass and
generalized spin in terms of differential operator constraints
imposed on the Fock space vectors. These constraints generate a
closed non-linear algebra of HS symmetry, which contains, with the
exception of $k$ basis generators in the Cartan subalgebra, a
system of first- and second-class constraints. The above algebra
coincides, modulo the isometry group generators, with its
Howe-dual $sp(2k)$ symplectic algebra. The construction
of a correct Lagrangian description requires the initial symmetry
algebra to be deformed to the algebra $\mathcal{A}_c(Y(k),AdS_d)$,
introducing the algebra $\mathcal{A}'(Y(k),AdS_d)$, different only
due to the respective quadratic parts given by (\ref{LiLjW})--(\ref{LiLj+bW})
and Tables~\ref{table}, \ref{table'}, with the use of Weyl ordering for
the right-hand sides.

We have generalized the method of constructing a Verma module
\cite{Vermamodule}, starting from Lie (super)algebras
\cite{BRbos, Liealgebra, 0001195, BurLeites} and the
quadratic algebra $\mathcal{A}'(Y(1),AdS_d)$ for totally symmetric
HS fields \cite{BurdikNavratilPasnev, BKL}, to the case of a
non-linear algebra underlying the mixed-symmetric HS bosonic
fields in an AdS space with a two-row Young tableaux. Theorem 1
presents our basic result in this respect. We show, as a
by-product of the Verma module construction, that the
Poincar\'{e}--Birkhoff--Witt theorem is valid in the case of the
algebra in question, thereby providing a lifting of the Verma
module for the Lie algebra $\mathcal{A}(Y(2), {R}^{1,d-1})$ --
isomorphic to $\left(T^2 \oplus T^{2*}\right) +
\hspace{-1em}\supset sp(4)$ -- to the Verma module for an algebra
quadratic in the deformation parameter $r$. The same is certainly
expected to hold true in the case of the general algebra
$\mathcal{A}'(Y(k),AdS_d)$, for which, we believe, it is pure
machinery to obtain an explicit form of the Verma module in a
recursive manner by means of some new \emph{primary and derived
block operators}, such as $\widehat{t}'_{12}$ (\ref{hattosc}) and
$\widehat{t}^{\prime+}_{12}$ (\ref{t'+F}).

We have obtained a representation for $15$ generators of the
algebra $\mathcal{A}'(Y(2),AdS_d)$ as a formal power series over
the Heisenberg--Weyl algebra with $6$ pairs of Grassmann-even
oscillators, which, in the flat-space limit $r=0$, takes a
polynomial form identical with familiar results, at least for
$m=0$ \cite{flatbos}, and appears to be novel in the massive case
\cite{flatbosm} for $k=2$ \cite{BRbos}. Theorem 2
finalizes our second basic result solving the Fock space
realization problem for $\mathcal{A}'(Y(2),AdS_d)$ by using the
approach (\ref{l'+ijF})--(\ref{l22fin}) of a generalized Verma
module construction.

On the basis of an exact BRST operator $Q'$ constructed for the
nonlinear algebra $\mathcal{A}_c(Y(2),AdS_d)$ of 15 converted
constraints $O_I$ (having a third order in the ghost coordinates),
by analyzing the corresponding structure of Jacobi identities, we
consistently develop a construction of gauge-invariant Lagrangian
descriptions for bosonic HS fields of a given spin
$\mathbf{s}=(s_1,s_2)$ and mass in an AdS${}_d$ space. To this
end, in the standard manner, we extract from $Q'$ (\ref{explQ'})
the related BRST operator $Q$ (\ref{Q}), which is associated with
the converted first-class constraints alone. The latter operator
is nilpotent only in those proper subspaces of the total Hilbert
space which have vanishing eigenvalues of the spin operators
$\vec{\sigma}_2 = (\sigma^1,\sigma^2)$ (\ref{sigmai}). The
corresponding Lagrangian formulation is, at most, a
$5$-th-stage-reducible abelian gauge theory, and is given by
(\ref{LEoM}), ({\ref{S}}), with a specific mass $m_0$ (\ref{m02}).
These last relations may be regarded as our main result in solving
the general problem of constructing a Lagrangian formulation for
non-Lagrangian initial AdS group irreducible representation
relations, which describe a bosonic HS field with two rows in the
Young tableaux (\ref{Eq-0b})--(\ref{Eq-3b}). It should be noted
that unconstrained Lagrangians for free mixed-symmetry HS fields
with two rows in the Young tableaux on AdS backgrounds, in both
the metric-like and frame-like formalisms have not been obtained
until now. We emphasize  that a Lagrangian description of massive
bosonic HS fields described by a two-row Young tableaux in (A)dS
spaces has been known only in the frame-like form
\cite{Zinoviev2}, with off-shell traceless and Young constraints.

We have used these results to study a deformation procedure in the
approach with a complete BRST operator, along the lines of
\cite{BuchbinderReshetnyak, BRmasscub, BRcub1}, so as to obtain an
interacting theory, in particular, a system of generating
equations (\ref{g1operV3}) for the cubic vertices of $3$ copies of
initial Lagrangian formulations, with a respective set of
constraints and oscillators. The above system is shown to be
specified by three equations (\ref{g1Lmod}), in the particular
case of coincident cubic vertices ${V}^3= \widetilde{V}_l^3 =
\widetilde{V}_{l+1}^3$ for $l=0,...,5$, so that a nontrivial
solution should be a BRST ($Q$)-closed three-vector with vanishing
values of spin $\vec{\sigma}_2$. By construction, all the
holonomic constraints (tracelessness and Young symmetry) are
included on equal footing with the remaining differential
constraints in the total BRST operator. This guarantees an equal
number of physical degrees of freedom in interacting and free
(undeformed) Lagrangian formulations, thus providing for the
interacting model to describe an irreducible interacting triple of
massive HS fields (see \cite{BRsuff} for the recent progress in
the necessary and sufficient conditions for an interacting model,
obtained with the use of the constrained BRST approach so as to
preserve irreducibility for interacting HS fields in flat spaces).
Unfortunately, there is no familiar application of Lagrangian
dynamics in an AdS space, even for a free HS field of a given mass
and spin, in the approach with an incomplete BRST operator, due to
the absence of off-shell BRST extended constraints.

In addition, we consider a number of simple examples of the
proposed Lagrangian descriptions: we deduce one for a totally
symmetric HS field obtained earlier in \cite{BKL} and another one
for an antisymmetric massive tensor field of spin $(1,1)$, whose
first-stage reducible unconstrained Lagrangian formulation with a
complete set of auxiliary fields and gauge parameters has been
obtained in Section~\ref{antsymm}. The final ungauged Lagrangian
is given by (\ref{S011}), after an application of partial
gauge-fixing and a resolution of some of the equations of motion,
and is shown to coincide with a Lagrangian given by a different
approach \cite{brst1}.

As a first by-product, we obtain polynomial deformations of the
$su(1,1)$ algebra related to the conversion analyzed in
Appendix~\ref{poltoy}. Second, we demonstrate that a BRST
Lagrangian formulation for an HS field of mass $m$ and spin
$(s_1,s_2)$ leads to equations of motions which are equivalent to
the initial irreducibility conditions (\ref{Eq-0b}--(\ref{Eq-3b}).
Third, we obtain a new Lagrangian formulation for HS field of mass
$m$ and spin $(s_1,s_2)$ in a $d$-dimensional flat space-time
starting from a general description for the same field in an AdS
space in the limit $r\to 0$, due to new representations for the
Verma and Fock modules given, respectively, by
(\ref{tfinL})--(\ref{l22totnVL}) and
(\ref{toscL})--(\ref{l22oscL}), with the gauge-invariant
Lagrangian action $\mathcal{S}^{m}_{0|\vec{s}_2} = \int d \eta_0
\; {}_{\vec{s}_2}\langle \chi |K^0_{\vec{s}_2} Q^0_{\vec{s}_2}|
\chi \rangle_{\vec{s}_2}$ for the BRST operator
$Q^0_{\vec{s}_2}=Q_{\vec{s}_2}|_{r=0}$ and
$K^0_{\vec{s}_2}=K_{\vec{s}_2}|_{r=0}$. This Lagrangian is
different from the one obtained in \cite{flatbosm} in view of a
dependence on two additional mass-like parameters $m_1, m_2$.

From the mathematical standpoint, the construction of a Verma
module for the algebra $\mathcal{A}'(Y(2)$, $AdS_d)$ opens a
possibility of analyzing the module structure and searching for
singular and subsingular vectors in it, so that, in principle,
this could then allow one to construct new (non-scalar)
infinite-dimensional representations of the given algebra.
Besides, the above results allow one, certainly, to understand the
problem of (generalized) Verma module construction for the HS
symmetry algebras and superalgebras underlying HS bosonic and
(respectively) fermionic fields in AdS spaces subject to a
multi-row Young tableaux.

We expect, first of all, to use the deformation procedure for
irreducible totally symmetric massless HS fields elaborated for
free fields in \cite{symint-adsmassless} as applied to AdS spaces,
and also anticipate the concept of BRST--BV quantization as
applied to deduce interacting models in the BRST approach, while
adapting the algorithm developed in flat spaces \cite{2010.15741},
albeit with a complete BRST operator, as in the case of a minimal
BRST--BV action \cite{BRST-BVR}. As a next problem to be solved,
we intend to apply our results to an explicit construction of a
generalized Verma module, a BRST operator, and an LF for
irreducible massive half-integer HS fields of spin
$(n_1+1/2,n_2+1/2)$ subject to $Y(n_1,n_2)$ in AdS spaces.

Among other directions of application and development of the
suggested approach, such as finding LFs implementing free theories
and cubic vertices for irreducible massive half-integer
higher-spin fields on AdS backgrounds, having in mind the
possibility of taking a flat-space limit
\cite{BoulangerLeclercqSundell} for numerous cubic vertices
\cite{FradkinVasiliev, FradkinVasiliev1}, one should also mention
the problems of constructing quartic and higher vertices, as well
as the related problems of locality (see the discussions in
\cite{T1, DT, T2, DGKV, Vasil, Didenko, Didenko1}), which, as we
believe, may be addressed using the BRST method. We also expect
that (ir)reducible HS fields can be employed as composite fields
in the recent approach \cite{MR2, MR3}.

\section*{Acknowledgments}

The authors are grateful to I.~Buchbinder, S.~Odintsov, V.~Dobrev,
V.~Krykhtin, P.~Lavrov, E.~Skvort\-sov, M.~Vasili\-ev and
Yu.~Zinoviev for useful discussions.
A.R. is indebted to D.~Francia for illuminating comments. The work
of A.R. has been partially supported by the Ministry of Education
of the Russian Federation under Project No.~QZOY-2023-0003.

\appendix
\section*{Appendix}

\section{Proposition proof}\label{proof}
\renewcommand{\theequation}{\Alph{section}.\arabic{equation}}
\setcounter{equation}{0}

In this appendix, we examine the correctness of our proposition in
Subsection~\ref{auxtheorem}. The analysis is based on deriving the
multiplication laws (\ref{addalg}), (\ref{conv-alg}),
(\ref{sumcoeff}) explicitly for the operator sets $\mathcal{A}'$
of $o'_I$ and $\mathcal{A}_c$ of $O_I$. Namely, from the r.h.s. of
(\ref{involrel}), we find (due to the commutativity of
$o_I$ with $o'_J$) the following equations that determine the
unknown structure functions ${F}_{IJ}^K({o}',{O})$:
\begin{eqnarray}\label{exprOoo'}
[\,{O}_I,{O}_J\}=[\,{o}_I,{o}_J\}+[\,o_I',o_J'\}
= \sum_{m=1}^nf_{ij}^{K_1\cdots K_m}\prod_{l=1}^{m}{o}_{K_l}
+ [\,o_I',o_J'\}.
\end{eqnarray}
Expressing in (\ref{exprOoo'}) the initial elements ${o}_{K_1},
\ldots, {o}_{K_n}$ through the enlarged $O_I$ and additional
$o'_I$ operators with the use of the $o'{O}$-ordering, we obtain
a sequence of relations for each power of $o_K$,
\begin{eqnarray}\label{ok1}
f_{IJ}^{K_1}{o}_{K_1}\hspace{-0.7em} &\hspace{-0.7em}
=\hspace{-0.7em}&\hspace{-0.7em} f_{IJ}^{K_1}{O}_{K_1} - f_{IJ}^{K_1}
{o}'_{K_1}\,, \\
f_{IJ}^{K_1K_2}{o}_{K_1}{o}_{K_2}
& = & f_{IJ}^{K_1K_2}{O}_{K_1}{O}_{K_2} -
(f_{IJ}^{K_1K_2}+ f_{IJ}^{K_2K_1}){o}'_{K_1}{O}_{K_2}
+ f_{IJ}^{K_2K_1}{o}'_{K_1}{o}'_{K_2}\, , \label{ok1k2}\\
\cdots   &\cdots & \cdots \quad \cdots \quad \cdots \quad \cdots \quad \cdots \quad \cdots \quad\cdots \quad, \nonumber\\
f_{IJ}^{K_1\cdots
K_n}\prod_{l=1}^{n}{o}_{K_l} & =  & f_{IJ}^{K_1\cdots
K_{n}}\prod\nolimits_{m=1}^{n}{O}_{K_m}
+\sum\nolimits_{s=1}^{n-1}(-1)^{s} f_{ij}^{\widehat{K_s\cdots
K_1}\widehat{K_{s+1}\cdots K_{n}}} \nonumber
\\
&\times& \prod\nolimits_{p=1}^{s}{o}'_{K_{p}}
\prod\nolimits_{m=s+1}^{n}{O}_{K_{m}}- (-1)^{
  n }f_{IJ}^{(n)K_{n}\cdots
K_1}\prod\nolimits_{s=1}^{n}{o}'_{K_s},\label{Ok1kn}
\end{eqnarray}
where the hats in $f_{ij}^{\widehat{K_s\cdots K_1}
\widehat{K_{s+1}\cdots K_{n}}}$ for the quantities
$f_{ij}^{{K_s\cdots K_1}{K_{s+1}\cdots K_{n}}}$ designate a set of
$C_s^n= \frac{n!}{s!(n-s)!}$ terms, obtained using
$f_{ij}^{\widehat{K_s\cdots K_1}\widehat{K_{s+1}\cdots K_{n}}}$ by
symmetrization, as shown explicitly in (\ref{sumcoeff}). First,
the above system (\ref{ok1})--(\ref{Ok1kn}) allows one, immediately,
to establish, from the above rightmost terms in
(\ref{exprOoo'})--(\ref{Ok1kn}), that the set of ${o}'_I$ forms a
polynomial algebra $\mathcal{A}'$ of order $n$, subject to the
algebraic relations (\ref{addalg}). Second, the remaining terms in
(\ref{exprOoo'})--(\ref{Ok1kn}) completely determine the structure
functions $F^{(m){}K}_{IJ}({o}',{O})$, $m=1,\ldots, n$, in the
form (\ref{expfunc}), and also show that the set of $O_I$ actually
determines the non-linear
algebra\footnote{The algebraic relations (\ref{conv-alg})
for the algebra $\mathcal{A}_c$ are different from those
for the polynomial algebra in view of a non-homogeneous character
of the structure functions $F^{(m){}K}_{IJ}({o}',{O})$ in $O_I$,
due to the presence of the elements $o'_I$.\label{prop}}
$\mathcal{A}_c$.

\section{Polynomial deformations of $su(1,1)$ algebra}\label{poltoy}
\renewcommand{\theequation}{\Alph{section}.\arabic{equation}}
\setcounter{equation}{0}

The polynomial algebra $\mathcal{A}^{n-1}$ of order $n-1$,
as a deformation of the algebra $su(1,1)$ with three bosonic elements,
$\{o_I\}= \{q_+, q_-, q_0\}$, $I=1,2,3$, and a central extension
\cite{1002.0667}, suggested in connection with quantum optics
and the Bethe ansatz systems, determines the non-vanishing
relations\footnote{In the case $n=1$, the corresponding algebra must be
of zero order, because of the relation (\ref{A1}), which has the form
$[\,q_+, q_{-}]=-1$; however, in this case, we can add the unity $1$
to the set of $o_I$, as has been done with the Heisenberg algebra $A_1$,
and then to remove the element $q_0$ from $o_I$, due to the possibility
of the representation $q_0 = q_-q_+$.}
\begin{eqnarray}
[\,q_0, q_{\pm}] &=& \pm q_{\pm},  \qquad  [\,q_+, q_{-}] \ =\ \phi^{(n)}(q_0)-\phi^{(n)}(q_0-1), \label{A0}\\
{} \phi^{(n)}(x) &=& -\prod_{l=1}^n\left(x +
\frac{ln-1}{n^2}\right)+ \prod_{l=1}^n\left( \frac{l-n}{n}-
\frac{1}{n^2}\right).\label{A1}
\end{eqnarray}
The second relation (\ref{A1}) can be recast in a form similar
to (\ref{polynom}), as follows:
\begin{eqnarray}
{} [\,q_+, q_{-}] &=& \sum_{l=0}^{n-1}f^{(l)}q_0^l, \quad
f^{(l)}= \frac{d^{(l)} \phi^{(n)}}{d x^l}_{(x=0)}- \frac{d^{(l)}
\phi^{(n)}}{d x^l}_{(x=-1)}.\label{Apolyn}
\end{eqnarray}
The study of \cite{1002.0667} has proposed a so-called one-mode
boson realization of the algebra $\mathcal{A}^{n-1}$, in terms
of a pair of creation and annihilation operators, $b^+, b$, $[b,b^+]=1$,
\begin{equation}\label{realizPolyn}
q_+ = \frac{(b^+)^n}{(n)^{\frac{n}{2}}}\,,\qquad q_- =
\frac{b^n}{(n)^{\frac{n}{2}}}\,,\qquad q_0 =
\frac{1}{n}\left(b^+b+\frac{1}{n}\right).
\end{equation}
The relations (\ref{A0}), (\ref{Apolyn}) for $\mathcal{A}^{n-1}$
show that we are within the conditions of our \emph{Proposition},
which implies the following composition law for the polynomial algebra
of order $n-1$, $\mathcal{A}^{\prime n-1}$ with the elements $q_+', q_-', q_0'$,
\begin{eqnarray}
  [\,q'_0, q'_{\pm}] &=& \pm q'_{\pm}\,,\qquad
  [\,q'_+, q'_{-}] \ =\  \sum_{l=0}^{n-1}(-1)^{l-1}f^{(l)}q_0^l\,,
  \label{A1'}
\end{eqnarray}
as we have chosen the value of the central extension for $\mathcal{A}^{\prime n-1}$
to be opposite to the one for $\mathcal{A}^{n-1}$, i.e., $(-f^{(0)})$, so that
the central extension of the non-homogeneous polynomial algebra
$\mathcal{A}_c^{n-1}$, $F^{(0)} = (f^{(0)}-f^{(0)})$ of order $n-1$
is zero. The composition law for the algebra $\mathcal{A}_c^{n-1}$
with the elements $Q_{\pm}=q_{\pm}+q'_{\pm},\ Q_0=q_0+q'_0$,
following the prescription (\ref{conv-alg}), (\ref{sumcoeff}), is determined by
\begin{eqnarray}
[\,Q_{0}, Q_{\pm}] &=& \pm Q_{\pm},  \label{A0c}\\
{} [\,Q_{+}, Q_{-}] &=& \sum_{l=1}^{n-1}f^{(l)}Q_{0}^{l} +
\sum_{l=2}^{n-1} f^{(l)}\sum_{s=1}^{l-1}(-1)^{s}
\frac{l!}{s!(l-s)!}({q'_0})^{s}Q_{0}^{l-s}.  \label{A1c}
\end{eqnarray}
The relations established for the polynomial algebras $\mathcal{A}_c^{n-1}$,
$\mathcal{A}_c^{n-1}$ can be applied to various objectives. From the number
of issues analyzed in this article, one can examine a Verma module and obtain
some new representations for each of the above algebras, as well as to construct
a BRST operator.

\section{Verma module construction for $\mathcal{A}'(Y(2),AdS_d)$}\label{addalgebra}
\renewcommand{\theequation}{\Alph{section}.\arabic{equation}}
\setcounter{equation}{0}

Here, we describe the details of a Verma module construction \cite{Vermamodule}
for the algebra
$\mathcal{A}'(Y(2), AdS_d)$, so as to validate the relations (\ref{t12fin})--(\ref{l22totnV})
of Theorem 1. Starting from (\ref{l'+lb}), (\ref{t'+b}), (\ref{l'0auxb}),
(\ref{t'auxb})--(\ref{l'22auxb}),
we formulate the following auxiliary

\noindent \textbf{\emph{{Lemma 1.}}} The $n$-th power in the action
of the operator
$\mathrm{ad}_{ l^{\prime +}_i}$ on
the operators\footnote{Explicitly, $W_{b}^{\prime{}12+}=
2r\left[l^{\prime + }_{12}(g_0^{\prime 2}-g_0^{\prime 1})-
l^{\prime +}_{11}t^{\prime +}_{12}+ l^{\prime +}_{22}
t^{\prime}_{12} \right]$, which follows from
$W_{b}^{\prime{}12}$ by hermitian conjugation.}
$W_{b}^{\prime{}12+}$, $X_b^{\prime{}12}$, $X_b^{\prime{}21}$
(\ref{lilj'})--(\ref{lilj+b'})
and the Casimir operator $\mathcal{K}_0(2)\equiv\mathcal{K}_0$
for the algebra $sp(4)$,
\begin{equation}\label{Casimirsb2}
\mathcal{K}_0 =
   \sum_{i}K_0^{0i} +
    2\mathcal{K}_0^{12}  =
    \sum_{i}\bigl((g_0^{\prime i})^2-2g_0^{\prime i} -4l^{\prime ii+}l^{\prime
ii}\bigr)+2
    \bigl(t^{\prime +}_{12}t'_{12}-
    4l^{\prime +}_{12}
      l^{\prime {12}} -
    g_0^{\prime 2} \bigr),
\end{equation}
denoted by $\bigl(\mathcal{K}^{i}_n, W_{bn}^{\prime{}12+i},X_{b i n}^{\prime{}12 }$,
$X_{b i n}^{\prime{}21}\bigr) \equiv\mathrm{ad}^n_{l^{\prime +}_i}
\bigl(\mathcal{K}_0,W_{b}^{\prime{}12+}, X_{b}^{\prime{}12 }$,
$X_{b}^{\prime{}21}\bigr)$,
yields the following set of relations for natural numbers $n$:
\begin{eqnarray}\label{Knb}
    \mathcal{K}^{i}_n  & = &  \left(-8rl^{\prime ii+}\right)^{\left[(n-1)/2\right]}
    \sum^{\left[(n+1)/2\right]}_{m=1}\left(\mathcal{K}^{i}_2
    \delta_{n,2m}+
    \mathcal{K}^{i}_1\delta_{n,2m-1}\right), \ n\geq 1,\\
    \mathcal{K}^{12{}i}_n & =&  \left(-2rl^{\prime ii+}\right)^{\left[(n-1)/2\right]}
    \sum^{\left[(n+1)/2\right]}_{m=1}\left(\mathcal{K}^{12{}i}_2
    \delta_{n,2m}+
    \mathcal{K}^{12{}i}_1\delta_{n,2m-1}\right),\ n\geq 1,\\
\label{W12+n}
    W_{bi{}n}^{\prime{}12+}  & = &  \left(-2rl^{\prime ii+}\right)^{
    \left[n/2\right]}
    \sum^{\left[(n+1)/2\right]}_{m=0}\left(W_b^{\prime{}12 +}
    \delta_{n,2m}+
    W_{b}^{\prime{}12 +i}\delta_{n,2m-1}\right),\\
\label{X12n}
    X_{b 2{}n}^{\prime{}12 }  & = &  \left(-2rl^{\prime 22+}\right)^{
    \left[n/2\right]}
    \sum^{\left[(n+1)/2\right]}_{m=0}\left(X^{\prime{}12 }_b
    \delta_{n,2m}+
    X_{b{}2}^{\prime{}12 }\delta_{n,2m-1}\right), \\
\label{X21n}
    X_{b 1{}n}^{\prime{}21 }  & = &  \left(-2rl^{\prime 11+}\right)^{
    \left[n/2\right]}
    \sum^{\left[(n+1)/2\right]}_{m=0}\left(X_b^{\prime{}21 }
    \delta_{n,2m}+
    X_{b{}1}^{\prime{}21 }\delta_{n,2m-1}\right),
\end{eqnarray}
where the symbol $[x]$ denotes the integer part of a real-valued
$x$; the operators $\mathcal{K}^{i}_k $, $\mathcal{K}^{12{}i}_k =
\mathrm{ad}^k_{l^{\prime +}_i}\mathcal{K}^{12}_0$, $k=1,2$,
$W_{b}^{\prime{}12 + i}$, $X_{b1}^{\prime{}21 }$,
$X_{b2}^{\prime{}21 }$, in accordance with the rule
(\ref{product}) for the operation $\mathrm{ad}^k_{l^{\prime
+}_i}$, and the action on the highest-weight vector $|0\rangle_V$,
are defined, respectively, by (\ref{K_ikb})--(\ref{X1221}) and
(\ref{K00012b})--(\ref{X12iV}) from Appendix~\ref{Ap-AddPartsb}.

The correctness of (\ref{Knb})--(\ref{X21n}) follows from
calculations proceeding by induction in $n$.

The results of \emph{Lemma 1} imply the validity of the following
rather technical

\noindent \textbf{\emph{{Lemma 2.}}} The $n$-th power
of the action of $\mathrm{ad}_{ l^{\prime +}_i}$ on any
$o'_I$, $o'_I \in \{ t^{\prime +}_{12}$, $l^{\prime +}_{2}$,
$l^{\prime }_0$, $\mathcal{E}^{+}_2\}$ is determined by the
relations (\ref{l_ii l_1+b})--(\ref{l0li+b}),
\begin{eqnarray}
\left[l^{\prime ii}, \textstyle\left(\frac{l^{\prime
i+}}{m_i}\right)^{n_i}\right]
 & = &
-\textstyle\frac{n_i}{m_i}\textstyle\left(\frac{l^{\prime
i+}}{m_i} \right)^{n_i-1}l^{\prime i} -
\textstyle\frac{n_i(n_i-1)}{2
m_i^2}\textstyle\left(\frac{l^{\prime i+}}{m_i}
\right)^{n_i-2}K_2^{\prime  0{}i} + 2
r\displaystyle\hspace{-0.5em}
\sum\limits^{\left[(n_i-1)/2\right]}_{m=1}\hspace{-0.5em}\left(
-8rl^{\prime ii
+}\right)^{m-1} \label{l_ii l_1+b}
\\
&&\times
 \textstyle\left(\frac{l^{\prime i+}}{m_i}
\right)^{n_i-2m-2}\hspace{-0.2em}\textstyle
\left(\hspace{-0.1em}
\frac{1}{m_i} \hspace{-0.1em}\right)^{2m+1}\hspace{-0.2em}\left\{
\textstyle\frac{l^{\prime i+}}{m_i}
C^{n_i}_{2m+1}\hspace{-0.2em}\left[\mathcal{K}_1^{i} -
\textstyle\frac{2}{4^m}\mathcal{K}_1^{12{}i}\right]\hspace{-0.2em}
\right.\nonumber\\
{}& & \ + \left.
\textstyle\frac{1}{m_i}C^{n_i}_{2m+2}\hspace{-0.2em}\left[\mathcal{K}_2^{i}
- \textstyle\frac{2}{4^m} \mathcal{K}_2^{12{}i}
\right]\right\},\nonumber
\end{eqnarray}
\begin{eqnarray}
 \left[l^{\prime i},\textstyle\left(\frac{l^{\prime i+}}{m_i}\right)^{n_i}\right] &
 =& \textstyle\frac{n_i}{m_i}\textstyle\left(\frac{l^{\prime i+}}{m_i}
\right)^{n_i-1}K_2^{\prime 0{}i} - 2
r\displaystyle\sum\limits^{\left[n_i/2\right]}_{m=1}\left(
-8rl^{\prime ii+}\right)^{m-1}\textstyle\left(\frac{l^{\prime
i+}}{m_i} \right)^{n_i-2m-1}\times \nonumber \\
& &\ \times \textstyle\left(\frac{1}{m_i}
\right)^{2m}\left\{\textstyle\frac{l^{\prime i+}}{m_i}
C^{n_i}_{2m}\left[\mathcal{K}_1^{i} - \textstyle\frac{2}{4^m}
\mathcal{K}_1^{12{}i}\right] +
\textstyle\frac{1}{m_i}C^{n_i}_{2m+1}\left[\mathcal{K}_2^{i} -
\textstyle\frac{2}{4^m}\mathcal{K}_2^{12{}i}\right]\right\},
\label{l_i l_i+b}\\
\left[t^{\prime }_{12}, \textstyle\left(\frac{l^{\prime
i+}}{m_i}\right)^{n_i}\right] &
 =& -\delta^{i2}\Bigl\{\textstyle\frac{n_2}{m_2}\textstyle\left(\frac{l^{\prime 2+}}{m_2}
\right)^{n_2-1}l^{\prime 1+} +
\displaystyle\sum\limits^{\left[n_2/2\right]}_{m=1}\left(
-2rl^{\prime 22+}\right)^{m-1}\textstyle\left(\frac{l^{\prime
2+}}{m_2} \right)^{n_2-2m-1}\times \nonumber \\
& & \quad\times\textstyle\left(\frac{1}{m_2}
\right)^{2m}\left[\textstyle\frac{l^{\prime 2+}}{m_2}
C^{n_2}_{2m}W_b^{\prime{}12 +} +
\textstyle\frac{1}{m_2}C^{n_2}_{2m+1}W_{b2}^{\prime{}12
+}\right]\Bigr\} ,\label{tl2+b}\\
\left[l^{\prime 1},\textstyle\left(\frac{l^{\prime
2+}}{m_2}\right)^{n_2}\right] &
 =&
\displaystyle\sum\limits^{\left[n_2-1/2\right]}_{m=0}\left(
-2rl^{\prime 22+}\right)^{m}\textstyle\left(\frac{l^{\prime
2+}}{m_2} \right)^{n_2-2m-2}\textstyle\left(\frac{1}{m_2}
\right)^{2m+1}\times \nonumber \\
& & \quad \times\left\{\textstyle\frac{l^{\prime 2+}}{m_2}
C^{n_2}_{2m+1}X_b^{\prime{}12 } +
\textstyle\frac{1}{m_2}C^{n_2}_{2m+2}X_{b2}^{\prime{}12 }\right\},
\label{l1 l2+b}\\
\left[l^{\prime 2},\textstyle\left(\frac{l^{\prime
1+}}{m_1}\right)^{n_1}\right] &
 =&
\displaystyle\sum\limits^{\left[n_1-1/2\right]}_{m=0}\left(
-2rl^{\prime 11+}\right)^{m}\textstyle\left(\frac{l^{\prime
1+}}{m_1} \right)^{n_1-2m-2}\textstyle\left(\frac{1}{m_1}
\right)^{2m+1}\times \nonumber \\
&& \quad \times \left\{\textstyle\frac{l^{\prime 1+}}{m_1}
C^{n_1}_{2m+1}X_b^{\prime{}21 } +
\textstyle\frac{1}{m_1}C^{n_1}_{2m+2}X_{b1}^{\prime{}21 }\right\}
,\label{l2 l1+b}
\\
\left[l^{\prime 12},\textstyle\left(\frac{l^{\prime
i+}}{m_i}\right)^{n_i}\right] &
 =&
-\textstyle \frac{n_i}{2m_i}\left(\frac{l^{\prime i+}}{m_i}
\right)^{n_i-1}l^{\prime\{1}\delta^{2\}i} - \delta^{i2}\frac{1}{2}
\displaystyle\sum\limits^{\left[n_2/2\right]}_{m=1}\left(
-2rl^{\prime 22+}\right)^{m-1}\textstyle\left(\frac{l^{\prime
2+}}{m_2} \right)^{n_2-2m-1}\textstyle\times \nonumber \\
& &\quad \times \left(\frac{1}{m_2}
\right)^{2m}\left\{\textstyle\frac{l^{\prime 2+}}{m_2}
C^{n_2}_{2m}X_b^{\prime{}12 } +
\textstyle\frac{1}{m_2}C^{n_2}_{2m+1}X_{b2}^{\prime{}12
}\right\}\nonumber \\
{}&& -
\delta^{i1}\frac{1}{2}\displaystyle\sum\limits^{\left[n_1/2\right]}_{m=
1}\left( -2rl^{\prime
11+}\right)^{m-1}\textstyle\left(\frac{l^{\prime 1+}}{m_1}
\right)^{n_1-2m-1}\textstyle\left(\frac{1}{m_1}
\right)^{2m}\times \nonumber \\
&& \quad \times \left\{\textstyle\frac{l^{\prime 1+}}{m_1}
C^{n_1}_{2m}X_b^{\prime{}21 } +
\textstyle\frac{1}{m_1}C^{n_1}_{2m+1}X_{b1}^{\prime{}21 }\right\}
\,,\label{l12 li+b}\\
\left[t^{\prime+}_{12},\textstyle\left(\frac{l^{\prime
i+}}{m_i}\right)^{n_i}\right] &
 =& -\delta^{i1}\Bigl\{\textstyle\frac{n_1}{m_1}\textstyle\left(\frac{l^{\prime 1+}}{m_1}
\right)^{n_1-1}l^{\prime 2+} -
\displaystyle\sum\limits^{\left[n_1/2\right]}_{m=1}\left(
-2rl^{\prime 11+}\right)^{m-1}\textstyle\left(\frac{l^{\prime
1+}}{m_1} \right)^{n_1-2m-1}\times \nonumber \\
& & \quad \times \textstyle\left(\frac{1}{m_1}
\right)^{2m}\left[\textstyle\frac{l^{\prime 1+}}{m_1}
C^{n_1}_{2m}W_b^{\prime{}12 +} +
\textstyle\frac{1}{m_1}C^{n_1}_{2m+1}W_{b1}^{\prime{}12
+}\right]\Bigr\} ,\label{t+l1+b}\\
\varepsilon_{ij}\left[l^{\prime
i+},\textstyle\left(\frac{l^{\prime j+}}{m_j}\right)^{n_j}\right]
&
 =&
-\displaystyle\sum\limits^{\left[n_j-1/2\right]}_{m=0}\left(
-2rl^{\prime jj+}\right)^{m}\textstyle\left(\frac{l^{\prime
j+}}{m_j} \right)^{n_j-2m-2}\textstyle\left(\frac{1}{m_j}
\right)^{2m+1}\times \nonumber \\
& & \quad\times\left\{\textstyle\frac{l^{\prime j+}}{m_j}
C^{n_j}_{2m+1}W_b^{\prime{}12 +} +
\textstyle\frac{1}{m_j}C^{n_j}_{2m+2}W_{bj}^{\prime{}12 +}\right\},\
\label{li+ lj+b}\\
 \left[{l}_0',\textstyle\left(\frac{l^{\prime i+}}{m_i}\right)^{n_i}\right] &
 =& - r\displaystyle\sum\limits^{\left[(n_i-1)/2\right]}_{m=0}\left(
-8rl^{\prime ii +}\right)^{m}\textstyle\left(\frac{l^{\prime
i+}}{m_i} \right)^{n_i-2m-2}\left(\frac{1}{m_i}
\right)^{2m+1}\times \nonumber \\
&& \quad\times \left\{\textstyle\frac{l^{\prime i+}}{m_i}
C^{n_i}_{2m+1}\mathcal{K}_1^{i} +
\textstyle\frac{1}{m_i}C^{n_i}_{2m+2}\mathcal{K}_2^{i}\right\}.
\label{l0li+b}
\end{eqnarray}
In the above formulae, we need to take into account that $C^n_{n+k}=
0$ for any $n, k \in \mathbf{N}$, and also, in (\ref{li+ lj+b}),
the fact that $\varepsilon_{ij}\varepsilon^{jl}=\delta_{i}^l$,
$\varepsilon_{21} = 1$.

It is then easy to see that the undetermined action of the
remaining operators from the subalgebra $\mathcal{E}^-_2 $,
$l^{\prime +}_{2}$,  $t^{\prime +}_{12}$ on the vector
$|\vec{{0}}_{ij}, \vec{n}_s\rangle_V$ in (\ref{l'+lb}),
(\ref{t'+b}) reduces to the action of
$t^{\prime}_{12}$ on the vector $|\vec{{0}}_{ij}, 0,0,
{n}_2\rangle_V = |{n}_2\rangle_V$. Explicitly, this is implied by
the relations\footnote{In (\ref{t+prom}), (\ref{l2+prom}) and
later on, for the sake of convenience, we use the following
notation for an arbitrary vector:
$\left|\vec{n}_{ij}+m\delta_{ij,11} -2\delta_{ij,12}, \vec{n}_s -
2m\delta_{s,1} + \delta_{s,12}\rangle_V\right.$ $\equiv$
$\left|\vec{n}_{ij}+ (m,-2,0), \vec{n}_s -
(2m,-1,0)\rangle_V\right.$ $\equiv$ $ \left|\vec{N}(2) + (m,-2,0,
- 2m,+1,0)\rangle_V\right.$.}
\begin{eqnarray}
\hspace{-1em}t^{\prime +}_{12}\left|\vec{0}_{ij},\vec{n}_s\rangle_V\right.
&\hspace{-1em} =&\hspace{-1em}
\sum^{\left[n_1/2\right]}_{m=0}\left(\frac{-2r}{m_1^2}\right)^m
\Bigl\{ C^{n_1}_{2m}\left|\vec{0}_{ij}+ m\delta_{ij,11}, \vec{n}_s
- 2m\delta_{s,1}+\delta_{s,12}\rangle_V\right.\nonumber
\\ {}&{}&-C^{n_1}_{2m+1} \frac{m_2}{m_1}\left|\vec{0}_{ij}+
m\delta_{ij,11}, \vec{n}_s -
(2m+1)\delta_{s,1}+\delta_{s,2}\rangle_V\right.\Bigr\} \nonumber
\\ {}&{}& -
\hspace{-0.3em}\sum^{\left[n_1/2\right]}_{m=1}\hspace{-0.3em}\left(\hspace{-0.1em}
\frac{-2r}{m_1^2}\right)^m \hspace{-0.3em}\Bigl\{\hspace{-0.1em}
\Bigl(C^{n_1}_{2m}(h^2-h^1+2p_{12}+n_2)-C^{n_1}_{2m+1}\Bigr)\nonumber\\
{}&&{} \times \left|\vec{0}_{ij}+
(m-1)\delta_{ij,11}+\delta_{ij,12},
\vec{n}_s - 2m\delta_{s,1}\rangle_V\right.+ C^{n_1}_{2m}p_{12}(h^1-h^2\nonumber\\
{}&{}& \quad -n_2-p_{12}+1)\left|\vec{0}_{ij}+
(m-1)\delta_{ij,11}+\delta_{ij,22}, \vec{n}_s -
2m\delta_{s,1}-\delta_{s,12}\rangle_V\right.
\Bigr\}\nonumber\\
{}&{}&\hspace{-1.0em} -
\sum^{\left[n_1/2\right]}_{m=1}\hspace{-0.3em}\left(\hspace{-0.2em}
\frac{-2r}{m_1^2}\right)^{m}\hspace{-0.3em}(l^{\prime
+}_{11})^{m-1}\left(\frac{l^{\prime
1+}}{m_1}\right)^{n_1-2m}\hspace{-0.3em}C^{n_1}_{2m}l^{\prime
22+}(t^{\prime + }_{12})^{p_{12}}t'_{12}|{n}_2\rangle_V, \label{t+prom}\\
\hspace{-1em}{l^{\prime +}_2|\vec{0}_{ij},\vec{n}_s\rangle_V}
&\hspace{-0.5em} =&\hspace{-0.5em}
m_2\hspace{-1.0em}\sum^{\left[n_1+1/2\right]}_{m=0}\hspace{-0.3em}
\left(\frac{-2r}{m_1^2} \right)^{m}
\hspace{-0.5em}C^{n_1}_{2m}\left|\vec{0}_{ij}+ m\delta_{ij,11},
\vec{n}_s - 2m\delta_{s,1}+\delta_{s,2}\rangle_V\right.
\nonumber \\
&& +
m_1\hspace{-1.0em}\sum^{\left[n_1-1/2\right]}_{m=0}\hspace{-0.3em}
\left(\hspace{-0.2em}\frac{-2r}{m_1^2} \right)^{m+1}
\hspace{-1.0em}\Bigl\{
 \Bigl(C^{n_1}_{2m+1}(h^2-h^1+2p_{12}+n_2)-C^{n_1}_{2m+2}\Bigr)\nonumber \\
&& \quad \times\left|\vec{0}_{ij}+
m\delta_{ij,11}+\delta_{ij,12}, \vec{n}_s -
(2m+1)\delta_{s,1}\rangle_V\right.+
C^{n_1}_{2m+1}\Bigl(p_{12}(h^1-h^2\nonumber
\\  {}&{}& \quad -n_2-p_{12}+1) \left|\vec{0}_{ij}+
m\delta_{ij,11}+\delta_{ij,22},
\vec{n}_s - (2m+1)\delta_{s,1}-\delta_{s,12}\rangle_V\right.\nonumber\\
{}&{}& \quad - \left|\vec{0}_{ij}+ (m+1)\delta_{ij,11}, \vec{n}_s
- (2m+1)\delta_{s,1}+\delta_{s,12}\rangle_V\right.\Bigr)
\Bigr\}\nonumber\\
{}&{}& \hspace{-1.0em}+
m_1\sum^{\left[n_1-1/2\right]}_{m=0}\hspace{-0.5em}
\left(\hspace{-0.3em}\frac{-2r}{m_1^2}\hspace{-0.2em}\right)^{m+1}
\hspace{-0.7em}(l^{\prime +}_{11})^{m} l^{\prime
+}_{22}\left(\frac{l^{\prime
1+}}{m_1}\right)^{n_1-2m-1}C^{n_1}_{2m+1}(t^{\prime
+ }_{12})^{p_{12}} t'_{12}\left|{n}_2\rangle_V\right..
\label{l2+prom}
\end{eqnarray}
In a manifest form, the action of $t'_{12}$ on the vector $|{n}_2\rangle_V$
is determined by

\vspace{1ex} \noindent \textbf{\emph{{Lemma 3.}}} The operator $t'_{12}$
in its action on $|{n}_2\rangle_V$ satisfies
the following \emph{recurrent relations}:
\begin{eqnarray}
 t'_{12}|{n}_2\rangle_V &=&
|A_{n_2}\rangle_V - \sum^{\left[n_2/2\right]}_{{}^1m=0}
\sum^{\left[n_2/2-{}^1m-1/2\right]}_{
{}^1l=0}\left(\frac{-2rl^{\prime
22+}}{m_2^2}\right)^{{}^1m+{}^1l+1}
 \nonumber\\
&& \times
 C^{n_2}_{2{}^1m+1}C^{n_2-2{}^1m-1}_{2{}^1l+1}t'_{12}\bigl|n_2-2({}^1m+{}^1l+1)
\rangle_V,  \label{recurrt'}
\end{eqnarray}
with a completely specified vector $|A_{n_2}\rangle_V$ from the Verma module,
\begin{eqnarray}
|A_{n_2}\rangle_V & =&
\sum^{\left[n_2/2\right]}_{m=1}\left(\frac{-2r}{m_2^2}\right)^m
\Bigl\{- C^{n_2}_{2m}\left|1,0,m-1,0, 1, {n}_2-2m\rangle_V\right.\nonumber\\
{}&&+ [C^{n_2}_{2m}(h^2-h^1)+C^{n_2}_{2m+1}]\left|0,1,m-1,0, 0,
{n}_2-2m\rangle_V\right.
\Bigr\}\nonumber
\end{eqnarray}
\begin{eqnarray}
&&-\frac{m_1}{m_2}\sum^{\left[n_2/2\right]}_{m=0}\left(\frac{-2r}{m_2^2}\right)^m
C^{n_2}_{2m+1}\left|0,0,m, 1, 0, {n}_2-2m-1\rangle_V\right.\nonumber\\
{}&& -\sum^{\left[n_2/2\right]}_{{}^1m=0}\sum^{
\left[n_2/2-{}^1m-1\right]}_{
{}^1l=0}\left(\frac{-2r}{m_2^2}\right)^{\hspace{-0.5em}{}^1m+{}^1l+1}\hspace{-1em}
C^{n_2}_{2{}^1m+1}\Biggl\{\Bigl[C^{n_2-2{}^1m-1}_{2{}^1l+1}\Bigl(h^2-h^1+n_2\nonumber\\
&&-2({}^1m+{}^1l+1)\Bigr)-C^{n_2-2{}^1m-1}_{2{}^1l+2}\Bigr]\left|0,1,{}^1m+{}^1l,
0, 0, {n}_2-2({}^1m+{}^1l+1)\rangle_V\right. \nonumber\\
&&-C^{n_2-2{}^1m-1}_{2{}^1l+1}
\left|1,0,{}^1m+{}^1l, 0, 1, {n}_2-2({}^1m+{}^1l+1)\rangle_V\right.\nonumber\\
&&+ \frac{m_1}{m_2}C^{n_2-2{}^1m-1}_{2{}^1l+2}\left|0,0,{}^1m+{}^1l+1,
1, 0, {n}_2-2({}^1m+{}^1l+1)-1\rangle_V\right.\Biggr\}. \label{An2}
\end{eqnarray}
Considering the representation (\ref{recurrt'}), we use the commutation relation
$t'_{12}$ with $\left(\frac{l^{\prime +}_2}{m_2}\right)^{n_2}$ (\ref{tl2+b}),
from which we obtain the first terms of $|A_{n_2}\rangle_V$ with single sums,
denoted below by S.S., the remaining term being a manifest example of resolving
the entanglement between the negative-root vectors $l^{\prime +}_2$, with
$l^{\prime +}_1$, due to the permuted structure opposite to that
of $|N(2)\rangle_V$ (\ref{VM2}),
\begin{equation}\label{entaglt12}
 t'_{12}|{n}_2\rangle_V =
 {\mathrm S.S.} - \sum^{\left[n_2/2\right]}_{{}^1m=0}
 \hspace{-0.3em}\left(\frac{-2rl^{\prime
 22+}}{m_2^2}\right)^{ \hspace{-0.2em}{}^1m}
 \hspace{-0.5em}C^{n_2}_{2{}^1m+1}
 \hspace{-0.3em}\underbrace{\left(\frac{l^{\prime
 +}_2}{m^2}\right)^{ \hspace{-0.2em}n_2-2{}^1m-1}
 \hspace{-0.3em}l^{\prime +}_1} |0 \rangle_V,
\end{equation}
In order to commute $\left(\frac{l^{\prime
+}_2}{m^2}\right)^{\hspace{-0.2em}n_2-2{}^1m-1}$ in
(\ref{entaglt12}) to the right through $l^{\prime+}_1$, one should
use the following relation, which is not difficult to establish,
\begin{eqnarray}
\left[\left(\frac{l^{\prime +}_2}{m_2}\right)^{n_2-2{}^1m-1},
 l^{\prime+}_1\right]
 &= &
\displaystyle\sum\limits^{\left[n_2-2{}^1m-2/2\right]}_{{}^1l=0}\left(
\frac{-2rl^{\prime 22+}}{m_2^2}\right)^{{}^1l}
\frac{1}{m_2}\Bigl\{
C^{n_2-2{}^1m-1}_{2{}^1l+1}W_b^{\prime{}12
+}\frac{l^{\prime +}_2}{m_2}\nonumber \\
& &  -
\frac{1}{m_2}C^{n_2-2{}^1m-1}_{2{}^1l+2}W_{b2}^{\prime{}12+}\Bigr\}
\left(\frac{l^{\prime +}_2}{m_2} \right)^{n_2-2({}^1m+
{}^1l+1)-1} ,\ \label{li+lj+b}
\end{eqnarray}
and which follows from an analogue of the commutation relation (\ref{product}),
\begin{equation}
\label{product*}
B^nA =
\sum^{n}_{k=0}C^n_k (\mathrm{\widehat{ad}}^k_B{}A) B^{n-k}\,, \ \ \
\mathrm{\widehat{ad}}_B{}A = [B, A].
\end{equation}
Rearranging the terms over the figure brackets in
(\ref{entaglt12}) by the rule (\ref{product*}), and also using
(\ref{lilj'2}), (\ref{W12i}), (\ref{X12iV}), we establish the
validity of the sought-for relations (\ref{recurrt'}),
(\ref{An2}).

From (\ref{entaglt12}), it follows that a single application of
the operator $t'_{12}$ to $|{n}_2\rangle_V$ implies, at least,
a reduction of the last component in the above vector by the value
of 2, i.e., a reduction to a vector with $|\vec{0}_{ij}
+ \delta_{i2}\delta_{j2}, 0,0,{n}_2-2\rangle_V$. Therefore,
applying the operator action from \emph{Lemma 3}
as many as $\left[\frac{(n_2-1)}{2}\right]$ times, we resolve
the recurrent relations (\ref{recurrt'}) in the form
\begin{eqnarray}
 t'_{12}|{n}_2\rangle_V
 & = &
 \sum\limits^{\left[(n_2-1)/2\right]}_{k=0}\Biggl\{\sum\limits^{
  \left[n_2/2\right]}_{{}^{1}m=0}
  \sum\limits^{\left[n_2/2-({}^{1}m + 1)\right]}_{{}^1l=0}\ldots\
\sum\limits^{\left[n_2/2-\sum\limits^{k-1}_{i=1}({}^{i}m + {}^il)
 -(k-1)\right]}_{{}^{k}m=0}
 \sum\limits^{\left[n_2/2-\sum\limits^{k-1}_{i=1}({}^{i}m + {}^il)-{}^{k}m
 -k \right]}_{{}^kl=0}
 (-1)^k
 \nonumber
 \end{eqnarray}
\begin{eqnarray}
 &&
\times\left(\frac{-2r}{m_2^2}\right)^{\sum_{i=1}^k({}^im+{}^il)+k}
 C^{n_2}_{2{}^1m+1}C^{n_2-2{}^1m-1}_{2{}^1l+1}\ldots
 C^{n_2-2(\sum_{i=1}^{k-1}({}^im+{}^il)+k-1)}_{2{}^km+1}
 \nonumber\\
 &&
 \times \hspace{0.3 em}  C^{n_2-2(\sum_{i=1}^{k-1}
 ({}^im+{}^il) +k-1) - 2{}^km -1}_{2{}^kl+1}\Big|A_{0,0,\sum_{i=1}^{k}({}^im+{}^il)+k,
 0,0,n_2-2[\sum\limits^k_{i=1}({}^{i}m + {}^il)
 +k]}\Big\rangle_V\Biggr\},\label{t'n2}
 \\
 &&
 \mathrm{where} \ \ \Big|A_{0,0,m,0,0,n_2}\Big\rangle_V\equiv
 (l^{\prime 22+})^m|A_{n_2}\rangle_V,\label{An2m}
\end{eqnarray}
so that we completely determine the action of $t'_{12}$
on the vector $|{n}_2\rangle_V$.

Due to Lemmas 2, 3, we consider the action of the operator ${t}'_{12}$
on the vector $|\vec{0}_{ij},0,0,{n}_2\rangle_V$ as a primary block operator
which can be used to obtain the action of all $o'_I$ on the vector
$\bigl|\vec{N}(2)\rangle_V$.

Note that, in the case of the flat-space limit $r=0$, it is only
the zero order in $r$ of (\ref{t'n2}), (\ref{An2m}) that survives,
so that we have a term which is not originated by the $sp(4)$ algebra:
\begin{eqnarray}
 && t'_{12}|\vec{0}_{ij},0,0,{n}_2\rangle_V\vert_{r=0} = -
\frac{m_1}{m_2} {n_2}|\vec{0}_{ij}, 1, 0, {n}_2-1\rangle_V .
\label{t12Mink}
\end{eqnarray}
Then the problem of finding the action of $t'_{12}$ on $|\vec{N}(2)\rangle_V$
in (\ref{t'auxb}) is readily solved in the form
\begin{eqnarray}
\label{t'ntot}
 t'_{12}\left|\vec{N}(2)\rangle_V\right. &=&
  p_{12}(h^1-h^2-n_2-p_{12}+1)\left|\vec{n}_{ij}, n_1, p_{12}-1, {n}_2\rangle_V
 \right. \nonumber \\
 \hspace{-1em} && \hspace{-1em}-\sum_l
  ln_{l2}\hspace{-0.2em}\left|{\vec{n}}_{ij}-\delta_{ij,l2}+
\delta_{ij,1l}, \vec{n}_s\rangle_V
 \right. +
 \sum\limits^{\left[(n_2-1)/2\right]}_{k=0}\Biggl\{\sum\limits^{
  \left[n_2/2\right]}_{{}^{1}m=0}
  \sum\limits^{\left[n_2/2-({}^{1}m + 1)\right]}_{{}^1l=0}\ldots\nonumber \\
\hspace{-1em}&&\hspace{-1em}
  \ldots\sum\limits^{\left[n_2/2-\sum\limits^{k-1}_{i=1}({}^{i}m + {}^il)
  -(k-1)\right]}_{{}^{k}m=0}
  \sum\limits^{\left[n_2/2-\sum\limits^{k-1}_{i=1}({}^{i}m + {}^il)-{}^{k}m
  -k \right]}_{{}^kl=0} (-1)^k\left(\frac{-2r}{m_2^2}\right)^{\sum_{i=1}^k({}^im+{}^il)+k}
\nonumber\\
\hspace{-1em}&&\hspace{-1em} \times C^{n_2}_{2{}^1m+1}C^{n_2-2{}^1m-1}_{2{}^1l+1}...
C^{n_2-2(\sum_{i=1}^{k-1}({}^im+{}^il)+k-1)}_{2{}^km+1}C^{n_2-2(
\sum_{i=1}^{k-1}(
{}^im+{}^il) +k-1)-2{}^km-1}_{2{}^kl+1}\nonumber\\
&&\ \ \times
\Big|\hat{A}_{\vec{n}_{ij}+[\sum\limits^k_{i=1}({}^{i}m + {}^il)
  +k]\delta_{i2}\delta_{j2},{n}_1,n, n_2-2[\sum\limits^k_{i=1}({}^{i}m + {}^il)
  +k]}\Big\rangle_V\Biggr\},
\end{eqnarray}
as we have extracted from $t'_{12}$ the action of the Lie part $t^{\prime Lie}_{12}$,
$sp(4)$, and also the part $\hat{t}^{\prime}_{12}$, being non-linear in $r$,
\begin{eqnarray}
\label{Liepart} t'_{12}\left|\vec{N}(2)\rangle_V\right. & =&
 p_{12}(h^1-h^2-n_2-p_{12}+1)\left|\vec{n}_{ij}, n_1, p_{12}-1, {n}_2\rangle_V
 \right. \nonumber \\
  && -\sum_l ln_{l2}\hspace{-0.2em}\left|{\vec{n}}_{ij} -
\delta_{ij,l2}+ \delta_{ij,1l}, \vec{n}_s\rangle_V
 \right.
+ \hat{t}'_{12} \left|\vec{N}(2)\rangle_V\right.,
  \end{eqnarray}
in order to prove the validity of (\ref{t12fin}). The vector
with a ``hat'', $|\hat{A}_{\ldots}\rangle_V$, in (\ref{t'ntot})
is constructed using the vector $\left|{A}_{\ldots}\rangle_V\right.$
of (\ref{An2}) by the rule
\begin{eqnarray}
 \left|\hat{A}_{\vec{N}(2)}\rangle_V\right. & =&
\displaystyle\prod_{i\leq j}^2\bigl(l^{\prime
+}_{ij}\bigr){}^{n_{ij}}\textstyle\left(\frac{l^{\prime
+}_1}{m_1}\right){}^{ n_1}\bigl(t^{\prime +}_{12}\bigr){}^{p_{12}}
\left|A_{\vec{0}_{ij},0,0,{n}_2}\rangle_V\right. \ =\  n_2p_{12} \left|\vec{N}(2) -\delta_{s,12}\rangle_V\right.
\nonumber
\end{eqnarray}
\begin{eqnarray}
&\hspace{-1em}&\hspace{-1em}  -\frac{m_1}{m_2}
\sum\limits^{\left[n_2/2\right]}_{{}^1m=0}\left(\frac{-2r}{
    m_2^2}\right)^{{}^1m}C^{n_2}_{2{}^1m +1}\left|\vec{N}(2)+{}^1m
    \delta_{ij,22}+\delta_{s,1}-(2{}^1m+1)\delta_{s,2}
    \rangle_V\right.  \nonumber\\
        &\hspace{-1em}&\hspace{-1em} -\sum\limits^{\left[n_2/2\right]}_{{}^1m=1}\left(\frac{-2r}{
    m_2^2}\right)^{{}^1m}\Biggl\{ C^{n_2}_{2{}^1m}\left|\vec{N}(2)+\delta_{ij,11} +({}^1m-1)
    \delta_{ij,22}+\delta_{s,12}-2{}^1m\delta_{s,2} \rangle_V\right.\nonumber\\
   &\hspace{-1em}&\hspace{-1em} -\Bigl[C^{n_2}_{2{}^1m}
     (h^2-h^1+2p_{12})+
     C^{n_2}_{2{}^1m+1}\Bigr]\left|\vec{N}(2)\hspace{-0.1em}+\hspace{-0.1em}
     \delta_{ij,12}\hspace{-0.1em}+\hspace{-0.1em}({}^1m-1)\delta_{ij,22}\hspace{-0.1em}
     -\hspace{-0.1em}
     2{}^1m\delta_{s,2}\rangle_V\right.\nonumber\\
     &\hspace{-1em}&\hspace{-1em} + p_{12}C^{n_2}_{2{}^1m}
     (h^2-h^1+p_{12}-1)\left|\vec{N}(2)+{}^1m\delta_{ij,22}-\delta_{s,12}-
     2{}^1m\delta_{s,2}\rangle_V\right.
 \Biggr\}\nonumber\\
&\hspace{-1em}&\hspace{-1em} -
\sum\limits^{\left[n_2/2\right]}_{{}^{1}m=0}
  \sum\limits^{\left[n_2/2-({}^{1}m + 1)\right]}_{{}^1l=0}
 \hspace{-0.3em}\left(\frac{-2r}{
    m_2^2}\right)^{{}^{1}m+{}^{1}l +1}\hspace{-0.7em}
    C^{n_2}_{2{}^{1}m+1}\Biggl\{\Bigl[
    C^{n_2-2{}^{1}m-1}_{2{}^1l+1}
    \bigl(h^2-h^1+ 2p_{12}+n_2\nonumber\\
&\hspace{-1em}&\hspace{-1em} - 2({}^{1}m
+{}^1l+1)\bigr)-C^{n_2-2{}^{1}m -1 }_{2{}^1l+2}
    \Bigr]
   \left|\vec{N}(2)+\delta_{ij,12}+({}^1m+{}^1l)\delta_{ij,22}-2({}^{1}m\right.
   \nonumber\\
&\hspace{-1em}&\hspace{-1em} \left.
+{}^{1}l+1)\delta_{s,2}\rangle_V\right.
-p_{12}C^{n_2-2{}^{1}m-1}_{2{}^1l+1}
    \bigl(h^2-h^1+ n_2-2({}^{1}m  +{}^1l+1)+p_{12}-1\bigr)\nonumber\\
    &\hspace{-1em}&\hspace{-1em}\times \left|\vec{N}(2)+({}^1m+{}^1l+1)\delta_{ij,22}-\delta_{s,12}-
2({}^{1}m +{}^{1}l+1)\delta_{s,2}\rangle_V\right.
\nonumber\\
  &\hspace{-1em}&\hspace{-1em}-
C^{n_2-2{}^{1}m-1}_{2{}^1l+1}\left|\vec{N}(2)+\delta_{ij,11}+({}^1m+{}^1l)\delta_{ij,22}+\delta_{s,12}-
2({}^{1}m +{}^{1}l+1)\delta_{s,2}\rangle_V\right.\nonumber\\
  &\hspace{-1em}&\hspace{-1em}
+
\frac{m_1}{m_2}C^{n_2-2{}^{1}m-1}_{2{}^1l+2}\left|\vec{N}(2)\hspace{-0.2em}
+\hspace{-0.2em}
({}^1m+{}^1l+1)\delta_{ij,22}\hspace{-0.2em}+\hspace{-0.2em}
\delta_{s,1}\hspace{-0.2em}-\hspace{-0.2em} 2({}^{1}m
+{}^{1}l+\textstyle\frac{3}{2})\delta_{s,2}\rangle_V\right.
      \Biggr\}.\label{Antot}
\end{eqnarray}
Thus, the action of the negative-root vectors $t^{ \prime+}_{12}$,
$l^{ \prime+}_{2}$ on $|\vec{0}_{ij},\vec{n}_s\rangle_V$
(\ref{t+prom}), (\ref{l2+prom}), due to \emph{Lemma~2}, is completely
determined, and the corresponding terms containing $t^{ \prime}_{12}$
can be presented as follows, in terms of the derived block operators
$\hat{t}^{\prime +}_{12}$, $\hat{l}^{\prime +}_{2}$ constructed
from the primary ones $\hat{t}^{ \prime}_{12}$, namely,
\begin{eqnarray}
{\hat{t}^{\prime +}_{12}|\vec{0}_{ij},\vec{n}_s\rangle_V}  &\equiv & {t}^{\prime +}_{12}|\vec{0}_{ij},\vec{n}_s\rangle_V
=...
- \sum^{\left[n_1/2\right]}_{m=1}\left(\frac{-2r}{m_1^2}\right)^{m}
C^{n_1}_{2m}\hat{t}'_{12}\nonumber\\
{}& &\times\left|\vec{0}_{ij}+ (m-1)\delta_{ij,11} +
\delta_{ij,22},
\vec{n}_s - 2m\delta_{s,1}\rangle_   V\right., \label{t+prom2}\\
{\hat{l}^{\prime  +}_2|\vec{0}_{ij},\vec{n}_s\rangle_V} &\equiv & {l^{\prime  +}_2|\vec{0}_{ij},\vec{n}_s\rangle_V} =...
m_1\sum^{\left[n_1-1/2\right]}_{m=0}\left(\frac{-2r}{m_1^2}\right)^{m+1}
C^{n_1}_{2m+1}\nonumber\\
{}& &\times \hat{t}'_{12}\left|\vec{0}_{ij}+ m\delta_{ij,11} +
\delta_{ij,22}, \vec{n}_s - (2m+1)\delta_{s,1}\rangle_V\right..
\label{l2+prom2}
\end{eqnarray}
Combining the results (\ref{l'+lb}), (\ref{t'+b}),
(\ref{t'ntot})--(\ref{l2+prom2}), we arrive at the final representation
for the action of $t^{ \prime+}_{12}$, $l^{\prime+}_{2}$ on an arbitrary
vector $|\vec{N}(2)\rangle_V$, in terms
of the primary block operator $\hat{t}^{\prime}_{12}\,$,
\begin{eqnarray}
{{t}^{\prime +}_{12}\left|\vec{N}(2)\rangle_V\right.} & = & -
 \sum_l (3-l) n_{1l}\left| {\vec{n}}_{ij}-\delta_{ij,1l} +\delta_{ij,l2},
 \vec{n}_s\rangle_V
 \right. \nonumber
 \end{eqnarray}
\begin{eqnarray}
{}&{}&
+\sum^{\left[n_1/2\right]}_{m=0}\left(\frac{-2r}{m_1^2}\right)^m
\Bigl\{ C^{n_1}_{2m}\left|\vec{N}(2)+ m\delta_{ij,11} -
2m\delta_{s,1}+\delta_{s,12}\rangle_V\right.\nonumber
\\  {}&{}&-C^{n_1}_{2m+1} \frac{m_2}{m_1}\left|\vec{N}(2) +
m\delta_{ij,11} -
(2m+1)\delta_{s,1}+\delta_{s,2}\rangle_V\right.\Bigr\}
  \nonumber
\\  {}&{}& -
\hspace{-0.3em}\sum^{\left[n_1/2\right]}_{m=1}\hspace{-0.3em}\left(\hspace{-0.1em}
\frac{-2r}{m_1^2}\right)^m \hspace{-0.3em}\Bigl\{
\Bigl[C^{n_1}_{2m}(h^2-h^1+2p_{12}+n_2)-C^{n_1}_{2m+1}\Bigr]\nonumber\\
{}&&{}\quad \times \left|{N}(2)+
(m-1)\delta_{ij,11}+\delta_{ij,12} - 2m\delta_{s,1}\rangle_V\right.+ C^{n_1}_{2m}p_{12}(h^1-h^2\nonumber\\
{}&{}& \quad -n_2-p_{12}+1)\left|{N}(2)+
(m-1)\delta_{ij,11}+\delta_{ij,22} -
2m\delta_{s,1}-\delta_{s,12}\rangle_V\right.
\Bigr\}\nonumber\\
  &&   -
\sum^{\left[n_1/2\right]}_{m=1}\left(\frac{-2r}{m_1^2}\right)^{m}
C^{n_1}_{2m}\hat{t}'_{12}\left|\vec{N}(2) + (m-1)\delta_{ij,11} +
\delta_{ij,22} - 2m\delta_{s,1}\rangle_V\right., \label{t+fin}\\
{{l}^{\prime +}_l\left|\vec{N}(2)\rangle_V\right.} & = & \delta^{l1}
m_1\left|\vec{N}(2) + \delta_{s,1}\rangle_V\right. \nonumber\\
 {}&{}&+
\delta^{l2}\Biggl\{m_2\hspace{-1.0em}\sum^{\left[n_1+1/2\right]}_{m=0}\hspace{-0.3em}
\left(\frac{-2r}{m_1^2} \right)^{m}
\hspace{-0.5em}C^{n_1}_{2m}\left|\vec{N}(2)+
m\delta_{ij,11}\vec{N}(2) -
2m\delta_{s,1}+\delta_{s,2}\rangle_V\right.
\nonumber \\
&& +
m_1\hspace{-1.0em}\sum^{\left[n_1-1/2\right]}_{m=0}\hspace{-0.3em}
\left(\hspace{-0.2em}\frac{-2r}{m_1^2} \right)^{m+1}
\hspace{-1.0em}\Bigl[
 \Bigl(C^{n_1}_{2m+1}(h^2-h^1+2p_{12}+n_2)-C^{n_1}_{2m+2}\Bigr)\nonumber \\
&& \quad \times\left|\vec{N}(2) + m\delta_{ij,11}+\delta_{ij,12}
- (2m+1)\delta_{s,1}\rangle_V\right.+
C^{n_1}_{2m+1}\Bigl(p_{12}(h^1-h^2\nonumber
\\  {}&{}& \quad -n_2-p_{12}+1) \left|\vec{N}(2) +
m\delta_{ij,11}+\delta_{ij,22} - (2m+1)\delta_{s,1}-\delta_{s,12}\rangle_V\right.\nonumber\\
{}&{}& \quad - \left|\vec{N}(2) + (m+1)\delta_{ij,11}-
(2m+1)\delta_{s,1}+\delta_{s,12}\rangle_V\right.\Bigr)
\Bigr]\nonumber\\
 {}&{}&\hspace{-0.5em}
+
m_1\hspace{-0.7em}\sum^{\left[n_1-1/2\right]}_{m=0}\hspace{-0.5em}
\left(\hspace{-0.3em}\frac{-2r}{m_1^2}\right)^{m+1}\hspace{-1.2em}
C^{n_1}_{2m+1}
\hat{t}'_{12}\left|\hspace{-0.2em}\vec{N}(2)\hspace{-0.1em}+
\hspace{-0.1em} m\delta_{ij,11} \hspace{-0.1em}+\hspace{-0.1em}
\delta_{ij,22}\hspace{-0.1em} -\hspace{-0.1em}
(2m+1)\delta_{s,1}\rangle_V\right.\hspace{-0.4em}\Biggr\}. \label{ll+fin}
\end{eqnarray}
Equivalently, the above representation for $t^{ \prime+}_{12}$,
$l^{ \prime+}_{2}$ can be written in terms of the derived blocks
$\hat{t}^{\prime+}_{12}$, $\hat{l}^{ \prime+}_{2}$, as in the
resulting formulae (\ref{t+fin2}), (\ref{ll+fin2}),
\begin{eqnarray}
 \hspace{-1em}&\hspace{-1em}& \hspace{-1em}
{{t}^{\prime +}_{12}\left|\vec{N}(2)\rangle_V\right.}  = -
 \sum_l (3-l) n_{1l}\left| {\vec{n}}_{ij}-\delta_{ij,1l} +\delta_{ij,l2},
 \vec{n}_s\rangle_V
 \right. +\hat{t}^{\prime +}_{12}\left|\vec{N}(2)\rangle_V\right., \label{t+fin2C}\\
\hspace{-1em}&\hspace{-1em}& \hspace{-1em} {{l}^{\prime
+}_l\left|\vec{N}(2)\rangle_V\right.}  = \delta^{l1}
m_1\left|\vec{N}(2) + \delta_{s,1}\rangle_V\right. +
\delta^{l2}\hat{l}^{\prime +}_{2}\left|\vec{N}(2)\rangle_V\right.,
\label{ll+fin2C}
\end{eqnarray}
with allowance for the notation
\begin{eqnarray}
  \hspace{-1em}&\hspace{-1em}&\hspace{-1em}
  \hat{B}\left|{\vec{N}(2)}\rangle_V\right.  =
\displaystyle\prod_{i\leq j}^2\bigl(l^{\prime
+}_{ij}\bigr){}^{n_{ij}}
B\left|{\vec{0}_{ij},{\vec{n}}_s}\rangle_V\right.,\ \
B\in\{{t}^{\prime +}_{12}, {l}^{\prime +}_{2}\},  \label{Bntot}
\end{eqnarray}
introduced by presenting the vectors $l^{\prime +}_{ij}$ in a
manner which does not affect the structure of $B|{\vec{0}_{ij},
{\vec{n}}_s}\rangle_V$; see (\ref{t+prom2}), (\ref{l2+prom2}).

\vspace{1ex} \noindent \textbf{\emph{{Lemma 4.}}}
The action of the operators $l'_0,l'_{l},l'_{l2}$, $l=1,2$ on the vector
$|\vec{0}_{ij},0,p_{12},{n}_2\rangle_V$ = $\bigl| p_{12},
{n}_2\rangle_V$ is given by
\begin{eqnarray}
\hspace{-1.5em}&\hspace{-1.5em}&\hspace{-1.5em}
l'_{0}\Big|p_{12},{n}_2\Big\rangle_V = m_0^2
\left|{\vec{0}_{ij},0,p_{12},{n}_2}\rangle_V\right.
-r\sum^{\left[n_2-1/2\right]}_{m=0}\left(
\frac{-8r}{m_2^2} \right)^{m}
\Bigl\{\Bigl[C^{n_2}_{2m+1}(2h^2+2p_{12}-1)\nonumber\\
&&  +
2C^{n_2}_{2m+2}\Bigr]\left|0,0,m,0, p_{12}, {n}_2-2m \rangle_V\right.
 +2\frac{m_1}{m_2}C^{n_2}_{2m+1}\left|0,0,m,1,
p_{12}+1, {n}_2-2m-1
\rangle_V\right.\Bigr\}\nonumber\\
&& +\frac{1}{2}
\sum^{\left[n_2-1/2\right]}_{m=0}\left(
\frac{-8r}{m_2^2} \right)^{m+1}
\hspace{-0.7em}C^{n_2}_{2m+2}\Bigl\{\Bigl(m_0^2-r\bigl[h^2(h^2-4)+h^1+p_{12}(2h^1-3-p_{12})\bigr]\Bigr)\nonumber\\
&&\quad \times\Big|0,0,m+1,0, p_{12}, {n}_2-2(m+1)\rangle_V
+r\Big|1,0,m,0, p_{12}+2, {n}_2-2(m+1)\rangle_V
\nonumber\\
&&\quad + 2r(h^1-p_{12}-2)\left.|0,1,m,0,
p_{12}+1, {n}_2-2(m+1) \rangle_V\right.  \Bigr\}\nonumber\\
&&
+2r\hspace{-0.4em}\sum^{\left[n_2-1/2\right]}_{m=0}
\sum^{\left[n_2/2-m-1\right]}_{l=0}\left(
\frac{-8r}{m_2^2} \right)^{m}
\left(\frac{-2r}{m_2^2}
\right)^{l+1}\hspace{-0.7em}C^{n_2}_{2m+1}\Biggl\{
\Bigl[C^{n_2-2m-1}_{2l+1}\bigl(h^2-h^1+2p_{12}+n_2
\nonumber\\
&& \quad -2(m+l)\bigr) - C^{n_2-2m-1}_{2l+2}\Bigr]\Big|0,1,m+l,0, p_{12}+1,
{n}_2-2(m+l+1)\rangle_V  \nonumber\\
&& - C^{n_2-2m-1}_{2l+1} \Bigl[\Big|1,0,m+l,0, p_{12}+2,
{n}_2-2(m+l+1)\rangle_V -\bigl(p_{12}+1\bigr)
 \nonumber\\
&&\quad \times\bigl[h^1-h^2 - n_2-p_{12} + 2(m+l+1)\bigr]\Big|0,0,m+l+1,0, p_{12},
{n}_2-2(m+l+1)\rangle_V   \Bigr]\nonumber\\
&& \quad
+  C^{n_2-2m-1}_{2l+2} \frac{m_1}{m_2}\Big|0,0,m+l+1,1,
p_{12}+1,
{n}_2-2(m+l+1)-1\rangle_V\Biggr\}\nonumber\\
&& \quad
+2r\sum^{\left[n_2-1/2\right]}_{m=0}\sum^{\left[n_2/2-m-1\right]}_{l=0}\left(\frac{-8r}{m_2^2}
\right)^{m} \left(\frac{-2r}{m_2^2} \right)^{l+1}C^{n_2}_{2m+1}C^{n_2-2m-1}_{2l+1}
\nonumber\\
&&\qquad
\times \widehat{t}'_{12}\Big|0,0,m+l+1,0, p_{12}+1,
{n}_2-2(m+l+1)\Big\rangle_V ,\label{l0nn2}\\
\hspace{-1.5em}&\hspace{-1.5em}&\hspace{-1.5em}
l'_{1}\Big|p_{12},{n}_2\Big\rangle_V =
- \frac{m_2}{2}
\sum^{\left[n_2-1/2\right]}_{m=0}\left(
\frac{-2r}{m_2^2} \right)^{m+1}
\Bigl\{C^{n_2}_{2m+1}(h^1+h^2-2)+
C^{n_2}_{2m+2}\Bigr\}\nonumber\\
&&\qquad\qquad \times\Big|0,0,m,0, p_{12}+1, {n}_2-2m-1\Big\rangle_V,
\label{l1nn2}\\
\hspace{-1.5em}&\hspace{-1.5em}&\hspace{-1.5em}
l'_{2}\Big|p_{12},{n}_2\Big\rangle_V = \sum^{\left[n_2/2\right]}_{m=0}
\left(\frac{-8r}{m_2^2} \right)^{m}\hspace{-0.5em}C^{n_2}_{2m+1}
\frac{[m_0^2-rh^2(h^2-3)]}{m_2} \Big|0,0,m,0, p_{12}, {n}_2-2m-1\Big\rangle_V
\nonumber\\
&&+\frac{m_2}{4}\hspace{-0.3em} \sum^{\left[n_2/2\right]}_{m=1}
\left(\frac{-8r}{m_2^2}
\right)^{m}
\Bigl\{C^{n_2}_{2m}(2h^2-1)  + 2C^{n_2}_{2m+1}\Bigr\}
\Big|0,0,m-1,0,
p_{12}, {n}_2-2m+1\Big\rangle_V\nonumber\\
 && \qquad - p_{12}
l'_{1}\Big|{\vec{0}_{ij},0,p_{12}-1,{n}_2}\Big\rangle_V
\nonumber
\\
&&
-  \sum^{\left[n_2/2\right]}_{m=0} \left(\frac{-2r}{m_2^2}
\right)^{m}(4^m-1)C^{n_2}_{2m}\Bigl\{\frac{m_1}{2}\Big|0,0,m-1,1,p_{12}+
1, {n}_2-2m\rangle_V \nonumber\\
&& \quad - p_{12}\frac{m_2}{2} \Big|0,0,m-1,0, p_{12},
{n}_2-2m+1\rangle_V\Bigr\} \nonumber
 \end{eqnarray}
  \begin{eqnarray}
  &&  + \frac{m_2}{2}
\sum^{\left[n_2/2\right]}_{m=0} \left(\frac{-2r}{m_2^2}
\right)^{m+1}(4^m-1)C^{n_2}_{2m+1}\Bigl\{\Bigl[(2p_{12}+1)h^1-h^2-3p_{12}-
p_{12}^2\Bigr]\nonumber
\\ && \quad \times\Big|0,0,m,0,
p_{12}, {n}_2-2m-1\rangle_V -
\Big|1,0,m-1,0,p_{12}+ 2,
{n}_2 -2m -1\Big\rangle_V \nonumber\\
{}&& \qquad  - 2(h^1- p_{12} -2)\Big|0,1,m-1,0,p_{12}+1, {n}_2 -2m -1
\Big\rangle_V
  \Bigr\} \nonumber
 \\
&&  - \frac{m_2}{2} \sum^{\left[n_2/2\right]}_{m=0}
\sum^{\left[n_2-2m-1/2\right]}_{l=0}\left(\frac{-2r}{m_2^2}
\right)^{m+l+1}\hspace{-1em}(4^m-1)C^{n_2}_{2m}\nonumber
\\ && \qquad \times\Biggl\{
\Bigl[C^{n_2-2m}_{2l+1}\bigl(h^2-h^1+2p_{12}+n_2-
2(m+l)+1\bigr) -C^{n_2-2m}_{2l+2}\Bigr]\nonumber\\
&& \qquad \times\Big|0,1,m+l-1,0, p_{12}+1, {n}_2 -2(m+l)
-1\Big\rangle_V\nonumber\\
&& \qquad - C^{n_2-2m}_{2l+1}\Bigl[\Big|1,0,m + l -1,0, p_{12}+2,
{n}_2 -2(m+l) -1\Big\rangle_V + \bigl(h^2-h^1+{n}_2
\nonumber\\
&& \qquad  -2(m+l)-1\bigr)\Big|0,0,m + l,0, p_{12},{n}_2
-2(m+l)-1
\Big\rangle_V\Bigr]\nonumber\\
&& \qquad + \frac{m_1}{m_2}C^{n_2-2m}_{2l+2}
\Big|0,0,m + l , 1,
p_{12} +1, {n}_2
-2(m+l+1)\Big\rangle_V \Biggr\}\nonumber\\
&&+  \sum^{\left[n_2/2\right]}_{m=0}
\sum^{\left[n_2-2m-1/2\right]}_{l=0}\left(\frac{-2r}{m_2^2}
\right)^{m+l+1}(4^m-1)C^{n_2}_{2m}
C^{n_2-2m}_{2l+1}\Biggl\{p_{12}\frac{m_2}{2}[h^2-h^1 \nonumber\\
&& \qquad +n_2+p_{12}- 2(m+l)]\Big|0,0,m + l ,0, p_{12},
{n}_2 -2(m+l) -1\Big\rangle_V\nonumber\\
&& \qquad- \frac{m_2}{2}  \widehat{t}'_{12}
 \Big|0,0,m + l , 0, p_{12}+1, {n}_2
-2(m+l)-1\Big\rangle_V\Biggr\}
,\label{l2nn2}\\
\hspace{-1.5em}&\hspace{-1.5em}&\hspace{-1.5em}
 l'_{12}\Big|p_{12},{n}_2\Big\rangle_V = \frac{1}{4}\sum^{\left[n_2/2\right]}_{m=1}\left(\frac{-2r}{m_2^2}
\right)^{m} \Bigl\{C^{n_2}_{2m}(h^1+h^2-2)+
C^{n_2}_{2m+1}\Bigr\}\nonumber\\
&&\qquad\qquad\quad\times\Big|0,0,m-1,0, p_{12}+1,
{n}_2-2m\Big\rangle_V,\label{l12nn2} \\
\hspace{-1.5em}&\hspace{-1.5em}&\hspace{-1.5em}
 l'_{22}\Big|p_{12},{n}_2\Big\rangle_V =
-2 {p_{12}}l'_{12}\Big| p_{12}-1, {n}_2\Big\rangle_V \nonumber\\
&&
-\hspace{-0.7em}\sum^{\left[(n_2-1)/2\right]}_{m=0}\left(\frac{-8r}{m_2^2} \right)^{m}
C^{n_2}_{2m+2}\frac{\bigl[m_0^2-r
h^2(h^2-3)\bigr]}{m_2^2}\Big|0,0,m,0,p_{12},{n}_2-2(m+1)
\Big\rangle_V
\nonumber\\
&&-\frac{1}{4}\sum^{\left[(n_2-1)/2\right]}_{m=1}
\left(\frac{-8r}{m_2^2}\right)^{m}
\Bigl\{C^{n_2}_{2m+1}(2h^2-1)
+
 2C^{n_2}_{2m+2}\Bigr\}
\Big|0, 0, m-1,
0,
p_{12},{n}_2-2m\Big\rangle_V
\nonumber
\\  {}&{}&+\frac{1}{2}\sum^{\left[(n_2-1)/2\right]}_{m=0}
\left(\frac{-2r}{m_2^2}
\right)^{m+1}\hspace{-1em}(4^{m}-1)\Biggl\{{C^{n_2}_{2m+2}}
\Bigl[\Big|1,0,m-1, 0,p_{12}+2,n_2-2(m+1)\Big\rangle_V\nonumber\\
{}&{}&\quad + 2\bigl((h^1-2)-p_{12}\bigr) \Big|0,1,m-1,
0,p_{12}+1,n_2-2(m+1)\Big\rangle_V
 \nonumber
 \end{eqnarray}
  \begin{eqnarray}
{}&{}& \quad + \bigl(h^2-h^1 -p_{12}(2h^1-p_{12}-3)\bigr)\Big|0,0,m,0, p_{12}, {n}_2-2(m+1)\Big\rangle_V\Bigr]\nonumber\\
{}&{}& \quad +{C^{n_2}_{2m+1}}\Bigl[
\frac{m_1}{m_2} \Big|0,0,m-1, 1,p_{12}+1,n_2-2m-1)\Big\rangle_V\nonumber\\
{}&{}& \qquad -p_{12} \Big|0,0,m-1,0, p_{12},
{n}_2-2m\Big\rangle_V \Bigr] \Biggr\}\nonumber\\
{}&{}&
 +
\frac{1}{2}\sum^{\left[(n_2-1)/2\right]}_{m=0}\sum^{\left[n_2/2
-m-1\right]}_{l=0}\left(\frac{-2r}{m_2^2}\right)^{m+l+1}(4^m-1)C^{n_2}_{2m+1}\Biggl\{
\Bigl[C^{n_2-2m-1}_{2l+1}(h^2-h^1+2p_{12}
 \nonumber\\
&{}& \quad  +n_2-2(m+l))- C^{n_2-2m-1}_{2l+2} \Bigr]\Big|0,1,m+l-1,
0,p_{12}+1,{n}_2-2(m+l+1)\Big\rangle_V  \nonumber\\
&& \quad +C^{n_2-2m-1}_{2l+1}\Bigl[
(p_{12}+1)\Bigl(h^1-h^2-p_{12}-{n}_2+2(m+l+1)\Bigr)\Bigr]\nonumber\\
&&\qquad\times\Big|{0},0,m+l,
0, p_{12},{n}_2-2(m+l+1)\Big\rangle_V \nonumber\\
&{}& - C^{n_2-2m-1}_{2l+1} \Big|1,0,m+l-1,0, p_{12}+2,
{n}_2-2(m+l+1)\Big\rangle_V
\nonumber\\
&{}&+ C^{n_2-2m-1}_{2l+2}\frac{m_1}{m_2}\Big|0,0,m+l,1, p_{12}+1,
{n}_2-2(m+l+1)-1\Big\rangle_V \Biggr\}\nonumber\\
&{}& +
\frac{1}{2}\sum^{\left[(n_2-1)/2\right]}_{m=0}\sum^{\left[n_2/2
-m-1\right]}_{l=0}\left(\frac{-2r}{m_2^2}\right)^{m+l+1}(4^m-1)
C^{n_2}_{2m+1}C^{n_2-2m-1}_{2l+1}
\nonumber\\
&&\qquad\times \widehat{t}'_{12}\Big|0,0,m+l,0, p_{12}+1,
{n}_2-2(m+l+1)\Big\rangle_V. \label{l22nn2}
\end{eqnarray}

The deduction of the above auxiliary formulae
(\ref{l0nn2})--(\ref{l22nn2}) is straightforward and relies on
(\ref{l_ii l_1+b}), (\ref{l_i l_i+b}),
(\ref{l1 l2+b})--(\ref{l0li+b}), (\ref{K00012b})--(\ref{X12iV}).

\vspace{1ex}

Using the results of \emph{Lemma~4}, we  introduce a {\it new
derived block operator}, by the following rule, similar to that
given by (\ref{Bntot}), for $l=1,2$:
\begin{eqnarray}
  \hspace{-1em}&\hspace{-1em}&\hspace{-1em}
  \hat{C}\left|{\vec{N}(2)}\rangle_V\right.  =
\displaystyle\prod_{i\leq j}^2\bigl(l^{\prime
+}_{ij}\bigr){}^{n_{ij}}\textstyle\left(\frac{l^{\prime
+}_1}{m_1}\right)^{n_1}
C\left|{\vec{0}_{ij},0,p_{12},{n}_2}\rangle_V\right.,\  C\in\{
{l}^{\prime }_{l2}, {l}^{\prime }_{l}, l'_0\}.  \label{Cntot}
\end{eqnarray}
This allows one to completely establish the operator action on the vectors
that remain undetermined by (\ref{l0finnV})--(\ref{l22totnV}),
due to the following

\noindent \textbf{\emph{{Lemma 5.}}} The action of the operators $l'_0, l'_{l}, l'_{ij}$,
for $i,j,l=1,2$ and $i\leq j$, on the vector $|{\vec{0}_{ij},{\vec{n}}_s}\rangle_V
= |{\vec{n}}_s\rangle_V$ is given by
\begin{eqnarray}
  \hspace{-1.5em}&\hspace{-1.5em}&\hspace{-1.5em}
  {l^{\prime  }_0\Big|\vec{n}_s\Big\rangle_V} =
\widehat{l}^{\prime  }_0\Big|{n}_s\rangle_V +
\frac{m_1}{2}\displaystyle\sum\limits^{\left[n_1-1/2\right]}_{m=0}\left(
\frac{-8r}{m_1^2}\right)^{m+1}C^{n_1}_{2m+1}\Bigl\{ \widehat{l}^{\prime
}_2\Big|(m,1,0),\vec{n}_s-(2m+1)\delta_{s,1}\Big\rangle_V\nonumber
\\
&&  +
\widehat{l}^{\prime
}_1\Big|(m+1,0,0),\vec{n}_s-(2m+1)\delta_{s,1}\Big\rangle_V
- \frac{m_2}{2}\widehat{t}^{\prime}_{12}
\Big|(m,0,0),\vec{n}_s-(2m+1)\delta_{s,1}+\delta_{s,2}
\Big\rangle_V
\nonumber\\
&&- \frac{m_2}{2}p_{12}(h^1-h^2-n_2-p_{12})
\Big|(m,0,0),\vec{n}_s-(2m+1,1,-1)\Big\rangle_V\Bigr\}
  \nonumber
 \\
&& - r
\displaystyle\sum\limits^{\left[n_1-1/2\right]}_{m=0}\left(
\frac{-8r}{m_1^2}\right)^{m}\Bigl\{
C^{n_1}_{2m+1}\bigl[2(h^1-p_{12})-3\bigr] +
2C^{n_1}_{2m+2}\Bigr\}\Big|(m,0,0),\vec{n}_s-2m\delta_{s,1}
\Big\rangle_V\nonumber
\end{eqnarray}
  \begin{eqnarray}
&& +
\frac{1}{2}\displaystyle\sum\limits^{\left[n_1-1/2\right]}_{m=0}\left(
\frac{-8r}{m_1^2}\right)^{m+1}C^{n_1}_{2m+2}
\Biggl\{\widehat{l}^{\prime
}_0\Big|(m+1,0,0),\vec{n}_s-(2m+2)\delta_{s,1}\Big\rangle_V
\nonumber
\\
&&
-r\Bigl([h^1-p_{12}][h^1-
p_{12}-2]-h^2
-p_{12} - n_2
+2p_{12} (h^1-
h^2-n_2 - p_{12}+1)\Bigr)\nonumber\\
&& \ \times
\Big|(m+1,0,0),\vec{n}_s-(2m+2)\delta_{s,1}
\Big\rangle_V
-2r\widehat{t}'_{12}\Big|(m+1,0,0),\vec{n}_s-
(2m+2,-1,0)\Big\rangle_V \nonumber\\
&& + 8r \widehat{l}^{\prime
}_{12}\Big|(m+1,1,0),\vec{n}_s\hspace{-0.2em}\hspace{-0.2em}-(2m\hspace{-0.2em}+2)\delta_{s,1}\hspace{-0.2em}
\Big\rangle_V+4r\widehat{l}^{\prime
}_{22}\Big|(m,2,0),\vec{n}_s\hspace{-0.2em}\hspace{-0.2em}-(2m\hspace{-0.2em}+2)\delta_{s,1}\hspace{-0.1em}
\Big\rangle_V\nonumber\\
&&
+2r(h^2+n_2+p_{12}-2)\widehat{t}'_{12}\Big|(m,1,0),\vec{n}_s-(2m+2)\delta_{s,1}
\Big\rangle_V
 \nonumber\\
&&+2rp_{12}(h^2\hspace{-0.2em}+n_2\hspace{-0.2em}+p_{12}\hspace{-0.2em}-2)
(h^1\hspace{-0.2em}-h^2\hspace{-0.2em}-n_2\hspace{-0.2em}-p_{12}
+1)\Big|(m,1,0),\vec{n}_s\hspace{-0.2em}-(2m+2,1,0)
\Big\rangle_V
 \Biggr\}
\nonumber \\
&& +
\frac{r}{2}\displaystyle\sum\limits^{\left[n_1-1/2\right]}_{m=0}\left(
\frac{-8r}{m_1^2}\right)^{m+1}C^{n_1}_{2m+2}
\Biggl\{p_{12}(p_{12}-1)(h^1-h^2-n_2-p_{12}+2)\nonumber\\
&& \times (h^1-h^2-n_2-p_{12}+1)\Big|(m,0,1),\vec{n}_s-(2m+2,2,0)
\Big\rangle_V \nonumber\\
&&+ p_{12}\textstyle(h^1-h^2-n_2-p_{12}+1)\widehat{t'}_{12}
\Big|(m,0,1),\vec{n}_s-(2m+2, 1,0) \Big\rangle_V
\nonumber\\
&&
+\sum\limits^{\left[(n_2-1)/2\right]}_{k=0}\Biggl[\sum\limits^{
  \left[n_2/2\right]}_{{}^{1}m=0}
  \sum\limits^{\left[n_2/2-({}^{1}m + 1)\right]}_{{}^1l=0}......
  \hspace{-1em}\sum\limits^{\left[n_2/2-\sum\limits^{k-1}_{i=1}({}^{i}m + {}^il)
  -(k-1)\right]}_{{}^{k}m=0}
  \sum\limits^{\left[n_2/2-\sum\limits^{k-1}_{i=1}({}^{i}m + {}^il)-{}^{k}m)
  -k \right]}_{{}^kl=0}\hspace{-0.7em}(-1)^k  \nonumber\\
  &&\quad \times
  \left(\frac{-2r}{m_2^2}\right)^{\sum_{i=1}^k({}^im+{}^il)+k}
C^{n_2}_{2{}^1m+1}C^{n_2-2{}^1m-1}_{2{}^1l+1}...\nonumber\\
  &&\quad \times
C^{n_2-2(\sum_{i=1}^{k-1}({}^im+{}^il)+k-1)}_{2{}^km+1}C^{n_2-2(\sum_{i=1}^
{k-1}({}^im+{}^il+k-1)-2{}^km-1}_{2{}^kl+1}\nonumber \\
  &&
   \times \widehat{t'_{12} \Big|A}_{(m,0,\sum_{i=1}^k({}^im+{}^il)+k+1),\vec{n}_s-(2m+2)\delta_{s,1}
 -2[\sum\limits^k_{i=1}({}^{i}m + {}^il)
  +k]\delta_{s,2}}\Big\rangle_V \Biggr]\Biggr\} ,\label{l0ns}\\
 \hspace{-1.5em}&\hspace{-1.5em}&\hspace{-1.5em}{l^{\prime}_1
 \Big|\vec{n}_s\Big\rangle_V} =
 \sum\limits^{\left[n_1/2\right]}_{m=0}
 \left(\frac{-8r}{m_1^2}\right)^{m}{C^{n_1}_{2m}}\widehat{l}^{\prime _1}\left|
 \vec{0}_{ij} +m\delta_{ij,11},\vec{n}_s-2m\delta_{s,1}\rangle_V\right.\nonumber\\
 && \qquad
 + \frac{1}{m_1}\displaystyle\sum\limits^{\left[n_1/2\right]}_{m=0}\left(
 \frac{-8r}{m_1^2}\right)^{m}{C^{n_1}_{2m+1}}\Biggl\{\widehat{l}^{\prime}_0
 \Big|(m,0,0), n_s-(2m+1, 0, 0)\Big\rangle_V \nonumber\\
 && \qquad -r \Bigl[(h^1-p_{12})(h^1-p_{12}-2) -
 h^2-p_{12}-n_2\Bigr]\Big|(m,0,0),\vec{n}_s-(2m+1,0,
 0) \Big\rangle_V\nonumber\\
 && \qquad
 -rp_{12}(h^1-h^2-n_2-p_{12}+1)\Big|(m,0,0),\vec{n}_s-(2m+1,0,
  0) \Big\rangle_V
 \nonumber\\
 && \qquad
 -r\widehat{t}'_{12}\Big|(m,0,0),\vec{n}_s-(2m+1, -1,0)
 \Big\rangle_V \Biggr\}\nonumber
 \\
 && - {4m_1}\displaystyle\sum\limits^{\left[n_1/2\right]}_{m=0}
 \left( \frac{-2r}{m_1^2}\right)^{m+1}
 {C^{n_1}_{2m+1}}\textstyle\left(4^m-\frac{1}{2}\right)
 \widehat{l}'_{12}\Big|(m,1,0),\vec{n}_s-(2m+1)\delta_{s,1}\Big\rangle_V\nonumber
 \end{eqnarray}
 \begin{eqnarray}
 &&
 + \frac{m_1}{4}\displaystyle\sum\limits^{\left[n_1/2\right]}_{m=1}\left(
 \frac{-8r}{m_1^2}\right)^{m}\Bigl\{{C^{n_1}_{2m}}(2(h^1-p_{12})-1)
 + 2{C^{n_1}_{2m+1}}\Bigr\}\nonumber\\
 && \qquad \times\Big|(m-1,0,0),\vec{n}_s-(2m-1,0,0)
 \Big\rangle_V\nonumber
\\
&& +
\frac{1}{2}\displaystyle\sum\limits^{\left[n_1/2\right]}_{m=1}\left(
\frac{-2r}{m_1^2}\right)^{m}(4^m-1){C^{n_1}_{2m}}\Bigl\{2
\widehat{l}^{\prime }_2\Big|(m-1,1,0),\vec{n}_s-(2m, 0, 0)
\Big\rangle_V
\nonumber \\
&&\ -
m_2\widehat{t}^{\prime}_{12}\Big|(m\hspace{-0.2em}-1,0,0),\vec{n}_s\hspace{-0.2em}-(2m,0,-1)
\Big\rangle_V-m_1\Big|(m\hspace{-0.2em}-1,0,0),\vec{n}_s-(2m-1)\delta_{s1}\Big\rangle_V\nonumber\\
&& \ -
m_2p_{12}(h^1-h^2-n_2-p_{12})\Big|(m-1,0,0),\vec{n}_s-(2m,1,-1)
\Big\rangle_V\Bigr\}
\nonumber\\
&&
+\frac{m_1}{2}\displaystyle\sum\limits^{\left[n_1/2\right]}_{m=1}
\left( \frac{-2r}{m_1^2}\right)^{m+1}(4^m-1)
  {C^{n_1}_{2m+1}}\Bigl\{p_{12}(h^1-h^2-n_2-p_{12}+1)
  \nonumber\\
  && \times \Big|(m,0,0),
\vec{n}_s-(2m+1)\delta_{s,1}\Big\rangle_V +
\widehat{t}'_{12}\Big|(m,0,0),
\vec{n}_s-(2m+1)\delta_{s,1}+\delta_{s,12}\Big\rangle_V\nonumber\\
&& -
  4\widehat{l}^{\prime }_{22}\Big|\vec{0}_{ij}+(m-1)\delta_{ij,11}+
2\delta_{ij,12},
\vec{n}_s-(2m+1)\delta_{s,1}\Big\rangle_V\nonumber\\
&&-2 (h^2+n_2+p_{12}-2)\Bigl[p_{12}(h^1-h^2-n_2-p_{12}+1)\nonumber\\
&& \quad
\times\Big|\vec{0}_{ij}+(m-1)\delta_{ij,11}+\delta_{ij,12},
\vec{n}_s-(2m+1)\delta_{s,1}-\delta_{s,12}\Big\rangle_V \nonumber\\
&& +
\widehat{t}'_{12}\Big|\vec{0}_{ij}+(m-1)\delta_{ij,11}+\delta_{ij,12},
\vec{n}_s-(2m+1)\delta_{s,1}\Big\rangle_V \Bigr]
  \Bigr\}\nonumber\\
&& -
\frac{m_1}{2}\displaystyle\sum\limits^{\left[n_1/2\right]}_{m=1}\left(
\frac{-2r}{m_1^2}\right)^{m+1}\hspace{-0.5em}(4^m-1)
  {C^{n_1}_{2m+1}}
\Biggl\{p_{12}(p_{12}-1)(h^1-h^2-n_2-p_{12}+2)\nonumber\\
&& \quad\times
(h^1-h^2-n_2-p_{12}+1)\Big|(m-1,0,1),\vec{n}_s-(2m+1)\delta_{s,1}
-2\delta_{s,12} \Big\rangle_V \nonumber\\
&&+ p_{12}(h^1-h^2-n_2-p_{12}+1)\widehat{t'}_{12}
\Big|(m-1,0,1),\vec{n}_s-(2m+1)\delta_{s,1}-\delta_{s,12}
\Big\rangle_V
\nonumber\\
&&
+\sum\limits^{\left[(n_2-1)/2\right]}_{k=0}\Biggl[\sum\limits^{
  \left[n_2/2\right]}_{{}^{1}m=0}
  \sum\limits^{\left[n_2/2-({}^{1}m + 1)\right]}_{{}^1l=0}... \nonumber\\
  &&
  ...\sum\limits^{\left[n_2/2-\sum\limits^{k-1}_{i=1}({}^{i}m + {}^il)
  -(k-1)\right]}_{{}^{k}m=0}
  \sum\limits^{\left[n_2/2-\sum\limits^{k-1}_{i=1}({}^{i}m + {}^il)-{}^{k}m)
  -k \right]}_{{}^kl=0}\hspace{-0.5em}(-1)^k
  \left(\frac{-2r}{m_2^2}\right)^{\sum_{i=1}^k({}^im+{}^il)+k}\nonumber\\
  &&
\times C^{n_2}_{2{}^1m+1}C^{n_2-2{}^1m-1}_{2{}^1l+1}...
C^{n_2-2(\sum_{i=1}^{k-1}({}^im+{}^il)+k-1)}_{2{}^km+1}C^{n_2-2(\sum_{i=1}^{k-1}(
{}^im+{}^il)+k-1)-2{}^km-1}_{2{}^kl+1}\nonumber \\
  &&
   \times \widehat{t'_{12} \Bigl|A}_{(m-1,0,\sum_{i=1}^k({}^im+{}^il)+k+1),\vec{n}_s-(2m+1)\delta_{s,1}
-2[\sum\limits^k_{i=1}({}^{i}m + {}^il)
  +k]\delta_{s,2}}\Bigr\rangle_V\Biggr]\Biggr\},\label{l1ns}
  \\
  \hspace{-1.5em}&\hspace{-1.5em}&\hspace{-1.5em}{l^{\prime  }_2\left|\vec{0}_{ij},\vec{n}_s\rangle_V\right.} =
\displaystyle\sum\limits^{\left[n_1+1/2\right]}_{m=0}\left(
\frac{-2r}{m_1^2}\right)^{m}C^{n_1}_{2m}\widehat{l}^{\prime 2
}\Big|\vec{0}_{ij}+m\delta_{ij,11},\vec{n}_s-2m\delta_{s,1}\Big\rangle_V \nonumber \\
&& -2m_1
\displaystyle\hspace{-0.3em}\sum\limits^{\left[n_1-1/2\right]}_{m=0}\left(
\frac{-2r}{m_1^2}\right)^{m+1}\hspace{-0.3em}
\Bigl\{C^{n_1}_{2m+1}\widehat{l}^{\prime
}_{12}\Big|\vec{0}_{ij}+(m+1)\delta_{ij,11},\vec{n}_s-(2m+1)
\delta_{s,1}
\Big\rangle_V\nonumber
\end{eqnarray}
\begin{eqnarray}
&& \quad+ C^{n_1}_{2m+1}\widehat{l}^{\prime
}_{22}\Big|\vec{0}_{ij}+m\delta_{ij,11}+\delta_{ij,12},
\vec{n}_s-(2m+1)\delta_{s,1}
\Big\rangle_V \nonumber \\
    &&\quad +\frac{1}{4}\Bigl[ C^{n_1}_{2m+1}(h^1+h^2+n_2 -
2)+
C^{n_1}_{2m+2}\Bigr]\widehat{t}'_{12}\Big|\vec{0}_{ij}+m\delta_{ij,11},
\vec{n}_s-(2m+1)\delta_{s,1}\Big\rangle_V   \nonumber\\
&& \quad + \frac{p_{12}}{4}\Bigl( C^{n_1}_{2m+1}(h^1+h^2+n_2 -
2)+  C^{n_1}_{2m+2}\Bigr)(h^1-h^2-n_2-p_{12}+1)\nonumber \\
  && \quad \times \Big|\vec{0}_{ij}+m\delta_{ij,11},
\vec{n}_s-(2m+1)\delta_{s,1}-\delta_{s,12}\Big\rangle_V
\Bigr\}
  ,\label{l2ns}\\
 \hspace{-1.5em}&\hspace{-1.5em}&\hspace{-1.5em}{l^{\prime}_{11}
 \left|\vec{0}_{ij},\vec{n}_s\rangle_V\right.} =
\frac{1}{2}\sum^{\left[n_1-1/2\right]}_{m=0}
 \left(\frac{-8r}{m_1^2}
\right)^{m+1}C^{n_1}_{2m+2}\widehat{l}^{\prime}_{12}
\Big|(m,1,0), \vec{n}_s -2(m+1)\delta_{s1}
\rangle_V\nonumber\\
&& -\frac{1}{8}\sum^{\left[n_1-1/2\right]}_{m=0}
 \left(\frac{-8r}{m_1^2}
\right)^{m+1}C^{n_1}_{2m+2} \Bigl\{p_{12}(h^1-h^2-p_{12}-n_2+1)+(h^1-p_{12}) \nonumber\\
&&  \quad \times(h^1-p_{12} -2)- h^2 -p_{12}-n_2\Bigr\}\Big|
\vec{0}_{ij}+m\delta_{ij,11},
\vec{n}_s-2(m+1)\delta_{s1}\rangle_V\nonumber\\
&&  -\frac{1}{m^2_1}\sum^{\left[n_1-1/2\right]}_{m=0}
\left(\frac{-8r}{m_1^2} \right)^{m}  C^{n_1}_{2m+2} \widehat{l}_0'
\left|\vec{0}_{ij}+m\delta_{ij,11}, \vec{n}_s-2(m+1)\delta_{s1}
\rangle_V\right.\nonumber\\
&& - \frac{1}{8}\sum^{\left[n_1-1/2\right]}_{m=0}
\left(\frac{-8r}{m_1^2} \right)^{m+1}
C^{n_1}_{2m+2}\widehat{t}'_{12}\left|
\vec{0}_{ij}+m\delta_{ij,11}, \vec{n}_s-2(m+1)\delta_{s1}+
\delta_{s,12}\rangle_V\right.\nonumber\\
&& - \frac{1}{4}\hspace{-0.4em}\sum^{\left[n_1-1/2\right]}_{m=1}
\hspace{-0.4em}\left(\hspace{-0.2em}\frac{-8r}{m_1^2}
\hspace{-0.2em}\right)^{m}
\hspace{-0.5em}\left\{C^{n_1}_{2m+1}(2(h^1-p_{12})-1)+2C^{n_1}_{2m+2}
\hspace{-0.2em}\right\}\nonumber\\
&& \qquad\times \left|\vec{0}_{ij}+(m-1)\delta_{ij,11},
\vec{n}_s-2m\delta_{s1} \rangle_V\right.
\nonumber\\
&&-\frac{1}{m_1}\sum^{\left[n_1-1/2\right]}_{m=0}\left(\frac{-8r}{m_1^2}
\right)^{m}C^{n_1}_{2m+1}\widehat{l}^{\prime
}_1\left|\vec{0}_{ij}+m \delta_{ij,11},
\vec{n}_s-(2m+1)\delta_{s1} \rangle_V\right.
\nonumber\\
&& -\frac{1}{m_1}
\sum^{\left[n_1-1/2\right]}_{m=0}\left(\frac{-2r}{m_1^2}
\right)^{m}(4^m-1) C^{n_1}_{2m+1}\widehat{l}^{\prime
}_2\Big|(m-1,1,0),\vec{n}_s-(2m+1)\delta_{s1} \rangle_V
\nonumber\\
&& +\frac{m_2}{2m_1}\hspace{-0.4em}
\sum^{\left[n_1-1/2\right]}_{m=0}\left(
\frac{-2r}{m_1^2} \right)^{m}(4^m-1)
C^{n_1}_{2m+1}\Bigl\{\widehat{t}^{\prime}_{12}
\Big|(m-1,0,0),\vec{n}_s
-(2m+1,0,-1)\rangle_V
\nonumber\\
&&\qquad +p_{12}(h^1-h^2-p_{12}-n_2+1)\Big|(m-1,0,0),\vec{n}_s-
(2m+1,1,-1) \rangle_V\Bigr\}
\nonumber\\
&&+\frac{1}{2}
\sum^{\left[n_1-1/2\right]}_{m=0}\left(\frac{-2r}{m_1^2}
\right)^{m}(4^m-1)
C^{n_1}_{2m+1}\Big|(m-1,0,0),\vec{n}_s-2m\delta_{s1}
\rangle_V
\nonumber
\\
&& +\frac{1}{2}
\sum^{\left[n_1-1/2\right]}_{m=0}
\hspace{-0.2em}\left(\frac{-2r}{m_1^2}
\right)^{m+1}(4^m-1)
pC^{n_1}_{2m+2}p_{12}(h^1-h^2-p_{12}-n_2+1)\nonumber\\
&&\quad
\times\Bigl\{2(h^2+n_2+p_{12}-2)\Big|
(m-1,1,0),\vec{n}_s-2(m+1)\delta_{s1}-\delta_{s,12}
\rangle_V
\nonumber\\
&&\quad -\Big|\vec{0}_{ij}+m\delta_{ij,11},\vec{n}_s-2(m+1)\delta_{s1}\rangle_V\Bigr\}
\nonumber\\
&&\hspace{-0.5em}-\frac{1}{2}\hspace{-0.2em}
\sum^{\left[n_1-1/2\right]}_{m=0}\left(
\frac{-2r}{m_1^2}
\right)^{m+1}\hspace{-0.3em}(4^m-1) C^{n_1}_{2m+2}
\Biggl\{\widehat{t}'_{12}\Big|(m,0,0),\vec{n}_s-2(m+1)
\delta_{s1}+ \delta_{s,12}\rangle_V
\nonumber
\end{eqnarray}
\begin{eqnarray}
&& -4 \widehat{l}^{\prime }_{12}\Big|(m,1,0),\vec{n}_s-2(m+1)
\delta_{s1}\rangle_V -4 \widehat{l}^{\prime
}_{22}\Big|(m-1,2,0),\vec{n}_s-2(m+1)
\delta_{s1}\rangle_V\nonumber\\
&& -2
(h^2+n_2+p_{12}-2)\widehat{t}'_{12}\Big|(m-1,1,0),\vec{n}_s-
2(m+1)\delta_{s1}\rangle_V\Biggr\}\nonumber\\
&& + \frac{1}{2}
\hspace{-0.3em}\sum^{\left[n_1-1/2\right]}_{m=0}\hspace{-0.2em}\left(
\frac{-2r}{m_1^2}
\right)^{m+1}\hspace{-0.5em}(4^m-1) C^{n_1}_{2m+2}
\hspace{-0.2em}\Biggl\{p_{12}(p_{12}-1)(h^1-h^2-n_2-p_{12}+2)\nonumber\\
&& \times (h^1-h^2-n_2-p_{12}+1)\Big|(m-1,0,1),\vec{n}_s-2(m+1)
\delta_{s1}-2\delta_{s,12}\rangle_V \nonumber\\
&&+ p_{12}\left(h^1-
h^2\hspace{-0.1em}-
n_2-p_{12}+1\right) \hspace{-0.2em}
\widehat{t'}_{12}\Big|(m\hspace{-0.2em}-1,0,1),\vec{n}_s\hspace{-0.2em}-2(m+1)
\delta_{s1}\hspace{-0.2em}-\delta_{s,12}\rangle_V\nonumber\\
&&
+\sum\limits^{\left[(n_2-1)/2\right]}_{k=0}\Biggl[\sum\limits^{
  \left[n_2/2\right]}_{{}^{1}m=0}
  \sum\limits^{\left[n_2/2-({}^{1}m + 1)\right]}_{{}^1l=0}... \nonumber\\
  &&
  ...\hspace{-1.5em}\sum\limits^{\left[n_2/2-\sum\limits^{k-1}_{i=1}({}^{i}m + {}^il)
  -(k-1)\right]}_{{}^{k}m=0}
  \sum\limits^{\left[n_2/2-\sum\limits^{k-1}_{i=1}({}^{i}m + {}^il)-{}^{k}m)
  -k \right]}_{{}^kl=0}\hspace{-1.0em}(-1)^k
    \hspace{-0.2em}\left(\hspace{-0.2em}\frac{-2r}{m_2^2}\hspace{-0.2em}\right)^{\sum_{i=1}^k({}^im+{}^il)+k}\nonumber\\
  &&\times\hspace{-0.2em}
C^{n_2}_{2{}^1m+1}\hspace{-0.1em}C^{n_2-2{}^1m-1}_{2{}^1l+1}...
\hspace{-0.1em}C^{n_2-2(\sum_{i=1}^{k-1}({}^im+{}^il)+k-1)}_{2{}^km+1}\hspace{-0.1em}C^{n_2-2(\sum_{i=1}^{k-1}(
{}^im+{}^il)+k-1 )-2{}^km-1}_{2{}^kl+1}\nonumber \\
  &&
   \times \widehat{t'_{12} \Bigl|A}_{(m-1,0,\sum_{i=1}^k({}^im+{}^il)+k+1),\vec{n}_s-2(m+1)
\delta_{s,1} -2[\sum\limits^k_{i=1}({}^{i}m + {}^il)
  +k]\delta_{s,2}}\Bigr\rangle_V\Biggr]\Biggr\},\label{l11ns}\\
  \hspace{-1.5em}&\hspace{-1.5em}&\hspace{-1.5em}{
  l^{\prime }_{12}\Big|\vec{n}_s\rangle_V} = -
\frac{1}{2m_1}\sum^{\left[n_1/2\right]}_{m=0}
\left(\frac{-2r}{m_1^2}
\right)^{m}C^{n_1}_{2m+1}\widehat{l}^{\prime
}_2\Big|\vec{0}_{ij} + m\delta_{ij,11},\vec{n}_s-(2m+1)\delta_{s1}\rangle_V\nonumber\\
&& + \sum^{\left[n_1/2\right]}_{m=0} \left(\frac{-2r}{m_1^2}
\right)^{m}C^{n_1}_{2m}\widehat{l}^{\prime
}_{12}\Big|\vec{0}_{ij}+m\delta_{ij,11},
\vec{n}_s-2m\delta_{s1}\rangle_V\nonumber\\
&& + \sum^{\left[n_1/2\right]}_{m=1}\left(\frac{-2r}{m_1^2}
\right)^{m} \Biggl\{C^{n_1}_{2m} \widehat{l}^{\prime
}_{22}\Big|\vec{0}_{ij}+(m-1)\delta_{ij,11}+\delta_{ij,12},
\vec{n}_s-2m\delta_{s1}\rangle_V
\nonumber\\
&& +\frac{1}{4}\Bigl[C^{n_1}_{2m}
(h^1+h^2+n_2-2)+C^{n_1}_{2m+1}\Bigr]\Bigl[
\widehat{t}'_{12}\Big|\vec{0}_{ij}+(m-1)\delta_{ij,11},
\vec{n}_s-2m\delta_{s1}\rangle_V  \nonumber \\
&&+
p_{12}(h^1-h^2-n_2-p_{12}+1)
\Big|\vec{0}_{ij}+(m-1)\delta_{ij,11},
\vec{n}_s-2m\delta_{s1}
-\delta_{s12}\rangle_V \Bigr]\Biggr\},\label{l12ns} \\
\hspace{-1em}&\hspace{-1em}&\hspace{-1em}{l^{\prime
}_{22}\Big|\vec{n}_s\rangle_V} =
\widehat{l}^{\prime
}_{22}\Big|\vec{n}_s\rangle_V.\label{l22ns}
\end{eqnarray}

In (\ref{l0ns})--(\ref{l22ns}), we use the block operators
introduced by the rule (\ref{Bntot}), (\ref{Cntot}) and the below
convention (\ref{t'totalAq}) for the vector ${\widehat{t'_{12}
\Bigl|A}_{\vec{n}_{ij},n_1,p_{12},q} \Bigr\rangle_V}$.
In establishing the validity of \emph{Lemma 5}, we have also intensely used
(\ref{l_ii l_1+b}), (\ref{l_i l_i+b}), (\ref{l1 l2+b})--(\ref{l0li+b}),
(\ref{K00012b})--(\ref{X12iV}). It is only the term
$(2rl^{\prime +}_{22}(t'_{12})^2$ $\left|\vec{0}_{ij},0,p_{12},n_2
\rangle_V\right.$, implied by the quantity $\mathcal{K}_2^{12{}i}$
for $i=1$ (\ref{K12_12ib}) and arising in the proof of the relations
(\ref{l0ns}), (\ref{l1ns}), (\ref{l11ns}), that calls for particular
attention. To this end, it is sufficient to use the result
of the action of ${t}'_{12}$, determined by (\ref{Liepart}),
on the vector $\Bigl|\widehat{A}_{\vec{0}_{ij}+q\delta_{ij,22},
0,p_{12},n_2-2q}\Bigr\rangle_V$ (\ref{Antot}), for
$q=[\sum_{i=1}^k({}^im+{}^il)+k]$, as in (\ref{l11ns}). This is
implied by the following representation, which is not difficult
to establish:
\begin{eqnarray}
\hspace{-1em}&\hspace{-1em}&\hspace{-1em}{(t'_{12})^2\Big| p_{12}, {n}_2
\rangle_V}\ = \
p_{12}(p_{12}-1)(h^1-h^2-n_2-p_{12}+2)\nonumber\\
&& \hspace{0.5em}\times(h^1-h^2-n_2-p_{12}+1)
\Big| p_{12}-2,
{n}_2 \rangle_V \nonumber\\
&&\hspace{0.5em}+ p_{12}(h^1-h^2-n_2-p_{12}+1)
\widehat{t}'_{12}\Big| p_{12}-1,
{n}_2 \rangle_V \nonumber\\
&&\hspace{0.5em}
+\sum\limits^{\left[(n_2-1)/2\right]}_{k=0}\Biggl\{\sum\limits^{
  \left[n_2/2\right]}_{{}^{1}m=0}
  \sum\limits^{\left[n_2/2-({}^{1}m + 1)\right]}_{{}^1l=0}... \nonumber\\
  &&\hspace{0.5em}
  ...\hspace{-1.5em}\sum\limits^{\left[n_2/2-\sum\limits^{k-1}_{i=1}({}^{i}m + {}^il)
  -(k-1)\right]}_{{}^{k}m=0}
  \hspace{-0.5em}
\sum\limits^{\left[n_2/2-\sum\limits^{k-1}_{i=1}({}^{i}m +
{}^il)-{}^{k}m)
  -k \right]}_{{}^kl=0}
  \hspace{-1.5em}(-1)^k
  \left(\frac{-2r}{m_2^2}\right)^{\sum_{i=1}^k({}^im+{}^il)+k}\times\nonumber\\
  &&
C^{n_2}_{2{}^1m+1}C^{n_2-2{}^1m-1}_{2{}^1l+1}...
C^{n_2-2(\sum_{i=1}^{k-1}({}^im+{}^il)+k-1)}_{2{}^km+1}
C^{n_2-2(\sum_{i=1}^{k-1}(
{}^im+{}^il+k-1)-2{}^km-1}_{2{}^kl+1}\nonumber \\
  &&
   \times
t'_{12}
\Big|\widehat{A}_{\vec{0}_{ij}+[\sum\limits^k_{i=1}({}^{i}m +
{}^il)
  +k]\delta_{ij,22},0,p_{12}, n_2-2[\sum\limits^k_{i=1}({}^{i}m +
{}^il)
  +k]}\Bigr\rangle_V\Bigr\}\Biggr\}.\label{t'2nn2}
\end{eqnarray}
To complete the proof, we present the final term explicitly, by using a linear
combination of the primary block $\widehat{t}'_{12}$ as acting on specific
vectors of the Verma module:
\begin{eqnarray}
 \hspace{-1em}&\hspace{-1em}& \hspace{-1em}{t'_{12}
\left|\widehat{A}_{\vec{0}_{ij}+q\delta_{ij,22},0,p_{12},n_2-2q}\rangle_V\right.}
=
-2q\left|\widehat{A}_{\vec{0}_{ij}+\delta_{ij,12}+(q-1)\delta_{ij,22},0,p_{12},n_2-2q}\rangle_V\right.\nonumber\\
&&  \qquad\qquad\qquad\qquad\qquad\qquad + (l^{\prime +}_{22})^q {t'_{12}
\left|\widehat{A}_{\vec{0}_{ij},0,p_{12},n_2-2q}\rangle_V\right.},\
\label{t'Arprom}\\
&& {t'_{12}
\left|\widehat{A}_{\vec{0}_{ij},0,p_{12},n_2-2q}\rangle_V\right.}
=  p_{12}(n_2-2q)\widehat{t}'_{12}\left|\vec{0}_{ij}, 0, p_{12}-1,
n_2-2q
\rangle_V\right.  \nonumber\\
{}&{}&- \sum^{\left[n_2/2-q\right]}_{m=1}
  \left(\frac{-2r}{m_2^2}\right)^m
\Bigl\{ C^{n_2-2q}_{2m}\widehat{t}'_{12}\Big|1,0,m-1,0, p_{12}+1,
n_2-2(q+m)
\rangle_V\nonumber\\
{}&{}& -  (C^{n_2-2q}_{2m}(h^2-h^1+2p_{12})+C^{n_2-2q}_{2m+1})
\widehat{t}'_{12}\Big|0,1,m-1,0, p_{12}, n_2-2(q+m)\rangle_V\nonumber\\
&&+p_{12}C^{n_2-2q}_{2m}(h^2-h^1+p_{12}-1)\widehat{t}'_{12}
\Big|0,0,m,0,
p_{12}-1, n_2-2(q+m)\rangle_V \Bigr\}\nonumber\\
&& -\frac{m_1}{m_2}\sum^{\left[n_2/2-q\right]}_{m=0}
\left(\frac{-2r}{m_2^2} \right)^m
C^{n_2-2q}_{2m+1}\widehat{t}'_{12} \Big|0,0,m, 1, p_{12},
n_2-2(q+m)-1\rangle_V
\nonumber\\
{}&{}& - \sum^{\left[n_2/2-q\right]}_{{}^1m=0 }\sum^{
\left[n_2/2-q-{}^1m-1\right]}_{{}^1l=0}
\hspace{-0.6em}\left(\frac{-2r}{m_2^2}\right)^{{}^1m+{}^1l+1}
\hspace{-1em}C^{n_2-2q}_{2{}^1m+1}\Bigl\{\Bigl[C^{n_2-2(q+{}^1m)-1}_{2{}^1l+1}
\bigl(h^2-h^1+n_2-2q +2p_{12}\nonumber
\\
&&-2({}^1m+{}^1l\hspace{-0.2em}+1)\bigr)
\hspace{-0.2em}-C^{n_2-2q-2{}^1m-1}_{2{}^1l+2}\Bigr]
\widehat{t}'_{12}\Big|0,
1,{}^1m+{}^1l, 0, p_{12},
n_2-2(q+{}^1m+{}^1l\hspace{-0.2em}
+1)\rangle_V \nonumber\\
&&- p_{12}C^{n_2-2q-2{}^1m-1}_{2{}^1l+1} \bigl(h^2-h^1+n_2-2q
+p_{12}
-2({}^1m+{}^1l+1)-1\bigr)  \nonumber\\
&& \times\widehat{t}'_{12}
 \Big|0, 0,{}^1m+{}^1l+1,
0, p_{12}-1,n_2-2(q+{}^1m+{}^1l+1)\Big\rangle_V \nonumber\\
&&- C^{n_2-2q-2{}^1m-1}_{2{}^1l+1}
\widehat{t}'_{12}\Big|1,0,{}^1m+{}^1l, 0, p_{12}+1,
n_2-2(q+{}^1m+{}^1l+1)\Big\rangle_V\nonumber
\end{eqnarray}
\begin{eqnarray}
&& +
\frac{m_1}{m_2}C^{n_2-2q-2{}^1m-1}_{2{}^1l+2}\widehat{t}'_{12}
\Big|0,0, {}^1m+{}^1l+1, 1, p_{12},
n_2-2(q+{}^1m+{}^1l+1)-1\rangle_V
\Bigr\}\nonumber\\
&&+ p_{12}(p_{12}-1)(n_2-2q)(h^1-h^2- n_2+2q
-p_{12}+2)\left|\vec{0}_{ij}, 0, p_{12}-2, n_2-2q
\rangle_V\right.\nonumber
\\
&&-\sum^{\left[n_2/2-q\right]}_{m=1}
  \left(\frac{-2r}{m_2^2}\right)^m
\Biggl\{
\Bigl[C^{n_2-2q}_{2m}\bigl[(p_{12}+1)(h^1-h^2-n_2+2(q+m)-p_{12})\nonumber\\
&& \quad + (h^2-h^1+2p_{12})\bigr] + C^{n_2-2q}_{2m+1}\Bigr]
\Big|1,0,m-1,0, p_{12}, n_2-2(q+m)\rangle_V\nonumber\\
&& \quad -2(m-1) C^{n_2-2q}_{2m}\Big|1,1,m-2,0, p_{12}+1,
n_2-2(q+m)\rangle_V
\nonumber\\
&& \quad -\Bigl(C^{n_2-2q}_{2m}(h^2-h^1+2p_{12})+ C^{n_2-2q}_{2m+1}\Bigr)
\Bigl[p_{12}(h^1-h^2-n_2+2(q+m)\nonumber\\
&& \quad-p_{12}+1)\Big|0,1,m-1,0, p_{12}-1,
 n_2-2(q+m)\rangle_V\nonumber\\
&& \quad -2(m-1)\Big|0,2,m-2,0, p_{12},
n_2-2(q+m)\rangle_V\Bigr]
\nonumber\\
&& \hspace{1em}+
p_{12}C^{n_2-2q}_{2m}(h^2-h^1+p_{12}-1)\Bigl[(p_{12}-1)(h^1-h^2-n_2+2(q+m)\nonumber\\
&& \quad -p_{12}+2)\Big|0,0,m,0, p_{12}-2,
n_2-2(q+m)\rangle_V \nonumber\\
&& \quad -2m
\Big|0,1,m-1,0,
p_{12}-1,
n_2-2(q+m)-1\rangle_V\Bigr]\Biggr\}\nonumber\\
&& +\frac{m_1}{m_2}
\sum^{\left[n_2/2-q\right]}_{m=0}\left(\frac{-2r}{m_2^2} \right)^m
C^{n_2-2q}_{2m+1}\Bigl\{2m
\Big|0,1,m - 1, 1, p_{12},
n_2-2(q+m)-\hspace{-0.2em}1\rangle_V\nonumber\\
{}&{}&-p_{12}[h^1-h^2-n_2+2(q+m)-p_{12}+2] \hspace{-0.2em}
\Big|0,0,m, 1, p_{12}-\hspace{-0.2em}1,
n_2-2(q+m) -\hspace{-0.2em}1 \rangle_V
\Bigr\}\nonumber\\
 {}&{}& - \sum^{\left[n_2/2-q\right]}_{{}^1m=0 }\sum^{
\left[n_2/2-q-{}^1m-1\right]}_{{}^1l=0}
\left(\frac{-2r}{m_2^2}\right)^{{}^1m+{}^1l+1}
\hspace{-0.7em}C^{n_2-2q}_{2{}^1m+1}\Biggl\{\Bigl[C^{n_2-2(q+{}^1m)-1}_{2{}^1l+1}
\bigl(h^2-h^1+n_2-2q \nonumber\\
&& \ +2p_{12}-2({}^1m+{}^1l +1)\bigr)
-C^{n_2-2q-2{}^1m-1}_{2{}^1l+2}\Bigr]\Bigl[
p_{12}(h^1-h^2-n_2+2(q+{}^1m+{}^1l\hspace{-0.2em}+1)\nonumber\\
&&\quad-p_{12}+1) \Big|0,
1,{}^1m\hspace{-0.2em}+{}^1l, 0, p_{12}-1,
n_2-2(q+{}^1m+{}^1l\hspace{-0.2em}
+1)\rangle_V \nonumber\\
&&\quad -\hspace{-0.2em}\left|1, 0,{}^1m\hspace{-0.2em}+{}^1l, 0,
p_{12},
n_2-2(q\hspace{-0.2em}+{}^1m\hspace{-0.2em}+{}^1l\hspace{-0.2em}
+1)\rangle_V\right. \nonumber\\
&&\quad-2 ({}^1m\hspace{-0.2em}+{}^1l)\hspace{-0.2em}\left|0,
2,{}^1m\hspace{-0.2em}+{}^1l-1, 0, p_{12},
n_2-2(q\hspace{-0.2em}+{}^1m\hspace{-0.2em}+{}^1l\hspace{-0.2em}
+1)\rangle_V\right.\Bigr]\nonumber
\\
&&- p_{12}C^{n_2-2q-2{}^1m-1}_{2{}^1l+1} \bigl(h^2-h^1+n_2-2q
+p_{12}
-2({}^1m+{}^1l+1)-1\bigr)  \nonumber\\
&& \times\Bigl[
(p_{12}-1)(h^1-h^2-n_2+2(q+{}^1m\hspace{-0.2em}+{}^1l\hspace{-0.2em}+1)\nonumber\\
&&\quad-p_{12}+2) \hspace{-0.2em}
 \left|0, 0,{}^1m+{}^1l+1,
0, p_{12}-2,n_2-2(q+{}^1m+{}^1l+1)\rangle_V\right. \nonumber\\
&&\quad-2 ({}^1m\hspace{-0.2em}+{}^1l+1)\hspace{-0.2em}
 \left|0, 1,{}^1m+{}^1l,
0, p_{12}-1,n_2-2(q+{}^1m+{}^1l+1)\rangle_V\right. \Bigr]\nonumber\\
&& - C^{n_2-2q-2{}^1m-1}_{2{}^1l+1} \Bigl[
(p_{12}+1)(h^1-h^2-n_2+2(q+{}^1m\hspace{-0.2em}+{}^1l\hspace{-0.2em}+1)\nonumber
\\
&&\quad-p_{12})\left|1,0,{}^1m+{}^1l, 0, p_{12},
n_2-2(q+{}^1m+{}^1l+1)\rangle_V\right.\nonumber\\
&&\quad-2
({}^1m\hspace{-0.2em}+{}^1l)\hspace{-0.2em}\left|1,1,{}^1m+{}^1l-1,
0, p_{12}+1,
n_2-2(q+{}^1m+{}^1l+1)\rangle_V\right.\Bigr]\nonumber
\end{eqnarray}
\begin{eqnarray}
&& + \frac{m_1}{m_2}C^{n_2-2q-2{}^1m-1}_{2{}^1l+2}\Bigl[
p_{12}(h^1-h^2-n_2+2(q+{}^1m\hspace{-0.2em}+{}^1l\hspace{-0.2em}+1)\nonumber\\
&&\quad  -p_{12}+2) \left|0,0, {}^1m+{}^1l+1, 1, p_{12}-1,
n_2-2(q+{}^1m+{}^1l+1)-1\rangle_V\right. \nonumber\\
&&\quad-2 ({}^1m\hspace{-0.2em}+{}^1l+1)\hspace{-0.2em}\left|0,1,
{}^1m+{}^1l, 1, p_{12},
n_2-2(q+{}^1m+{}^1l+1)-1\rangle_V\right.\Bigr]\Biggr\},
\label{t't+Ar}
\end{eqnarray}
as we have assumed $q=\sum\limits^k_{i=1}({}^{i}m + {}^il)+k$.

Finally, adopting notation based on (\ref{t'Arprom}), (\ref{t't+Ar}) and using
the conventions (\ref{Bntot}), (\ref{Cntot}),
\begin{equation}\label{t'totalAq}
 \hspace{-0.5em} {\widehat{t'_{12} \Bigl|A}_{\vec{n}_{ij}+q\delta_{ij,22},n_1,p_{12},n_2-2q}
 \Bigr\rangle_V} \equiv
 \displaystyle\prod_{i\leq j}^2\bigl(l^{\prime +}_{ij}\bigr){}^{n_{ij}}
 \textstyle\hspace{-0.2em}\left(\frac{l^{\prime +}_1}{m_1}
 \hspace{-0.2em}\right)^{n_1}
 \hspace{-0.7em}{t'_{12}\hspace{-0.2em}\left|\widehat{A}_{\vec{0}_{ij}
 + q\delta_{ij,22},0,p_{12},n_2-2q}\rangle_V\right.},
\end{equation}
we arrive at the validity of \emph{Lemma~5}, and therefore also at the validity
of \emph{Theorem 1}.

In the flat-space limit $r=0$, we obtain a new Verma module realization
for the above Lie algebra $\mathcal{A}(Y(2), R^{1,d-1})$, being different
from that for $sp(4)$ in \cite{flatbos}, as well as in
\cite{BRbos} for $k=2$,
\begin{eqnarray}
 \hspace{-1em} &\hspace{-1em}& \hspace{-1ex}
t'_{12}\left|\vec{N}(2)\rangle_V\right. =
  p_{12}(h^1-h^2-n_2-p_{12}+1)\left|\vec{n}_{ij}, \vec{n}_s- \delta_{s,12}
\rangle_V \right. \nonumber \\
  && -\sum_l ln_{l2}\hspace{-0.2em}
\left|{\vec{n}}_{ij}-\delta_{ij,l2}+\delta_{ij,1l},
\vec{n}_s\rangle_V
 \right. -
\frac{m_1}{m_2}
{n_2}\left|\vec{N}(2)+ \delta_{s,1}-\delta_{s,2}\rangle_V\right.,\label{tfinL}\\
{}&\hspace{-1em}& \hspace{-1em} {t}^{\prime
+}_{12}\left|\vec{N}(2)\rangle_V\right.  = -
 \sum_l (3-l) n_{1l}
\left| {\vec{n}}_{ij}-\delta_{ij,1l} +\delta_{ij,l2},
 \vec{n}_s\rangle_V\right. \nonumber\\
{}&{}& + \left|\vec{N}(2)+ \delta_{s,12}\rangle_V\right.-{n_1}
\frac{m_2}{m_1}\left|\vec{N}(2)  -
\delta_{s,1}+\delta_{s,2}\rangle_V\right. , \label{t+finL}\\
&\hspace{-1em}& \hspace{-1em} l^{\prime+
}_l\left|\vec{N}(2)\rangle_V\right.
 = m_l\left|\vec{N}(2)+ \delta_{s,l}\rangle_V\right.,\ l=1,2,   \label{llfinnVL}\\
 &\hspace{-1em}& \hspace{-1em} l^{\prime  }_0\left|\vec{N}(2)\rangle_V\right.
 = m^2_0\left|\vec{N}(2)\rangle_V\right.,   \label{l0finnVL}\\
{} &\hspace{-1em}&\hspace{-1em} l^{\prime
}_1\left|\vec{N}(2)\rangle_V\right.
 =
n_1\frac{m_0^2}{m_1}\left|\vec{N}(2)-\delta_{s,1}\rangle_V\right.
  -m_1 n_{11}\left|\vec{n}_{ij}-\delta_{ij,11},
  \vec{n}_s+\delta_{s,1}\rangle_V\right. \nonumber \\
{}&{}&-
\frac{m_2n_{12}}{2}\left|\vec{n}_{ij}-\delta_{ij,12},\vec{n}_s+\delta_{s,2}\rangle_V\right.
, \label{l1finnVL}
\\
&\hspace{-1em}&\hspace{-1em} {l^{\prime
}_2\left|\vec{N}(2)\rangle_V\right.}
 = n_2\frac{m_0^2}{m_2}\left|\vec{N}(2)-\delta_{s,2}\rangle_V\right. -m_1 \frac{n_{12}}{2}
 \left| {\vec{n}}_{ij}-\delta_{ij,12},
   \vec{n}_s +\delta_{s,1}\rangle_V\right.\nonumber \\
{}&{}& - {m_2n_{22}}
\left|\vec{n}_{ij}-\delta_{ij,22},\vec{n}_s+\delta_{s,2}\rangle_V\right.,
\label{l2finnVL}
\\
&\hspace{-1em}&\hspace{-1em} {l^{\prime
}_{11}\left|\vec{N}(2)\rangle_V\right.} =
  {n}_{11}({n}_{11}+{n}_{12}+n_1- p_{12} -1 + h^1)
   \left|\vec{n}_{ij} -\delta_{ij,11}, \vec{n}_s\rangle_V
 \right. \nonumber\\
 && + \frac{{n}_{12}({n}_{12}-1)}{4}\hspace{-0.2em}
 \left| \vec{n}_{ij} \hspace{-0.2em}-2\delta_{ij,12}\hspace{-0.2em}
+\delta_{ij,22}, \vec{n}_s\rangle_V
\right.\hspace{-0.2em}-\frac{m_0^2}{2m_1^2}{{n}_{1}({n}_{1}\hspace{-0.2em}-1)}
\hspace{-0.2em}\left|\vec{N}(2)\hspace{-0.2em}-2\delta_{s,1}
\rangle_V\right.
 \nonumber \\
 &&
 - \frac{{n}_{12}}{2}\Bigl\{\hspace{-0.1em}\left|\vec{N}(2)\hspace{-0.2em}
-\delta_{ij,12}\hspace{-0.2em}+
\delta_{s,12}\rangle_V\right.\hspace{-0.2em}-{n_1}
\frac{m_2}{m_1}\left|\vec{N}(2)\hspace{-0.2em}
-\delta_{ij,12}\hspace{-0.2em} -
\delta_{s,1}\hspace{-0.2em}+\delta_{s,2}\rangle_V\right.\Bigr\},
 \label{l11nVfinL}
\\
&\hspace{-1em}&\hspace{-1em} {l^{\prime
}_{12}\left|\vec{N}(2)\rangle_V\right.} =\frac{{n}_{12}}{4}
\Bigl({n}_{12}+
 \displaystyle\sum_{l}\hspace{-0.1em}(2{n}_{ll}+n_l+h^l)-1
\Bigr)\left|
 {\vec{n}}_{ij} - \delta_{ij,12},  \vec{n}_s\rangle_V
 \right.\nonumber
 \\
&& + \frac{1}{2} p_{12} n_{11}(h^2-h^1 + n_2+ p_{12} - 1)\left|
 {\vec{n}}_{ij}-\delta_{ij,11}, n_1, p_{12}-1, {n}_2\rangle_V \right.
 \nonumber
 \end{eqnarray}
\begin{eqnarray}
 &&  + n_{11}n_{22}\left|
 {\vec{n}}_{ij}-\delta_{ij,11} -\delta_{ij,22}+ \delta_{ij,12}, \vec{n}_s\rangle_V
 \right. -\frac{n_1n_2m_0^2}{2m_1m_2}
\left|\vec{N}(2)-\delta_{s,1}-\delta_{s,2}\rangle_V\right.\nonumber\\
&& -\frac{n_{22}}{2}\Bigl\{ \left|\vec{N}(2)-\delta_{ij,22}+
\delta_{s,12}\rangle_V\right.-{n_1}
\frac{m_2}{m_1}\left|\vec{N}(2)-\delta_{ij,22}  -
\delta_{s,1}+\delta_{s,2}\rangle_V\right.\Bigr\}\nonumber\\
{}&{}& +\frac{m_1}{m_2}\frac{n_2n_{11}}{2}
\left|\vec{N}(2)-\delta_{ij,11}+
\delta_{s,1}-\delta_{s,2}\rangle_V\right.,
 \label{l12nVfinL}
\\
&\hspace{-1em}&\hspace{-1em}{l^{\prime
}_{22}\left|\vec{N}(2)\rangle_V\right.} =  n_{22}(n_{12} + p_{12}
+ n_2 + n_{22} -1 + h^2)
 \left|\vec{n}_{ij} -\delta_{ij,22}, \vec{n}_s\rangle_V
 \right. \nonumber\\
  && +  \frac{n_{12}p_{12}}{2}(p_{12}-1 + h^2 - h^1+n_2)
\left| \vec{n}_{ij}-\delta_{ij,12},
 \vec{n}_s -\delta_{s,12}\rangle_V
 \right.
\nonumber \\
 && +  \frac{n_{12}(n_{12}-1)}{4}\left|\vec{n}_{ij}+\delta_{ij,11}
-2\delta_{ij,12}, \vec{n}_s\rangle_V\right. -
\frac{n_2(n_2-1)}{2}\frac{m_0^2}{{m_2^2}}\left|\vec{N}(2)-2\delta_{s,2}
\rangle_V\right. \nonumber\\
&& +\frac{m_1}{m_2}\frac{{n_2}n_{12}}{2}
\left|\vec{N}(2)-\delta_{ij,12}+
\delta_{s,1}-\delta_{s,2}\rangle_V\right. \label{l22totnVL} ,
\end{eqnarray}
whereas the action of the negative-root vectors $l^{\prime +}_{ij}$
and that of the Cartan generators $g^{\prime i}_0$ on $|\vec{N}(2)\rangle_V$
remains unchanged and is given by (\ref{l'+ijb}), (\ref{g'0ib}).

\subsection{Auxiliary calculation for $\mathcal{A}'(Y(2),AdS_d)$}
\label{Ap-AddPartsb}

Here, we list the formulae for the operator quantities from \emph{Lemma 1}:
\begin{eqnarray}\label{K_ikb}
    \mathcal{K}^{i}_n & = & K_n^{0{}i} +  2\mathcal{K}_n^{12{}i}
    ,\qquad \left(K_n^{0{}i},\mathcal{K}_n^{12{}i}\right)
 \equiv \mathrm{ad}_{l^{\prime i +}}
    \left(K_{n-1}^{0{}i},\mathcal{K}_{n-1}^{12{}i}\right),\
    n\in \mathbf{N}\,,\\
    \mathcal{K}^{i}_1 & = & K_1^{0{}i}  + 2\mathcal{K}_1^{12{}i}
\nonumber\\
&=& \left\{4l^{\prime ii+}l^{\prime i}+  l^{\prime i +}(2g^{\prime
i}_0-1)\right\}+2 \left\{2l^{\prime  +}_{12}l^{\prime \{1}\delta^{2\}i
}- \delta^{1i}l^{\prime 2+}t'_{12} - \delta^{2i}l^{\prime
1+}t^{\prime +}_{12}\right\},\\
 K_2^{0{}i} & = &
  4l^{\prime ii+}K_2^{\prime 0
  {}i}+ 2{l^{\prime i+}}^2, \label{K_2ib} \\
   \mathcal{K}_2^{12{}i}  &=& 2l^{\prime 12+}X_b^{\prime\{1i}\delta^{2\}i} +
 W^{\prime 21+}_b(\delta^{1i}t'_{12}
  - \delta^{2i}t^{\prime +}_{12}) \nonumber\\
   && =
 2rl^{\prime
 12+}\Bigl\{
  \delta^{i1}\Bigl[ 4l^{\prime
 12+}l^{\prime 22}+ 2(g_0^{\prime 2} -1)t'_{12}\Bigr] +\delta^{i2}\Bigl[ 4l^{\prime
 12+}l^{\prime 11} + 2(g_0^{\prime 1} -
 1)t^{\prime +}_{12}\Bigr]\Bigr\}
\nonumber \\
 {}&{}& -2rl^{\prime ii+}
 \left(\mathcal{K}_0^{12} + g_0^{\{1}\delta^{2\}i}\right)
+ 2r\Bigl\{\delta^{i1}l^{\prime
 22+}(t'_{12})^2 +\delta^{i2}l^{\prime
 11+}(t^{\prime +}_{12})^2 \Bigr\} \,,\label{K12_12ib}
 \\
K_2^{\prime 0 i } & = & X_b^{\prime ii} = l_0' -  r
\Bigl(K^{\prime 0i}_0 + \mathcal{K}^{\prime 12}_0\Bigl),
\label{K02ib}\\
W_{b}^{\prime{}12+i} & =& [W_b^{\prime{}12 +},l^{\prime i+}]
 =  2r\Bigl\{ \bigl[l^{\prime  +}_{12}l^{\prime 2 +} -
l^{\prime  +}_{22}l^{\prime 1 +}\bigr]\delta^{i2} - \bigl[l^{\prime
+}_{12}l^{\prime 1 +} - l^{\prime +}_{11}l^{\prime 2 +}\bigr]\delta^{i1}
\Bigr\} ,\label{W12i}\\
\Bigl(X_{b{}2}^{\prime{}12 },{}X_{b{}1}^{\prime{}21 }\Bigr) &=&
\Bigl([X_b^{\prime{}12 },l^{\prime 2+}],{}[X_b^{\prime{}21 },
l^{\prime 1+}]\Bigr) = r\Bigl\{-2l^{\prime  +}_{22}l^{\prime 1 }
+t^{\prime +}_{12}l^{\prime 2+},{}  -2l^{\prime  +}_{11}l^{\prime 2 }
+l^{\prime 1+}t^{\prime }_{12}\Bigr\}.\label{X1221}
\end{eqnarray}
The non-vanishing action of the operators (\ref{Casimirsb2}), (\ref{K_ikb})--(\ref{K02ib})
and that of $\{-W_b^{\prime{}12 +}, X_b^{\prime{}12 }\}$, as acting on the
highest-weight vector $|0\rangle_V$, are given by
\begin{eqnarray}
  \Bigl( \mathcal{K}_0 ,\mathcal{K}^{12{}i}_1\Bigr)|0\rangle_V
  &=& \Bigl(\Bigl[\sum\nolimits_i h^i(h^i-2) -2h^2\Bigr]|0\rangle_V,-m_1\delta^{i2}
  \left|\vec{0}_{lm},1,1,0\rangle_V\right.\Bigr),
  \label{K00012b}
  \\
  \mathcal{K}_1^i|0\rangle_V & = & 2m_i h^i\left|\vec{0}_{lm},
\delta^{i1},0,\delta^{i2}\rangle_V\right.  -
2m_1\delta^{i2}\left|\vec{0}_{lm}, 1,1,0\rangle_V\right. , \label{K10012b}\\
 K_2^{\prime 0{}i}|0\rangle_V & = & M_i^2|0\rangle_V , \qquad  M_i^2
=
  {m}_0^2 -
r{h^i}({h^i}-2 - \delta^{i2}) ,
\\
K_2^{0{}i} |0\rangle_V &=& 4
M_i^2\left|\delta^{i1},0,\delta^{i2},\vec{0}_s\rangle_V\right. + 2m_i^2\left|
\vec{0}_{lm},
2\delta^{i1},0,2\delta^{i2}\rangle_V\right.\,, \label{K20ib}
 \end{eqnarray}
\begin{eqnarray}
\mathcal{K}^{12{}i}_2|0\rangle_V &=&
2r\delta^{2i}\Bigl\{(h^2-h^{1})
\left|0,0,\delta^{i2},\vec{0}_s\rangle_V\right. +
2(h^1-2)\delta^{i2}\left|0,1,0,0,1,{0}\rangle_V\right.\nonumber \\
 &&+ \delta^{i2}\left|1,{0},0,0,2,{0}\rangle_V\right.
\Bigr\},  \label{K122iVb}
\\
\mathcal{K}^{i}_2 |0\rangle_V &=& \Bigl\{4
M_i^2+4r\delta^{2i}(h^2-h^{1})\Big\})
\left|\delta^{i1},0,\delta^{i2},\vec{0}_s\rangle_V\right. +
2m_i^2\left|
\vec{0}_{lm}, 2\delta^{i1},0,2\delta^{i2}\rangle_V\right. \nonumber\\
&& +8r(h^1-2)\delta^{i2}\left|0,1,0,0,1,{0}\rangle_V\right. + 4r
\delta^{i2}\left|1,{0},0,0,2,{0}\rangle_V\right. , \label{Kf22iVb}\\
W_b^{\prime{}12 +}  |0\rangle_V &=&
2r\Bigl\{(h^2-h^1)\left|{0},1,0,\vec{0}_s\rangle_V\right. -
\left.|1,0,0,0,1,0\rangle_V\right. \Bigr\},\label{W12iV}
\\
X_b^{\prime{}12 }|0\rangle_V&=&
-2r(h^2+h^1-{2})\left|\vec{0}_{lm},0,1,0\rangle_V\right.\,.\label{X12iV}
\end{eqnarray}

\section{Oscillator realization of additional parts: new Fock space}\label{oscVM2}
\renewcommand{\theequation}{\Alph{section}.\arabic{equation}}
\setcounter{equation}{0}

In this appendix, we examine the correctness of \emph{Theorem 2} for a bosonic
realization of the non-linear algebra $\mathcal{A}'(Y(2), AdS_d)$ over the
Heisenberg algebra $A_6$ of $(b_i,b_i^+, b_{ij}, b^+_{ij}, d_{12}, d_{12}^+)$.
To this end, it is sufficient to show how the most involved of the Verma module
relations can be transformed to an oscillator form. Thus, the vector
$\left|A_{n_2}\rangle_V\right.$ (\ref{An2}) reads
\begin{eqnarray}
\hspace{-1.0em}
\left|A_{n_2}\rangle_V\right.
&\longleftrightarrow &
 \Biggl\{
   -\frac{m_1}{m_2}\sum_{m=0}\left(\frac{-2r}{m_2^2}\right)^m
    \frac{(b_{22}^+)^m}{(2m+1)!}b_1^+
    b_2^{2m+1}\nonumber \\ &&\hspace{-1.0em} +\sum_{m=1}
  \left(\frac{-2r}{m_2^2}\right)^m
   (b_{22}^+)^{m-1}\Biggl[b_{12}^+
   \left\{
    \frac{h^2-h^1}{(2m)!}+\frac{b^+_2b_2}{(2m+1)!}\right\}- \frac{b_{11}^+d^+_{12}}{(2m)!}\Biggr]b^{2m}_2 \nonumber \\
   \hspace{-1.0em}   &&  \hspace{-1.0em} - \sum_{m=0}\sum_{l=0}\left(\frac{-2r}{m_2^2}\right)^{m+l+1}
   \frac{1}{(2m+1)!}(b_{22}^+)^{m+l}\Biggl[ b_{12}^+
   \Bigl\{\frac{h^2-h^1+b_2^+b_2}{(2l+1)!}\nonumber\\
 \hspace{-1.0em}   &&  \hspace{-1.0em} -
   \frac{b_2^+b_2}{(2l+2)!}
   \Bigr\}
   -\frac{b_{11}^+d^+_{12}}{(2l+1)!}  +
   \frac{m_1}{m_2}\frac{b_{22}^+}{(2l+2)!}b_1^+b_2
   \Biggr]  b_2^{2(m+l+1)}
   \Biggr\}\left|\vec{0}_{ij},0,0,n_2\rangle\right. . \label{exmap}
\end{eqnarray}
This relation allows one to establish the form of the
primary block operator $\hat{t}'_{12}$ in (\ref{hattosc}).
Let us specify the operator related to the action of
$(t'_{12})^2$ in (\ref{t't+Ar}). For instance, there is an obvious
one-to-one correspondence for the following term in
$l'_0\left|\vec{0}_{ij},\vec{n}_s\rangle_V\right.$ (\ref{l0ns}):
\begin{eqnarray}
&&\frac{r}{2}\displaystyle\sum\limits^{\left[n_1-1/2\right]}_{m=0}\left(
\frac{-8r}{m_1^2}\right)^{m+1}C^{n_1}_{2m+2}(l^{\prime +}_{11})^m
l^{\prime +}_{22}\left(\frac{l^{\prime
+}_{1}}{m_1}\right)^{n_1-2(m+1)}\underbrace{(t'_{12})^2}\left|\vec{0}_{ij},0,p_{12},n_2\rangle_V\right.\nonumber \\
&& \equiv
\frac{r}{2}\displaystyle\sum\limits^{\left[n_1-1/2\right]}_{m=0}\left(
\frac{-8r}{m_1^2}\right)^{m+1}C^{n_1}_{2m+2}
\Biggl[p_{12}(p_{12}-1)(h^1-h^2-n_2-p_{12}+2)\nonumber\\
&& \times (h^1-h^2-n_2-p_{12}+1)\left|(m,0,1),\vec{n}_s-(2m+2,2,0)
\rangle_V\right. \nonumber\\
&&+ p_{12}\textstyle(h^1-h^2-n_2-p_{12}+1)\widehat{t'}_{12}
\left|(m,0,1),\vec{n}_s-(2m+2, 1,0) \rangle_V\right.
\nonumber\\
&&
+\sum\limits^{\left[(n_2-1)/2\right]}_{k=0}\Biggl\{\sum\limits^{
  \left[n_2/2\right]}_{{}^{1}m=0}
  \sum\limits^{\left[n_2/2-({}^{1}m + 1)\right]}_{{}^1l=0}... \nonumber
   \end{eqnarray}
\begin{eqnarray}
  &&
  ...\sum\limits^{\left[n_2/2-\sum\limits^{k-1}_{i=1}({}^{i}m + {}^il)
  -(k-1)\right]}_{{}^{k}m=0}
  \sum\limits^{\left[n_2/2-\sum\limits^{k-1}_{i=1}({}^{i}m + {}^il)-{}^{k}m)
  -k \right]}_{{}^kl=0}(-1)^k
  \textstyle
  \left(\frac{-2r}{m_2^2}\right)^{\sum_{i=1}^k({}^im+{}^il)+k}
\nonumber\\
  &&
\times C^{n_2}_{2{}^1m+1}C^{n_2-2{}^1m-1}_{2{}^1l+1}...
C^{n_2-2(\sum_{i=1}^{k-1}({}^im+{}^il)+k-1)}_{2{}^km+1}C^{n_2-2(\sum_{i=1}^{
k-1}(
{}^im+{}^il+k-1)-2{}^km-1}_{2{}^kl+1}\nonumber \\
  &&
  \times \widehat{t'_{12} \Bigl|A}_{(m,0,\sum_{i=1}^k({}^im+{}^il)+k+1),\vec{n}_s-(2m+2)\delta_{s,1}
 -2[\sum\limits^k_{i=1}({}^{i}m + {}^il)
  +k]\delta_{s,2}}\Bigr\rangle_V \Biggr\}\Biggr]
 \nonumber\\
 &
 &
 \longleftrightarrow \frac{r}{2}\displaystyle\sum\limits^{}_{m\geq 0}\left(
 \frac{-8r}{m_1^2}\right)^{m+1}\frac{(b_{11}^+)^mb_{22}^+}{(2m+2)!}
  \Biggl[\Bigl\{(h^1-h^2-b^+_2b_2-d^+_{12}d_{12}) d_{12}
 + \widehat{t'}_{12}\Bigr\}
\nonumber\\
&&
\times(h^1\hspace{-0.2em}-h^2\hspace{-0.2em}-b^+_2b_2\hspace{-0.2em}-d^+_{12}d_{12})
d_{12}\hspace{-0.2em} + {t'}_{12}\widehat{t'}_{12}
\Biggr](b_{1})^{2(m+1)}(b_{1}^+)^{n_1}\hspace{-0.3em}\left|\vec{0}_{ij},0,p_{12},{n}_2\rangle\right.
\hspace{-0.2em}. \label{t'2nn2''}
\end{eqnarray}
In particular, (\ref{t'2nn2''}) implies an oscillator
representation for the action of $({t'}_{12})^2$ on the vector
$\left|\vec{0}_{ij},0,p_{12},{n}_2\rangle_V\right. $, due to an
associative composition of the operator ${t}'_{12}$,
$\hat{t}'_{12}$ given by (\ref{t'Lf}), (\ref{hattosc}),
\begin{eqnarray}\label{t'2oscnn2}
&& ({t'}_{12})^2
\left|\vec{0}_{ij},0,p_{12},{n}_2\rangle_V\right.\longleftrightarrow
   \Biggl[\Bigl\{(h^1-h^2-b^+_2b_2-d^+_{12}d_{12}) d_{12} + \widehat{t'}_{12}
   \Bigr\}
\nonumber\\
&& \qquad
\times(h^1\hspace{-0.2em}-h^2\hspace{-0.2em}-b^+_2b_2\hspace{-0.2em}-d^+_{12}d_{12})
d_{12}\hspace{-0.2em} + {t'}_{12}\widehat{t'}_{12}
\Biggr]\hspace{-0.3em}\left|\vec{0}_{ij},0,p_{12},{n}_2\rangle\right..
\end{eqnarray}
By using the arguments on the oscillator realization of the quadratic terms
in ${t}'_{12}$ and the structure of the operator related to the vector
$\left|A_{n_2}\rangle_V\right.$, all of the operators $o'_I$ can be restored
in an oscillator form, given by (\ref{l'+ijF})--(\ref{l22fin}), by virtue of

\vspace{1ex} \noindent \textbf{\emph{{Theorem 2.}}} An oscillator
realization of the non-linear second-order algebra $\mathcal{A}'(Y(2), AdS_d)$
over the Heisenberg algebra $A_6$ does exist in terms of a formal power series
in the degrees of creation and annihilation operators, is given by the relations
(\ref{l'+ijF})--(\ref{t+totF}), (\ref{l2'+F}), (\ref{l0fin})--(\ref{l22fin}),
and expressed using the \emph{primary block operator} $\widehat{t}'_{12}$
(\ref{hattosc}) and the \emph{derived block operators}
$\widehat{t}^{\prime +}_{12}, \widehat{l}^{\prime }_{0} ,
\widehat{l}^{\prime }_{m}, \widehat{l}^{\prime }_{m2}, m=1,2$
(\ref{t'+F}), (\ref{l0oscnn2})--(\ref{l22oscnn2}) as follows.

First, for trivial negative-root vectors, we have
\begin{eqnarray}
 && l^{\prime  +}_1  =  m_1b_1^+
  \,,\quad l^{\prime+}_{ij}  =  b_{ij}^+\,,
  \quad g_0^{\prime i} =
 2b_{ii}^+b_{ii}+b_{12}^+b_{12} + (-1)^id^+_{12}d_{12} + b^+_ib_i+ h^i .
 \label{l'+ijF}
\end{eqnarray}
Second, for the operator ${t}'_{12}(B,B^+)$ and the \emph{primary block operator}
$\widehat{t}'_{12}(B,B^+)$,
we obtain\footnote{In (\ref{hattosc}) for $k=0$, there are no
double sums. Besides, the products $\prod_{i=1}^0...$ are equal to
$1$, and the terms inside the internal brackets,
$\big(b_2^+d_{12}b_2-\frac{m_1}{m_2}
\sum_{m=0}\left(\frac{-2r}{m_2^2}\right)^m\frac{(b_{22}^+)^m}{(2m+1)!}
b_1^+ b_2^{2m+1} +\sum_{m=1}...\big)$, are the only ones to survive.}
{\begin{eqnarray}
   {t}'_{12} &=&
   \left(h^1-h^2 - b^+_2b_2 - d^+_{12}d_{12}\right)d_{12} -
   b_{11}^+b_{12}
   -2b_{12}^+b_{22} +\widehat{t}'_{12} \,,
 \label{t'Lf}  \\
\widehat{t}'_{12} & =&
\sum_{k=0}\Biggl[\sum_{{}^1m=0}\hspace{-0.2em}
   \sum_{{}^1l=0} ... \sum_{{}^km=0}\hspace{-0.2em}\sum_{{}^kl=0}
   (-1)^k\hspace{-0.2em}
   \left(\hspace{-0.2em}\frac{-2r}{m_2^2}\hspace{-0.2em}\right)^{\hspace{-0.2em}\sum_{i=1}^k({}^im+{}^il)+k}
   \hspace{-0.2em}\prod_{i=1}^k
   \frac{1}{(2{}^im+1)!}\frac{1}{(2{}^il+1)!}\nonumber
\\
   && \times \bigl(b_{22}^+\bigr)^{\sum_{i=1}^k({}^im+{}^il)+k}\Biggl\{
   b_2^+d_{12}b_2 -\frac{m_1}{m_2}\sum_{m=0}\left(\frac{-2r}{m_2^2}\right)^m
    \frac{(b_{22}^+)^m}{(2m+1)!}b_1^+
    b_2^{2m+1}\nonumber \\
   && +\sum_{m=1}\left(\frac{-2r}{m_2^2}\right)^m
   (b_{22}^+)^{m-1}\Bigl[b_{12}^+
   \Bigl\{
    \frac{h^2-h^1+2d^+_{12}d_{12}}{(2m)!}+\frac{b^+_2b_2}{(2m+1)!}\Bigr\}
   \nonumber
   \end{eqnarray}
   \begin{eqnarray}
  &&  -
\frac{b_{11}^+d^+_{12}}{(2m)!} -
\frac{b_{22}^+}{(2m)!}(h^2-h^1+d^+_{12}d_{12})d_{12}\Bigr]b^{2m}_2 \nonumber \\
     &&   - \sum_{m=0}\sum_{l=0}\left(\frac{-2r}{m_2^2}\right)^{m+l+1}
   \frac{1}{(2m+1)!}(b_{22}^+)^{m+l}\Bigl[ b_{12}^+
   \Bigl\{\frac{h^2-h^1+2d^+_{12}d_{12}+b_2^+b_2}{(2l+1)!}\nonumber
   \\
    && -
   \frac{b_2^+b_2}{(2l+2)!}
   \Bigr\}
   -\frac{b_{11}^+d^+_{12}}{(2l+1)!}  -\frac{b_{22}^+}{(2l+1)!}(h^2-h^1+
   b_2^+b_2+d^+_{12}d_{12})d_{12} \nonumber\\
      &&  +
   \frac{m_1}{m_2}\frac{b_{22}^+}{(2l+2)!}b_1^+b_2
   \Bigr]  b_2^{2(m+l+1)}
   \Biggr\}(b_2)^{2(\sum_{i=1}^k({}^im+{}^il)+k)}\Biggr].\label{hattosc}
\end{eqnarray}}
Third, for the operators $t^{\prime+}_{12}, l^{\prime+}_2$ and the derived
block operator $\widehat{t}^{\prime+}_{12}$, we have
\begin{eqnarray}
 t^{\prime+}_{12} &=&
 - 2 b_{12}^+b_{11}-b_{22}^+b_{12}
 +\widehat{t}^{\prime+}_{12}, \label{t+totF} \\
   \widehat{t}^{\prime+}_{12}  & = &
 \sum_{m=0}\left(\frac{-2r}{m_1^2}\right)^m(b_{11}^+)^m\Bigl\{
\frac{d^+_{12}}{(2m)!}
-\frac{m_2}{m_1}\frac{b_2^+b_1}{(2m+1)!}\Bigr\}b_1^{2m}\nonumber\\
 && \hspace{1.0em}
-\hspace{-0.2em}\sum_{m=1}\hspace{-0.2em}\left(\hspace{-0.2em}
\frac{-2r}{m_1^2}\hspace{-0.2em}\right)^m\hspace{-0.2em}
(b_{11}^+)^{m-1}\Biggl\{
b_{12}^+\hspace{-0.2em}\left[\frac{(h^2\hspace{-0.2em}-h^1\hspace{-0.2em}+2d^+_{12}d_{12}\hspace{-0.2em}+b^+_2b_2)}{(2m)!}
\hspace{-0.2em}-\frac{b_1^+b_1}{(2m+1)!}\hspace{-0.2em}\right] \nonumber\\
 && \hspace{1.0em} +b_{22}^+\hspace{-0.2em}
\frac{(h^1\hspace{-0.2em}-h^2\hspace{-0.2em}-d^+_{12}d_{12}
\hspace{-0.2em}-b^+_2b_2)}{(2m)!}d_{12} + \frac{b_{22}^+}{ (2m)!}\
\widehat{t}'_{12}\ \Biggr\}b_1^{2m} , \label{t'+F}\\
 l^{\prime+}_2 &= &
m_1\sum_{m=0}\hspace{-0.2em}\left(\hspace{-0.2em}\frac{-2r}{m_1^2}\hspace{-0.2em}
\right)^{m+1}\hspace{-0.5em}(b_{11}^+)^{m}\Biggl\{
b_{12}^+\left[\frac{(h^2-h^1+2d^+_{12}d_{12}+b^+_2b_2)}{(2m+1)!}
-\frac{b_1^+b_1}{(2m+2)!}\right]  \nonumber\\
 &&
 + b_{22}^+
\frac{(h^1-h^2-d^+_{12}d_{12}-b^+_2b_2)}{(2m+1)!}d_{12}-
b_{11}^+\frac{d^+_{12}}{(2m+1)!}\Biggr\}b_1^{2m+1}
\nonumber\\
&& +m_2\sum_{m=0}\left(\frac{-2r}{m_1^2}\right)^m
\frac{(b_{11}^+)^m b_2^+}{(2m)!} b_1^{2m}
+m_1\sum_{m=0}\left(\frac{-2r}{m_1^2}\right)^{m+1}\frac{(b_{11}^+)^{m}
b_{22}^+}{(2m+1)!}\ \widehat{t}'_{12}\ b_{1}^{2m+1}. \label{l2'+F}
\end{eqnarray}
Fourth, for the Cartan-like vector ${l^{\prime }_0}$, the following
representation holds true:
 \begin{eqnarray}
 {l^{\prime  }_0} &=& \widehat{l}^{\prime  }_0 +
 \frac{m_1}{2}\displaystyle\sum\limits_{m=0}\left(
 \frac{-8r}{m_1^2}\right)^{m+1}\frac{(b_{11}^+)^m}{(2m+1)!}\Biggl\{
 b_{12}^+ \ \widehat{l}^{\prime }_2\  +\ b_{11}^+ \
 \widehat{l}^{\prime }_1\nonumber \\
 \hspace{-1.0em} && \quad - \frac{m_2}{2}
 [\widehat{t}^{\prime}_{12} +(h^1-h^2-b^{2+}b_2 -
 d^+_{12}d_{12})d_{12}]\ b_2^+ \Biggr\}b_1^{2m+1} \nonumber
 \\
 &&  - r \displaystyle\sum\limits_{m=0}\left(
  \frac{-8r}{m_1^2}\right)^{m}(b_{11}^+)^m b_1^+\left\{
 \frac{2(h^1-d^+_{12}d_{12})-3}{(2m+1)!} +
 \frac{2b_1^+b_1}{(2m+2)!}\right\}b_1^{2m+1}\nonumber \\
 &&  +
 \frac{1}{2}\hspace{-0.2em}\displaystyle\sum\limits_{m=0}\hspace{-0.2em}
 \left(\hspace{-0.2em}
 \frac{-8r}{m_1^2}\hspace{-0.2em}\right)^{m+1}\hspace{-1.0em}
 \frac{(b_{11}^+)^{m}}{ (2m+2)!}\Biggl\{b_{11}^+\
 \widehat{l}^{\prime }_0\
 \hspace{-0.2em}-rb_{11}^+\Bigl([h^1\hspace{-0.2em} - d^+_{12}d_{12}][h^1 \hspace{-0.2em}
 - d^+_{12}d_{12} \hspace{-0.2em}- 2]\nonumber
\\
&&  \quad - h^2 - d^+_{12}d_{12} - b^+_2b_2 +
 2d^+_{12}(h^1 - h^2 - b^+_2b_2 - d^+_{12}d_{12})d_{12}\Bigr)
\nonumber
\end{eqnarray}
\begin{eqnarray}
&&  \quad -2r b_{11}^+\widehat{t}'_{12}\ d^+_{12}
+ 8r b_{11}^+ b_{12}^+ \widehat{l}^{\prime}_{12} \ +4r
(b_{12}^+)^2\ \widehat{l}^{\prime}_{22} + 2rb_{12}^+ \Bigl[\
(h^1\hspace{-0.2em} - h^2
\nonumber
\\
&& \quad  - b^+_2b_2 - d^+_{12}d_{12})d_{12}+
\widehat{t}'_{12}\Bigr] \ \Bigl[ h^2 + b^+_2b_2 + d^+_{12}d_{12} -
2\Bigr]\Biggr\}b_1^{2m+2}
 \nonumber\\
 &&  +
\frac{r}{2}\displaystyle\sum\limits_{m=0}\left(
\frac{-8r}{m_1^2}\right)^{m+1}\frac{(b_{11}^+)^m}{(2m+2)!}b_{22}^+\
\Biggl\{\Bigl[(h^1-h^2-b^+_2b_2-d^+_{12}d_{12}) d_{12} +
\widehat{t'}_{12}
   \Bigr]\nonumber
\\
 && \qquad
\times(h^1\hspace{-0.2em}-h^2\hspace{-0.2em}-b^+_2b_2\hspace{-0.2em}-d^+_{12}d_{12})
d_{12}\hspace{-0.2em} + t'_{12}\widehat{t}'_{12} \Biggr\}\
b_1^{2m+2}
 \label{l0fin}.
\end{eqnarray}
Fifth, the operators $l^{\prime }_m$, $l^{\prime }_{lm}$, $l,m=1,2, l\leq m$,
read as follows:
\begin{eqnarray}
l^{\prime }_1 & =& -m_1b^+_1b_{11}
 -\frac{1}{2}{l}^{\prime 2+}b_{12}  +\sum\limits_{m=0}
\left(\frac{-8r}{m_1^2}\right)^{m}\frac{(b_{11}^+)^m}{(2m)!}
\widehat{l}^{\prime 1}b_1^{2m}  \nonumber\\
&& + \frac{1}{m_1}\displaystyle\sum\limits_{m=0}\left(
\frac{-8r}{m_1^2}\right)^{m}{\frac{(b_{11}^+)^m}{(2m+1)!}}
\Biggl\{\widehat{l}^{\prime }_0  -r
\Bigl[(h^1-d^+_{12}d_{12})(h^1-d^+_{12}d_{12} -2) -
h^2\nonumber\\
&&  -d^+_{12}d_{12}-b^+_{2}b_{2}\Bigr] -
rd^+_{12}\Bigl[h^1-h^2-b_2^+b_2 -d^+_{12}d_{12}\Bigr] d_{12}
 - r\ \widehat{t}'_{12}\ d^+_{12} \Biggr\}b_1^{2m+1}\nonumber \\
&& - 4{m_1}\displaystyle\sum\limits_{m=0} \left(
\frac{-2r}{m_1^2}\right)^{m+1}{\frac{(b_{11}^+)^m
b_{12}^+}{(2m+1)!}}
\textstyle(4^m-\frac{1}{2})\ \widehat{l}'_{12}\ b_1^{2m+1}\nonumber\\
&&
 + \frac{m_1}{2}\displaystyle\sum\limits_{m=1}\left(
\frac{-8r}{m_1^2}\right)^{m}(b_{11}^+)^{m-1}b_1^+
\left\{\frac{1}{2}\frac{(2(h^1-d^+_{12}d_{12})-1)}{(2m)!} +
\frac{b_1^+b_1}{(2m+1)!}\right\}b_1^{2m} \nonumber\\
&& + \frac{1}{2}\displaystyle\sum\limits_{m=1}\left(
\frac{-2r}{m_1^2}\right)^{m}(4^m-1)
{\frac{(b_{11}^+)^{m-1}}{(2m)!}}\Bigl\{2b_{12}^+
\widehat{l}^{\prime 2}   - m_1b_1^+\nonumber\\
&& - m_2 (\widehat{t}^{\prime }_{12}+
h^1-h^2-b_2^+b_2-d^+_{12}d_{12})d_{12}b_2^+ \Bigr\}b_1^{2m}
\nonumber\\
&& +\frac{m_1}{2}\displaystyle\sum\limits_{m=1}\left(
\frac{-2r}{m_1^2}\right)^{m+1}(4^m-1)
  {\frac{(b_{11}^+)^{m-1}}{(2m+1)!}}
  \Biggl\{ b_{11}^+\ \widehat{t}'_{12}\ d^+_{12} -
  4(b_{12}^+)^2\widehat{l}^{\prime }_{22} \nonumber
 \\
  && \qquad -2 b_{12}^+
  \Bigl[(h^1-h^2- b_2^+b_2 - d^+_{12}d_{12})d_{12}  +
  \widehat{t}'_{12} \Bigr]
  (h^2 + b_2^+b_2 + d^+_{12}d_{12} - 2)\nonumber\\
&& \qquad+ b_{11}^+d^+_{12}(h^1-h^2-b_2^+b_2-d^+_{12}d_{12})d_{12}
    \Biggr\}b_1^{2m+1}\nonumber
    \\
&& -
\frac{m_1}{2}\displaystyle\sum\limits_{m=1}\hspace{-0.2em}\left(
\frac{-2r}{m_1^2}\hspace{-0.2em}\right)^{m+1}\hspace{-1.0em}(4^m-1)
  {\frac{(b_{11}^+)^{m-1} }{(2m+1)!}}b_{22}^+
  \Biggl\{\Bigl[(h^1-h^2-b^+_2b_2-d^+_{12}d_{12}) d_{12} \nonumber\\
&&\qquad  + \widehat{t}'_{12}
   \Bigr](h^1\hspace{-0.2em}-h^2\hspace{-0.2em}-b^+_2b_2\hspace{-0.2em}-d^+_{12}d_{12})
d_{12}\hspace{-0.2em} + {t}'_{12}\widehat{t}'_{12}
\Biggr\}b_1^{2m+1}, \label{l1'F}
\\
{l^{\prime }_2}
 &= &
\displaystyle\sum\limits_{m=0}\hspace{-0.2em}\left(
\hspace{-0.2em}\frac{-2r}{m_1^2}\hspace{-0.2em}\right)^{m}\hspace{-0.2em}\frac{(b_{11}^+)^m}{2m!}
\widehat{l}^{\prime }_2\
b_1^{2m}\hspace{-0.2em}-2{m_1}\hspace{-0.2em}\displaystyle\sum\limits_{m=0}
 \hspace{-0.2em}\left( \hspace{-0.2em}\frac{-2r}{m_1^2}\hspace{-0.2em}\right)^{m+1}
\frac{(b_{11}^+)^{m}}{(2m+1)!}\sum_kb_{1k}^+ \widehat{l}^{\prime
}_{k2}
b_1^{2m+1}\nonumber
 \end{eqnarray}
    \begin{eqnarray}
&& -2m_1 \displaystyle\sum\limits_{m=0}\left(
\frac{-2r}{m_1^2}\right)^{m+1}
\hspace{-0.2em}\frac{(b_{11}^+)^m}{4} \widehat{t}'_{12}
\hspace{-0.2em}\left\{
  \frac{(h^1+h^2 + b_2^+b_2 -
2)}{(2m+1)!} + \frac{b_1^+b_1}{(2m+2)!}\right\}
 b_1^{2m+1}\nonumber\\
&&-\frac{m_1}{2}\displaystyle\sum\limits_{m=0} \left(
\frac{-2r}{m_1^2}\right)^{m+1}(b_{11}^+)^{m}\Biggl\{
\frac{(h^1+h^2+b_2^+b_2 - 2)}{(2m+1)!} +
\frac{b_1^+b_1}{(2m+2)!}\Biggr\}\times\nonumber\\
&& \quad \times(h^1-h^2-b_2^+b_2-d_{12}^+d_{12})d_{12}b_1^{2m+1} -
\frac{m_1}{2}b^+_1b_{12} -  {l}^{\prime+}_2\ b_{22}, \label{l2fin}
\\
{l'_{11}}  &=&
  ({b}^+_{11}b_{11}+{b}^+_{12}{b}_{12}+b^+_1b_1 -d^+_{12} d_{12}
   + h^1)b_{11}
    + \frac{b_{22}^+}{4}b_{12}^2
 - \frac{1}{2}\widehat{t}^{\prime
+}_{12}b_{12} \nonumber\\
&& + \frac{1}{2}\sum_{m=0}  \left(\frac{-8r}{m_1^2}
\right)^{m+1}\frac{(b_{11}^+)^{m}}{(2m+2)!}\Biggl\{\frac{1}{4r}
\widehat{l}_0'\ + b_{12}^+\ \widehat{l}^{\prime }_{12} -
\frac{1}{4}\Bigl[d^+_{12}(2-h^1-h^2-b^+_2b_2)d_{12} \nonumber\\
&&\quad + h^1(h^1-2) - h^2-b^+_2b_2+\widehat{t}'_{12} d^+_{12}  \Bigr]
\Biggr\} b_1^{2m+2} \nonumber\\
&& - \frac{1}{2}\sum_{m=1} \left(\frac{-8r}{m_1^2} \right)^{m}
(b_{11}^+)^{m-1}b_1^+\left\{\frac{h^1-d^+_{12}d_{12}-\frac{1}{2}}{(2m+1)!}
+\frac{b_1^+b_1}{(2m+2)!}
\right\}b_1^{2m+1}\nonumber\\
&&-\frac{1}{m_1}\sum_{m=0}\Biggl\{\left(\frac{-8r}{m_1^2}
\right)^{m}\frac{(b_{11}^+)^{m}}{(2m+1)!}\ \widehat{l}^{\prime
}_1\
  - \left(\frac{-2r}{m_1^2}
\right)^{m}(4^m-1) \frac{(b_{11}^+)^{m-1}b_{12}^+}{(2m+1)!}\
\widehat{l}^{\prime }_2\Biggr\}\ b_{1}^{2m+1}
\nonumber\\
&& + \frac{1}{2} \sum_{m=0}\left(\frac{-2r}{m_1^2}
\right)^{m}(4^m-1)
(b_{11}^+)^{m-1}\frac{b_1^+b_1}{(2m+1)!}b_{1}^{2m}
\nonumber\\
&& +\frac{m_2}{2m_1}\hspace{-0.2em}
\sum_{m=0}\hspace{-0.2em}\left(\hspace{-0.2em}\frac{-2r}{m_1^2}\hspace{-0.2em}
\right)^{m}\hspace{-0.5em}(4^m-1)
\frac{(b_{11}^+)^{m-1}}{(2m+1)!}\ \Bigl\{
{t}^{\prime}_{12}\hspace{-0.2em}  + \sum_l lb^+_{1l}b_{l2}\Bigr\}
b_2^+ b_{1}^{2m+1}
\nonumber\\
&& -\frac{1}{2}
\hspace{-0.2em}\sum_{m=0}\hspace{-0.2em}\left(\hspace{-0.2em}\frac{-2r}{m_1^2}
\hspace{-0.2em}\right)^{m+1}\hspace{-0.7em}(4^m-1)
\frac{(b_{11}^+)^{m-1}}{(2m+2)!}\left[b_{11}^+d^+_{12}\hspace{-0.2em}
-2b_{12}^+( h^2+b_2^+b_2+d_{12}^+d_{12}-1)\right]\times
\nonumber\\
&& \qquad \times\Bigl(h^1-h^2-d_{12}^+d_{12}-b^+_2b_2\Bigr)d_{12}
b_{1}^{2m+2}
\nonumber\\
&&-\frac{1}{2} \sum_{m=0}\left(\frac{-2r}{m_1^2}
\right)^{m+1}(4^m-1) \frac{(b_{11}^+)^{m-1}}{(2m+2)!} \Bigl\{\
b_{11}^+\ \widehat{t}'_{12}\ d^+_{12} -4b_{11}^+b_{12}^+\
\widehat{l}^{\prime }_{12}\ \nonumber\\
&&\qquad -4(b_{12}^+)^2\ \widehat{l}^{\prime }_{22}\
 -2 b_{12}^+ \ \widehat{t}'_{12}\ (h^2+b^+_2b_2+d^+_{12}d_{12}-2)\Bigr\}b_{1}^{2m+2}
 \nonumber
 \\
&& + \frac{1}{2} \sum_{m=0}\left(\frac{-2r}{m_1^2}
\right)^{m+1}(4^m-1) \frac{(b_{11}^+)^{m-1}}{(2m+2)!} b_{22}^+ \
\left\{\Bigl[(h^1-h^2-b^+_2b_2-d^+_{12}d_{12})
d_{12}+\widehat{t}'_{12}
   \Bigr] \times\right.\nonumber\\
&&\qquad \left.\times \bigl(h^1- h^2 -b^+_2b_2
-d^+_{12}d_{12}\bigr) d_{12} + {t}'_{12}\widehat{t}'_{12} \right\}
b_{1}^{2m+2},
 \label{l11fin}
\\
 l'_{12} & = &
\frac{1}{4}\Bigl( b_{12}^+b_{12}+
 \displaystyle\sum_{m}(2b_{mm}^+b_{mm}+b_m^+b_m+h^m)\Bigr)
 b_{12}+ b_{12}^+b_{11}b_{22}
-\frac{1}{2} \left(\widehat{t}^{\prime +}_{12}\ b_{22} +
\widehat{t}^{\prime }_{12}\ b_{11}\right)
 \nonumber \\
 && + \frac{1}{2} (h^2-h^1 + b_2^+b_2 + d^+_{12}d_{12} )b_{11}d_{12}
 - \frac{1}{2m_1}\hspace{-0.2em}\sum_{m=0}
\hspace{-0.2em}\left(\hspace{-0.2em}\frac{-2r}{m_1^2}\hspace{-0.2em}
\right)^{m}\hspace{-0.3em}\frac{(b_{11}^+)^m}{(2m+1)!}
\widehat{l}^{\prime
}_2b_1^{2m+1}\nonumber\\
&& + \sum_{m=0}  \left(\frac{-2r}{m_1^2}
\right)^{m}\frac{(b_{11}^+)^m}{(2m)!} \widehat{l}^{\prime
}_{12}b_1^{2m}+ \sum_{m=1}\left(\frac{-2r}{m_1^2} \right)^{m}
\frac{(b_{11}^+)^{m-1}b_{12}^+}{(2m)!}\ \widehat{l}^{\prime
 }_{22}b_1^{2m}\nonumber
 \end{eqnarray}
    \begin{eqnarray}
&& + \frac{1}{4}\sum_{m=1}\left(\frac{-2r}{m_1^2} \right)^{m}
(b_{11}^+)^{m-1} \Biggl\{\widehat{t}'_{12}\
\left[\frac{(h^1+h^2+b_2^+b_2-2)}{(2m)!}
+\frac{b_1^+b_1}{(2m+1)!}\right] \nonumber
 \\
&& +\left[\frac{h^1+h^2+b_2^+b_2-2}{(2m)!}
+\frac{b_1^+b_1}{(2m+1)!} \right]
(h^1-h^2-b^+_2b_2-d_{12}^+d_{12})d_{12} \Biggr\} b_1^{2m},
 \label{l12fin}
 \\
l^{\prime }_{22}  & = & \widehat{l}^{\prime }_{22}\ + \bigl(
b^+_{12}b_{12} + d^+_{12}d_{12} + b^+_2b_2 + b^+_{22}b_{22}+
h^2\bigr)b_{22}  +  \frac{1}{4}b_{11}^+b_{12}^2
   -\frac{1}{2}\ \widehat{t}'_{12}\ b_{12}
 \nonumber\\
  &&+\frac{1}{2}( h^2 - h^1 + d^+_{12}d_{12} + b^+_2b_2)d_{12}b_{12}.
  \label{l22fin}
\end{eqnarray}
In (\ref{l0fin})--(\ref{l22fin}), the derived block operators
$\widehat{l}^{\prime}_0$, $\widehat{l}^{\prime}_m$,
$\widehat{l}^{\prime}_{m2}$, for $m=1,2$, are written as
follows:\footnote{These operators are immediately determined by their action
on the vector $|\vec{0}_{ij},0,p_{12},n_2\rangle_V$ in
(\ref{l0nn2})--(\ref{l22nn2}).}
\begin{eqnarray}
{\widehat{l}^{\prime  }}_0  & = & m_0^2
-r\sum_{m=0}\left(\frac{-8r}{m_2^2} \right)^{m}(b_{22}^+)^{m}
\Biggl\{b_2^+\left[\frac{1}{(2m+1)!}(2h^2+2d^+_{12}d_{12} -
1)\right. \nonumber
\\
&&\quad \left.+ \frac{2b_2^+b_2}{(2m+2)!}\right]b_2
-2\frac{m_1}{m_2}\frac{b_1^+d^+_{12}}{(2m+1)!}b_2\Biggr\}b_2^{2m}\nonumber\\
\hspace{-1.0em} && +\frac{1}{2} \sum_{m=0}\left(\frac{-8r}{m_2^2}
\right)^{m+1}
\frac{(b_{22}^+)^{m}}{(2m+2)!}\Bigl\{b_{22}^+\Bigl(m_0^2-
r\bigl[h^2(h^2-4)+h^1-(d^+_{12})^2d_{12}^2\nonumber\\
\hspace{-1.0em} && \quad   +2d^+_{12}d_{12}(h^1-2)\bigr]\Bigr)+
2rb_{12}^+d^+_{12} (h^1 - d^+_{12}d_{12} - 2) +
rb_{11}^+(d^+_{12})^2\Bigr\}b_2^{2m+2}\nonumber\\
\hspace{-1.0em} && +2r\sum_{m=0}\sum_{l=0}\left(\frac{-8r}{m_2^2}
\right)^{m} \left(\frac{-2r}{m_2^2}
\right)^{l+1}\frac{(b_{22}^+)^m}{(2m+1)!}
\frac{(b_{22}^+)^{l+1}}{(2l+1)!} \ \widehat{t}'_{12}\ d^+_{12}
b_2^{2(m+l+1)}\nonumber\\
 && +2r\sum_{m=0}\sum_{l=0}\left(
\frac{-8r}{m_2^2}\right)^{m} \left(\frac{-2r}{m_2^2}
\right)^{l+1}\frac{(b_{22}^+)^{m+l}}{(2m+1)!}\Biggl\{
b_{12}^+d^+_{12}\hspace{-0.2em}\Biggl[\hspace{-0.2em}\frac{[h^2\hspace{-0.2em}-h^1
\hspace{-0.2em}+2d^+_{12}d_{12}\hspace{-0.2em}+b^+_2b_2\hspace{-0.2em}
+2]}{(2l+1)!}  \nonumber\\
\hspace{-1.0em} &&
 - \frac{b_2^+ b_2}{(2l+2)!}\Biggr] -
\frac{b_{11}^+(d^+_{12})^2}{(2l+1)!} +
\frac{b_{22}^+(d^+_{12}d_{12} +1) }{(2l+1)!}  [h^1 - h^2 -
b_2^+b_2 - d^+_{12}d_{12}]\nonumber\\
\hspace{-1.0em} && \quad +
\frac{m_1}{m_2}\frac{b_{22}^+b_1^+d^+_{12}}{(2l+2)!}
b_2\Biggr\}b_2^{2(m+l+1)}, \label{l0oscnn2}
\\
 \widehat{l}^{\prime  }_1  &=&
-\frac{m_2}{2}\sum_{m=0}\left( \frac{-2r}{m_2^2} \right)^{m+1}
(b_{22}^+)^{m} d^+_{12}\left\{\frac{(h^1+h^2-2)}{(2m+1)!}+
\frac{b_2^+b_2}{(2m+2)!}\right\}b_2^{2m+1} ,
\label{l1oscnn2}\\
\widehat{l}^{\prime }_2 & =& - \widehat{l}^{\prime }_1 d_{12}
-\sum_{m=0} \left(\frac{-8r}{m_2^2}
\right)^{m}\frac{(b_{22}^+)^m}{(2m+1)!}
\frac{[m_0^2-rh^2(h^2-3)]}{m_2}b_2^{2m+1} \nonumber\\
 &&  + \frac{m_2}{4} \sum_{m=1}
\left(\frac{-8r}{m_2^2}
\right)^{m}(b_{22}^+)^{m-1}b_2^+\left\{\frac{(2h^2-1)}{(2m)!} +
2\frac{b_2^+b_2}{(2m+1)!}\right\}b_2^{2m}\nonumber\\
 &&
 +
\frac{m_2}{2}\sum_{m=1}\left(\hspace{-0.2em}\frac{-2r}{m_2^2}
\right)^{m+1} (b_{22}^+)^m
d^+_{12}\left\{\frac{(h^1+h^2-2)}{(2m+1)!}+
\frac{b_2^+b_2}{(2m+2)!}\right\}d_{12} b_2^{2m+1} \nonumber
\\
 &&  -  \frac{1}{2}\sum_{m=0}
\left(\frac{-2r}{m_2^2} \right)^{m}(4^m-1)(b_{22}^+)^{m-1}
d^+_{12}\left\{\frac{{m_1}b_1^+}{(2m)!}-\frac{{m_2}b_2^+d_{12}}{(2m)!}
\right\}
b_2^{2m} \nonumber
\end{eqnarray}
\begin{eqnarray}
 &&  + \frac{m_2}{2} \sum_{m=0}
\left(\frac{-2r}{m_2^2}
\right)^{m+1}(4^m-1)\frac{(b_{22}^+)^{m-1}}{(2m+1)!}\Bigl\{b_{22}^+
\Bigl[2(h^1-2)d^+_{12}d_{12}
 \nonumber
 \\
 && \quad +h^1- h^2-(d^+_{12})^2d_{12}^2\Bigr]
-2b_{12}^+d^+_{12}(h^1-d^+_{12}d_{12}-2)- b_{11}^+ (d^+_{12})^2
\Bigr\}b_2^{2m+1}
\nonumber
\\
 &&  - \frac{m_2}{2} \sum_{m=0}
\sum_{l=0}\left(\frac{-2r}{m_2^2}
\right)^{m+l+1}\hspace{-1.0em}(4^m-1)\frac{(b_{22}^+)^{m+l-1}}{(2m)!}\Biggl\{\frac{m_1}{m_2}\frac{b_{22}^+b_1^+d^+_{12}}{(2l+2)!}b_2
 \nonumber
\\
 && \hspace{1.0em}
+b_{12}^+d^+_{12}\left[\frac{(h^2 - h^1 + 2d^+_{12}d_{12}+
b_2^+b_2 +2}{(2l+1)!}  - \frac{b_2^+b_2}{(2l+2)!}\right]
 \nonumber \\
 && \hspace{1.0em}
-\frac{\Bigl(b_{11}^+(d^+_{12})^2 +b_{22}^+\bigl[h^2-h^1+b_2^+b_2
\bigr]\Bigr)}{(2l+1)!} \Biggr\}b_2^{2(m+l)+1}
\nonumber\\\
 &&  + \frac{m_2}{2} \hspace{-0.2em}\sum_{m=0}
\sum_{l=0}\left(\frac{-2r}{m_2^2}
\right)^{m+l+1}(4^m-1)\frac{(b_{22}^+)^{m}}{(2m)!}
\frac{(b_{22}^+)^{l}}{(2l+1)!}\times\nonumber\\
 &&  \qquad \times d_{12}^+\left\{h^2-h^1+b_2^+b_2 +
d^+_{12}d_{12} +2\right\}d_{12}b_2^{2(m+l)+1}
\nonumber\\
 &&  - \frac{m_2}{2} \sum_{m=0}
\sum_{l=0}\left(\frac{-2r}{m_2^2}
\right)^{m+l+1}(4^m-1)\frac{(b_{22}^+)^{m}}{(2m)!}\frac{(b_{22}^+)^{l}}{
(2l+1)!}  \ \widehat{t}'_{12}\ d^+_{12} b_2^{2(m+l)+1},
 \label{l2oscnn2}\\
 \widehat{l}^{\prime}_{12} &= &
\frac{1}{4}\sum_{m=1}\left(\frac{-2r}{m_2^2}
\right)^{m}(b_{22}^+)^{m-1}d^+_{12}\left\{\frac{(h^1+h^2-2)}{(2m)!}+
\frac{b_2^+b_2}{(2m+1)!}\right\} b_2^{2m},
 \label{l12oscnn2}
\\
\widehat{l}^{\prime }_{22}  & = & - 2
\widehat{l}^{\prime}_{12}d_{12} -
\sum_{m=0}\left(\frac{-8r}{m_2^2} \right)^{m}
\frac{(b_{22}^+)^m}{m_2^2(2m+2)!}\bigl\{m_0^2-r
h^2(h^2-3)\bigr\}b_2^{2(m+1)}\nonumber\\
\hspace{-1.0em} && -\frac{1}{2}\sum_{m=1} \left(\frac{-8r}{m_2^2}
\right)^{m}
(b_{22}^+)^{m-1}b_2^+\left\{\frac{1}{2}\frac{(2h^2-1)}{(2m+1)!} +
\frac{b_2^+b_2}{(2m+2)!}\right\}b_2^{2m+1} \nonumber
\\
 &&  + \frac{1}{2}\sum_{m=0}
\left(\frac{-2r}{m_2^2}
\right)^{m+1}\hspace{-1.0em}(4^{m}-1)(b_{22}^+)^{m-1}
\left\{{\frac{b_{12}^+d^+_{12}}{(2m+2)!}}\bigl[2(h^1-
2)-2d^+_{12}d_{12}\bigr]
b_2\right.  \nonumber\\
 &&  \left. \  + {\frac{b_{22}^+}{(2m+2)!}} \Bigl[h^2-h^1
-2d^+_{12}d_{12}(h^1-2)+ (d^+_{12})^2d_{12}^2\Bigr]b_2 +
{\frac{b_{11}^+(d^+_{12})^2 b_2}{(2m+2)!}}\nonumber\right.
\\
 && \left. \
 -{\frac{b_2^+d^+_{12}d_{12}}{(2m+1)!}} \right\}b_2^{2m+1} \ + \ \frac{m_1}{2m_2}\sum_{m=0}
\left(\frac{-2r}{m_2^2}
\right)^{m+1}{\frac{(b_{22}^+)^{m-1}b_1^+d^+_{12}}{(2m+1)!}}(4^{m}-1)
b_2^{2m+1}\nonumber\\
 &&
 +
\frac{1}{2}\sum_{m=0}\sum_{l=0}\left(\frac{-2r}{m_2^2}\right)^{m+l+1}
\hspace{-1.0em}(4^m-1)\frac{(b_{22}^+)^{m+l-1}}{(2m+1)!}\Biggl\{
\frac{m_1}{m_2}\frac{b_{22}^+b_1^+d^+_{12}}{(2l+2)!} b_2-
\frac{b_{11}^+(d^+_{12})^2}{(2l+1)!}
 \nonumber\\
 && \
+\frac{b_{22}^+}{(2l+1)!}\bigl(d^+_{12}d_{12}+1\bigr)
\bigl(h^1-h^2-b_2^+b_2-d^+_{12}d_{12}\bigr)\nonumber\\
 && \ + b_{12}^+d^+_{12}\left[\frac{(h^2 - h^1 +
2d^+_{12}d_{12} + b_2^+b_2+2)}{(2l+1)!} - \frac{b_2^+b_2
}{(2l+2)!} \right]
\Biggr\}b_2^{2(m+l+1)}\nonumber\\
 &&  +
\frac{1}{2}\sum_{m=0}\sum_{l=0}\left(\frac{-2r}{m_2^2}\right)^{
m+l+1}\hspace{-1.0em}(4^m-1)
\frac{(b_{22}^+)^{m+l}}{(2m+1)!(2l+1)!}\ \widehat{t}'_{12} \
d^+_{12}b_2^{2(m+l+1)}.
\label{l22oscnn2}
\end{eqnarray}

As a consequence, in the flat-space case $r=0$, the oscillator
realization of the second-order algebra $\mathcal{A}'(Y(2),
AdS_d)$ is reduced to a $2$-parametric (with the parameters
$m_l, l=1,2$) polynomial realization of the Lie algebra
$\mathcal{A}(Y(2), R^{1,d-1})$, with the central extension $m_0^2$
given by (\ref{l'+ijb}), (\ref{g'0ib}), and the following relations:
\begin{eqnarray}
 \hspace{-1em} &\hspace{-1em}& \hspace{-1ex}
t'_{12} =\left(h^1-h^2  - d^+_{12}d_{12}\right)d_{12} -\sum_l l
b^+_{1l}b_{l2}- \frac{m_1}{m_2}
b_1^+{b_2},\label{toscL}\\
{}&\hspace{-1em}& \hspace{-1em} {t}^{\prime +}_{12}  = d^+_{12}-
 \sum_l (3-l) b^+_{l2}b_{1l}
- \frac{m_2}{m_1}b_2^+{b_1}
  , \qquad \qquad l^{\prime  }_0 = m_0^2,\label{t+oscL}\\
&\hspace{-1em}& \hspace{-1em} l^{\prime+ }_l
 = m_lb_l^+,\qquad l^{\prime }_l =
\frac{m_0^2}{m_l}b_l   -\frac{m_1 }{l}b_1^+b_{1l} -
\frac{lm_2}{2}b_2^+b_{l2}, \ l=1,2,  \label{lloscL}\\
&\hspace{-1em}&\hspace{-1em} {l^{\prime }_{11}} =
({b}^+_{11}b_{11}+{b}^+_{12}{b}_{12}+b^+_1b_1 -d^+_{12} d_{12}
   + h^1)b_{11}
    + \frac{b_{22}^+}{4}b_{12}^2 -\frac{m_0^2}{m_1^2}
    {b}_{1}^2\nonumber\\
    && \qquad
     - \frac{1}{2}\Bigl\{\hspace{-0.1em}{d}^+_{12}
      -
\frac{m_2}{m_1}b^+_2b_1\Bigr\}{b}_{12},
 \label{l11oscL}
\\{}&\hspace{-1em}&\hspace{-1em} {l^{\prime }_{12}}
=\frac{1}{4}\Bigl( b_{12}^+b_{12}+
 \displaystyle\sum_{k}(2b_{kk}^+b_{kk}+b_k^+b_k+h^k)\Bigr)
 b_{12}+ b_{12}^+b_{11}b_{22} -
 \frac{m_0^2}{2m_1m_2}b_1b_2
\nonumber \\
 &&\quad  + \frac{1}{2} (h^2-h^1 + d^+_{12}d_{12}
 )b_{11}d_{12}
 -\frac{1}{2}\Bigl\{ d^+_{12}-
\frac{m_2}{m_1}b^+_{2}b_{1}\Bigr\}b_{22}
+\frac{m_1}{2m_2}b_1^+b_2b_{11} ,
 \label{l12oscL}
\\
&\hspace{-1em}&\hspace{-1em}{l^{\prime }_{22}} = ( b^+_{12}b_{12}
+ d^+_{12}d_{12} + b^+_2b_2 + b^+_{22}b_{22}+ h^2)b_{22}+
\frac{b_{11}^+}{4}b_{12}^2
 \nonumber\\
  &&+\frac{1}{2}( h^2 - h^1+d^+_{12}d_{12})d_{12}b_{12}
  -
\frac{m_0^2}{2{m_2^2}}b_2^2 +\frac{m_1}{2m_2}{b_1^+{b_2}b_{12}}.
 \label{l22oscL}
\end{eqnarray}

\section{Reduction to initial irreducible relations}\label{reductionC}
\renewcommand{\theequation}{\Alph{section}.\arabic{equation}}
\setcounter{equation}{0}

Here, we demonstrate that the equations of motion (\ref{Eq-0b})--(\ref{Eq-3b}),
or, equivalently, (\ref{l0}) for $\tilde{l}_0 = l_0+m^2$ and (\ref{lilijt}),
can be obtained from the action (\ref{S}), after gauge-fixing and removing
the auxiliary fields by using part of the equations of motion.
Let us start by gauge-fixing.

\subsection{Gauge-fixing}

Let us consider the field $|\chi^l \rangle$, for $l=0, 1,...,6$,
at certain fixed values of spin $(s_1, s_2)$. In this section,
we omit the subscripts associated with the eigenvalues
of the $\sigma^i$-operators (\ref{sigmai}). Then we extract the dependence
of $Q$ (\ref{Q}) on the zero-mode ghosts $\eta_0$ and $\mathcal{P}_0$,
\begin{eqnarray}\label{DecQ}
Q &=& \eta_0\widetilde{L}_0 + {\imath}\sum_m\eta_m^+\eta^m {\cal{}P}_0+ \Delta
Q
\\
\label{DQ}
\Delta Q &=& \eta_i^+L_i +\sum\limits_{l\leq m}\eta_{lm}^+L_{lm} +
\vartheta^+_{12}T_{12}- \vartheta_{12}^+\sum\nolimits_{n}(1+\delta_{1n})\eta^+_{1n}
\mathcal{P}_{n2}\\
&& + \vartheta_{12}^+
\sum\nolimits_{n}(1+\delta_{n2})\eta_{n2} \mathcal{P}^+_{1n}+
\textstyle\frac{1}{2}\sum_{n }\eta^+_{n2}\eta_{1n}\lambda_{12}
\nonumber\\
&&
 - \textstyle\frac{1}{2}\sum\limits_{l\leq
m}(1+\delta_{lm})\eta_m\eta_{lm}^+\mathcal{P}_l  -
\bigl[\vartheta_{12} \eta^+_{2}\mathcal{P}_1 +\vartheta^+_{12} \eta^+_{1}\mathcal{P}_2 \bigl]
\nonumber
\\
 &&   +r\left\{ - \textstyle\frac{1}{2}\sum_{i,j}\eta^+_i \eta^+_{j}
\varepsilon^{ij}\Bigl[\sum_m(-1)^m\mathcal{G}_0^m \mathcal{P}_{12}
- \bigl( \mathcal{T}_{12} \mathcal{P}_{11} +
\mathcal{L}_{11}\lambda_{12}\bigr)  + \mathcal{T}^+_{12}
\mathcal{P}_{22} + \mathcal{L}_{22}\lambda^+_{12} \Bigr]\right. \nonumber \\
&  & \left. +\textstyle 2\eta^+_i \eta_{j}\Bigl[\sum_{m}
\mathcal{L}^+_{jm}\mathcal{P}_{im} -\frac{1}{8}\bigl(
\mathcal{T}^+_{12} \lambda_{12} +\mathcal{T}_{12}\lambda^+_{12}
\bigr)\delta^{ij} + \frac{1}{4}\sum_{m}\mathcal{G}_0^m
\lambda_{12} \delta^{j1}\delta^{i2} \Bigr] \right\} +h.c., \nonumber
\end{eqnarray}
where
\begin{eqnarray}\label{L0ext}
&& \widetilde{L}_0 := \textstyle\frac{1}{2}{L}_0 + r\sum\nolimits_{i}\eta^+_i\Bigl[
 2\mathcal{L}_{ ii}\mathcal{P}_i^+ + 2\mathcal{L}^+_{ i}\mathcal{P}_{ii}
  + \mathcal{G}^{i}_0  \mathcal{P}_{ i }+2
  \bigl(\mathcal{\mathcal{L}}_{12 }
  \mathcal{P}^+_{\{1} +
\mathcal{L}^+_{ \{1}\mathcal{P}_{12 }\bigr)\delta_{2\}i }
  \\
&& \quad -\delta^{1i}\bigl(\mathcal{L}_{2}\lambda^+_{12}
 + \mathcal{T}^+_{12}\mathcal{P}_{2}\bigr) - \delta^{2i}
\bigl(\mathcal{L}_{ 1}\lambda_{12} +
\mathcal{T}_{12}\mathcal{P}_{1}\bigr)\Bigr]\nonumber\\
&&  \quad  +r^2\hspace{-0.3em}\left\{\sum\nolimits_{i,j}\hspace{-0.3em}\eta_i\eta_j\varepsilon^{ij}\Bigl[
\textstyle \frac{1}{2}\sum_m\Bigl(G^m_0\bigl[\lambda_{12}
\mathcal{P}^{+}_{22}
 - \lambda_{12}^+
\mathcal{P}^{+}_{11}\bigr]
    + 4L_{mm}\mathcal{P}^{+}_{m2}
\mathcal{P}^{+}_{1m}\Bigr)  +2
L_{12}\mathcal{P}^{+}_{22}\mathcal{P}^{+}_{11}\Bigr]
 \right. \nonumber\\
&& \left. \quad\textstyle + \sum_{i,j}\eta^+_i\eta_j\Bigl[
2\sum_{m}\bigl(
L^{+}_{22}\mathcal{P}_{22}-L_{11}\mathcal{P}^{+}_{11})\bigr)
\lambda_{12}\delta^{1j} \delta^{2i}
-2T_{12}\mathcal{P}_{11}\mathcal{P}^{+}_{22}\delta^{1i}\delta^{2j}\right.
 \nonumber\\
&& \hspace{1em} \left. + 2\varepsilon^{\{1j}\delta^{2\}i}\Bigl(\bigl(
L_{12}\lambda_{12}-T_{12}\mathcal{P}_{12}\bigr)\mathcal{P}^{+}_{22}
  + \bigl(T^+_{12}\mathcal{P}_{12}-L_{12}\lambda_{12}^+\bigr)
\mathcal{P}^{+}_{11} \Bigr)\right.
\nonumber\\
&& \left.\hspace{1em} - \textstyle
2\sum_{m}(-1)^m\hspace{-0.2em}\Bigl(G^m_0\mathcal{P}_{11}\delta^{1i}\delta^{2j}-
G^m_0\mathcal{P}_{22} \delta^{2i}\delta^{1j} \Bigr)
\mathcal{P}^{+}_{12}
  \Bigr] \right\}+ h.c.,\nonumber
\end{eqnarray}
and do the same with the fields and gauge parameters:
\begin{equation}\label{decompeta0}
|\chi^l\rangle=|S^l\rangle +\eta_0|B^l\rangle.
\end{equation}
Then the equations of motion and gauge transformations (\ref{LEoM})
can be recast as follows:
\begin{eqnarray}\label{regthr}
 \delta |S^{l - 1} \rangle &=& \Delta Q |S^l \rangle -
\sum_m\eta_m^+\eta^m |B^l \rangle ,
\\
 \delta |B^{l - 1} \rangle & = & \widetilde{L}_0 |S^l \rangle - \Delta Q|B^l
\rangle \, , \qquad \delta |\chi^{-1} \rangle \equiv 0,
\end{eqnarray}
for $l=0, \ldots , 6$.

As a next step, we examine the lowest-level gauge transformation
\begin{eqnarray}
\label{lowgtr}
\delta |S^{5} \rangle = \Delta Q|S^{6} \rangle,
&\qquad& \delta |B^{5} \rangle = \widetilde{L}_0 |S^{6} \rangle,
\end{eqnarray}
where, due to the ghost number condition, we have taken account
of $|B^{6}\rangle\equiv 0$. Extracting the dependence of the gauge
parameters, and that of the operator $\Delta{}Q$ (\ref{DQ}), on
the ghost coordinate and momentum $\eta_{11}$, $\mathcal{P}_{11}^+$,
\begin{eqnarray}
\label{chil}
|\chi^l\rangle=|\chi^l_0\rangle+P_{11}^+|\chi^l_{1}\rangle,
&\qquad& \Delta Q=\Delta
Q_{11}+\eta_{11}T_{11}^++U_{11}\mathcal{P}_{11}^+,
\end{eqnarray}
where the quantities $|\chi^l_0\rangle$, $|\chi^l_{1}\rangle$,
$T_{11}^+$, $U_{11}$, $\Delta Q_{11}$ do not depend on
$\eta_{11}$, $\mathcal{P}_{11}^+$, we obtain the gauge
transformation of $|S^{5}_0\rangle$, with the same decomposition
for $|S^{5}\rangle$, $|S^{5}\rangle  = |S^{5}_0\rangle$ $+
\mathcal{P}_{11}^+|S^{5}_1\rangle$ as the one for $|\chi^l\rangle$
(\ref{chil}),
\begin{eqnarray}
\delta |S^{5}_0 \rangle & = & T_{11}^+ |S^{6}_1
\rangle. \label{dSk(k+1)}
\end{eqnarray}
Here, we have used the fact that $|S^{6}_0\rangle\equiv0$, due to
the ghost number condition. Since $T_{11}^+ = L_{11}^+ + O
(\mathcal{C}) = b_{11}^++\ldots$, as implied by the structure
of $\Delta Q$ in (\ref{DQ}), we can remove the dependence of
$|S^{5}_0\rangle$ on the operator $b_{11}^+$, using all of the
degrees of freedom in $|S^{6}_1\rangle$. Therefore, after
gauge-fixing at the lowest level of gauge transformations, we find
the following conditions for $|S_0^{5}\rangle$:
\begin{eqnarray}
b_{11}|S^{5}_0\rangle=0 &\Longleftrightarrow&
b_{11}\mathcal{P}_{11}^+|\chi^{5}\rangle=0,
\label{gk(k+1)-1}
\end{eqnarray}
so that the theory becomes a $4$-th-stage reducible gauge theory.

Let us turn to the next level of gauge transformations. Extracting
the explicit dependence of the gauge parameters and $\Delta{}Q$
on $\eta_{11}$, $\mathcal{P}_{11}^+$, $\eta_{12}$, $\mathcal{P}_{12}^+$,
and using similar arguments as those at the previous level of
gauge transformations, we find that $|\chi^{4}\rangle$ can be
subject to the gauge
\begin{eqnarray}\label{gk(k+1)-2}
b_{11}\mathcal{P}_{11}^+|\chi^{4}\rangle=0, &\qquad&
b_{12}\mathcal{P}_{11}^+\mathcal{P}_{12}^+|\chi^{4}\rangle=0.
\end{eqnarray}
To obtain these gauge conditions, one has to use all the degrees
of freedom in the gauge parameters $|\chi^{5}\rangle$, restricted
by (\ref{gk(k+1)-1}).

Proceeding with the above, step by step, first of all, for $l=3$, we obtain
\begin{eqnarray}\label{gk(k+1)-3}
 b_{11}\mathcal{P}_{11}^+|\chi^{l}\rangle=0, \quad
 b_{12}\mathcal{P}_{11}^+\mathcal{P}_{12}^+|\chi^{l}\rangle=0,
 \quad b_{22}\mathcal{P}_{11}^+\mathcal{P}_{12}^+\mathcal{P}_{22}^+|\chi^{l}\rangle=0.
 \end{eqnarray}
As we define the following set of operators used in
(\ref{gk(k+1)-1})--(\ref{gk(k+1)-3}),
\begin{eqnarray}\label{Al}
 [\mathcal{A}^l] = \Bigl(b_{11}\mathcal{P}_{11}^+,
 b_{12}\prod_i^2\mathcal{P}_{1i}^+, b_{22}\mathcal{P}_{22}^+
 \prod_i^{2}\mathcal{P}_{1i}^+\Bigr), \ \ \ l=1,2,3
\end{eqnarray}
(where, for instance, the second component of the set $[\mathcal{A}^l]$
is equal to $\mathcal{A}^{2} = b_{12}\prod_i^2\mathcal{P}_{1i}^+$),
we can equivalently represent the gauge conditions
(\ref{gk(k+1)-1})--(\ref{gk(k+1)-3}), and all of the subsequent ones,
which are based on decomposing the gauge parameters in all
of the ghost momenta $P_{ij}^+, i\leq j$, namely,
\begin{eqnarray}\label{Apgauge}
[\mathcal{A}^l]|\chi^{6-l}\rangle=0, \ \ \ \mathrm{for} \ \ \ l = 1,2,3.
\end{eqnarray}
Next, we apply the same procedure as above, albeit starting
from the gauge parameter $|\chi^{2}\rangle$, and carry out extraction
from this parameter and from the operator $\Delta{}Q$ (\ref{DQ})
of the ghost coordinates and momenta $\eta_{ij}$, $\mathcal{P}_{ij}^+$,
$i\leq j$, as well as $\eta_1$, $\mathcal{P}_{1}^+$. As a result, we obtain
a set of gauge conditions for the parameter $|\chi^{2}\rangle$:
\begin{equation}\label{gk(k+1)2-1}
\Bigl([\mathcal{A}^{3}],b_1\mathcal{P}_1^+\prod_{i,j=1, i\leq j}^{2}
\mathcal{P}_{ij}^+\Bigr)|\chi^{2}\rangle =0 .
\end{equation}
In the process of further extraction, starting with the ghosts $\eta_1$, $\eta_2$,
$\mathcal{P}_{1}^+$, $\mathcal{P}_{2}^+$, we obtain two sets of gauge
conditions for the parameters $|\chi^{3-m}\rangle$, $m=1,2$,
\begin{eqnarray}\label{gk(k-1)2}
&& \Bigl([\mathcal{A}^{3}],
b_1\mathcal{P}_1^+\prod_{i,j=1, i\leq j}^{2} \mathcal{P}_{ij}^+,
b_2\prod_m^2\mathcal{P}_m^+\prod_{i,j=1, i\leq j}^{2}
\mathcal{P}_{ij}^+ \Bigr)|\chi^{1}\rangle =0.
\end{eqnarray}
Finally, implementing the same algorithm as the one above, albeit
initiated by the field $|\chi^{0}\rangle$, and extracting from this field,
and also from the operator $\Delta{}Q$ (\ref{DQ}), the ghost coordinates
and momenta $\eta_m$, $\mathcal{P}_{m}^+$, $\eta_{ij}$,
$\mathcal{P}_{ij}^+$, $i\leq j$, $\vartheta_{12}$, $\lambda_{12}^+$,
we deduce a set of gauge conditions for the field $|\chi^{0}\rangle$:
\begin{equation}
\Bigl([\mathcal{A}^{3}],
b_1\mathcal{P}_1^+\prod_{i,j=1, i\leq j}^{2} \mathcal{P}_{ij}^+,
b_2\prod_m^2\mathcal{P}_m^+\prod_{i,j=1, i\leq j}^{2}
\mathcal{P}_{ij}^+,
d_{12}\lambda_{12}^+\prod_m^2\mathcal{P}_m^+\prod_{i,j=1, i\leq
j}^{2} \mathcal{P}_{ij}^+ \Bigr)|\chi^{0}\rangle =0. \label{G1}
\end{equation}

Let us now turn to removing the auxiliary fields by using
the equations of motion.

\subsection{Removal of auxiliary fields by resolving equations of motion}

As a first step, we decompose the field $|S^0\rangle$ in the form
\begin{eqnarray}\label{Bdecomp}
|S^0 \rangle & =& |S^0_0 \rangle + \mathcal{P}_{11}^+ |S^0_{1}
\rangle,
\qquad \qquad |S^0_{(0)_{3}} \rangle =
|S^0_{(0)_{3}0} \rangle + \mathcal{P}_{1}^+
|S^0_{(0)_{3}1}\rangle,
\\
|S^0_{0}\rangle &=& |S^0_{00} \rangle + \mathcal{P}_{12}^+
|S^0_{01} \rangle,  \qquad\quad
|S^0_{(0)_{3}0} \rangle
= |S^0_{(0)_{5}}
\rangle + \mathcal{P}_{2}^+
|S^0_{(0)_{4}1}\rangle,\nonumber
\\
|S^0_{(0)_2} \rangle &=& |S^0_{(0)_20} \rangle +
\mathcal{P}_{22}^+ |S^0_{(0)_21} \rangle,
\qquad |S^0_{(0)_{5} }\rangle =
|S^0_{(0)_{5}0} \rangle + \lambda_{12}^+
|S^0_{(0)_{5}1} \rangle,\nonumber
\end{eqnarray}
and do the same with the vector $|B^0\rangle$,
\begin{eqnarray}\label{Bdecomp0}
|B^0 \rangle &=& \mathcal{P}_{11}^+ |B^0_{1} \rangle +
\mathcal{P}_{12}^+|B^0_{01}\rangle + \mathcal{P}_{22}^+
|B^0_{(0)_{2}1}\rangle + \mathcal{P}_{1}^+ |B^0_{(0)_{3}1}\rangle
+
\mathcal{P}_{2}^+ |B^0_{(0)_{4}1} \rangle
+\lambda_{12}^+ |B^0_{(0)_{5}1} \rangle , \nonumber
\end{eqnarray}
where the term independent of the ghost momenta is absent due to
the spin value and the ghost number condition $gh(|B^0\rangle)=-1$.
Notice that, due to the spin value and the fact that
$gh(|S^0\rangle)=0$, the vector $|S^0_{(0)_{6}}\rangle$ does not
depend on the ghost coordinates and momenta, as a consequence
of the gauge conditions (\ref{G1}), $|S^0_{(0)_{6}}\rangle=|\Phi\rangle$,
with $|\Phi\rangle$ being the physical field (\ref{PhysState}) for $k=2$.

Next, similar to the fields, we extract from $\Delta{}Q$ (\ref{DQ})
the dependence on $\eta_{11}$, $\mathcal{P}_{11}^+$, $\eta_{12}$,
$\mathcal{P}_{12}^+$, $\eta_{22}$, $\mathcal{P}_{22}^+$, and then
the dependence on $\eta_{l}$, $\mathcal{P}_{l}^+$, $l=1,2$,
as well as on $\vartheta_{12}$, $\lambda_{12}^{+}$, respectively.

Substituting these $6$ decompositions into the equation of motion
\begin{eqnarray}
\tilde{L}_0 |S^0 \rangle - \Delta Q|B^0 \rangle=0,
\end{eqnarray}
and using the gauge conditions (\ref{G1}), we can show that
$|B^0_{(0)_{5}1}\rangle\hspace{-0.1em}= 0$, and then
$|B^0_{(0)_{4}1}\hspace{-0.1em} \rangle$ = $0$, etc., until
$|B^0_{1}\rangle=0$, which implies
\begin{eqnarray}
\tilde{L}_0|S^0\rangle=0, &\qquad& |B^0\rangle=0. \label{E1}
\end{eqnarray}

In a similar way, we consider the second equation of motion
\begin{eqnarray}
\Delta Q |S^0 \rangle=0, \label{E2}
\end{eqnarray}
where $|B^0\rangle=0$ has been used. After the same decomposition,
we deduce, step-by-step, the facts that
\begin{eqnarray}
|S^0_{(0)_{5}1}\rangle = |S^0_{(0)_{4}1}\rangle
= \ldots = |S^0_{01}\rangle = |S^0_1\rangle=0. \label{E3}
\end{eqnarray}
The relations (\ref{E1}) and (\ref{E3}) imply that all of the auxiliary fields
are zero, and then, as a result, we have $|\chi^0\rangle_{(s)_k}=|\Phi\rangle$.
The operator ${\widetilde{L}}_0$ reduces to $L_0$, with no dependence
on the auxiliary oscillators when acting on  $|\Phi\rangle$, so that
the equations of motion (\ref{Eq-0b}), (\ref{Eq-1b})--(\ref{Eq-3b})
hold true, with allowance for the mass term $m_0^2$ (\ref{m02}). Therefore,
we have proved that the space of cohomologies for the BRST operator $Q$ (\ref{Q})
with a vanishing ghost number is determined only by the constraints related
to an irreducible AdS group representation.

\medskip


\end{document}